\documentclass[aps,prd,twocolumn,preprintnumbers,showpacs,nofootinbib,amssymb]{revtex4}
\usepackage{graphicx}
\usepackage{amsmath}
\usepackage{amssymb}
\usepackage{amsfonts}
\usepackage{bm}
\usepackage{color}

\def\be{\begin{equation}}
\def\ee{\end{equation}}
\def\bea{\begin{eqnarray}}
\def\eea{\end{eqnarray}}

\begin{document}

\title{Phase transitions between dilute and dense axion stars}
\author{Pierre-Henri Chavanis}
\affiliation{Laboratoire de Physique Th\'eorique,
Universit\'e Paul Sabatier, 118 route de Narbonne  31062 Toulouse, France}

\begin{abstract}

We study the nature of phase transitions between dilute and dense axion stars
interpreted as self-gravitating Bose-Einstein condensates. We develop a
Newtonian model based on the
Gross-Pitaevskii-Poisson equations for a complex
scalar field with a self-interaction potential $V(|\psi|^2)$ involving an
attractive $|\psi|^4$
term and a repulsive $|\psi|^6$ term. Using a Gaussian ansatz for the wave
function, we analytically obtain the mass-radius relation of dilute and dense
axion stars for arbitrary values of the self-interaction parameter
$\lambda\le 0$. 
We show the existence of a critical point $|\lambda|_c\sim
(m/M_P)^2$, where $m$ is the axion mass and
$M_P$ the Planck mass, above which a first order phase transition takes
place. We
qualitatively estimate general relativistic corrections on the mass-radius
relation of axion stars. For weak
self-interactions $|\lambda|<|\lambda|_c$, a system of self-gravitating axions
forms a stable dilute axion star below a general relativistic maximum mass
$M_{\rm max,GR}^{\rm dilute}\sim M_P^2/m$ and collapses into a black hole above
that mass. For strong self-interactions $|\lambda|>|\lambda|_c$, a
system of
self-gravitating axions forms a stable dilute axion star below a Newtonian
maximum mass $M_{\rm max,N}^{\rm dilute}=5.073 M_P/\sqrt{|\lambda|}$  [P.H.
Chavanis, Phys. Rev. D {\bf
84}, 043531 (2011)],  collapses into a dense axion star above that mass,
and collapses into a black hole above a
general relativistic maximum mass $M_{\rm max,GR}^{\rm dense}\sim
\sqrt{|\lambda|}M_P^3/m^2$.
Dense axion stars explode below a Newtonian minimum mass $M_{\rm min,N}^{\rm
dense}\sim m/\sqrt{|\lambda|}$ and form dilute axion stars of
large size or disperse away. We
determine the phase diagram of
self-gravitating axions and show the existence of a triple point
$(|\lambda|_*,M_*/(M_P^2/m))$
separating dilute axion stars, dense axion stars, and black holes. We make
numerical applications  for QCD axions and
ultralight axions. Our approximate analytical results are in good agreement with
the exact numerical results of Braaten {\it et al.} [Phys. Rev. Lett. {\bf 117},
121801 (2016)] for Newtonian dense axion stars. They are also qualitatively
similar to those obtained by Helfer {\it et al.} [JCAP {\bf 03}, 055 (2017)] for
general relativistic axion stars but
they differ quantitatively for weak self-interactions due presumably to the use
of a different self-interaction potential $V(|\psi|^2)$. We point out analogies
between the evolution of self-gravitating axions (bosons) at zero temperature
evolving
from dilute axion stars to dense axion stars and black
holes,  and the evolution of compact degenerate (fermion) stars at zero
temperature evolving from white
dwarfs to 
neutron stars and black holes. We also discuss some analogies between the phase
transitions of axion stars at zero temperature and the phase transitions of
self-gravitating fermions at finite temperature. Finally, we suggest
that a dense axionic nucleus may form at the center
of dark matter halos through
the collapse of
a dilute axionic core (soliton) passing above  the maximum
mass 
$M_{\rm max,N}^{\rm dilute}$. It would have a mass 
 $1.11\times 10^9 (f/m)\, M_{\odot}$, a radius $0.949/(mf^{1/3})\, {\rm
pc}$, a density $2.10\times 10^{-8} (m^2f^2)\, {\rm g/m^3}$, a
pulsation period $8.24/(mf^{1/3})\, {\rm yrs}$ and an
energy $-5.59\times 10^{62} (f/m)\, {\rm ergs}$, where the axion mass $m$ is
measured in units of $10^{-22}\, {\rm eV/c^2}$ and the axion decay constant $f$
is measured in units of $10^{15}{\rm GeV}$.  This
dense axionic nucleus could be the remnant of a bosenova associated with the 
emission of a characteristic radiation [Levkov {\it et al.},
Phys. Rev. Lett. {\bf 118}, 011301 (2017)].

\end{abstract}


\maketitle

\section{Introduction}
\label{sec_introduction}

The nature of dark matter (DM) is still unknown and constitutes one of the
greatest
mysteries of modern cosmology. The  cold dark
matter (CDM) model in which DM is assumed to be made of weakly
interacting massive particles (WIMPs) works remarkably well at large
(cosmological) scales but encounters some problems at small (galactic) scales.
These problems are known as the cusp problem \cite{cusp}, the missing
satellite
problem \cite{satellites}, and the too big to fail problem \cite{tbtf}. In
order to
solve this ``CDM crisis'', it has been proposed to take the quantum nature of
the
particles into account. For example, it has been suggested that DM may be
made of bosons in the form of Bose-Einstein condensates (BECs) at
absolute zero
temperature \cite{baldeschi,khlopov,membrado,sin,jisin,leekoh,schunckpreprint,
matosguzman,sahni,
guzmanmatos,hu,peebles,goodman,mu,arbey1,silverman1,matosall,silverman,
lesgourgues,arbey,fm1,bohmer,fm2,bmn,fm3,sikivie,mvm,lee09,ch1,lee,prd1,prd2,
prd3,briscese,
harkocosmo,harko,abrilMNRAS,aacosmo,velten,pires,park,rmbec,rindler,lora2,
abrilJCAP,mhh,lensing,glgr1,ch2,ch3,shapiro,bettoni,lora,mlbec,madarassy,
abrilph,playa,stiff,guth,souza,freitas,alexandre,schroven,pop,cembranos,
schwabe,fan,calabrese,marsh,bectcoll,chavmatos,hui,zhang,abrilphas,
chavtotal,shapironew,suarezchavanisprd3} (see the Introduction of
\cite{prd1} for a short historic of this model). In this model, DM halos are
interpreted as gigantic boson stars described by a scalar field (SF) that may
represent the wavefunction $\psi$ of the BEC. The mass of the DM boson has to
be very small (see below) for quantum mechanics to manifest itself at galactic
scales. The
bosons may be noninteracting or have a repulsive or attractive
self-interaction. Different names have been given to this model such as SFDM,
fuzzy dark matter (FDM), BECDM, and $\psi$DM. In the fully general relativistic
context, the evolution of the SF is described by the Klein-Gordon-Einstein (KGE)
equations. At the scale of DM halos, Newtonian
gravity can be used so the evolution of the wave function of the
self-gravitating BEC is governed by the Gross-Pitaevskii-Poisson (GPP)
equations.

Using the transformations introduced by Madelung 
\cite{madelung} and de Broglie \cite{broglie1927a,broglie1927b,broglie1927c},
hydrodynamic representations of the GPP and KGE equations have been introduced
in
\cite{bohmer,sikivie,abrilMNRAS,prd1,prd2,prd3,aacosmo,rindler,abrilph,playa,
chavmatos} (see the Introduction of \cite{chavmatos} for a short historic of the
hydrodynamic interpretation of quantum mechanics). The
quantum Euler equations
are similar to the hydrodynamic equations of CDM except
that
they include a quantum force arising from the Heisenberg uncertainty principle
and a pressure force due to the self-interaction of the bosons measured by their
scattering length $a_s$.   In the BEC model, DM
halos are stable stationary solutions of the KGE, GPP, or quantum
Euler equations.
They satisfy a condition of hydrostatic equilibrium corresponding to the balance
between the
gravitational attraction, the repulsive quantum force (Heisenberg's uncertainty
principle), and the pressure force due to the
self-interaction (scattering) that can be repulsive ($a_s>0$) or
attractive ($a_s<0$).  The
mass-radius
relation of self-gravitating BECs at $T=0$ has been
obtained numerically
(exactly) and analytically (approximately) in \cite{prd1,prd2} for any value
(positive or negative) of
the scattering length $a_s$ of the bosons.\footnote{This mass-radius relation
describes ultracompact dwarf DM halos like Fornax
that are completely condensed (ground state), and the solitonic core of large
BECDM
halos. It does not describe the virialized envelope  of large DM halos that has
a
Navarro-Frenk-White (NFW) profile (see, e.g., \cite{ch2,chavtotal} for more
details).} For bosons with a positive scattering length ($a_s> 0$),
corresponding to a repulsive self-interaction, this study makes the link between
the noninteracting limit  $GM^2m a_s/\hbar^2\ll 1$ in which the pressure due to
the scattering can be neglected and the Thomas-Fermi (TF)
limit $GM^2ma_s/\hbar^2\gg 1$ in which the
quantum potential can be neglected. In that case, an equilibrium state exists
for any mass. For bosons with a negative scattering
length ($a_s< 0$), corresponding to an attractive self-interaction, this study
shows that stable halos can exist only below a maximum mass.
The cosmological evolution of a relativistic self-interacting complex SF
described by the KGE equations has been studied in \cite{shapiro,abrilphas}.
The gravitational instability of an infinite homogeneous complex SF
described by the KGE and GPP equations (quantum Jeans problem) has been treated
through the hydrodynamical picture  in
\cite{khlopov,sikivie,abrilMNRAS,prd1,aacosmo,abrilph,suarezchavanisprd3}.

One possible DM candidate is the axion
\cite{kc}. Axions are hypothetical pseudo-Nambu-Goldstone bosons of the
Peccei-Quinn \cite{pq} phase transition associated with a $U(1)$ symmetry that
solves the strong charge parity (CP) problem of quantum chromodynamics (QCD).
The axion is a spin-$0$ particle with a very
small
mass  $m=10^{-4}\,
{\rm eV}/c^2$ and an extremely weak self-interaction $a_s=-5.8\times
10^{-53}\, {\rm m}$  arising
from
nonperturbative effects in QCD. Their role in cosmology has been first
investigated in \cite{preskill,abbott,dine,davis}. They are produced in the
early universe by
non-thermal
mechanisms, either by vacuum misalignment \cite{preskill,abbott,dine} or cosmic
string decay \cite{davis}.
 Axions are extremely nonrelativistic and have
huge occupation numbers, so they can be described by a classical field.
Recently, it has been proposed that gravitational
interactions can thermalize the axions, so axionic DM can form a BEC
during
the radiation-dominated era \cite{sikivie1,sikivie2}.   Axions can thus be
described by a
relativistic quantum field theory with a real scalar field $\varphi$ whose
evolution is governed by the KGE equations. In the
nonrelativistic
limit, they can
be described by an effective field theory with a complex scalar field $\psi$
whose evolution is governed  by the GPP equations.
Therefore, axions are good candidates for the BECDM scenario.  One
particularity of the QCD axion is to have a negative scattering length ($a_s<0$)
corresponding
to an attractive self-interaction.

The formation of structures in an axion-dominated Universe was investigated by
Hogan and Rees \cite{hr} and Kolb and Tkachev \cite{kt}. In the very early
Universe, the axions are relativistic but self-gravity can be neglected with
respect to their attractive self-interaction. These authors found that the
attractive self-interaction of the axions generates  very dense structures
corresponding to pseudo-soliton configurations  that they called ``axion
miniclusters'' \cite{hr} or ``axitons'' \cite{kt}. These axitons have a mass
$M_{\rm
axiton}\sim 10^{-12}\, M_{\odot}$ and a radius $R_{\rm
axiton}\sim 10^{9}\, {\rm m}$. 
At later times, self-gravity must be taken into account. Kolb and Tkachev
\cite{kt} mentioned the possibility to form
boson stars\footnote{Boson stars, that are the solutions of the 
KGE equations, were introduced by Kaup \cite{kaup} and
Ruffini and Bonazzola \cite{rb} in the case where the bosons have no
self-interaction. Boson stars in which the bosons have a repulsive
self-interaction were considered
later by Colpi {\it et al.} \cite{colpi}, Tkachev \cite{tkachev} and Chavanis
and Harko \cite{chavharko}. These
authors showed that boson stars can exist only below a maximum mass
(see Appendixes \ref{sec_cs} and \ref{sec_is}) due to
general relativistic effects.} by Jeans instability through Bose-Einstein
relaxation in the gravitationally bound clumps of axions. In other words,
axitons are expected to collapse into
boson stars due to Jeans instability when self-gravity becomes important.
This possibility was originally proposed by Tkachev \cite{tkachev,tkachevrt} who
introduced the names ``gravitationally bound axion condensates''
\cite{tkachev} and ``axionic Bose stars'' \cite{tkachevrt}, becoming later
``axion stars''. It is important to stress, however, that Tkachev
\cite{tkachev,tkachevrt} and Kolb and Tkachev \cite{kt} considered 
boson stars with a repulsive self-interaction ($a_s>0$). Since axions have an
attractive self-interaction ($a_s<0$), one cannot directly apply the standard
results of boson stars \cite{kaup,rb,colpi,tkachev,chavharko} to axion stars.

The case of self-interacting boson stars with an attractive
self-interaction ($a_s<0$), possibly representing axion stars, has been
considered only
recently \cite{gu,bb,prd1,prd2,braaten,davidson,cotner,bectcoll,ebybosonstars,
braatenEFT,ebycollapse,bbb,tkachevprl,braatenR,helfer,ebycollisions,ebydecay}.
The analytical expression  of the maximum mass of
Newtonian self-gravitating BECs
with an attractive self-interaction was first obtained in \cite{prd1,prd2}. For
QCD
axions, this leads to a maximum mass $M_{\rm max}^{\rm
exact}=6.46\times 10^{-14}\,
M_{\odot}=1.29\times 10^{17}\, {\rm kg}=2.16\times 10^{-8}\,
M_{\oplus}$ and a minimum radius $(R_{99}^*)^{\rm exact}=3.26\times 10^{-4}\,
R_{\odot}=227\, {\rm km}=3.56\times 10^{-2}\, R_{\oplus}$. This is the maximum
mass of dilute QCD axion stars. These values correspond to the typical size of
asteroids.
Obviously, QCD axions cannot form giant BECs with the
dimension of DM halos. By contrast, they may form mini axion stars that could be
the constituents of DM halos under the form of mini massive compact halo objects
(mini MACHOs) \cite{bectcoll}.\footnote{The
mass $M$ of the mini MACHOs must be less than $2\times 10^{-9}\, M_{\odot}$ to
be consistent with the observational constraints imposed by
gravitational microlensing \cite{eros,griest}. The maximum mass $M_{\rm
max}^{\rm
exact}=6.46\times
10^{-14}\, M_{\odot}$ of QCD axion stars satisfies this
constraint. Coincidentally, these mini axion stars have a mass comparable
to the mass of the axitons \cite{kt} but their mechanism of formation is
completely different. In particular, the maximum mass of a mini
axion star is obtained by balancing the quantum pressure against the
gravitational and self-interaction forces whereas the mass of a minicluster is
determined by the volume of the horizon when the axion starts to
oscillate. } These mini axion
stars are Newtonian self-gravitating BECs
of QCD axions with an attractive self-interaction stabilized
by the quantum pressure (Heisenberg uncertainty principle). They may cluster
into structures similar to standard CDM
halos but made of scalar field MACHOs instead of WIMPs. However, mini axion
stars (MACHOs)  behave essentially as
CDM and do not solve the small-scale crisis of CDM.

Other types of axions with a very small mass may exist and are called
ultralight axions (ULA) \cite{marsh}. Such axions appear in string theory
\cite{witten} leading to the notion of string axiverse \cite{axiverse}.
ULAs can form very  big objects of the size of DM halos. For an ULA with a mass 
$m=2.19\times 10^{-22}\, {\rm eV}/c^2$  and a very small attractive
self-interaction $a_s=-1.11\times 10^{-62}\, {\rm fm}$, one finds 
\cite{bectcoll} that the maximum mass and the minimum radius of an
axionic DM halo are $M_{\rm max}^{\rm exact}=10^8\, M_{\odot}$ and  $R=1\, {\rm
kpc}$, comparable to the dimensions of dwarf spheroidal galaxies (dSphs) like
Fornax.\footnote{See Appendix D of \cite{abrilphas} for a
more detailed discussion about the mass of ULAs, including noninteracting bosons
and bosons with a repulsive or an attractive self-interaction.} Therefore, ULAs
can form giant
BECs with the dimensions of DM halos.  

The previous discussion shows that QCD axions and ULAs behave very differently.
QCD axions form mini axion stars of the asteroid size which behave like
CDM. On the other hand, ULAs form DM halos which can solve the small-scale
problems of CDM. In the following, to simplify the discussion, we shall always
speak of ``axion stars'' but we stress that this term can refer either
to  ``mini axion stars'' (QCD axions) or ``axionic DM halos'' (ULAs). Our
formalism is
general and can handle these two situations.

One may wonder what happens when the mass of an axion
star is larger than the maximum mass $M_{\rm max}$  \cite{prd1,prd2}.
In that case, there is no equilibrium state
and
the star  undergoes gravitational collapse. Similarly, white
dwarf
stars
\cite{chandra31}, neutron stars \cite{ov} and boson stars
\cite{rb,kaup,colpi,tkachev,chavharko} exist only below a  maximum mass
and collapse when they are too massive. The collapse of massive stars
leads to black holes. We note, however, that the
maximum mass of the axion stars is a Newtonian nonrelativistic result
(essentially due to
the attractive self-interaction of the axions) while the maximum mass of white
dwarf stars,
neutron stars and boson stars arises from special and/or general
relativity. As a result, the collapse of  axion
stars is expected to be very different from the collapse of white dwarfs,
neutron stars and boson stars.

The collapse of dilute axion stars  above $M_{\rm max}$ was first
discussed in \cite{bectcoll}. In this paper, we determined the general
expression of the collapse time $t_{\rm coll}(M,R_0)$ as a function of the mass
$M\ge M_{\rm max}$ of the axion star and its initial radius
$R_0$.\footnote{The main purpose of our paper was to obtain the expression of
the collapse time $t_{\rm coll}(M,R_0)$, not to study accurately
the collapse dynamics of the axion star or determine what is the final outcome
of the collapse. We mentioned the possibility to form black holes, but we also
stressed the limitation of our approach and mentioned other scenarios when a
more realistic self-interaction potential is considered. We
note that the
order of magnitude of the collapse time $t_{\rm coll}(M,R_0)$ obtained in
\cite{bectcoll} is totally
insensitive to the precise nature of the final object that is formed as a
result of the collapse as long
as it is sufficiently small (see Appendix \ref{sec_tco}).} We
used an
approximation based on a  Gaussian ansatz which is expected to be very good
close to the critical point $M_{\rm max}$.  However, in order to determine the
final fate
of an axion star experiencing
gravitational collapse, a more precise study is necessary.

Let us first consider, as in \cite{bectcoll}, axion stars with a purely
attractive $|\psi|^4$
self-interaction potential in the framework of Newtonian gravity. The dynamical
evolution above $M_{\rm max}$ of a nonrelativistic self-gravitating BEC with a
purely attractive $|\psi|^4$ interaction described by the GPP equations is
interesting in its own right even if this model may not be realistic on a
physical point of view (see below). In that case,
we expect that the collapse will ultimately lead to a Dirac peak $\rho({\vec
x})=M\delta({\vec x})$ because the effective
potential is not bounded from below (see Appendix \ref{sec_dpbh}). However, the
evolution of the system is not trivial. In the
early
stage of the collapse, self-gravity must be taken into account. By contrast, in
the
late stages of the collapse, the attractive self-interaction of the bosons
dominates, and self-gravity can be neglected. The collapse of a
nongravitational BEC with an attractive self-interaction ($a_s<0$) has been
studied extensively and is well-known \cite{sulem,zakharov}. After an initial
smooth
evolution where the BEC density profile is close to a Gaussian (if it is
Gaussian initially), the system undergoes a self-similar collapse (called
``wave collapse'') leading to a
finite
time singularity: at $t=t_{\rm coll}(M,R_0)$,\footnote{This collapse time
corresponds precisely to the function $t_{\rm coll}(M,R_0)$ studied in
\cite{bectcoll}. It cannot be obtained from the self-similar
solution of the GP equations because it appears as an undetermined constant of
integration \cite{sulem,zakharov}. Therefore, a complementary study, such as the
one performed in \cite{bectcoll}, is necessary to obtain $t_{\rm coll}(M,R_0)$.}
the
central density $\rho_0(t)$ becomes
infinite and the core radius $r_0(t)$ tends to zero leading to a singular
density profile $\rho(r,t=t_{\rm coll})\propto r^{-2}$ diverging at the origin.
During the self-similar evolution, the central density
increases as $\rho_0(t)=\rho(0,t)\sim (t_{\rm coll}-t)^{-1}$ and the core radius
decreases as $r_0(t)\sim (t_{\rm coll}-t)^{1/2}$. Note that the
singularity at $t=t_{\rm coll}$ contains no central mass:
$M(r)=4\pi\int_0^{r}\rho(r')r'^{2}dr'\rightarrow 0$ as $r\rightarrow 0$.
However, the evolution continues in the post collapse regime $t>t_{\rm
coll}$, where a Dirac peak  $\rho({\vec
x},t)=M_D(t)\delta({\vec x})$ with an increasing mass $M_D(t)$ forms at $r=0$
by accreting particles around it.\footnote{As far as we
know, this postcollapse regime has not been studied in detail for the GP
equation with an attractive self-interaction. Therefore, our discussion should
be confirmed, or infirmed, by numerical simulations. Our claim that a Dirac peak
is formed in the postcollapse regime comes from the analogy between the GP
equation with an attractive self-interaction and  the Smoluchowski-Poisson
equations describing self-gravitating Brownian
particles \cite{sp1,sp2,sp3,sp4,sp5,sp6}.}
If we take general relativity into account, using the KGE equations  with a
purely attractive $\phi^4$ interaction, we expect that the Dirac peak will be
replaced by a  black hole.  Of course, the
postcollapse dynamics of the
GPP and KGE equations is
difficult to study mathematically and numerically because one needs to cross the
$r^{-2}$ singularity at $t=t_{\rm coll}$. It
would be interesting to study this model in detail at a mathematical level even
if it is purely academic on a physical point of view for the reasons explained
below.

First, the previous scenario assumes that the system
remains spherically symmetric and compact during the collapse. Recently, Cotner
\cite{cotner} has
numerically solved the GPP equations with a purely attractive self-interaction.
When $M>M_{\rm max}$, he found that the system collapses and first
forms an extremely dense core surrounded by a fluctuating halo of scalar waves
(suggesting a large amount of interference). Then, he observed that the core
fragments into several stable pieces (axion ``drops'') of mass $M'<M_{\rm
max}$,
thereby
avoiding its catastrophic collapse into a singularity. This
fragmentation process was previously suggested in \cite{davidson}.

On the other hand, even in the case of a spherical collapse, new
physical processes can come into play   when the system
becomes dense enough and can 
prevent the formation of a finite time singularity at  $t=t_{\rm coll}$. In
particular,
when the system
becomes dense, the   $|\psi|^4$ approximation is not valid anymore and we have
to take into account higher order terms in the expansion of the SF potential
(or, better, consider the exact axionic self-interaction potential). These
higher order terms, which can
be repulsive (unlike the $\phi^4$ term), can account for strong collisions
between axions. These collisions may have important consequences on the collapse
dynamics. Three possibilities have been considered in the recent literature:

(i) The first possibility is to form {\it dense axion stars}. Indeed, when
repulsive
terms are taken into account, the SF potential becomes bounded from
below. In that case, the collapse is halted by the repulsion of the bosons that
becomes dominant at high densities. An
equilibrium state exists in
which the
self-gravity is balanced by
the repulsive self-interaction. This
possibilty was first proposed by Braaten {\it et al.} \cite{braaten} who
calculated numerically the equilibrium configurations of dense axion stars.

(ii) The second possibility is a {\it bosenova} phenomenon. The collapse of the
axion star may be accompanied by a burst of relativistic axions
(radiation) produced by
inelastic reactions when the density reaches high values.
Relativistic
collisions between
axions can stop the collapse and even drive a re-expansion of the system. In
that case, the collapse (implosion) is followed by an explosion. This phenomenon
was shown 
experimentally by Donley {\it et al.} \cite{donley} for
nongravitational relativistic BECs with an attractive self-interaction. Braaten
{\it et al.} \cite{braaten}  and
Eby {\it et al.} \cite{ebycollapse} argued that axion stars, as
they collapse, emit many highly energetic free axions. This could cause the
partial, or even complete, disappearance  of the axion star.
Recently, the validity of this scenario (bosenova) for
relativistic axion stars
was demonstrated
by Levkov {\it
et al.} \cite{tkachevprl} from direct
numerical
simulations of the KGE
equations with the exact axionic potential taking collisions into account. In
these simulations, the system does not reach an
equilibrium state (i.e. it does not form a dense axion star) but undergoes a
series of collapses and
explosions. Multiparticle relativistic interactions in
the dense center create an outgoing stream of mildly relativistic particles
which carries away an essential part of the star mass (about 30$\%$). The
explosion is a period of violent oscillations accompanied by strong emission
of outgoing high-frequency waves with a characteristic spectrum (see Fig. 3 in
 \cite{tkachevprl}).

(iii) The third possibility, in the case where general
relativity is taken into account, is the formation of a  black hole if the mass
of the axion star is sufficiently large or if the self-interaction is
sufficiently weak. This possibility has been demonstrated numerically by Helfer
{\it et al.} \cite{helfer}. For small masses or strong self-interactions, they
obtained either a stable axion star or a dispersion phenomenon similar to that
reported by Levkov {\it et al.} \cite{tkachevprl}.  For large masses or weak
self-interactions, a black hole was formed.

Other works have been considered in relation to axion stars.  In particular, it
has been proposed that Fast radio bursts (FRBs), whose origin is one of the
major mysteries of high energy astrophysics,  could be caused by axion
stars that can engender bursts when undergoing conversion into photons during
their collision with the magnetosphere of neutron stars (magnetars),  during
their collision with
the
magnetized
accretion disk of a black hole, or
during their collapse above the maximum mass. We refer to
\cite{tkachev2015,iwazaki,raby,iwazakinew,bai} for the suggestion of this
scenario
and to
\cite{pshirkov} for an interesting critical discussion.

For all these reasons, it is important to study the phase transitions
between dilute and dense axion stars in detail. The aim
of this paper is to perform an exhaustive analytical study of this problem and
to isolate characteristic mass, length, density and energy scales that play an
important role. Even if our analytical approach is approximate, these typical
scales are fully relevant and determine the gross characteristic features of the
system. In general, in astrophysics, we are more interested by
orders of magnitude than by exact values. Our study provides the
correct orders of magnitude. Furthermore, when it was possible to compare our
approximate analytical results with
exact numerical results \cite{prd2,braaten}, we found that our study proves to
be relatively accurate. 

The paper is organized as follows. In Sec. \ref{sec_hydro},
starting from the KGE equations for a real SF, and considering the
nonrelativistic limit, we derive the hydrodynamic equations describing axionic
dark matter in an expanding universe taking into account the self-interaction
of the axions.
In Sec. \ref{sec_eos}, we determine
the general equation of state of axionic dark matter and introduce the
simplified equation of state studied in this paper. In Sec. \ref{sec_maxmass},
we recall important results concerning the maximum mass of dilute axion stars
and express it with different parameters that are relevant to our problem. In
Sec. \ref{sec_mrdd}, we derive the analytical mass-radius relation of dilute and
dense axion stars within the Gaussian ansatz. In Sec. \ref{sec_tf}, we consider
the TF approximation which is particularly well-suited to describe dense axion
stars. In Sec. \ref{sec_eap}, we derive the energy and pulsation of dilute
and dense axion stars and derive
a simplified version of Poincar\'e's turning point argument based on the
Gaussian ansatz. In Sec. \ref{sec_r}, we derive the radius of the axion star
resulting from a collapse or an explosion at a critical point. In Sec.
\ref{sec_sapt}, we study the stability of axion stars and the nature of phase
transitions
between dilute and dense configurations. In Sec. \ref{sec_qgr}, we take into
account general relativistic effects in a qualitative manner and obtain an
estimate of the maximum mass of axion stars above which they collapse into a
black hole. This allows us to obtain a phase diagram exhibiting a triple point
separating dilute axion stars, dense axion stars, and black holes. In Sec.
\ref{sec_summary}, we summarize our main results and discuss analogies and
differences between axion stars and fermion stars. The Appendixes present
important complementary results giving, for example, the fundamental scalings of
mass, radius, density, energy... or providing exact results (going beyond the
Gaussian ansatz) in particular limits of the problem.

\section{Hydrodynamic equations of axionic dark matter in an expanding universe}
\label{sec_hydro}

In this section, we derive the hydrodynamic equations describing axionic DM in
an expanding universe in the nonrelativistic limit $c\rightarrow +\infty$,
starting from the KGE equations for a real SF. The
nonrelativistic approximation is extremely well justified for axions. We
compare the resulting equations with those obtained in our previous papers
\cite{abrilph,playa} for a complex SF.

\subsection{The KGE equations}
\label{sec_kge}

We consider the weak field gravity limit and work with the 
Newtonian gauge which is a
perturbed form
of the Friedmann-Lema\^itre-Robertson-Walker  (FLRW) line element \cite{ma}. We
use the simplest form of 
Newtonian gauge, only taking
into account scalar perturbations which are the ones that contribute to
the formation of structures in cosmology. Vector perturbations (which are
supposed to be always small) vanish during cosmic inflation and tensor
perturbations (which account for gravitational waves) are neglected \cite{mfb}.
We also neglect anisotropic stresses. We finally assume that the Universe is
flat in agreement with the observations of the cosmic microwave background
(CMB). Under these conditions, the line element is given by
\begin{equation}
ds^2=c^2\left(1+2\frac{\Phi}{c^2}\right)dt^2-a(t)^2\left(1-2\frac{\Phi}{c^2}
\right)\delta_{ij}dx^idx^j,
\label{conf1}
\end{equation}
where ${\Phi}/{c^2}\ll 1$. In this metric, $\Phi({\vec x},t)$ represents the
gravitational potential of
classical Newtonian gravity and $a(t)$ is the scale factor. Working with this
metric enables us
to obtain exact equations at the
order  $O(\Phi/c^2)$ taking into account both relativistic and gravitational
contributions inside an expanding Universe.

We consider a real SF $\varphi$ with a potential $V(\varphi^2)$ whose evolution
is
governed by the KGE equations. In the weak field approximation $\Phi/c^2\ll 1$,
using the Newtonian gauge, the KGE equations write 
\begin{eqnarray}
\frac{1}{c^2}\frac{\partial^2\varphi}{\partial
t^2}+\frac{3H}{c^2}\frac{\partial\varphi}{\partial
t}-\frac{1}{a^2}\left(1+\frac{4\Phi}{c^2}\right)\Delta\varphi
-\frac{4}{c^4}\frac{\partial\Phi}{\partial t}\frac{\partial\varphi}{\partial
t}\nonumber\\
+\left (1+\frac{2\Phi}{c^2}\right) \frac{m^2
c^2}{\hbar^2}\varphi+2\left
(1+\frac{2\Phi}{c^2}\right)\frac{dV}{d\varphi^2}\varphi
=0,
\label{kge1}
\end{eqnarray}
\begin{eqnarray}
\frac{\Delta\Phi}{4\pi G a^2}&=&\frac{1}{2c^4}\left (1-\frac{2\Phi}{c^2}\right
)\left (\frac{\partial\varphi}{\partial t}\right )^2\nonumber\\
&+&\frac{1}{2a^2c^2}\left (1+\frac{2\Phi}{c^2}\right
)(\vec\nabla\varphi)^2+\frac{m^2}{2\hbar^2}\varphi^2\nonumber\\
&+&\frac{1}{c^2}V(\varphi^2)-\frac{3H^2}{8\pi G}+\frac{3H}{4\pi Gc^2}\left
(\frac{\partial\Phi}{\partial t}+H\Phi\right ),\nonumber\\
\label{kge2}
\end{eqnarray}
where $H=\dot a/a$ is the Hubble parameter. They can be compared to the KGE
equations (12) and (18) of \cite{playa} for a complex SF.

\subsection{The GPE equations}
\label{sec_gpe}

To ultimately obtain the nonrelativistic limit $c\rightarrow +\infty$, we make
the
transformation
\begin{equation}
\label{gpe1}
\varphi=\frac{1}{\sqrt{2}}\frac{\hbar}{m}\left\lbrack \psi({\vec
x},t)e^{-imc^2t/\hbar}+\psi^*({\vec
x},t)e^{imc^2t/\hbar}\right\rbrack.
\end{equation}
This is the counterpart of the Klein transformation  for a complex SF $\varphi$
(see Eq.
(24) of
\cite{playa}).
Equation (\ref{gpe1})
introduces a complex function $\psi({\vec
x},t)$ that we shall call the pseudo wave
function. This transformation allows us to separate the fast
oscillations of the SF with pulsation $\omega=mc^2/\hbar$
caused by its rest
mass from the slow evolution of $\psi({\vec
x},t)$. Neglecting the oscillatory
terms that average out to zero in the fast
oscillation regime $\omega \gg H$, we find
that\footnote{This approximation may lead to
incorrect results in the
relativistic regime but it becomes exact in the nonrelativistic limit
$c\rightarrow +\infty$ considered after Eq. (\ref{gpe5}).
Note that there is no such approximation for a
complex SF since the relation $|\varphi|^2=(\hbar^2/m^2)|\psi|^2$ is always
valid in that case
(see Eq. (24) of \cite{playa}).}
\begin{eqnarray}
\varphi^2\simeq \frac{\hbar^2}{m^2}|\psi|^2.
\label{gpe2}
\end{eqnarray}
We define the pseudo rest-mass density
by
\begin{eqnarray}
\rho= |\psi|^2.
\label{gpe3}
\end{eqnarray}
We stress that it is only in the nonrelativistic limit $c\rightarrow +\infty$
that $\psi$ has the interpretation of a wave function and that $\rho$
has the interpretation of a rest-mass density (see below). In the relativistic
regime,
$\psi$ and $\rho$ do not have a clear physical interpretation but they can
always be defined as convenient notations \cite{abrilph,playa}.

Substituting Eq. (\ref{gpe1}) into Eqs. (\ref{kge1}) and (\ref{kge2}), and
neglecting the the oscillatory terms again (see footnote 8),  we
obtain the GPE equations
\begin{eqnarray}
i\hbar\frac{\partial\psi}{\partial t}&-&\frac{\hbar^2}{2m
c^2}\frac{\partial^2\psi}{\partial t^2}-\frac{3}{2}H\frac{\hbar^2}{m
c^2}\frac{\partial\psi}{\partial t}\nonumber\\
&+&\frac{\hbar^2}{2 m a^2}\left
(1+\frac{4\Phi}{c^2}\right )\Delta\psi\nonumber\\
&-&m\Phi \psi-\left(1+\frac{2\Phi}{c^2}\right
)m\frac{dV}{d|\psi|^2}\psi+\frac{3}{2}i\hbar H\psi\nonumber\\
&+&\frac{2\hbar^2}{m c^4}\frac{\partial\Phi}{\partial t}\left
(\frac{\partial \psi}{\partial t}-\frac{i m c^2}{\hbar}\psi\right )=0,
\label{gpe4}
\end{eqnarray}
\begin{eqnarray}
\frac{\Delta\Phi}{4\pi G a^2}&=&\left (1-\frac{\Phi}{c^2}\right
)|\psi|^2+\frac{\hbar^2}{2m^2c^4}\left (1-\frac{2\Phi}{c^2}\right
)\left|\frac{\partial\psi}{\partial t}\right |^2\nonumber\\
&+&\frac{\hbar^2}{2a^2m^2c^2}\left
(1+\frac{2\Phi}{c^2}\right
)|\vec\nabla\psi|^2+\frac{1}{c^2}V(|\psi|^2)\nonumber\\
&-&\frac{\hbar}{m c^2}\left(1-\frac{2\Phi}{c^2}\right ){\rm Im} \left
(\frac{\partial\psi}{\partial t}\psi^*\right )
-\frac{3H^2}{8\pi G}\nonumber\\
&+&\frac{3H}{4\pi G c^2}\left
(\frac{\partial\Phi}{\partial t}+H\Phi\right ).
\label{gpe5}
\end{eqnarray}
These equations look similar to those obtained for a complex SF without making
any approximation (see Eqs.
(25) and (26) of \cite{playa}). However, there is a crucial difference. In
the present case, $V(|\psi|^2)$ is an effective potential deduced from
$V(\varphi^2)$ by substituting $\varphi$ from Eq. (\ref{gpe1}) and neglecting
the oscillatory terms. Therefore, it is different from the potential that
one would get by directly substituting $\varphi$ (already averaged) from Eq.
(\ref{gpe2}) into $V(\varphi^2)$, as we can do when $\varphi$ is complex
\cite{playa}. The instantonic potential of
axions is worked out explicitly in Sec. \ref{sec_inst}
following \cite{ebycollapse}.

In the nonrelativistic limit $c\rightarrow
+\infty$,\footnote{Although we take the $c\rightarrow
+\infty$ limit in the field equations, we keep
relativistic contributions in the SF potential (see Sec. \ref{sec_inst}).} the
GPE
equations 
(\ref{gpe4}) and (\ref{gpe5}) reduce to the GPP
equations
\begin{eqnarray}
i\hbar\frac{\partial\psi}{\partial
t}+\frac{3}{2}i\hbar
H\psi=-\frac{\hbar^2}{2 m a^2}\Delta\psi+m\Phi \psi
+m\frac{dV}{d|\psi|^2}\psi,
\label{gpe6}
\end{eqnarray}
\begin{eqnarray}
\frac{\Delta\Phi}{4\pi G a^2}=|\psi|^2 -\frac{3H^2}{8\pi G},
\label{gpe7}
\end{eqnarray}
similar to Eqs. (29) and (30) of \cite{playa} (with the same warning as before
 concerning the effective potential).
When $V=0$, they reduce to the Schr\"odinger-Poisson
equations in an expanding Universe. The GP equation  (\ref{gpe6}) can
be rewritten as
\begin{eqnarray}
i\hbar\frac{\partial\psi}{\partial
t}+\frac{3}{2}i\hbar
H\psi=-\frac{\hbar^2}{2 m a^2}\Delta\psi
+m\lbrack\Phi
+h(|\psi|^2)\rbrack\psi
\label{gpe8}
\end{eqnarray}
where
\begin{eqnarray}
\label{gpe9}
h(|\psi|^2)=\frac{dV}{d|\psi|^2}, \qquad {\rm i.e.}\qquad h(\rho)=V'(\rho).
\end{eqnarray}

{\it Remark:} In the full relativistic theory, the axions are
represented by a real SF $\varphi$ and the particle number is not conserved.
However, in the nonrelativistic limit, they are just spinless particles
described by a complex wavefunction $\psi$ and their number $N=\frac{1}{m}\int
|\psi|^2\, d{\bf r}$ is conserved. Physically, the particle number is conserved
because, by removing the fast oscillating terms, we have eliminated the particle
number violating processes that are energetically forbidden for
nonrelativistic particles.

\subsection{The hydrodynamic representation}
\label{sec_mad}

Using the Madelung \cite{madelung} transformation, we can write the GPP
equations under the form of hydrodynamic equations. To that purpose, we
write the wave function $\psi$ as
\begin{eqnarray}
\psi(\vec{x},t)&=&\sqrt{\rho(\vec{x},t)} e^{iS(\vec{x},t)/\hbar},
\label{mad1}
\end{eqnarray}
where $\rho=|\psi|^2$ is the density
and $S=(1/2)i\hbar\ln(\psi^*/\psi)$ is the real action. Following
Madelung, we also define a velocity field and an energy field by
\begin{eqnarray}
\vec{v}(\vec{x},t)&=&\frac{\vec\nabla S}{ma},\qquad
E(\vec{x},t)=-\frac{\partial S}{\partial t},
\label{mad2}
\end{eqnarray}
where the scale factor $a$ has been introduced in the velocity field in order to
take into account the
expansion of the Universe. Since the
velocity is potential, the flow is
irrotational: $\nabla\times {\bf
u}={\bf 0}$. Substituting Eq. (\ref{mad1}) into Eq. (\ref{gpe8}) and separating
real and imaginary parts, we obtain the quantum Euler-Poisson (EP) equations
in an
expanding background \cite{aacosmo,playa,chavmatos}:
\begin{eqnarray}
\frac{\partial\rho}{\partial t}+3H\rho+\frac{1}{a}\vec\nabla\cdot (\rho
{\vec v})=0,
\label{mad3}
\end{eqnarray}
\begin{eqnarray}
\frac{\partial S}{\partial t}+\frac{(\vec\nabla S)^2}{2 m
a^2}=\frac{\hbar^2}{2
m a^2}\frac{\Delta\sqrt{\rho}}{\sqrt{\rho}}
-m\Phi-mh(\rho),
\label{mad4}
\end{eqnarray}
\begin{eqnarray}
\frac{\partial {\vec v}}{\partial t}+H{\vec v}+\frac{1}{a}({\vec v}\cdot
\vec\nabla){\vec v}&=&
\frac{\hbar^2}{2m^2a^3}\vec\nabla\left( \frac{\Delta\sqrt{\rho}}{\sqrt{
\rho}} \right )\nonumber\\
&-&\frac{1}{a}
\vec\nabla\Phi-\frac{1}{\rho a}\vec\nabla P,
\label{mad5}
\end{eqnarray}
\begin{eqnarray}
\frac{\Delta\Phi}{4\pi G a^2}=\rho-\frac{3H^2}{8\pi
G}.
\label{mad6}
\end{eqnarray}

The hydrodynamic equations  (\ref{mad3})-(\ref{mad6}) have a clear physical
interpretation. Equation (\ref{mad3}), corresponding to the imaginary
part of the GP equation, is the continuity equation. It accounts for the local
conservation of mass $M=\int \rho\, d{\bf r}$. Equation (\ref{mad4}),
corresponding to the real part of the GP equation, is the  quantum 
Hamilton-Jacobi equation. It can also be interpreted as a generalized Bernoulli
equation for a potential flow. It involves a quantum potential
\begin{equation}
\label{mad7}
Q=-\frac{\hbar^2}{2ma^2}\frac{\Delta
\sqrt{\rho}}{\sqrt{\rho}}=-\frac{\hbar^2}{4ma^2}\left\lbrack
\frac{\Delta\rho}{\rho}-\frac{1}{2}\frac{(\nabla\rho)^2}{\rho^2}\right\rbrack
\end{equation}
which takes into account the Heisenberg uncertainty
principle. Equation (\ref{mad5}),
obtained by taking the gradient of  Eq. (\ref{mad4}), is the quantum Euler
equation.  It involves a quantum force (per unit of mass)
${\bf F}_Q=-(1/ma)\nabla Q$. It also involves
a pressure force $-(1/\rho a)\nabla P$
where the pressure is given by a barotropic equation of state $P(\rho)$
determined by the nonlinearity $h(\rho)$ in the GP equation
(\ref{gpe8}) through
the relation
\begin{equation}
\label{mad8}
h'(\rho)=\frac{P'(\rho)}{\rho}.
\end{equation}
This equation shows that $h$ can be interpreted as an enthalpy in the
hydrodynamic representation \cite{prd1}.  Equation (\ref{mad8}) can be
integrated
into \cite{playa,chavmatos}:
\begin{equation}
\label{mad9}
P(\rho)=\rho h(\rho)-V(\rho)=\rho V'(\rho)-V(\rho).
\end{equation}
The speed of sound is 
\begin{eqnarray}
c_s^2=P'(\rho)=\rho V''(\rho).
\label{mad10}
\end{eqnarray}
 Finally, Eq. (\ref{mad5}) involves the gravitational force $-\nabla\Phi$, where
$\Phi$ is determined by  the Poisson equation (\ref{mad6}).

{\it Remarks:} We stress that the hydrodynamic
equations (\ref{mad3})-(\ref{mad6})  do not involve viscous terms because they
are equivalent to the GPP equations (\ref{gpe6}) and (\ref{gpe7}). As a
result, they describe a superfluid. We also note that the hydrodynamic equations
(\ref{mad3})-(\ref{mad6}) with the quantum
potential (proportional to $\hbar^2$)  neglected provide a TF, or
semiclassical, description of
nonrelativistic SFs. When $\hbar=P=0$, one recovers the
classical equations of CDM.

\subsection{The spatially homogeneous SF}
\label{sec_hom}

For a spatially homogeneous SF ($\rho(\vec x,t)=\rho_b(t)$,
$\vec v_b(\vec x,t)=\vec 0$, $\Phi_b(\vec x,t)=0$, $S_b(\vec x,t)=S_b(t)$), the
fluid equations (\ref{mad3})-(\ref{mad6}) reduce to
\cite{aacosmo,playa,chavmatos}:
\begin{equation}
\frac{d\rho_b}{dt}+3H\rho_b=0,\qquad E=mV'(\rho_b),\qquad
{H^2}=\frac{8\pi G}{3}\rho_b,
\label{hom1}
\end{equation}
where $E(t)=-dS_b/dt$ is the time-dependent energy of the spatially
homogeneous SF in
the comoving frame. The wave function of the SF is $\psi_b(\vec
x,t)=\psi_b(t)=\sqrt{\rho_b(t)}e^{-(i/\hbar)\int E(t)\, dt}$. These equations
govern the evolution of the cosmological background. We find that $\rho_b\propto
a^{-3}$, $a\propto t^{2/3}$, $H=2/3t$ and $\rho_b=1/6\pi
Gt^2$,  corresponding to the Einstein-de Sitter (EdS) solution. Therefore,
in the nonrelativistic era, and for the homogeneous background, BECDM behaves
as CDM.
Using Eq. (\ref{hom1}), the Poisson equation (\ref{mad6}) can be rewritten as
\begin{eqnarray}
\Delta\Phi=4\pi G a^2 (\rho-\rho_b),
\label{hom2}
\end{eqnarray}
where $\rho_b(t)$ is the background density.

\subsection{The equations for the density contrast}
\label{sec_dc}

Let us write the density in the form $\rho(\vec x,t)=\rho_b(t)[1+\delta(\vec
x,t)]$,
where $\rho_b\propto 1/a^3$ and $\delta(\vec x,t)$ is the density contrast.
Substituting this expression into Eqs. (\ref{mad3})-(\ref{mad6}), we
obtain the following equations for the density contrast \cite{aacosmo}:
\begin{eqnarray}
\frac{\partial\delta}{\partial t}+\frac{1}{a}\vec\nabla\cdot [(1+\delta)
{\vec v}]=0,
\label{dc1}
\end{eqnarray}
\begin{eqnarray}
\frac{\partial {\vec v}}{\partial t}+H{\vec v}+\frac{1}{a}({\vec v}\cdot
\vec\nabla){\vec v}=
\frac{\hbar^2}{2m^2a^3}\vec\nabla\left( \frac{\Delta\sqrt{1+\delta}}{\sqrt{
1+\delta}} \right )\nonumber\\
-\frac{1}{a}
\vec\nabla\Phi-\frac{c_s^2}{(1+\delta) a}\vec\nabla \delta,
\label{dc2}
\end{eqnarray}
\begin{eqnarray}
\Delta\Phi=4\pi G \rho_b a^2 \delta,
\label{dc3}
\end{eqnarray}
where $c_s^2=P'(\rho)=P'[\rho_b(1+\delta)]$ is the square of the speed of
sound in the inhomogeneous Universe. We stress that these equations are exact
in the nonrelativistic limit, i.e. they do not rely on the assumption $\delta\ll
1$. The linearized
quantum EP equations have been studied in \cite{aacosmo,abrilph} in the context
of
structure formation for BECDM.

\section{The equation of state of axions}
\label{sec_eos}

\subsection{The instantonic potential of axions}
\label{sec_inst}

The instantonic potential of the axions is \cite{pq,wittenV,vv}:
\begin{equation}
\label{inst1}
V(\varphi)=\frac{m^2cf^2}{\hbar^3}\left\lbrack 1-\cos\left
(\frac{\hbar^{1/2}c^{1/2}\varphi}{f}\right
)\right\rbrack-\frac{m^2c^2}{2\hbar^2}\varphi^2,
\end{equation}
where $m$ is the mass of the axion and $f$ is the axion decay
constant.\footnote{Another,
more relevant, axionic potential is the chiral potential \cite{vv,chvv}.} 
Expanding the cosine
term in Taylor series, we get
\begin{equation}
\label{inst2}
V(\varphi)=-\frac{m^2cf^2}{\hbar^3}\sum_{n=2}^{+\infty}\frac{(-1)^n}{
(2n)!}\left (\frac{\hbar^{1/2}c^{1/2}\varphi}{f}\right
)^{2n}. 
\end{equation}
If we keep only the first two terms of this expansion, we obtain the
$\varphi^6$ potential
\begin{equation}
\label{inst3}
V(\varphi)=-\frac{m^2c^3}{24 f^2\hbar}\varphi^4+\frac{m^2c^4}{720
f^4}\varphi^6.
\end{equation}
The effective potential  $V(|\psi|^2)$ appearing in the GP
equation
(\ref{gpe6}) can be
obtained
from the following heuristic procedure introduced by Eby {\it et
al.} \cite{ebycollapse}. Raising
$\varphi$ in Eq.
(\ref{gpe1}) to the power $2n$ thanks to the binomial
formula and
keeping only terms that do not oscillate, we obtain
\begin{equation}
\label{inst4}
\varphi^{2n}=\frac{1}{2^n}\left (\frac{\hbar}{m}\right )^{2n}C_{2n}^n
|\psi|^{2n},
\end{equation}
where $C_{2n}^n=(2n)!/(n!)^2$. Substituting this expression into Eq.
(\ref{inst2}), we get
\begin{equation}
\label{inst5}
V(|\psi|^2)=-\frac{m^2cf^2}{\hbar^3}\sum_{n=2}^{+\infty}\frac{(-1)^n}{
(n!)^2}\left (\frac{\hbar^{3}c|\psi|^2}{2f^2m^2}\right
)^{n}. 
\end{equation}
The series can be performed analytically leading to 
an effective potential of the form
\begin{equation}
\label{inst6}
V(|\psi|^2)=\frac{m^2cf^2}{\hbar^3}\left\lbrack
1-\frac{\hbar^3 c}{2f^2m^2}|\psi|^2-J_0\left
(\sqrt{\frac{2\hbar^{3}c|\psi|^2}{f^2m^2}}\right
)\right\rbrack,
\end{equation}
where $J_0$ is the Bessel function of zeroth order. If we keep only the first
two
terms in the expansion of Eq. (\ref{inst5}), we obtain the $|\psi|^6$ potential
\begin{equation}
\label{inst7}
V(|\psi|^2)=-\frac{\hbar^3
c^3}{16f^2m^2}|\psi|^4+\frac{\hbar^6c^4}{288f^4m^4}|\psi|^6.
\end{equation}
We note that $V(|\psi|^2)$ is different from the expression
obtained by substituting Eq. (\ref{gpe2}) into $V(\varphi)$. The
difference is apparent already in the first term of the expansion of the
potential which involves a coefficient $-1/16$ [see Eq. (\ref{inst7})] instead
of $-1/24$ [see Eq. (\ref{inst3})].

In the nonrelativistic limit $c\rightarrow +\infty$, or for dilute systems
satisfying $|\psi|^2\ll f^2m^2/\hbar^3c$, the potential
$V(|\psi|^2)$ is dominated by the $|\psi|^4$ term. A $|\psi|^4$ potential is
usually written as 
\begin{equation}
\label{inst7b}
V(|\psi|^2)=\frac{2\pi
a_s\hbar^2}{m^3}|\psi|^4,
\end{equation}
where $a_s$ is the s-scattering length of the bosons \cite{revuebec}. Comparing
this expression with the
first
term of Eq.
(\ref{inst7}), we obtain
\begin{equation}
\label{inst8}
a_s=-\frac{\hbar c^3m}{32\pi f^2}.
\end{equation}
Using this relationship, the potential of Eq. (\ref{inst7}) can be rewritten as
\begin{equation}
\label{inst9}
V(|\psi|^2)=\frac{2\pi
a_s\hbar^2}{m^3}|\psi|^4+\frac{32\pi^2\hbar^4a_s^2}{9m^6c^2}|\psi|^6.
\end{equation}
We note that the first term is nonrelativistic while the second
term is relativistic in nature since it involves the speed of light $c$.
It provides a relativistic correction of order $1/c^2$ to the  $|\psi|^4$
potential.

For dilute configurations satisfying  $|\psi|^2\ll m^3c^2/|a_s|\hbar^2$, it is
sufficient to keep only the first term in
the expansion of the
potential $V(|\psi|^2)$ given by Eq. (\ref{inst5}). Since the axions have a
negative scattering length ($a_s<0$), we see from Eq. (\ref{inst9}) that the
$|\psi|^4$ term corresponds to an attraction between axions. This term is
sufficient to describe dilute axion stars below the maximum mass $M_{\rm max}$
\cite{prd1,prd2}. However, for $M>M_{\rm max}$, dilute axion stars undergo
gravitational collapse and their density increases. When the star
becomes dense enough, we
need to take into account higher order terms in the expansion of the potential.
We see that the $|\psi|^6$ term from Eq. (\ref{inst9}) corresponds to a
repulsion between axions. Therefore, this term can stop the collapse and lead to
equilibrium states corresponding to dense axion stars \cite{braaten}. This
repulsion has a relativistic origin. In
this paper, we will
consider a simplified model where we only retain the first two terms in the
expansion of the potential, i.e., we will use Eq. (\ref{inst9}) as an
approximation of the potential  $V(|\psi|^2)$ given by Eq. (\ref{inst6}).
Because of this approximation, our study may present differences with
respect to the studies of \cite{braaten,ebycollapse}
based on the
potential (\ref{inst6}). However, we will see that these differences are small.

{\it Remark:} the expression of the effective potential $V(|\psi|^2)$ given by
Eq. (\ref{inst6}),  obtained from the heuristic procedure of Eby {\it et al.}
\cite{ebycollapse}, is not exact. Braaten {\it et al.}
\cite{braatenEFT} have shown that the first two terms in the expansion
of the exact potential are given by
\begin{equation}
\label{inst10}
V_{\rm exact}(|\psi|^2)\simeq -\frac{\hbar^3
c^3}{16f^2m^2}|\psi|^4-\frac{\hbar^6c^4}{256f^4m^4}|\psi|^6+...
\end{equation}
The exact potential (\ref{inst10})  differs from Eq. (\ref{inst7})  at the level
of the $|\psi|^6$ term. Not only the value of the prefactor is different but its
sign also is different. For the potential of  Eq. (\ref{inst7}), the $|\psi|^6$ 
term corresponds to a repulsion while for the exact potential  of  Eq.
(\ref{inst10}) it corresponds to an attraction. Therefore, our study based on
the potential (\ref{inst9}), and the studies of 
\cite{braaten,ebycollapse} based on the
potential (\ref{inst6}), may present quantitative differences with
respect to the studies of \cite{tkachevprl,helfer} based on the exact axionic
potential
(\ref{inst1}).

\subsection{The general equation of state}
\label{sec_g}

\begin{figure}
\begin{center}
\includegraphics[clip,scale=0.3]{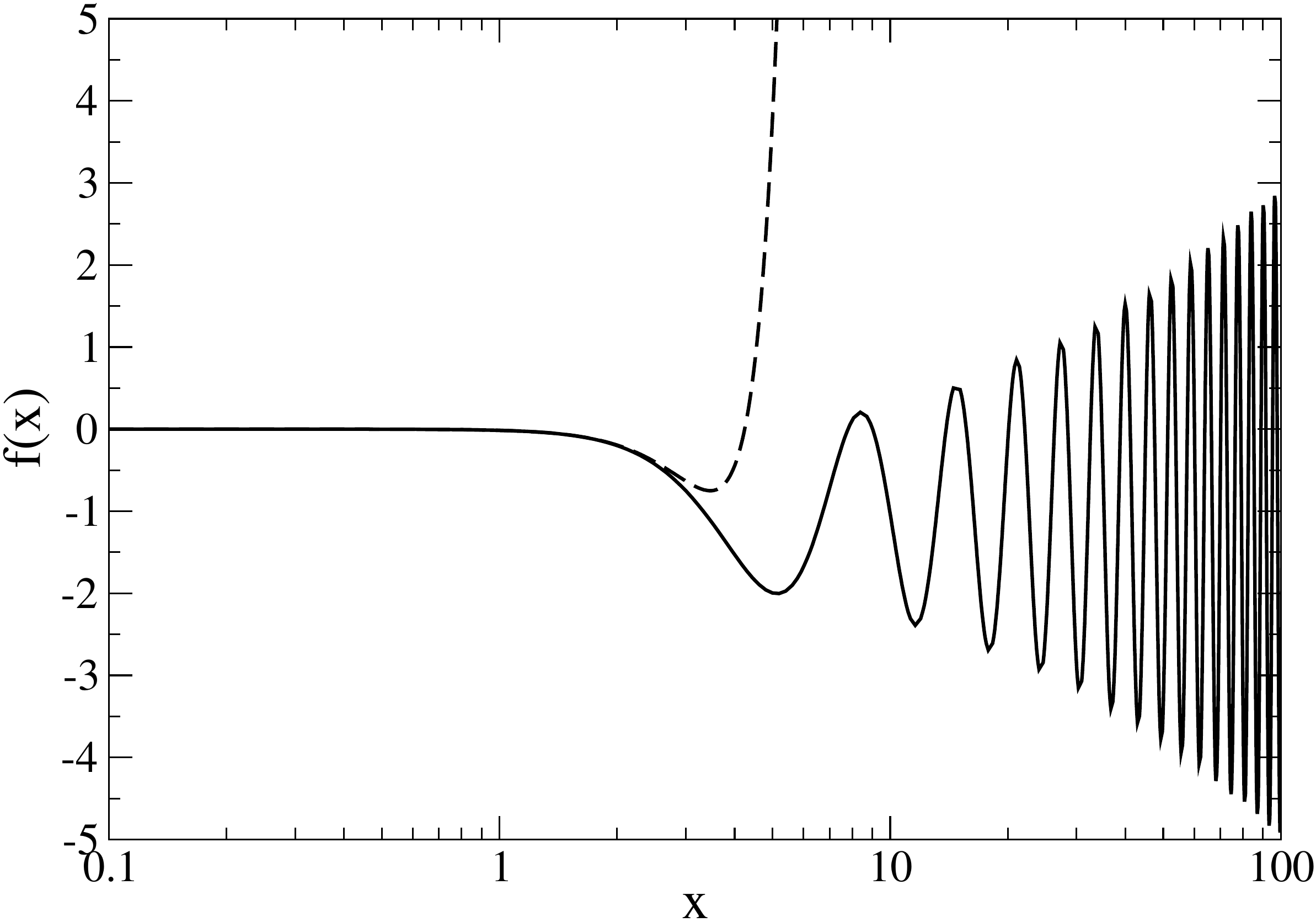}
\caption{The function $f(x)$ characterizing the equation of state of axions. For
comparison, we have plotted
in dashed line the function obtained by keeping only the two first terms in the
expansion of $f(x)$ for $x\rightarrow 0$ [see Eq. (\ref{g5})]. It corresponds
to the
simplified equation of state of Sec. \ref{sec_s} studied in this paper.}
\label{f}
\end{center}
\end{figure}

The general equation of state of an axion star is given by Eq. (\ref{mad9})
with the potential
\begin{equation}
\label{g1}
V(\rho)=\frac{m^2cf^2}{\hbar^3}\left\lbrack
1-\frac{\hbar^3 c}{2f^2m^2}\rho-J_0\left
(\sqrt{\frac{2\hbar^{3}c\rho}{f^2m^2}}\right
)\right\rbrack.
\end{equation}
Using $J_0'(x)=-J_{1}(x)$, we obtain 
\begin{eqnarray}
\label{g3}
P(\rho)=\frac{m^2cf^2}{\hbar^3}\biggl\lbrack
\frac{1}{2}\sqrt{\frac{2\hbar^{3}c\rho}{f^2m^2}}J_1\left
(\sqrt{\frac{2\hbar^{3}c\rho}{f^2m^2}}\right )\nonumber\\
+J_0\left
(\sqrt{\frac{2\hbar^{3}c\rho}{f^2m^2}}\right
)-1\biggr\rbrack.
\end{eqnarray}
This equation of state involves the function 
\begin{equation}
\label{g4}
f(x)=\frac{x}{2}J_1(x)+J_0(x)-1
\end{equation}
that is plotted in Fig. \ref{f}. 
For $x\rightarrow 0$,
\begin{equation}
\label{g5}
f(x)\simeq -\frac{x^4}{64}+\frac{x^6}{1152}.
\end{equation}
For $x\rightarrow +\infty$,
\begin{equation}
\label{g6}
f(x)\sim \sqrt{\frac{x}{2\pi}}\cos\left (x-\frac{\pi}{2}-\frac{\pi}{4}\right
).
\end{equation}

The equation of state (\ref{g3}) has a complicated oscillatory
behavior for $\rho\rightarrow +\infty$. To understand the meaning of these
oscillations, let us progressively increase the
density of axion stars, from dilute axion stars to dense axion stars. At low
densities, the pressure force is attractive (because $a_s<0$) and adds to the
gravitational attraction. A dilute axion star can resist this attraction
thanks to the repulsive quantum force arising from Heisenberg's uncertainty
principle. However, above a critical
density,  corresponding to the maximum mass $M_{\rm max}$ \cite{prd1,prd2}, the
axion
star becomes unstable and collapses. During the
collapse, its density increases. At sufficiently high densities the
pressure force becomes repulsive and counteracts the gravitational attraction.
An equilibrium state is reached,
corresponding to a dense axion star \cite{braaten}. If we keep increasing the
density, the
pressure force becomes attractive again. The dense axion star becomes unstable
and collapses until, at even higher densities, the pressure force becomes
repulsive and stops the collapse. This
process of stabilization/destabilization occurs periodically at higher and
higher densities.  As a result, the mass-radius relation of axion stars should
exhibit a branch of dilute axion stars and several branches of dense
axion stars. In this paper, we shall only consider the
first branch of dense axion stars. Therefore, it is sufficient to consider the
approximation of the equation of state for sufficiently low densities until
the first stabilization takes place. This is dicussed in the
following section.

\subsection{The simplified equation of state}
\label{sec_s}

In our simplified model, the
equation of state of an axion star is given
by Eq. (\ref{mad9}) with the potential
\begin{equation}
\label{s1}
V(\rho)=\frac{2\pi
a_s\hbar^2}{m^3}\rho^2+\frac{32\pi^2\hbar^4a_s^2}{9m^6c^2}\rho^3.
\end{equation}
This yields
\begin{equation}
\label{s2}
P(\rho)=\frac{2\pi
a_s\hbar^2}{m^3}\rho^2+\frac{64\pi^2\hbar^4a_s^2}{9m^6c^2}\rho^3.
\end{equation}
The equation of state can be written as
\begin{equation}
\label{s3}
P(\rho)=K_2\rho^2+K_3\rho^3,
\end{equation}
where $K_2={2\pi
a_s\hbar^2}/{m^3}$ and $K_3={64\pi^2\hbar^4a_s^2}/{9m^6c^2}$. This is the sum of
two polytropic equations of state of index $\gamma=2$
($n=1$) and $\gamma=3$ ($n=1/2$). The first equation of state has an attractive
effect ($K_2<0$) and the second equation of state has a repulsive
effect ($K_3>0$).

The equation of state (\ref{s2}) can also be obtained from the exact equation of
state (\ref{g3}) by keeping only the first two terms
in its expansion for $\rho\rightarrow 0$. The approximate equation of state
(\ref{s2}) is plotted in
dashed line in Fig. \ref{f} and compared with the exact equation of state
(\ref{g3}). We see that the approximate equation of state has a behavior
similar to the
exact equation of state at low densities. Starting from $P=0$, the pressure
first decreases (being therefore negative), reaches a minimum and
finally increases up to infinity. At low densities, corresponding to dilute
axion stars, the two equations of state coincide. At higher densities, when the
repulsive self-interaction comes into play, they differ quantitatively but
present nevertheless the same qualitative behaviour, both reaching a minimum
before increasing. This is the minimum requirement for the validity of our
study.

In this paper, we will
study in detail the simplified equation of state (\ref{s2}), corresponding to
the
complex SF potential (\ref{inst9}) involving an attractive $|\psi|^4$ term and a
repulsive 
$|\psi|^6$ term. This polynomial potential is interesting in its own right even
if it does
not exactly correspond to the potential of axions. A more precise treatment
of axion stars should consider the general equation of state (\ref{g3}),
associated with the effective potential (\ref{inst6}), as in Braaten {\it et
al.} \cite{braaten} and Eby {\it et al.} \cite{ebycollapse}. An even more
precise treatment should be based on the
instantonic potential \cite{pq,witten,vv} [see Eq. (\ref{inst1})], or on the
chiral potental \cite{vv,chvv}, as
in Levkov {\it et al.} \cite{tkachevprl} and Helfer {\it et al.} \cite{helfer}.
An interest of our model is that it leads to a  fully analytical study that 
allows us to identify characteristic mass, length and density scales of
fundamental importance. Furthermore, as we shall see, it gives results that
are in good agreement with the exact numerical results of Braaten {\it et al.}
\cite{braaten} based on the potential (\ref{inst6}).

\section{The maximum mass of dilute axion stars}
\label{sec_maxmass}

The mass-radius relation of a nonrelativistic self-gravitating  BEC/SF  at
$T=0$ described
by a purely $|\psi|^4$ potential
[see Eq. (\ref{inst7b})] has been obtained analytically (approximately) and
numerically (exactly) in \cite{prd1,prd2}. When
$a_s>0$, the short-range interaction between the bosons is repulsive. When
$a_s<0$, it is attractive. In this paper, we restrict ourselves to attractive
self-interactions.

\subsection{Maximum mass, minimum radius, and maximum average density}
\label{sec_maxmr}

When the self-interaction between the bosons is attractive ($a_s<0$), like in
the case of axions, an equilibrium state exists
only below a maximum mass \cite{prd1,prd2}:
\begin{eqnarray}
\label{maxmr1}
M_{\rm max}^{\rm exact}=1.012\frac{\hbar}{\sqrt{Gm|a_s|}}.
\end{eqnarray}
This is the maximum mass of stable dilute Newtonian axion stars. The radii of
the stable configurations satisfy $R_{99}\ge (R_{99}^*)^{\rm
exact}$ where 
\cite{prd1,prd2}:
\begin{eqnarray}
\label{maxmr2}
(R_{99}^*)^{\rm exact}=5.5\left (\frac{|a_s|\hbar^2}{Gm^3}\right )^{1/2}
\end{eqnarray}
is the radius corresponding to the maximum mass. This is the minimum radius
of stable dilute Newtonian axion stars. The
subscript $99$ means that $R_{99}$ is the radius containing $99\%$
of the mass (the density profile has not a compact support but extends to
infinity so the radius of the axion star is formally
infinite). The minimum radius is related to the maximum mass by the relation
\begin{eqnarray}
R_{99}^*=5.57\frac{\hbar^2}{GM_{\rm max}m^2}.
\end{eqnarray}
The
maximum average density of stable dilute Newtonian axion stars,
defined by
$\overline{\rho}_{\rm max}=3M_{\rm max}/4\pi (R_{99}^*)^3$, is
\begin{eqnarray}
\label{maxmr2b}
\overline{\rho}_{\rm max}^{\rm exact}=1.45\times
10^{-3}\frac{Gm^4}{a_s^2\hbar^2}.
\end{eqnarray}

For $M\ll
M_{\rm max}$ and $R_{99}\gg R^*_{99}$, the mass-radius relation is given by
\cite{membrado,prd1,prd2}:
\begin{eqnarray}
\label{maxmr2bb}
R_{99}^{\rm exact}=9.946\frac{\hbar^2}{GMm^2},
\end{eqnarray}
corresponding to the noninteracting limit. This branch is stable.

For $M\ll
M_{\rm max}$ and $R_{99}\ll R^*_{99}$, the mass-radius relation is given by
\cite{prd1,prd2}:
\begin{eqnarray}
\label{maxmr2cc}
R_{99}^{\rm exact}=3.64\frac{|a_s|}{m}M,
\end{eqnarray}
corresponding to the nongravitational limit. This branch is unstable. 

{\it Remark:} For future reference, we recall that when
the self-interaction between the bosons is repulsive ($a_s>0$), an
equilibrium state exists, in the TF approximation, at a unique radius
\cite{leekoh,goodman,arbey,bohmer,prd1}:
\begin{eqnarray}
\label{maxmr2pos}
R^{\rm exact}=\pi\left (\frac{a_s\hbar^2}{Gm^3}\right )^{1/2},
\end{eqnarray}
independent of the mass. We note that its scaling coincides with that of
$(R_{99}^*)^{\rm exact}$ in Eq. (\ref{maxmr2}).

\subsection{Alternative expressions}
\label{sec_alt}

Alternative expressions of the maximum mass of axion stars can
be given.
Introducing the
dimensionless self-interaction constant (see, e.g.,
Appendix A of \cite{bectcoll}):
\begin{eqnarray}
\label{lambda}
\lambda=\frac{8\pi a_s mc}{\hbar},
\end{eqnarray}
we get
 \cite{prd1,prd2}:
\begin{eqnarray}
\label{maxmr3}
M_{\rm max}^{\rm
exact}=5.073\frac{M_P}{\sqrt{|\lambda|}},
\end{eqnarray}
where $M_{\rm P}=(\hbar c/G)^{1/2}=2.18\times 10^{-5}\, {\rm
g}$ is the Planck mass.\footnote{This
scaling can be compared to the scaling $M_{\rm max}=0.384 M_P^3/m^2$ of the
maximum mass of
fermion stars \cite{ov}, to the scaling  $M_{\rm max}=0.633 M_P^2/m$ of
the maximum mass of noninteracting
boson stars \cite{kaup,rb}, and to the scaling $M_{\rm max}=0.0612
\sqrt{\lambda}
M_P^3/m^2$ of
the maximum mass of 
self-interacting boson stars \cite{colpi,tkachev,chavharko}. We emphasize,
however, that
the
maximum mass of dilute axion stars given by Eq.
(\ref{maxmr3}) is a Newtonian result contrary to the other limits that
come from general relativity. The usually very small mass-radius ratio
of dilute axion stars (see below) justifies {\it a posteriori} why a
Newtonian treatment is sufficient to describe them.} Similarly,
the minimum radius
can be written as \cite{prd1,prd2}:
\begin{eqnarray}
\label{maxmr4}
(R_{99}^*)^{\rm
exact}=1.1\sqrt{|\lambda|}\frac{M_P}{m}
\lambda_C ,
\end{eqnarray}
where  $\lambda_C=\hbar/mc$ is the Compton
wavelength of the bosons. The maximum average density is 
\begin{eqnarray}
\label{maxmr2c}
\overline{\rho}_{\rm max}^{\rm exact}=0.917\frac{1}{\lambda^2}\left
(\frac{m}{M_P}\right )^4\frac{M_P^2}{m\lambda_C^3}.
\end{eqnarray}

We can also express the maximum mass, minimum radius, and maximum average
density of axion stars in
terms of the axion decay constant [see Eq. (\ref{inst8})]:
\begin{equation}
\label{decay}
f=\left (\frac{\hbar c^3m}{32\pi |a_s|}\right )^{1/2}.
\end{equation}
We obtain
\begin{equation}
\label{maxmr5}
M_{\rm
max}^{\rm
exact}=10.15 \frac{f}{M_P c^2}
\frac{M_{\rm P}^2}{m},
\end{equation}
\begin{equation}
\label{maxmr6}
(R_{99}^*)^{\rm exact}=0.55
\frac{M_Pc^2}{f}\lambda_C,
\end{equation}
\begin{eqnarray}
\label{maxmr2d}
\overline{\rho}_{\rm max}^{\rm exact}=14.7 \left (\frac{f}{M_P c^2}\right
)^4\frac{M_P^2}{m\lambda_C^3}.
\end{eqnarray}

{\it Remark:} We note the relation $f=mc^2/2|\lambda|^{1/2}$. We also note that
the scalings from Eqs.
(\ref{maxmr3})-(\ref{maxmr2d}) artificially involve the speed of
light $c$ because of the definition of
$\lambda$ and $f$. Actually, $c$ should not appear in the equations since the
maximum mass, minimum radius, and maximum average
density of dilute axion stars are nonrelativistic results, as shown in the
scalings of Eqs. (\ref{maxmr1})-(\ref{maxmr2b}).

\subsection{Maximum scattering length}
\label{sec_maxsl}

Instead of expressing the maximum mass as a function of the scattering length,
we can express the maximum scattering length  of the bosons (in absolute value)
as a function
of the mass of the star. Self-gravitating BECs with an attractive
self-interaction
can be at equilibrium only if the scattering length of the bosons is less than a
maximum value \cite{prd1,prd2}:
\begin{eqnarray}
|a_s|_{\rm max}^{\rm exact}=1.024\frac{\hbar^2}{GmM^2}.
\label{maxsl1}
\end{eqnarray}
This corresponds to
\begin{eqnarray}
|\lambda|_{\rm max}^{\rm exact}=25.74\left (\frac{M_P}{M}\right )^2,
\label{maxsl2b}
\end{eqnarray}
\begin{eqnarray}
\frac{f_{\rm min}^{\rm exact}}{M_P c^2}=9.855\times 10^{-2} \frac{mM}{M_P^2}.
\label{maxsl2}
\end{eqnarray}

\subsection{Numerical applications}
\label{sec_na}

Let us make numerical applications for different types of axions. 

\subsubsection{QCD axions}
\label{sec_na1}

Considering QCD
axions with  $m=10^{-4}\,
{\rm eV}/c^2$ and
$a_s=-5.8\times 10^{-53}\, {\rm m}$ \cite{kc}, corresponding to
$\lambda=-7.39\times
10^{-49}$ and $f=5.82\times 10^{19}\, {\rm eV}=4.77\times 10^{-9} M_P c^2$, we
obtain $M_{\rm
max}^{\rm
exact}=6.46\times 10^{-14}\,
M_{\odot}=1.29\times 10^{17}\, {\rm kg}=2.16\times 10^{-8}\,
M_{\oplus}$, $(R_{99}^*)^{\rm exact}=3.26\times 10^{-4}\,
R_{\odot}=227\, {\rm km}=3.56\times 10^{-2}\, R_{\oplus}$, and
$\overline{\rho}_{\rm max}=2.62\times 10^3\, {\rm
g/m^3}$ (the maximum number of bosons is $N_{\rm max}=M_{\rm
max}/m=7.21\times 10^{56}$). These values correspond to the typical size of
asteroids. QCD axions cannot form DM halos of relevant mass
and size.
However, they
can form mini boson stars (mini axion stars or dark matter stars) of very low
mass
which are stable gravitationally bound BECs. They might play a role as DM
components (i.e. DM halos could be made of mini axion
stars interpreted as MACHOs) if they exist in the universe in
abundance.

{\it Remark:} For QCD axions, the product $m f\equiv(\Lambda_{\rm QCD}/c)^2$ of
the mass
and decay constant is fixed to the value $\Lambda_{\rm QCD}=7.6\times 10^7\,
{\rm
eV}$ \cite{kc}. Astrophysical and cosmological constraints restrict $f$ to
the interval $0.5\times 10^{18} {\rm eV}<f<7.6\times 10^{21}{\rm eV}$. The axion
mass lies therefore in the interval $7.7\times 10^{-7}\, {\rm
eV/c^2}<m<1.2\times 10^{-2}\, {\rm eV/c^2}$. According to Eq.  (\ref{inst8}), we
have
\begin{equation}
\label{na3}
\frac{|a_s|}{m^3}=\frac{\hbar c^3}{32\pi f^2m^2}=\frac{\hbar
c^7}{32\pi\Lambda_{\rm QCD}^4}.
\end{equation}
Since $mf$ is fixed for QCD axions, then $|a_s|/m^3$ is fixed. From Eq.
(\ref{maxmr2}),
this implies that the minimum radius $R_{99}^*$ is fixed to the value 
$(R_{99}^*)^{\rm exact}=0.55(M_P c^2/\Lambda_{\rm QCD})^2l_P=227\, {\rm km}$,
independently of
the
axion mass $m$ (here $l_P=(\hbar G/c^3)^{1/2}=1.62\times 10^{-35}\, {\rm m}$ is
the Planck
length). By contrast, the maximum mass $M_{\rm
max}=10.15(\Lambda_{\rm QCD}/M_Pc^2)^2M_P^3/m^2$ from Eq.
(\ref{maxmr1}) scales with the mass of the axion as $m^{-2}$.

\subsubsection{Ultralight axions}
\label{sec_na2}

Considering ULAs with $m=2.19\times 10^{-22}\, {\rm eV}/c^2$
and
$a_s=-1.11\times 10^{-62}\, {\rm fm}$ corresponding to
$\lambda=-3.10\times
10^{-91}$ and $f=1.97\times 10^{23}\, {\rm eV}=1.61\times 10^{-5} M_P c^2$
(see Appendix D of
\cite{abrilphas}), we obtain $M_{\rm
max}^{\rm exact}=10^{8}\,
M_{\odot}$, $(R_{99}^*)^{\rm exact}=1\, {\rm kpc}$, and
$\overline{\rho}_{\rm max}=1.62\times 10^{-18}\, {\rm g/m^3}$
(the maximum number of bosons is $N_{\rm max}=5.09\times 10^{95}$). These
values are typical of dwarf DM halos like Fornax, or typical of
 the solitonic core of large DM halos (see footnote 1). 

{\it Remark:} We note that a halo of mass $M=10^{8}\, M_{\odot}$ made of ULAs
with mass 
$m=2.19\times 10^{-22}\, {\rm eV}/c^2$ is stable if $a_s=0$ but becomes
unstable, and collapses, if
the scattering length of the axions is larger (in absolute value) than
$|a_s|_{\rm max}=1.11\times 10^{-62}\, {\rm fm}$
\cite{bectcoll}. This corresponds to $|\lambda|_{\rm max}=3.10\times
10^{-91}$ and $f_{\rm min}=1.97\times 10^{23}\, {\rm eV}=1.61\times 10^{-5} M_P
c^2$. Therefore $\lambda=0$ is
very different from $\lambda=-3.10\times
10^{-91}$ (!) The extraordinarily small value of  $|\lambda|_{\rm
max}=3.10\times
10^{-91}$ was stressed in our previous papers
\cite{prd2,bectcoll,abrilphas}. 

\subsubsection{White dwarfs and neutron stars}
\label{sec_wdns}

It is interesting to compare these results with other systems exhibiting a
maximum mass like white dwarfs \cite{chandra31} and neutron stars \cite{ov}. For
ideal $^4_2{\rm He}$ white dwarfs treated within the framework of general
relativity \cite{chandratooper}, we have $M_{\rm max}=1.39\, M_{\odot}$,
$R_{\rm min}=1.02\times
10^{3}\, {\rm km}$ and $(\rho_0)_{\rm max}=2.35\times 10^{16}\,
{\rm g/m^3}$. For ideal neutron stars \cite{ov}, we have $M_{\rm
max}=0.710\, M_{\odot}$, $R_{\rm min}=9.16\,
{\rm km}$ and 
$(\rho_0)_{\rm max}=3.54\times 10^{21}\,
{\rm g/m^3}$.

\subsection{Validity of the Newtonian treatment for dilute
axion stars}
\label{sec_val}

\begin{figure}
\begin{center}
\includegraphics[clip,scale=0.3]{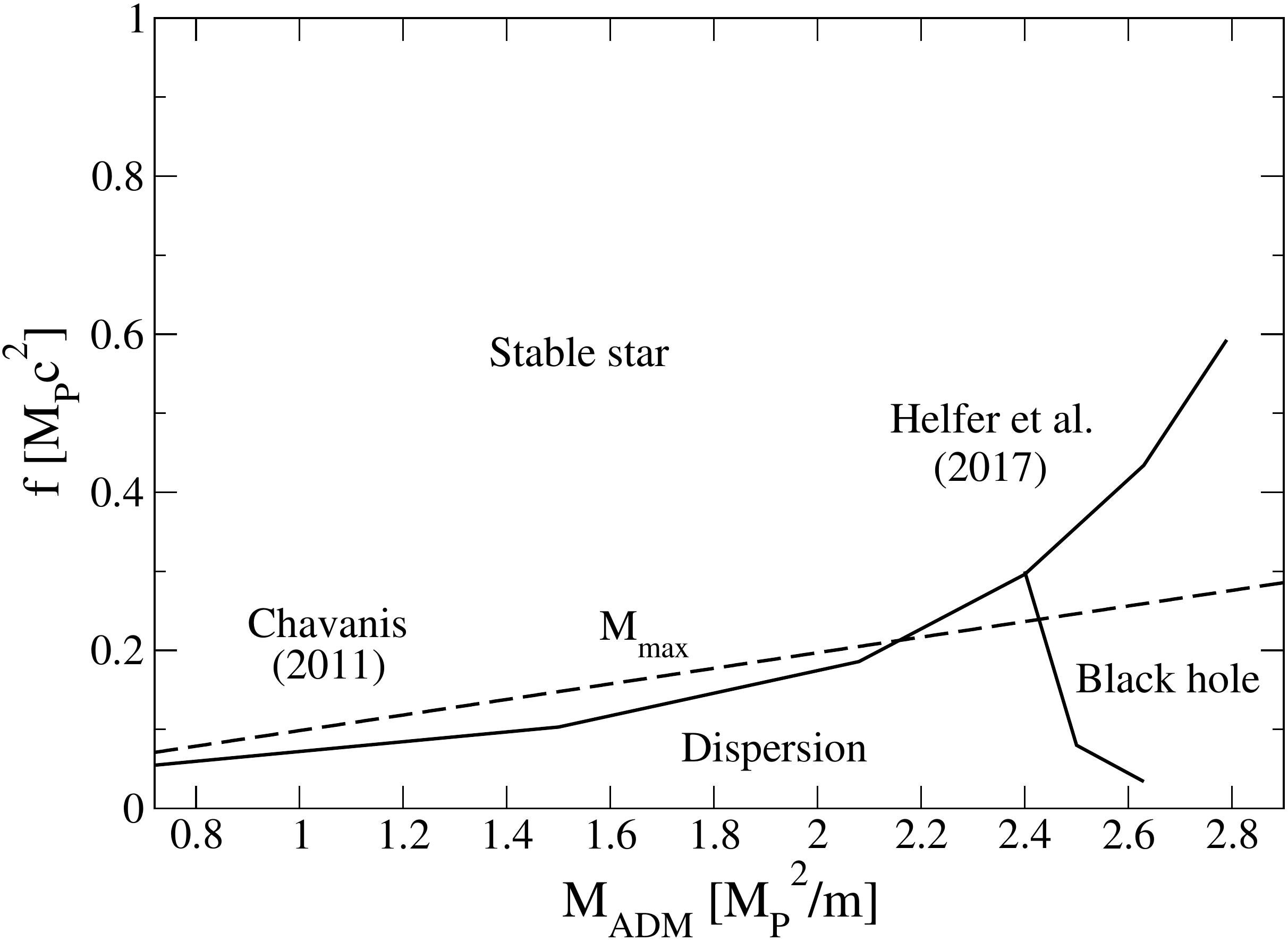}
\caption{Sketch of the phase diagram obtained by  Helfer {\it et al.}
\cite{helfer} by numerically solving the KGE equations with the
instantonic potential (\ref{inst1}). The dashed line corresponds to the
Newtonian maximum mass of dilute axion stars given by Eq.
(\ref{val1}) obtained in \cite{prd1}. It is valid for $f\ll M_P c^2$ (strong
self-interactions). We see
that it gives a fair agreement with the numerical results of \cite{helfer}
up to the
triple point at $(f,M)\sim (0.3 M_P c^2, 2.4 M_P^2/m)$.}
\label{marsh}
\end{center}
\end{figure}

The previous results are valid for Newtonian dilute axion stars. We note that
the maximum mass given by Eq. (\ref{maxmr1}) increases as the self-interaction
decreases. This result is expected to be valid for sufficiently small
masses, i.e., for sufficiently  strong self-interactions (in a sense made
precise below). When the mass becomes
large, i.e., for weak self-interactions, general relativity must be
taken
into account. Helfer {\it et al.} \cite{helfer} numerically solved the
general relativistic KGE equations with the instantonic potential 
(\ref{inst1})
and studied the formation of black holes from axion stars. They obtained
the phase diagram sketched in Fig. \ref{marsh}. For a sufficiently strong
self-interactions (small $f$), the maximum mass of dilute axion stars is given
by
Eq. (\ref{maxmr1}). It can be rewritten as
\begin{equation}
\label{val1}
\frac{M_{\rm
max}^{\rm
exact}}{M_{\rm P}^2/m}=10.15 \frac{f}{M_{\rm P}c^2}.
\end{equation}
This formula  gives a relatively good agreement with the numerical results of
Helfer {\it et al.} \cite{helfer} up to about $f\sim 0.25
M_P c^2$ and $M\sim 2.54 M_P^2/m$ (corresponding approximately to the position
of the triple point in their phase
diagram). For weaker
self-interactions (larger $f$),  the general relativistic results of
Helfer {\it et al.} \cite{helfer} start to substantially deviate from the
expression of the Newtonian maximum mass given by Eq. (\ref{val1}).

We can estimate the validity of our Newtonian description of
dilute axion stars as follows. The
Newtonian theory of gravitation is valid if the radius
$R$ of a  star of mass $M$ is much larger than its Schwarzschild radius
\begin{equation}
\label{sw}
R_S=\frac{2GM}{c^2}.
\end{equation}
Using Eq. (\ref{maxmr1}), the
Schwarzschild radius of a dilute axion star with the maximum mass can be written
as
\begin{equation}
\label{swb}
R_S^{\rm exact}=2.024\left (\frac{G\hbar^2}{|a_s|mc^4}\right )^{1/2}.
\end{equation}
We note the alternative expressions
\begin{equation}
\label{swc}
R_S^{\rm exact}=10.146\frac{l_P}{\sqrt{|\lambda|}}=20.30\frac{f}{M_P
c^2}\frac{M_P}{m}l_P.
\end{equation}
If axion stars with the
maximum mass can be treated with Newtonian gravity, then all the branch of
dilute axion stars can be treated with Newtonian gravity (see Appendix E of
\cite{bectcoll}). Therefore, the condition of validity of the Newtonian
treatment for dilute axion stars is
\begin{equation}
\label{val2}
\frac{R_{99}^*}{R_S}=\frac{R_{99}^*
c^2}{2GM_{\rm max}}\gg 1.
\end{equation}
Using Eqs. (\ref{maxmr1}) and (\ref{maxmr2}), we get
\begin{equation}
\label{val3}
\frac{R_{99}^*
c^2}{2GM_{\rm max}}\sim \frac{|a_s|c^2}{2Gm},
\end{equation}
where we have not written the prefactors since Eq. (\ref{val2}) just gives an
estimate of the validity of the Newtonian approximation. We note the exact
relations
\begin{equation}
\label{val4}
\frac{|a_s|c^2}{Gm}=\frac{2|a_s|}{r_S}=\frac{1}{32\pi}\left (\frac{M_P
c^2}{f}\right
)^2=\frac{|\lambda|}{8\pi}\left (\frac{M_P}{m}\right )^2,
\end{equation}
where
\begin{equation}
\label{val5}
r_S=\frac{2Gm}{c^2}
\end{equation}
is the effective Schwarzschild radius of a particle of mass $m$ (we
have introduced this quantity in \cite{abrilphas}). As a
result, the criterion (\ref{val2}) expressing the validity of the Newtonian
approach at the critical point
can be written in the equivalent forms
\begin{equation}
\label{val6}
|a_s|\gg r_S,\qquad f\ll M_Pc^2,\qquad |\lambda|\gg \left
(\frac{m}{M_P}\right )^2.
\end{equation}
The Newtonian approach is therefore valid for sufficiently strong
self-interactions. By contrast, when $|a_s|\sim r_S$, $f\sim M_Pc^2$ and
$|\lambda|\sim
({m}/{M_P})^2$, general
relativity must be taken into account. In that case, the maximum mass and
minimum radius of axion stars, obtained by substituting these equivalents into
Eqs. (\ref{maxmr1})-(\ref{maxmr2d}), become of the order of
\begin{equation}
\label{val7}
M_{\rm max}\sim \frac{M_P^2}{m},\qquad R_{*}\sim \lambda_C,
\end{equation}
corresponding to the Kaup scales (see Appendix
\ref{sec_cs}). A more detailed
general relativistic treatment is performed in  Sec. \ref{sec_qgr}.

Let us make some numerical applications to complete those reported in Sec.
\ref{sec_na}.

For QCD axions with $m=10^{-4}\,
{\rm eV}/c^2$ and
$a_s=-5.8\times 10^{-53}\, {\rm m}$, we find $R_S^{\rm exact}=1.91\times
10^{-10}\, {\rm m}$, $r_S=2.65\times 10^{-67}\, {\rm m}$, $M_P
c^2=1.22\times 10^{28}\, {\rm eV}$ and $(m/M_P)^2=6.71\times 10^{-65}$, implying
$(R_{99}^*/R_S)_{\rm exact}=1.19\times 10^{15}$.

For ULAs with $m=2.19\times 10^{-22}\,
{\rm eV}/c^2$ and
$a_s=-1.11\times 10^{-62}\, {\rm fm}$, we find $R_S^{\rm exact}=9.57\times
10^{-6}\, {\rm pc}$, $r_S=1.88\times 10^{-101}\, {\rm pc}$, $M_P
c^2=1.22\times 10^{28}\, {\rm eV}$ and $(m/M_P)^2=3.22\times 10^{-100}$,
implying
$(R_{99}^*/R_S)_{\rm exact}=1.04\times 10^{8}$.

In the following sections, we shall assume that the condition
$f\ll M_Pc^2$ 
(strong self-interaction) is
fulfilled so that the Newtonian treatment is valid for dilute axion stars and
for dense axion stars whose masses
are not too large. 
General
relativistic corrections for large
values of $f$ (weak self-interaction), or for large masses,
will
be obtained qualitatively in Sec. \ref{sec_qgr}. We
see from the previous numerical applications that QCD axions and
ULAs are in the strongly self-interacting regime. Therefore, the Newtonian
approximation generally provides an excellent description of axion stars made of
QCD axions and ULAs, except if they have very large masses.

\section{Mass-radius relation of dilute and dense axion stars within the
Gaussian ansatz}
\label{sec_mrdd}

When $M>M_{\rm max}$, dilute axion stars undergo gravitational collapse
\cite{prd1,prd2}. As
discussed in the Introduction, one possibility is that the collapse ends with
the formation of dense axion stars \cite{braaten}. In this section, we study
phase transitions between dilute and dense axion stars and establish their
mass-radius relation. We consider a simplified model corresponding to the
polynomial SF
potential  (\ref{inst9}) leading  to the equation of state (\ref{s2}). We
develop an
analytical approach based on a Gaussian ansatz. This is a  generalization of our
previous work \cite{prd1} for dilute axion stars. This ansatz usually
proves to be accurate to study the
equilibrium configurations of self-gravitating BECs \cite{prd1,prd2}. We shall
compare our analytical results with the exact numerical results obtained by  
Braaten {\it et al.} \cite{braaten} and find good agreement.

\subsection{The effective potential}
\label{sec_ep}

Using a Gaussian ansatz for the wave function \cite{prd1,bectcoll,chavtotal},
one
can show that the total energy of an axion star of mass $M$ and radius $R$
described by the equation of state (\ref{s2}) is given by 
\begin{eqnarray}
E_{\rm tot}=\frac{1}{2}\alpha
M\left (\frac{dR}{dt}\right
)^2+V(R)
\label{ep1}
\end{eqnarray}
with the effective potential
\begin{eqnarray}
V(R)=\sigma\frac{\hbar^2M}{m^2R^2}
-\nu\frac{GM^2}{R}+\zeta\frac{2\pi
a_s\hbar^2M^2}{m^3R^3}\nonumber\\
+\zeta\frac{32\pi^2\hbar^4a_s^2M^3}{9m^6c^2R^6}
.
\label{ep2}
\end{eqnarray}
The coefficients are $\alpha=3/2$, $\sigma=3/4$, $\zeta=1/(2\pi)^{3/2}$ and
$\nu=1/\sqrt{2\pi}$.\footnote{Other types of ansatz give the same results 
with slightly different values of the coefficients. For the sake of
generality, we will express our results in terms of $\alpha$, $\sigma$, $\zeta$
and $\nu$ so they can be applied to more general situations if necessary.} The
first term in Eq. (\ref{ep1}) is the classical
kinetic energy $\Theta_c$ and the second term is the
potential energy $V$. The potential energy includes the contribution of the
quantum
kinetic energy $\Theta_Q$ (or quantum potential), the gravitational energy $W$,
the internal energy $U_2$ due to the repulsive self-interaction (corresponding
to a polytropic equation of state of index $\gamma=2$ and polytropic
constant $K_2<0$), and the internal energy
$U_3$ due to the attractive self-interaction (corresponding to a polytropic
equation of state of index $\gamma=3$ and  polytropic
constant $K_3>0$). The general formalism needed to
obtain the
expression (\ref{ep2}) of the effective potential can be found in
\cite{chavtotal}. 
Writing $\dot E_{\rm tot}=0$ expressing the conservation of energy, we find
that the
dynamical equation satisfied by
the radius of the BEC is
\begin{eqnarray}
\alpha M \frac{d^2R}{dt^2}=-V'(R).
\label{ep3}
\end{eqnarray}
This equation is similar to the equation of motion of a fictive particle of mass
$\alpha
M$ moving in an effective potential $V(R)$.

\subsection{Maximum mass of dilute axion stars}

Within the Gaussian ansatz, the maximum mass and the minimum radius of
dilute axion stars are  \cite{prd1}:
\begin{eqnarray}
\label{cmmm1}
M_{\rm max}=\left
(\frac{\sigma^2}{6\pi\zeta\nu}\right
)^{1/2}\frac{\hbar}{\sqrt{Gm|a_s|}}=1.085\,  M_a,
\end{eqnarray}
\begin{eqnarray}
\label{cmmm2}
R_{*}=\left (\frac{6\pi \zeta}{\nu}\right )^{1/2}\left
(\frac{|a_s|\hbar^2}{Gm^3}\right )^{1/2} =1.732\, R_a,
\end{eqnarray}
where the scales $M_a$ and $R_a$ are defined in Appendix \ref{sec_cis}.
For a Gaussian density profile, the relation between the radius $R$ and the
radius $R_{99}$ containing $99\%$
of the mass is $R_{99}=2.38167 R$  \cite{prd1}.
The expressions (\ref{cmmm1}) and (\ref{cmmm2}) can be compared with the exact
results (\ref{maxmr1}) and (\ref{maxmr2}).

We define a density scale, a pressure scale, an energy scale
and a dynamical time scale by
\begin{eqnarray}
\label{cmmm3}
\rho_{0}=\frac{\sigma
\nu}{(6\pi\zeta)^2}\frac{Gm^4}{a_s^2\hbar^2}=0.209\, \rho_a,
\end{eqnarray}
\begin{eqnarray}
\label{cmmm4}
P_{0}=\frac{2\pi\sigma^2\nu^2}{(6\pi\zeta)^4}\frac{G^2m^5}{|a_s|^3\hbar^2}
=0.274 \, P_a,
\end{eqnarray}
\begin{eqnarray}
V_0=\frac{\sigma^2\nu^{1/2}}{(6\pi\zeta)^{3/2}}\frac{\hbar
m^{1/2}G^{1/2}}{|a_s|^{3/2}}=0.271 \, E_a,
\label{cmmm5}
\end{eqnarray}
\begin{eqnarray}
t_D=\frac{6\pi\zeta}{\nu}\left
(\frac{\alpha}{\sigma}\right
)^{1/2}\frac{|a_s|\hbar}{Gm^2}=4.24 \, t_a,
\label{cmmm6}
\end{eqnarray}
where the scales $\rho_a$, $P_a$, $E_a$ and $t_a$ are defined
in Appendix \ref{sec_cis}.
We note the
identities
\begin{eqnarray}
M_{\rm max}=\frac{\sigma}{\nu}\frac{\hbar^2}{Gm^2 R_*},\qquad
\rho_{0}=\frac{M_{\rm max}}{R_{*}^3},
\label{cmmm7}
\end{eqnarray}
\begin{eqnarray}
V_0=\nu \frac{GM_{\rm max}^2}{R_*},\qquad t_D=\left (\frac{\alpha}{\nu}\right
)^{1/2}\frac{1}{\sqrt{G\rho_0}}
\label{cmmm8}
\end{eqnarray}

Considering QCD axions with  $m=10^{-4}\,
{\rm eV}/c^2$ and
$a_s=-5.8\times 10^{-53}\, {\rm m}$, we obtain
$M_{\rm max}=6.92\times 10^{-14}\,
M_{\odot}=1.37\times 10^{17}\, {\rm kg}=2.31\times 10^{-8}\,
M_{\oplus}$, $N_{\rm max}=7.72\times 10^{56}$, $R_*=1.03\times
10^{-4}\,
R_{\odot}=71.5\, {\rm km}=1.12\times 10^{-2}\, R_{\oplus}
$, $\rho_{0}=3.76\times
10^{5}\, {\rm g/m^3}$, $V_0=7.06 \times 10^{25}\, {\rm ergs}$, and
$t_D=1.22\times 10^4\, {\rm s}=3.40\, {\rm hrs}$.

Considering ultralight axions (ULAs) with $m=2.19\times 10^{-22}\,
{\rm eV}/c^2$ and
$a_s=-1.11\times 10^{-62}\, {\rm fm}$, we obtain $M_{\rm
max}=1.07\times 10^{8}\,
M_{\odot}$, $N_{\rm max}=5.45\times 10^{95}$,
$R_*=0.313\, {\rm kpc}$, $\rho_{0}=2.36\times
10^{-16}\, {\rm g/m^3}$,  $V_0=1.25 \times 10^{54}\, {\rm ergs}$, and
$t_D=4.88\times 10^{14}\, {\rm s}=15.5\, {\rm
Myrs}$.

\subsection{Dimensionless variables}

We introduce the dimensionless variables
\begin{eqnarray}
\hat M=\frac{M}{M_{\rm max}},\qquad \hat R=\frac{R}{R_*},\qquad  \hat
\rho=\frac{\rho}{\rho_{0}},
\label{dep1}
\end{eqnarray}
\begin{eqnarray}
 \hat P
=\frac{P}{P_{0}},\qquad \hat
V=\frac{V}{V_0}, \qquad \hat t=\frac{t}{t_D},\qquad \hat\omega=\omega t_D.
\label{dep2}
\end{eqnarray}
We shall work with these
dimensionless variables but, from now on, we forget the ``hats'' in order to
simplify the notations.

\begin{figure}
\begin{center}
\includegraphics[clip,scale=0.3]{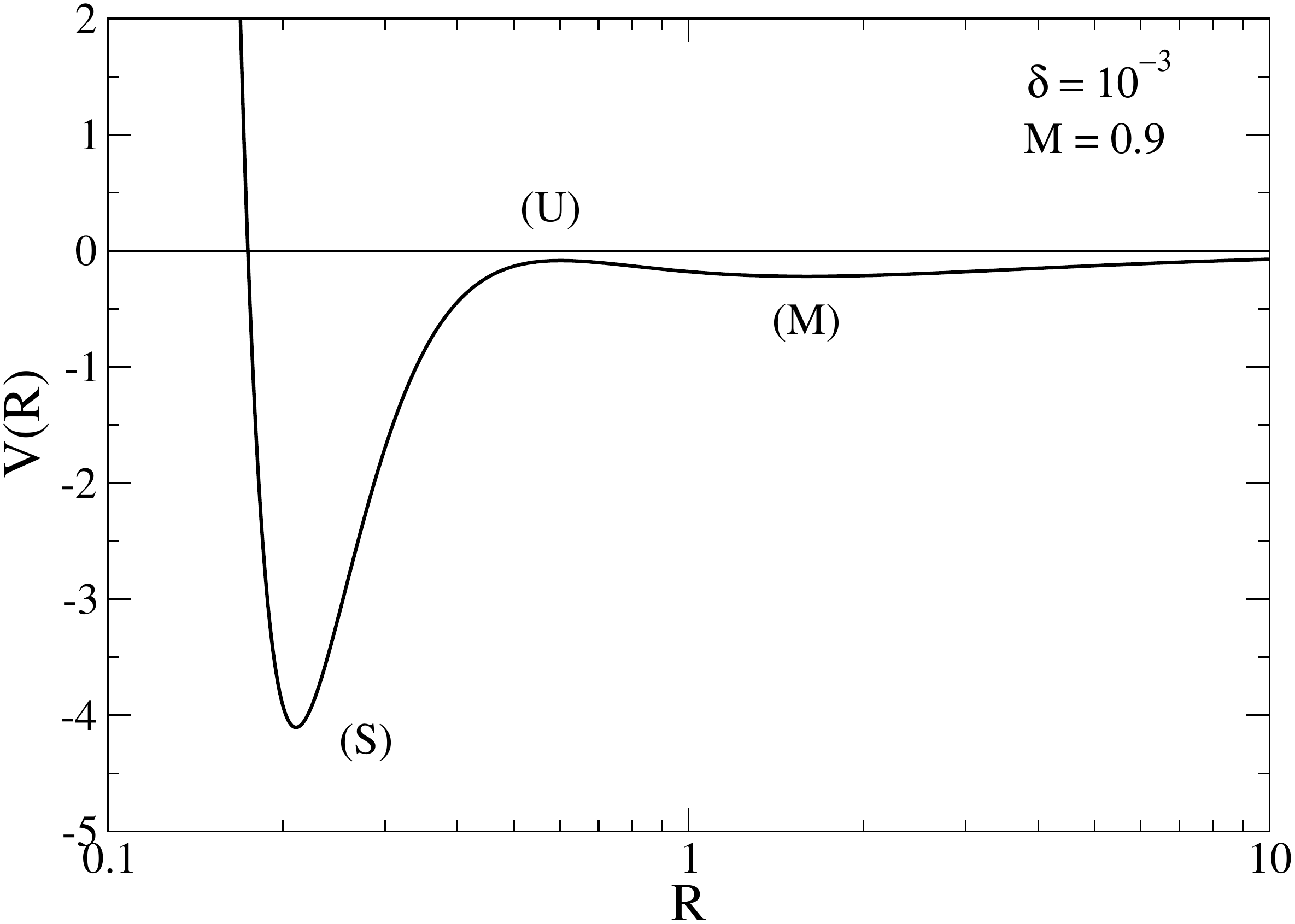}
\caption{Effective potential $V(R)$ as a function of the radius $R$
for $\delta=10^{-3}$ and
$M=0.9<M_{\rm max}=1$. It presents a local minimum (M) corresponding to a
metastable dilute axion star, a global minimum (S) corresponding to a fully
stable
dense axion star, and a local maximum (U) corresponding to an unstable axion
star.}
\label{rv}
\end{center}
\end{figure}

With these dimensionless variables, the average density and the total
energy write
\begin{equation}
\rho=\frac{3M}{4\pi R^3}
\label{dep5}
\end{equation} 
and 
\begin{eqnarray}
E_{\rm tot}=\frac{1}{2}M\left (\frac{dR}{dt}\right )^2+V(R).
\label{ep6}
\end{eqnarray}
The effective potential is given by 
\begin{equation}
\label{ep7}
V(R)=\frac{M}{R^2}-\frac{M^2}{R}-\frac{M^2}{3R^3}+\delta\frac{M^3}{R^6},
\end{equation}
where
\begin{equation}
\label{ep8}
\delta=\frac{16\pi\nu\sigma}{27(6\pi\zeta)^2}
\frac{Gm}{ |a_s|c^2}=0.389 \frac{Gm}{
|a_s|c^2 }
\end{equation}
is a dimensionless interaction parameter. For
illustration, the
effective
potential $V(R)$ is plotted in Fig. \ref{rv} for $\delta=10^{-3}$ and $M=0.9$.
The dynamical equation satisfied by the radius of the BEC is
\begin{eqnarray}
M \frac{d^2R}{dt^2}=-V'(R).
\label{ep9}
\end{eqnarray}

{\it Remark:} In terms of the dimensionless variables, the general equation of
state (\ref{g3}) writes
\begin{equation}
P=\frac{4}{729\delta^2}f(\sqrt{108\delta\rho})
\label{dep3}
\end{equation}
and the simplified equation of state (\ref{s2}) writes
\begin{equation}
P=-\rho^2+6\delta\rho^3.
\label{dep4}
\end{equation}

\subsection{The interaction parameter $\delta$}
\label{sec_ip}

Using Eq. (\ref{val4}), we can write the interaction parameter $\delta$ as
\begin{equation}
\label{ip1}
\delta=\frac{16\pi\nu\sigma}{27(6\pi\zeta)^2}\frac{r_S}{2|a_s|}=0.194\frac{r_S}{
|a_s|},
\end{equation}
\begin{equation}
\label{ip2}
\delta=\frac{16\pi\nu\sigma}{27(6\pi\zeta)^2} 32\pi\left (\frac{f}{M_P
c^2}\right
)^2=39.1 \left (\frac{f}{M_P
c^2}\right
)^2,
\end{equation}
\begin{equation}
\label{ip3}
\delta=\frac{16\pi\nu\sigma}{27(6\pi\zeta)^2}
\frac{8\pi}{|\lambda|}\left (\frac{m}{M_P}\right
)^2=9.77\frac{1}{|\lambda|}\left (\frac{m}{M_P}\right )^2.
\end{equation}
From the discussion of Sec. \ref{sec_val}, we note that the validity of the
Newtonian treatment at the critical point expressed by the
condition (\ref{val2}) can be written as
\begin{equation}
\delta\ll 1.
\end{equation}

Considering QCD axions with  $m=10^{-4}\,
{\rm eV}/c^2$ and
$a_s=-5.8\times 10^{-53}\, {\rm m}$, we obtain
 $\delta=8.88\times
10^{-16}$. 

Considering ultralight axions (ULAs) with $m=2.19\times 10^{-22}\,
{\rm eV}/c^2$ and
$a_s=-1.11\times 10^{-62}\, {\rm fm}$, we obtain 
$\delta=1.02\times
10^{-8}$.

These numerical applications confirm that the Newtonian approach is valid for
axion stars made of QCD
axions and ULAs, except if the mass $M$ is large.

\subsection{The mass-radius relation}
\label{sec_mr}

A stable equilibrium state corresponds to a minimum of the effective potential 
$V(R)$. Its first derivative is
\begin{equation}
\label{mr1}
V'(R)=-\frac{2M}{R^3}+\frac{M^2}{R^2}+\frac{M^2}{R^4}-6\delta\frac{M^3}{R^7}.
\end{equation}
The condition $V'(R)=0$ leads to the mass-radius relation
\begin{equation}
\label{mr2}
-\frac{2M}{R^3}+\frac{M^2}{R^2}+\frac{M^2}{R^4}-6\delta\frac{M^3}{R^7}=0,
\end{equation}
which can be rewritten as
\begin{equation}
\label{mr3}
6\delta M^2-R^3(1+R^2)M+2R^4=0.
\end{equation}
This is a second degree equation whose solutions are
\begin{equation}
\label{mr4}
M=\frac{R^3(1+R^2)\pm\sqrt{R^6(1+R^2)^2-48\delta R^4}}{12\delta}.
\end{equation}
The mass-radius relation $M(R)$ has two branches (+) and (-) determined by the
sign in front
of the square root.

In the general case, the equilibrium state
results from the balance between the gravitational attraction, the
repulsion due to the quantum potential arising from the Heisenberg uncertainty
principle, the attractive self-interaction and the repulsive self-interaction.
All these contributions are present in the mass-radius relation (\ref{mr4}).

\begin{figure}
\begin{center}
\includegraphics[clip,scale=0.3]{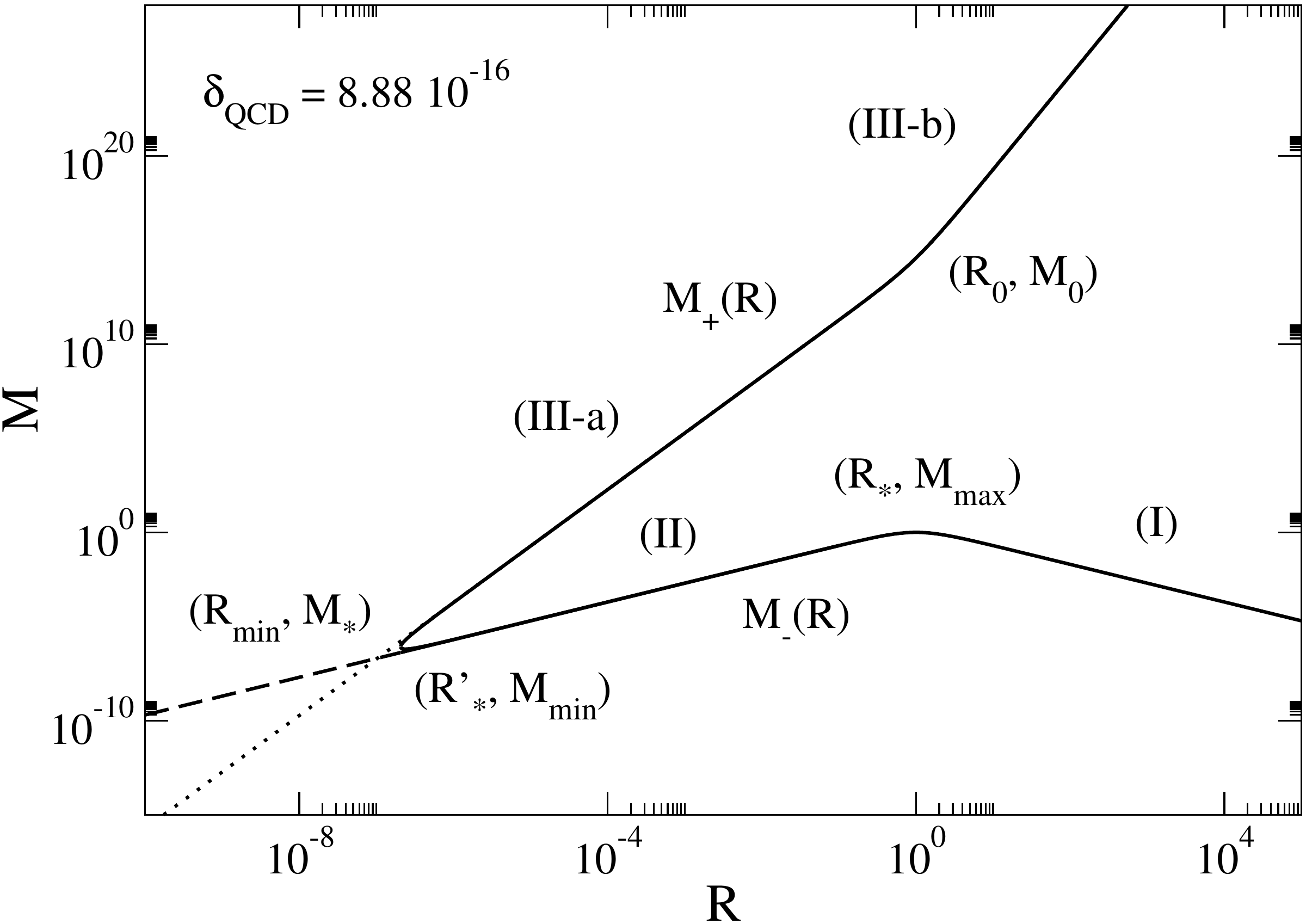}
\caption{Mass-radius relationship of QCD axion stars ($\delta=8.88\times
10^{-16}$). The dashed line corresponds to the mass-radius relation of
dilute axion stars with $\delta=0$ \cite{prd1}. The dotted line corresponds to
the
TF approximation of dense axion stars (see Sec. \ref{sec_tf}).
}
\label{mrQCD}
\end{center}
\end{figure}

\begin{figure}
\begin{center}
\includegraphics[clip,scale=0.3]{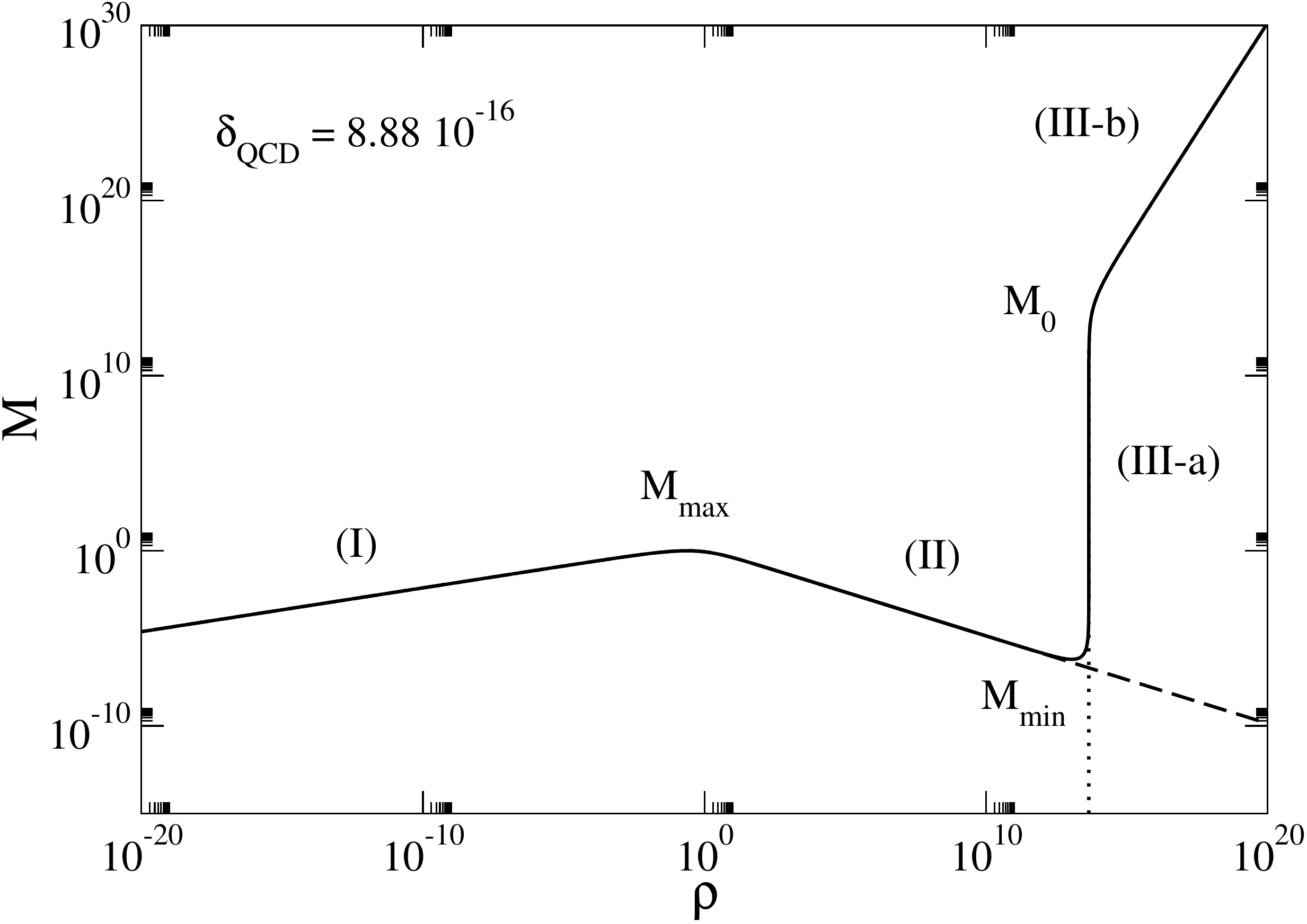}
\caption{Mass-density relationship of QCD axion stars ($\delta=8.88\times
10^{-16}$).}
\label{rhomass}
\end{center}
\end{figure}

The mass-radius relation (\ref{mr4}) is plotted in Fig. \ref{mrQCD} for
$\delta=8.88\times 10^{-16}$ corresponding to QCD axions. We have also plotted
the mass-density relation in Fig. \ref{rhomass}, where
$\rho$ represents the average density defined by Eq. (\ref{dep5}).
These curves present
different critical points that will be studied
specifically in the following sections. There is a local maximum mass $M_{\rm
max}(\delta)$ at a radius $R_*(\delta)$  and a  local minimum
mass $M_{\rm min}(\delta)$ at a radius $R'_*(\delta)$. There is
also a minimum radius  $R_{\rm min}(\delta)$ corresponding to a mass
$M_*(\delta)$.

The mass-radius relation $M(R)$ defines a series of equilibria (see Appendix
\ref{sec_p}). There can be
several solutions with the same mass $M$. A stable equilibrium state must be a
minimum of $V(R)$ so it must satisfy $V''(R)>0$. Computing the second derivative
of $V(R)$ from Eq. (\ref{ep7}), we get 
\begin{equation}
\label{mr7}
V''(R)=\frac{6M}{R^4}-\frac{2M^2}{R^3}-\frac{4M^2}{R^5}+42\delta\frac{M^3}{R^8}.
\end{equation}
We can distinguish three
branches in the curve $R(M)$ delimited by the maximum mass $M_{\rm
max}(\delta)$ and the minimum
mass $M_{\rm min}(\delta)$. As will be shown below (see Secs. \ref{sec_turning}
and \ref{sec_sapt}),
the branch (I) corresponds to stable dilute axion stars, the branch (III)
corresponds to stable dense axion stars, and the intermediate
branch (II)
corresponds to unstable axion stars. The change of stability in the series of
equilibria takes place at the maximum and minimum masses $M_{\rm max}(\delta)$ 
and  $M_{\rm min}(\delta)$ in agreement with the
Poincar\'e theory of 
linear series of equilibria \cite{poincare}, the $M(R)$ theorem of Wheeler
introduced in the physics of compact objects (white dwarfs and neutron stars)
\cite{htww}, and the catastrophe (or bifurcation) theory \cite{catastrophe} (see
Appendix \ref{sec_p}).\footnote{To be complete, we also
quote the  necessary Vakhitov-Kolokolov condition of stability $dM/d\rho>0$
\cite{zakharov,vk} mentioned in \cite{tkachevprl}. Applied to the mass-density
relationship of Fig. \ref{rhomass} this criterion indicates that the branch 
(II) is unstable.}

{\it Remark:} In the nongravitational limit, corresponding to
$R\ll 1$, the mass-radius relation (\ref{mr4}) can be
simplified by neglecting $R^2$ in front of $1$.  This approximation describes
the branches (II) and (III-a) for $\delta\ll 1$.  However, since this
approximation does not significantly simplify the equations, we shall not
consider this limit specifically.

\subsection{The mass-radius relation of dilute axion stars}
\label{sec_mrd}

When $\delta=0$, we recover the mass-radius relation of
dilute axion stars \cite{prd1}:
\begin{equation}
\label{mrd1}
M=\frac{2R}{1+R^2}\qquad {\rm or}\qquad R=\frac{1\pm\sqrt{1-M^2}}{M}.
\end{equation}
This relation also provides a good approximation of the branches (I) and (II) of
the general mass-radius relation
(\ref{mr4}) for $\delta\ll 1$ (the approximation being better and better as
$\delta\rightarrow 0$).\footnote{More precisely, when
$\delta\rightarrow 0$, the branches (I) and (II) tend towards the mass-radius
relation of dilute axion stars ($\delta=0$) while the branch (III-a) is
pushed towards the vertical axis $R=0$ and the branch (III-b) is rejected at
infinity. Therefore, the general mass-radius relation (\ref{mr4}) tends
towards the mass-radius relation of dilute axion stars (\ref{mrd1}) except for
small radii (see Appendix \ref{sec_dpbh} for a more detailed description of the
differences between the limit $\delta\rightarrow 0$ and the nonrepulsive case
$\delta=0$).} This
amounts to neglecting the self-repulsion
(nonrepulsive limit). The relation (\ref{mrd1}) describes the equilibrium
between the
self-gravity, the quantum potential, and the attractive self-interaction. This
curve exhibits a maximum mass $M_{\rm max}=1$ at $R_{*}=1$ with a maximum
average density $\rho_{\rm max}=3/4\pi$.

For $R\gg 1$, we obtain  
\begin{equation}
\label{mrd3}
M\sim \frac{2}{R}.
\end{equation}
This is also the asymptotic behavior of the general mass-radius relation 
(\ref{mr4}) for
$R\rightarrow +\infty$ and $M\rightarrow 0$. In that limit, the self-interaction
is negligible and Eq. (\ref{mrd3}) describes a stable equilibrium between the
self-gravity and the quantum potential. 
This corresponds to the noninteracting limit. The density is given by  $\rho\sim
3/2\pi
R^4$ so  the mass-density relation is $M\sim
2(2\pi/3)^{1/4}\rho^{1/4}$. Since  $\rho\rightarrow 0$ for $R\rightarrow
+\infty$, this branch corresponds to dilute axion
stars. Coming back to
dimensional variables, we get
\begin{eqnarray}
M=\frac{2\sigma}{\nu}\frac{\hbar^2}{GRm^2}=3.76 \frac{\hbar^2}{GRm^2}.
\end{eqnarray}
This relation can be compared with the exact result (\ref{maxmr2bb}).

For $R\ll 1$, we obtain  
\begin{equation}
\label{mrd4}
M\sim 2R.
\end{equation}
In that limit, the self-gravity is negligible and Eq. (\ref{mrd4}) describes an
unstable equilibrium between the quantum potential and the attractive
self-interaction. This corresponds to the
nongravitational limit. The density is given by  $\rho\sim
3/(2\pi
R^2)$ so  the mass-density relation is $M\sim
2(3/2\pi\rho)^{1/2}$. Since  $\rho\rightarrow +\infty$ for $R\rightarrow 0$,
this branch corresponds to unstable dense axion
stars. Coming back to
dimensional variables, we get
\begin{eqnarray}
M=\frac{2\sigma}{6\pi\zeta}\frac{mR}{|a_s|}=1.25\frac{mR}{|a_s|}.
\end{eqnarray}
This relation can be compared with the exact result (\ref{maxmr2cc}).

\subsection{The minimum radius}
\label{sec_mir}

\begin{figure}
\begin{center}
\includegraphics[clip,scale=0.3]{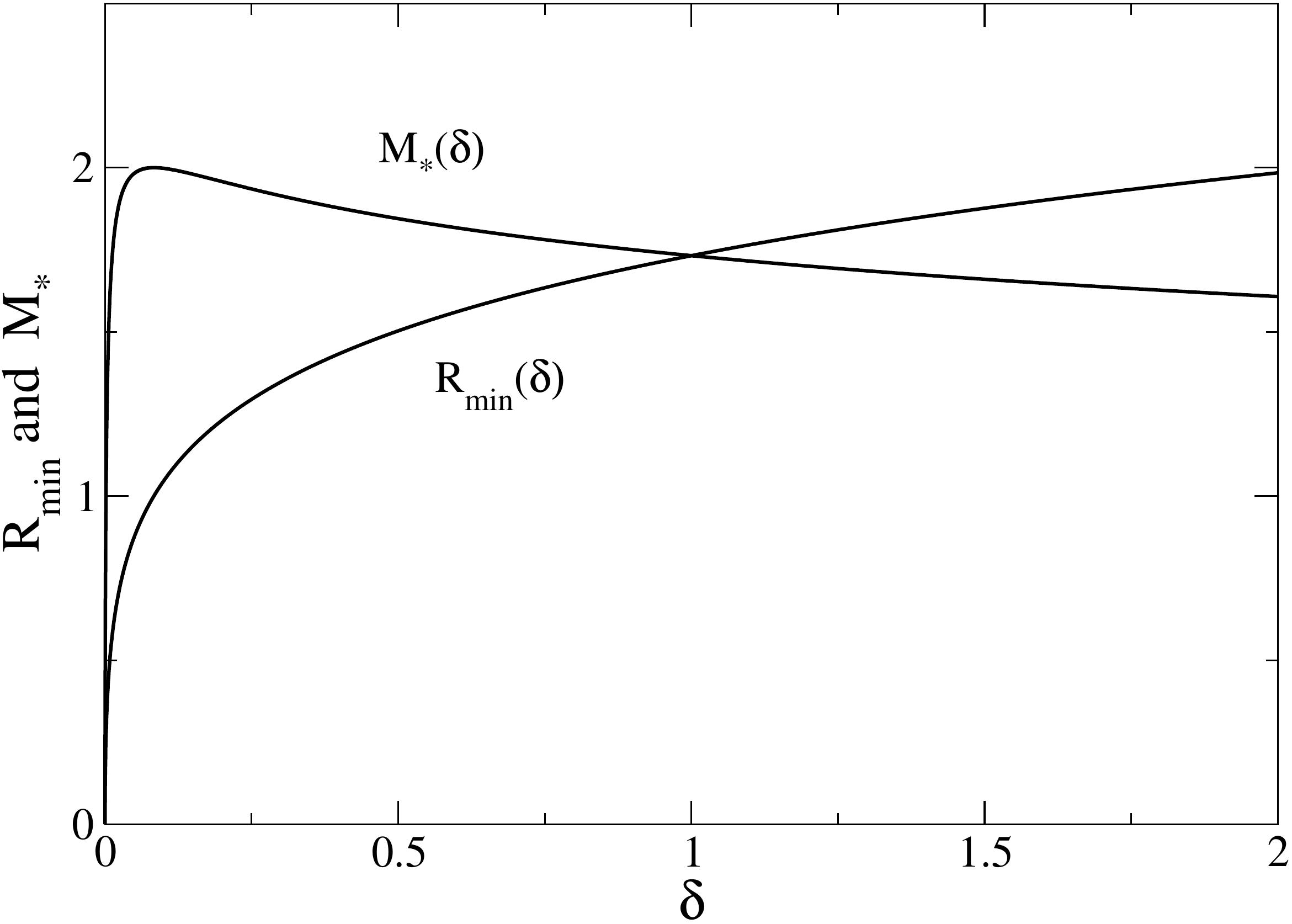}
\caption{The minimum radius $R_{\rm min}(\delta)$ and the corresponding mass
$M_*(\delta)$ as a function of $\delta$.
We note, parenthetically, that these curves cross each other
at  $\delta=1$ (at that point $R_{\rm min}=M_*=\sqrt{3}$).}
\label{deltaRmin}
\end{center}
\end{figure}

\begin{figure}
\begin{center}
\includegraphics[clip,scale=0.3]{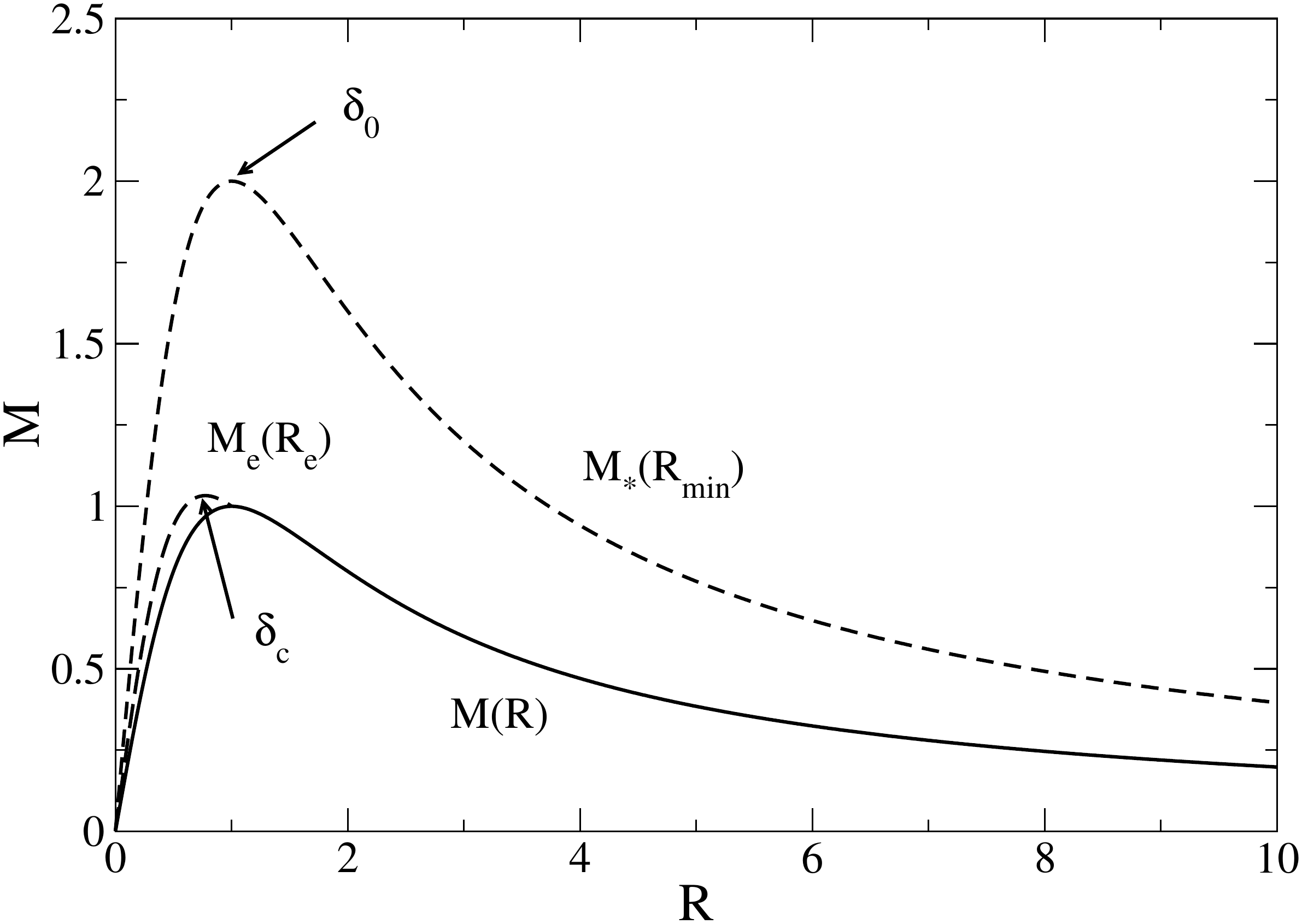}
\caption{The curve $M_*(R_{\rm min})$ (short-dashed) and the curve $M_e(R_{e})$
(long-dashed).  For
comparison, we have also plotted the mass-radius relation (\ref{mrd1}) of dilute
axion stars with $\delta=0$ in full lines.}
\label{RMskelette}
\end{center}
\end{figure}

Equilibrium states only exist above a minimum radius $R_{\rm min}(\delta)$
where the branches (+) and (-) merge. The minimum radius $R_{\rm min}(\delta)$,
corresponding to the vanishing of the discriminant of Eq. (\ref{mr3}), is
determined by the equation 
\begin{equation}
\label{mir1}
\delta=\frac{R_{\rm min}^2}{48}(1+R_{\rm min}^2)^2.
\end{equation}
This is a third degree equation
whose physical solution is 
\begin{eqnarray}
\label{mir2}
R_{\rm min}=\frac{1}{18^{1/3}}\left
(9\sqrt{48\delta}+\sqrt{12+3888\delta}\right
)^{1/3}\nonumber\\
-\left (\frac{2}{3}\right )^{1/3}\left
(9\sqrt{48\delta}+\sqrt{12+3888\delta}\right
)^{-1/3}.
\end{eqnarray}
The mass $M_*(\delta)$ at the minimum radius is determined by the equation
\begin{equation}
\label{mir3}
M_*=\frac{4R_{\rm min}}{1+R_{\rm min}^2},
\end{equation}
obtained by substituting for $\delta$ from Eq. (\ref{mir1}) into
Eq. (\ref{mr4}).
For $\delta\rightarrow 0$:
\begin{equation}
\label{mir4}
R_{\rm min}\sim 4\sqrt{3\delta},\quad M_*\sim
16\sqrt{3\delta},\quad M_*\sim 4R_{\rm min}.
\end{equation}
For $\delta\rightarrow +\infty$:
\begin{equation}
\label{mir5}
R_{\rm min}\sim 2^{2/3}(3\delta)^{1/6},\quad M_*\sim
\frac{2^{4/3}}{(3\delta)^{1/6}},\quad M_*\sim \frac{4}{R_{\rm min}}.
\end{equation}
The functions $R_{\rm min}(\delta)$ and $M_*(\delta)$ are plotted in Fig.
\ref{deltaRmin}. 
The mass $M_*(\delta)$ reaches a maximum
$M_*=2$ at $\delta=\delta_0=1/12=0.083333...$. At that point, $R_{\rm min}=1$.
The function $M_*(R_{\rm min})$ given by Eq. (\ref{mir3}) is plotted in Fig.
\ref{RMskelette}.

Coming back to dimensional variables, we get for $|a_s|\gg
r_S$:
\begin{equation}
\label{mir6}
R_{\rm min}\sim 16\sqrt{3}\left (\frac{\pi}{27}\right )^{1/2}\left
(\frac{\sigma}{6\pi\zeta}\right )^{1/2} \frac{\hbar}{mc}\sim 7.48\, R_i,
\end{equation}
\begin{equation}
\label{mir7}
M_{\rm *}\sim 64\sqrt{3}\left (\frac{\pi}{27}\right )^{1/2}\left
(\frac{\sigma}{6\pi\zeta}\right )^{3/2}\frac{\hbar}{|a_s|c}\sim
18.7 M_i,
\end{equation}
\begin{equation}
\label{mir8}
\rho\sim \frac{27}{256\pi^2}\frac{m^3c^2}{|a_s|\hbar^2}\sim
1.07\times 10^{-2}\rho_i,
\end{equation}
where the scales $M_i$, $R_i$ and $\rho_i$  are defined in Appendix
\ref{sec_ngsi}. We note that the gravitational constant $G$ does not appear
in these scales since they are obtained in the nongravitational limit $R\ll
1$ (see the Remark at the end of Sec. \ref{sec_mr}).

For QCD axions with   $\delta=8.88\times
10^{-16}$, we obtain $R_{\rm min}=2.06\times 10^{-7}$, $M_*=8.26\times
10^{-7}$, and $\rho\sim 1/(16\pi\delta)=2.26\times 10^{13}$ (in terms of
dimensional variables, this
corresponds to $R_{\rm
min}=1.47\, {\rm cm}$, $M_*=5.72\times
10^{-20}\, {\rm M_{\odot}}$, and $\rho=8.48\times 10^{18}\, {\rm g/m^3}$).

For ULAs with $\delta=1.02\times
10^{-8}$, we obtain $R_{\rm min}=7.00\times 10^{-4}$, $M_*=2.80\times
10^{-3}$, and $\rho\sim 1/(16\pi\delta)=1.95\times 10^{6}$ (in terms of
dimensional variables, this
corresponds to $R_{\rm
min}=0.219\, {\rm pc}$, $M_*=3.00\times
10^{5}\, {\rm M_{\odot}}$, and $\rho=4.60\times 10^{-10}\, {\rm g/m^3}$
).

\subsection{The maximum and minimum masses}
\label{sec_mmm}

The maximum mass $M_{\rm max}(\delta)$ corresponding to the radius
$R_{*}(\delta)$ and the minimum mass $M_{\rm min}(\delta)$ corresponding to the
radius $R_{*}'(\delta)$ are located on the branch $(-)$. They are
determined by the condition $M'(R_{e})=0$. Taking the derivative of Eq.
(\ref{mr4}) with respect to $R$, and setting it equal to zero, we obtain after
simplification the equation
\begin{equation}
\label{mmm1}
\delta=\frac{R_{e}^2}{192}(3+2R_{e}^2-5R_{e}^4).
\end{equation}
This equation determines $R_{*}(\delta)$ and $R_{*}'(\delta)$. Substituting for
$\delta$ from Eq. (\ref{mmm1}) into Eq. (\ref{mr4}) we obtain after some
algebra  
\begin{equation}
\label{mmm2}
M_{e}=\frac{8R_{e}}{3+5R_{e}^2}.
\end{equation}
This equation can be obtained more rapidly by differentiating Eq.
(\ref{mr3})
and setting $dM=0$. Equation (\ref{mmm2}) determines $M_{\rm max}(R_*)$ and
$M_{\rm min}(R'_*)$. Equations (\ref{mmm1}) and (\ref{mmm2}) determine $M_{\rm
max}(\delta)$ and $M_{\rm min}(\delta)$ in parametric form with parameter
$R_e$.

For $\delta\rightarrow 0$:
\begin{equation}
\label{mmm3}
R'_*\sim 8\sqrt{\delta},\qquad M_{\rm min}\sim
\frac{64}{3}\sqrt{\delta},\qquad M_{\rm min}\sim \frac{8}{3}R'_*,
\end{equation}
\begin{equation}
\label{mmm4}
R_*\simeq  1-12\delta,\qquad M_{\rm max}\simeq 1+3\delta.
\end{equation}

\begin{figure}
\begin{center}
\includegraphics[clip,scale=0.3]{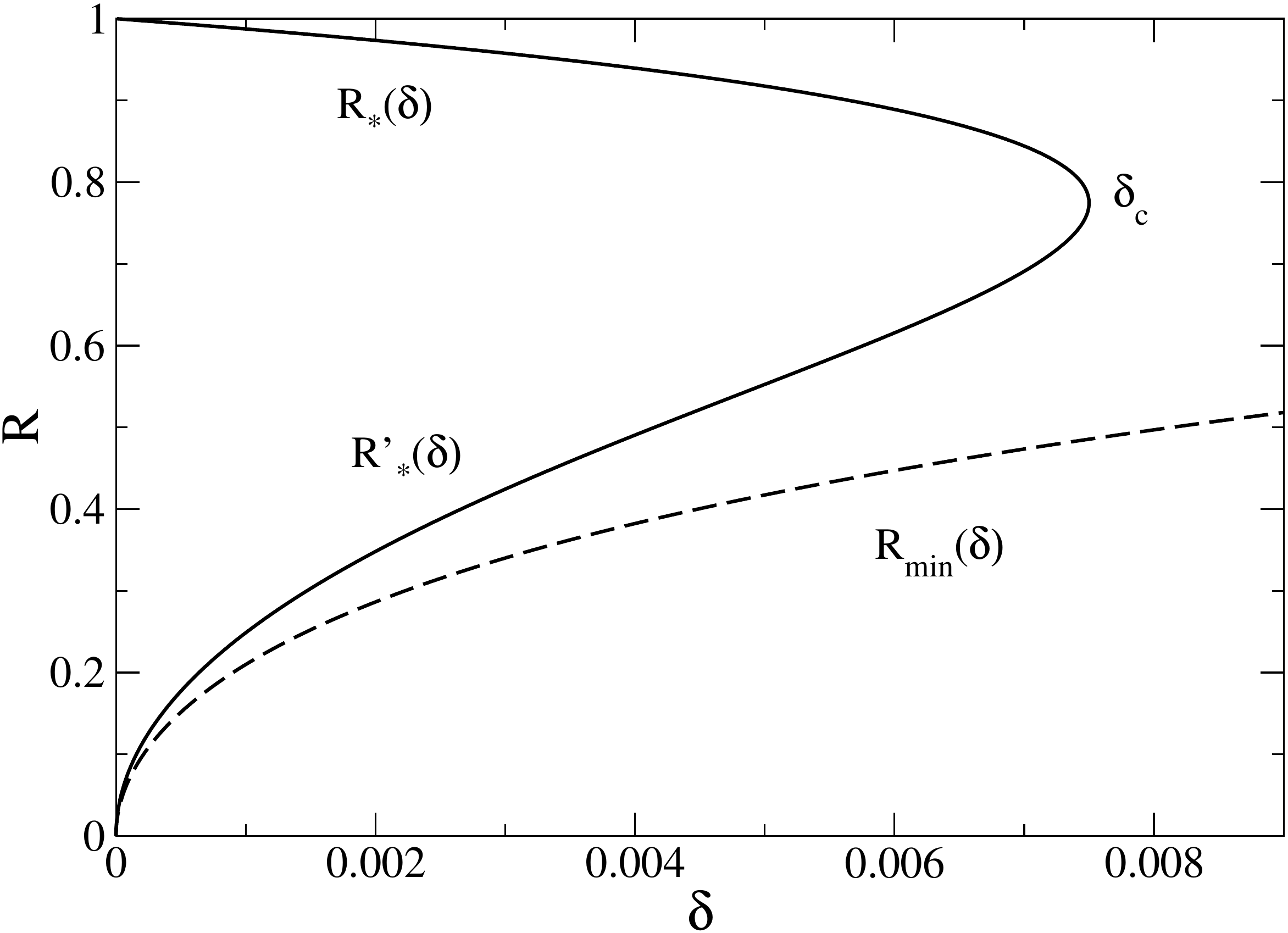}
\caption{The radius $R_*(\delta)$ corresponding to the
maximum mass $M_{\rm max}(\delta)$ and the radius $R'_{*}(\delta)$ corresponding
to the minimum mass $M_{\rm min}(\delta)$ as a function of
$\delta$. They coincide at the critical point $\delta_c$ (see Sec.
\ref{sec_cp}). We have also plotted the minimum radius $R_{\rm
min}(\delta)$ for comparison.}
\label{deltaRextrema}
\end{center}
\end{figure}

\begin{figure}
\begin{center}
\includegraphics[clip,scale=0.3]{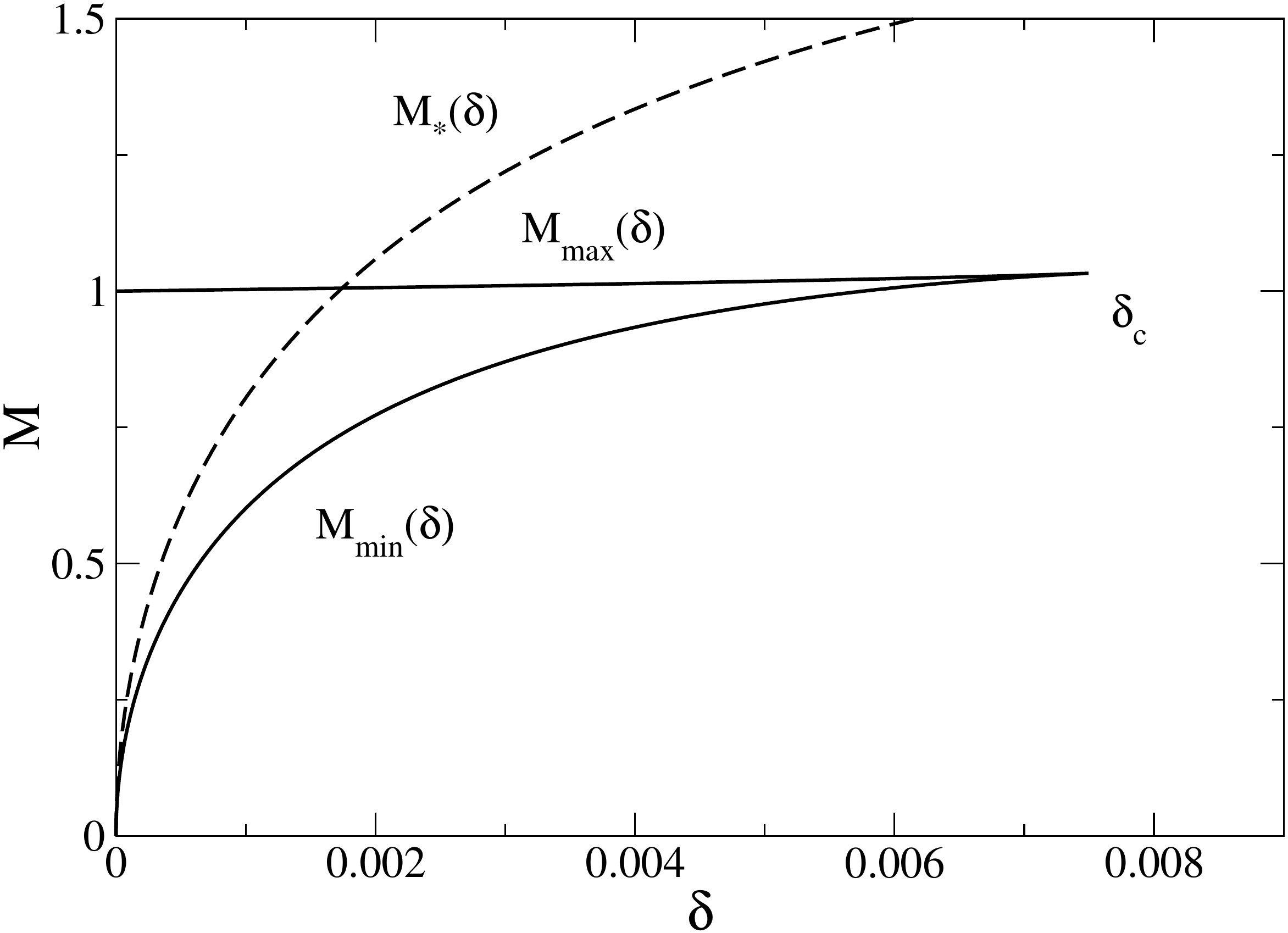}
\caption{The maximum mass $M_{\rm max}(\delta)$ and the minimum mass $M_{\rm
min}(\delta)$ as a function of $\delta$.  They coincide at the critical point
$\delta_c$ (see Sec.
\ref{sec_cp}). We have also plotted the mass 
$M_{*}(\delta)$ corresponding to the minimum radius for comparison (we note,
parenthetically, that $M_{*}=M_{\rm max}$ for $\delta=1.74\times 10^{-3}$ and
$M=1.005$). Since
the functions $\delta(R_e)$ and $M(R_e)$ are maximum at the critical point (see
Sec. \ref{sec_cp}) the curve $M_e(\delta)$ presents a spike at
$\delta=\delta_c$.}
\label{deltaMextrema}
\end{center}
\end{figure}

The functions $R_*(\delta)$ and $R'_*(\delta)$  are plotted in
Fig.
\ref{deltaRextrema}. The functions  $M_{\rm max}(\delta)$  and 
$M_{\rm min}(\delta)$ are plotted in Fig.
\ref{deltaMextrema}. The functions $M_{\rm max}(R_*)$ and  
$M_{\rm
min}(R'_*)$, corresponding to the right and left
branches  of the curve $M_e(R_e)$ defined by Eq. (\ref{mmm2}), 
are plotted in Fig.
\ref{RMskelette}.

Coming back to dimensional variables, we get for $|a_s|\gg r_S$:
\begin{equation}
\label{mmm5}
M_{\rm min}\sim \frac{256}{3}\left (\frac{\pi}{27}\right )^{1/2}\left
(\frac{\sigma}{6\pi\zeta}\right )^{3/2}\frac{\hbar}{|a_s|c}\sim
14.4 M_i,
\end{equation}
\begin{equation}
\label{mmm6}
R'_{*}\sim 32\left (\frac{\pi}{27}\right )^{1/2}\left
(\frac{\sigma}{6\pi\zeta}\right )^{1/2} \frac{\hbar}{mc}\sim 8.64\, R_i,
\end{equation}
\begin{equation}
\rho\sim \frac{27}{512\pi^2}\frac{m^3c^2}{|a_s|\hbar^2}\sim
5.34\times 10^{-3}\rho_i.
\end{equation}

For QCD axions with   $\delta=8.88\times
10^{-16}$, we obtain $R_{*}=1$, $M_{\rm max}=1$, $\rho=3/(4\pi)=0.239$,
$R'_{*}=2.38\times 10^{-7}$, $M_{\rm
min}=6.36\times
10^{-7}$ and $\rho\sim 1/(32\pi\delta)=1.12\times 10^{13}$ (in terms of
dimensional variables, this corresponds to $R_*=71.5\, {\rm
km}$,  $M_{\rm max}=6.92\times 10^{-14}\, M_{\odot}$, $\rho=8.98\times 10^4\,
{\rm g/m^3}$, $R'_{*}=1.70\, {\rm cm}$, $M_{\rm
min}=4.40\times
10^{-20}\, {\rm M_{\odot}}$ and $\rho=4.21\times 10^{18}\, {\rm g/m^3}$).

For ULAs with $\delta=1.02\times
10^{-8}$, we obtain $R_{*}=1$, $M_{\rm max}=1$,
$\rho=3/(4\pi)=0.239$, $R'_{*}=8.08\times 10^{-4}$, $M_{\rm min}=2.15\times
10^{-3}$ and  $\rho\sim 1/(32\pi\delta)=9.75\times 10^5$ (in terms of
dimensional variables, this corresponds to $R_*=0.313\, {\rm kpc}$, $M_{\rm
max}=1.07\times 10^8\, M_{\odot}$,  $\rho=5.63\times 10^{-17}\,
{\rm g/m^3}$, $R'_{*}=0.253\, {\rm pc}$, $M_{\rm min}=2.30\times
10^{5}\, {\rm M_{\odot}}$ and $\rho=2.30\times 10^{-10}\, {\rm g/m^3}$).

{\it Remark:} The exact result of Braaten {\it et al.} \cite{braaten} for the
radius of the dense axion star with the minimum mass is
$R'_{*}=9.2(\hbar/mc)$, which is very close to our
estimate $R'_{*}=8.64(\hbar/mc)$ from Eq. (\ref{mmm6}). Other
comparisons between our approximate results and the exact results of Braaten
{\it et al.}
\cite{braaten} are made in the caption of Fig. \ref{axionsqcd}.

\subsection{The critical point}
\label{sec_cp}

\begin{figure}
\begin{center}
\includegraphics[clip,scale=0.3]{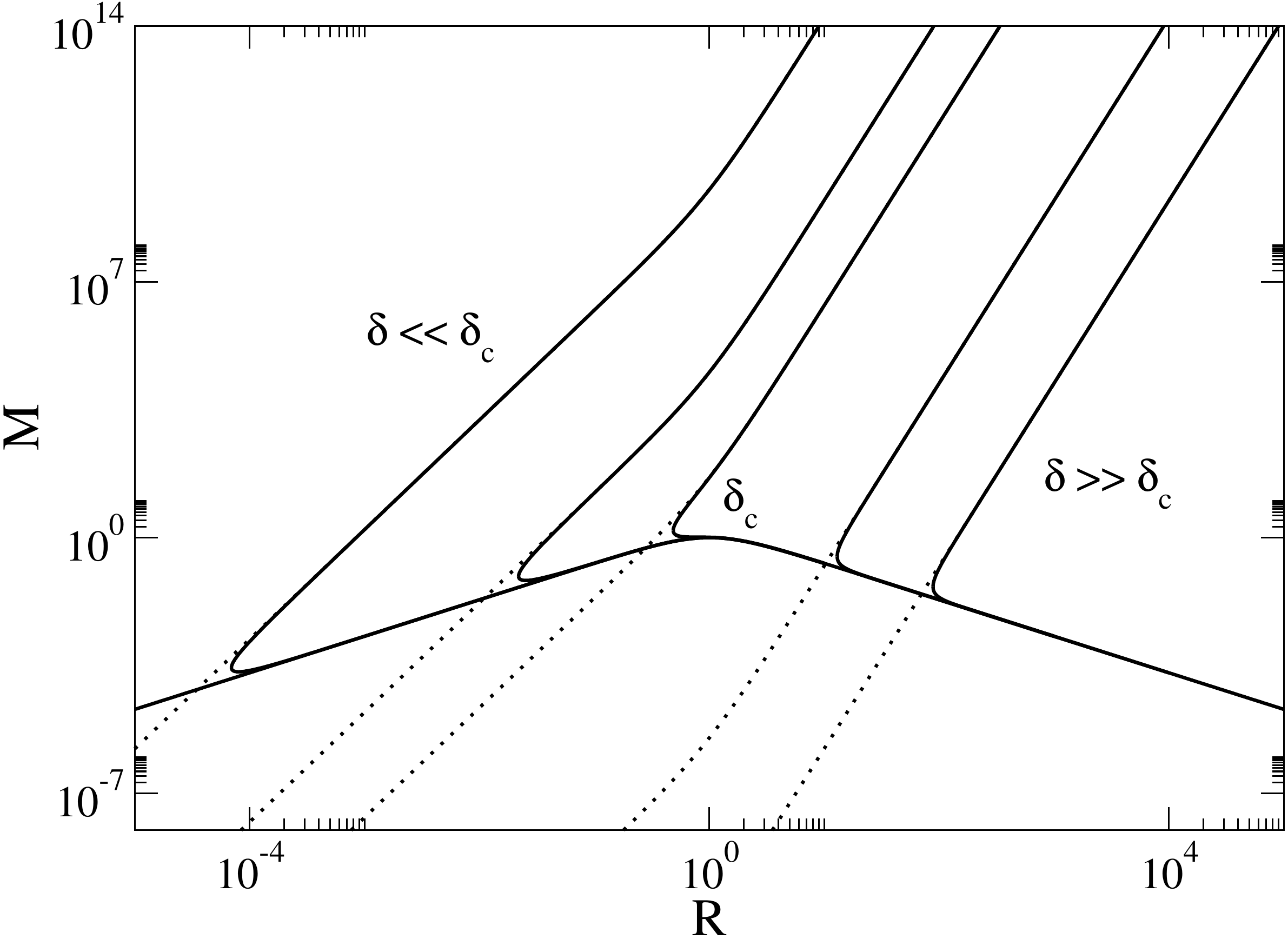}
\caption{Mass-radius relationship for different values of the 
interaction parameter ($\delta=10^{-10}, 10^{-5}, \delta_c,
10^5, 10^{10}$). }
\label{mr}
\end{center}
\end{figure}

There is a critical point when the maximum mass and the minimum mass coincide.
This corresponds to the maximum of the function $M_e(R_e)$ or, equivalently,  to
the maximum of the function $\delta(R_e)$. We find
\begin{equation}
\label{cp1}
\delta_c=\frac{3}{400}=0.0075.
\end{equation}
For this critical  value of $\delta$, the mass-radius relation $M(R)$ presents
an inflexion point at
\begin{equation}
\label{cp2}
R_c=\sqrt{\frac{3}{5}}=0.774597...\qquad M_c=\frac{4}{\sqrt{15}}=1.0327955...
\end{equation}
We also have
\begin{equation}
\label{cp3}
(R_{\rm min})_c=0.485537...,\qquad (M_*)_c=1.57164...
\end{equation}
The mass-radius relation is plotted in Fig. \ref{mr} for different values of
$\delta$ including the critical value $\delta_c$. The curve $R(M)$ is
multivalued for $\delta<\delta_c$ and univalued for $\delta>\delta_c$.

Using Eqs. (\ref{ip1})-(\ref{ip3}), this critical point
corresponds to
\begin{equation}
\label{gq1}
|a_s|_c=\frac{3200\pi\nu\sigma}{81(6\pi\zeta)^2}r_S=25.9\, r_S,
\end{equation}
\begin{equation}
\label{gq2}
f_c=\frac{9(6\pi\zeta)}{320\pi\sqrt{2\nu\sigma}}M_Pc^2=1.38\times 10^{-2}\,
M_Pc^2,
\end{equation}
\begin{equation}
\label{gq3}
|\lambda|_c=\frac{51200\pi^2\nu\sigma}{81(6\pi\zeta)^2}\left(\frac{m}{M_P}\right
)^2=1.30\times 10^{3} \left
(\frac{m}{M_P}\right
)^2.
\end{equation}

We will see that a first order phase transition between dilute
axion stars and dense axion stars occurs for $\delta<\delta_c$. This
condition can be written as  $|a_s|>|a_s|_c$, $f<f_c$, or
$|\lambda|>|\lambda|_c$. We will comment on the typical values of 
$|a_s|_c$, $f_c$ and $|\lambda|_c$ in Sec. \ref{sec_qgr} where we take
general relativity into account and introduce
similar quantities  $|a_s|_*$, $f_*$ and $|\lambda|_*$ separating the strongly
self-interacting regime from the weakly self-interacting regime.

\section{The TF approximation for dense axion stars}
\label{sec_tf}

In the TF approximation where the quantum potential energy (first
term in Eq. (\ref{ep7})) can be neglected, the mass-radius relation reduces to
\begin{equation}
\label{tf1}
M=\frac{R^3(1+R^2)}{6\delta}.
\end{equation}
The curve $M(R)$ is monotonically increasing. It is plotted in dotted line in
Figs. \ref{mrQCD} and \ref{mr}. This curve almost coincides with the branch
(III) of stable dense axion stars. Therefore, dense axion stars are very
well-described by the TF
approximation. It is only close to the minimum mass $M_{\rm min}$ that the
effect of the quantum potential is important. Two regimes can be evidenced.

\subsection{Nongravitational limit}

For $R\ll 1$, we obtain  
\begin{equation}
\label{tf2}
M\sim\frac{R^3}{6\delta}.
\end{equation}
In that limit, self-gravity is negligible and Eq. (\ref{tf2}) describes the
equilibrium between the attractive
self-interaction and the repulsive self-interaction (see Appendix
\ref{sec_tfng}). This corresponds to the nongravitational limit defining the
branch (III-a) in
Fig. \ref{mrQCD}. In that limit, the axion
stars
all have the same density
(independent of $M$ or $R$) given by
\begin{equation}
\label{tf2b}
\rho_{\rm dense}=\frac{1}{8\pi\delta}.
\end{equation}
The constancy of the density is clearly visible in Fig. \ref{rhomass}.
Since $\rho\gg 1$ for $\delta\ll 1$, the axion stars on the branch (III-a) are
dense. Coming back to dimensional variables, we
get
\begin{equation}
\label{tf2c}
M\sim \frac{9}{32\pi}\frac{m^3c^2}{|a_s|\hbar^2}R^3=8.95\times
10^{-2}\frac{m^3c^2}{|a_s|\hbar^2}R^3
\end{equation}
and
\begin{equation}
\rho_{\rm dense}=\frac{27}{128\pi^2}\frac{m^3c^2}{|a_s|\hbar^2}=2.14\times
10^{-2}\rho_i.
\end{equation}
These relations can be compared
with the exact results from Eqs. (\ref{tfng3rho}) and (\ref{tfng5}).

For QCD axions with   $\delta=8.88\times
10^{-16}$, we obtain $\rho_{\rm dense}=4.48\times 10^{13}$ (in terms of
dimensional variables, this corresponds to
$\rho_{\rm dense}=1.68\times 10^{19}\, {\rm g/m^3}$).

For ULAs with $\delta=1.02\times
10^{-8}$, we obtain  $\rho_{\rm dense}=3.90\times 10^6$ (in terms of
dimensional variables, this corresponds to
$\rho_{\rm dense}=9.21\times 10^{-10}\, {\rm g/m^3}$).

{\it Remark:} If we
consider the intersection between the nongravitational limit of dilute
axion stars ($M\sim 2R$ for $R\ll
1$) and the  nongravitational limit of dense
axion stars ($M\sim R^3/6\delta$ for $R\ll
1$), we obtain a point with radius $R\sim (12\delta)^{1/2}$ and mass
$M\sim
2(12\delta)^{1/2}$ whose scaling with $\delta$ agrees with $(R_{\rm
min},M_*)$ and $(R'_*,M_{\rm min})$ for $\delta\rightarrow 0$.

\subsection{Nonattractive limit}

For $R\gg 1$, we obtain
\begin{equation}
\label{tf3}
M\sim\frac{R^5}{6\delta}.
\end{equation}
This is also the asymptotic behavior of the general mass-radius relation
(\ref{mr4}) for 
$R\rightarrow +\infty$ and $M\rightarrow +\infty$. In that limit, the attractive
self-interaction is negligible and Eq. (\ref{tf3}) describes the equilibrium
between the self-gravity and the repulsive
self-interaction.
This corresponds to the nonattractive limit defining the branch (III-b) in
Fig. \ref{mrQCD}. In
that limit, the axion stars are equivalent to polytropes of index $n=1/2$ (see
Appendix \ref{sec_pol}). The density is given by $\rho\sim R^2/8\pi\delta$ so 
the mass-density relation is $M\sim (1/6\delta)(8\pi\delta\rho)^{5/2}$. Since 
$\rho\gg 1$ for $\delta\ll 1$ and/or $R\gg 1$, the axion stars on
the branch (III-b) are dense. Coming back to dimensional variables, we get
\begin{equation}
\label{gw1}
M\sim \frac{9}{32\pi}\frac{\nu}{6\pi\zeta}\frac{G
m^6c^2}{a_s^2\hbar^4}R^5=2.98\times 10^{-2}\frac{G
m^6c^2}{a_s^2\hbar^4}R^5.
\end{equation}
This relation can be compared with the exact result (\ref{pol3}) for a
polytrope $n=1/2$.

{\it Remark:} If we
consider the intersection between the noninteracting limit of
dilute
axion stars ($M\sim 2/R$ for $R\gg
1$) and the  nonattractive limit of dense
axion stars ($M\sim R^5/6\delta$ for $R\gg
1$), we obtain a point with radius $R\sim (12\delta)^{1/6}$ and mass
$M\sim
2/(12\delta)^{1/6}$ whose scaling with $\delta$ agrees with $(R_{\rm
min},M_*)$ for $\delta\rightarrow +\infty$.

\subsection{Transition between the
nongravitational and nonattractive TF regimes}
\label{sec_tran}

Dense axion stars are well-described by the mass-radius relation of Eq.
(\ref{tf1}) obtained in the TF approximation. We have evidenced two
regimes: a regime (III-a), corresponding to $R\ll 1$, where self-gravity is
negligible [see Eq. (\ref{tf2})] and a regime (III-b), corresponding to $R\gg
1$, where the attractive self-interaction is negligible [see Eq. (\ref{tf3})]. 
Qualitatively, the transition between these two
regimes
occurs at
\begin{equation}
\label{tran1}
R_0=1,\qquad M_0=\frac{1}{6\delta}.
\end{equation}
Coming back to dimensional variables, we get
\begin{equation}
\label{tran2}
R_0=\left (\frac{6\pi\zeta}{\nu}\right )^{1/2}\left
(\frac{|a_s|\hbar^2}{Gm^3}\right )^{1/2}=1.73 R_r=R_*,
\end{equation}
\begin{equation}
\label{tran3}
M_0=\frac{9}{32\pi}\left (\frac{6\pi\zeta}{\nu}\right
)^{3/2}\left
(\frac{|a_s|\hbar^2 c^4}{G^3m^3}\right )^{1/2}=0.465 M_r,
\end{equation}
where the scales $R_r$ and $M_r$ are defined in Appendix \ref{sec_is}. These
relations can be compared with the exact results from Eqs. (\ref{ext1}) and
(\ref{ext2}).

The regime (III-a) exists provided that $R_0>R_{\rm min}$, implying
$\delta<\delta_0$ with
\begin{equation}
\delta_0=\frac{1}{12}.
\end{equation}
At that point, $M_0=M_{\rm *}=2$ (see Sec. \ref{sec_mir}).
When $\delta<\delta_0$ and $M<M_0$, dense axion stars are in the
regime (III-a) where  self-gravity is negligible. In that case,
a dense axion
star of
mass $M$ has a radius 
\begin{equation}
\label{ray1}
R\sim (6M\delta)^{1/3}.
\end{equation}
When $\delta<\delta_0$ and $M>M_0$, or when  $\delta>\delta_0$, dense axion
stars are in the
regime (III-b) where  the attractive self-interaction is negligible. In that
case, a
dense axion star of mass $M$ has a radius 
\begin{equation}
\label{ray2}
R\sim (6M\delta)^{1/5}.
\end{equation}
We
will see in Sec. \ref{sec_fri}
that the solutions (III-b) are unphysical, or at most only marginally physical,
because general relativistic effects become important precisely when $M\sim
M_0$. However, for mathematical completeness, we will continue to discuss these
solutions in
the Newtonian framework, but keep this limitation in mind.

For QCD axions with   $\delta=8.88\times
10^{-16}$, we obtain $R_0=1$, $M_0=1.88\times 10^{14}$ and
$\rho_{\rm dense}=1/(8\pi\delta)=4.48\times 10^{13}$ (in terms of
dimensional variables, this corresponds to $R_0=71.5\, {\rm km}$,
$M_0=13.0 M_{\odot}$, and $\rho_{\rm dense}=1.68\times 10^{19}\, {\rm g/m^3}$).

For ULAs with $\delta=1.02\times
10^{-8}$, we obtain $R_0=1$, $M_0=1.63\times 10^7$  and
$\rho_{\rm dense}=1/(8\pi\delta)=3.90\times 10^{6}$ (in terms of
dimensional variables, this corresponds to $R_0=0.313\, {\rm kpc}$,
$M_0=1.75\times 10^{15}
M_{\odot}$, and $\rho_{\rm dense}=9.21\times 10^{-10}\, {\rm g/m^3}$).

\section{Energy and pulsation}
\label{sec_eap}

\subsection{The total energy}
\label{sec_te}

\begin{figure}
\begin{center}
\includegraphics[clip,scale=0.3]{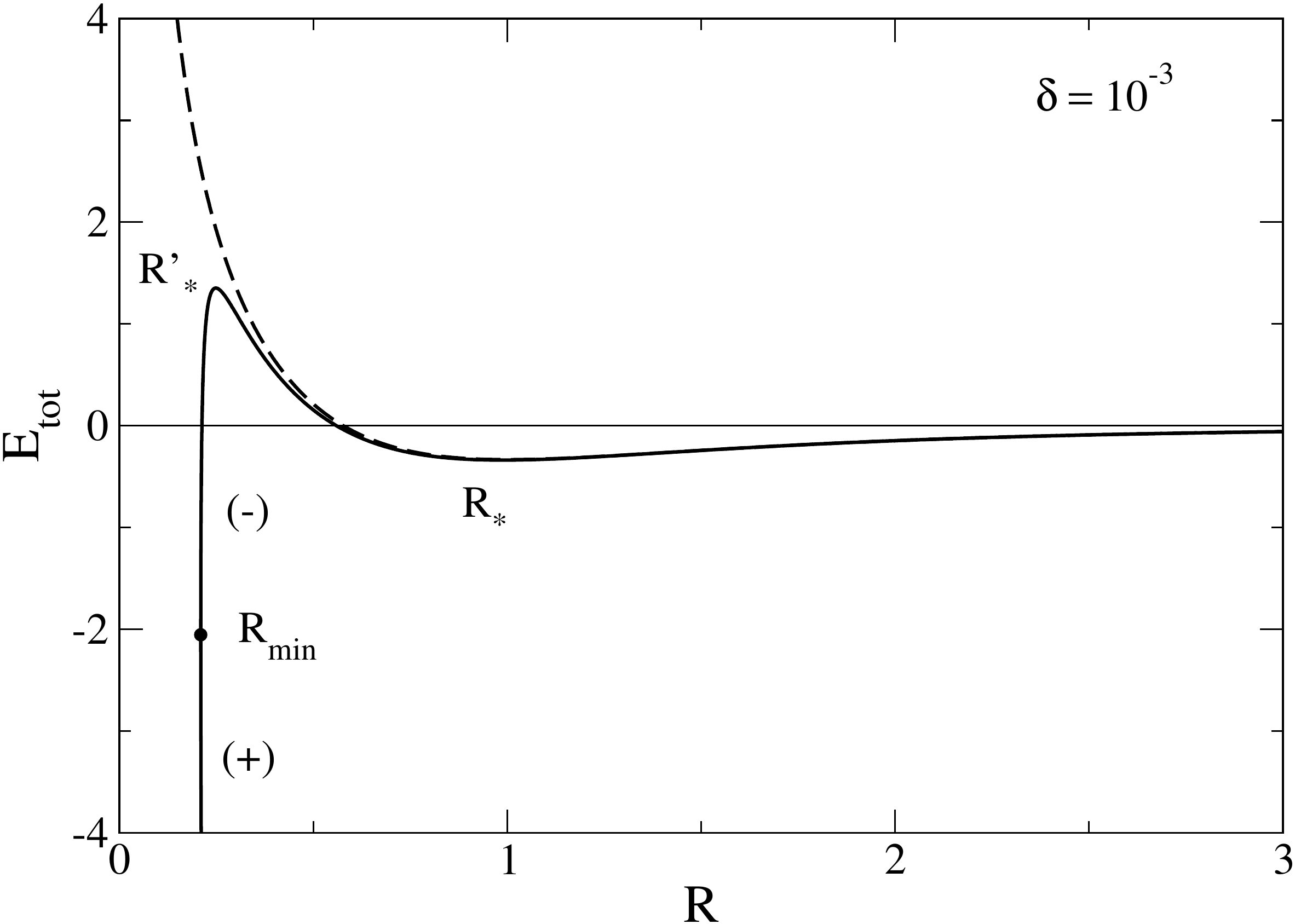}
\caption{Total energy $E_{\rm tot}$ as a function of the radius $R$ for 
axion stars with $\delta=10^{-3}$ (for illustration, we have taken a relatively
large
value of $\delta$ to facilitate the reading of the figure). The dashed line
corresponds to dilute axion
stars with $\delta=0$.}
\label{retot}
\end{center}
\end{figure}

\begin{figure}
\begin{center}
\includegraphics[clip,scale=0.3]{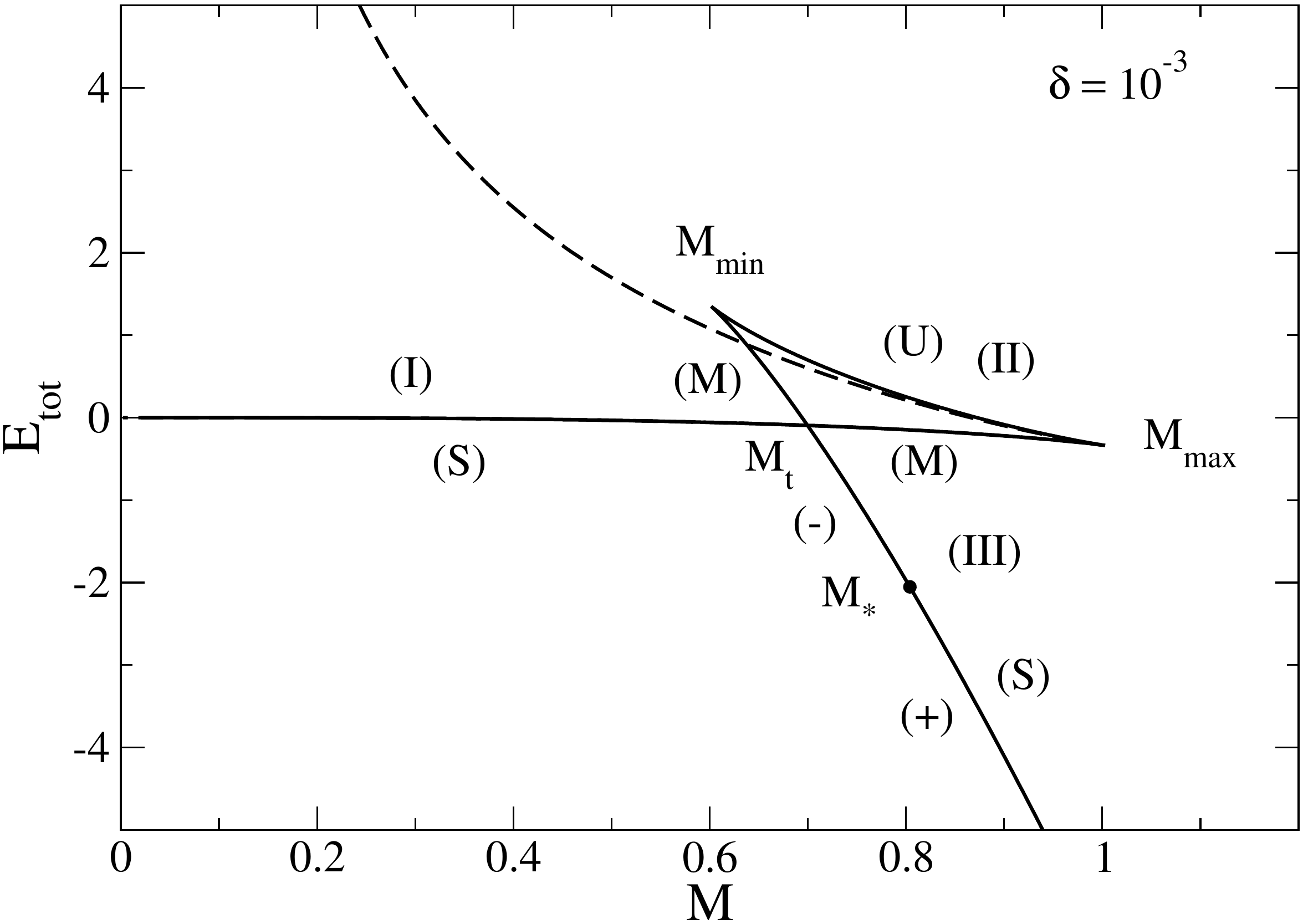}
\caption{Total energy $E_{\rm tot}$ as a function of the
mass $M$ for 
axion stars with $\delta=10^{-3}$. The dashed line corresponds to dilute axion
stars with $\delta=0$.}
\label{metot}
\end{center}
\end{figure}

At equilibrium ($\dot R=0$), the total energy of an axion star 
is given by
$E_{\rm tot}=V(R)$. Therefore
\begin{equation}
\label{te1}
E_{\rm tot}=\frac{M}{R^2}-\frac{M^2}{R}-\frac{M^2}{3R^3}+\delta\frac{M^3}{R^6}.
\end{equation}
We can obtain the functions $E_{\rm tot}(R)$ and $E_{\rm tot}(M)$ by combining
Eqs.
(\ref{mr4}) and (\ref{te1}). These functions are plotted in Figs. 
\ref{retot} and \ref{metot}.

The functions $M(R)$ and $E_{\rm tot}(R)$ achieve their extrema at the same
points.\footnote{The intrinsic reason is explained in
Appendix \ref{sec_p}. See also the Remark at the end of this section.} The
local minimum energy
$E_{\rm
tot}^{\rm min}(\delta)$ is
reached at the local maximum mass point $(M_{\rm max}(\delta),R_*(\delta))$ and
the local maximum energy $E_{\rm tot}^{\rm max}(\delta)$ is reached at
the local minimum mass point $(M_{\rm min}(\delta),R'_*(\delta))$. As a
consequence, the function  $E_{\rm tot}(M)$
presents a spike at the local maximum and minimum masses (see Fig. \ref{metot}).

For dilute axion stars with $\delta=0$ at the critical point
($M=R=1$), we get  
\begin{equation}
E_{\rm tot}=-\frac{1}{3}.
\end{equation}
Coming back to dimensional variables, we obtain 
\begin{equation}
\label{etotcrit}
E_{\rm tot}=-\frac{1}{3}\frac{\sigma^2\nu^{1/2}}{(6\pi\zeta)^{3/2}}\frac{\hbar
m^{1/2}G^{1/2}}{|a_s|^{3/2}}=-9.03\times 10^{-2}\, E_a,
\end{equation}
where the scale $E_a$ is defined in Appendix \ref{sec_cis}.
For QCD axions with  $m=10^{-4}\,
{\rm eV}/c^2$ and
$a_s=-5.8\times 10^{-53}\, {\rm m}$, we get $E_{\rm
tot}=-2.35\times
10^{25}\, {\rm ergs}$. For ULAs with $m=2.19\times 10^{-22}\, {\rm eV}/c^2$
and
$a_s=-1.11\times 10^{-62}\, {\rm fm}$, we get $E_{\rm
tot}=-4.17\times
10^{53}\, {\rm ergs}$.

In the TF approximation, the first term in Eq. (\ref{te1}) can be neglected.
Using Eq. (\ref{tf1}), the total energy can be written as
\begin{equation}
\label{te2}
E_{\rm tot}=-\frac{R^3(1+R^2)^2}{216\delta^2}(5R^2+1).
\end{equation}

For $R\ll 1$ (nongravitational limit):
\begin{equation}
\label{te3}
E_{\rm tot}\sim -\frac{R^3}{216\delta^2},\qquad E_{\rm tot}\sim
-\frac{M}{36\delta}.
\end{equation}
Coming back to dimensional variables, we obtain
\begin{equation}
\label{te3a}
E_{\rm tot}\sim
-\frac{81\zeta}{1024\pi}\frac{m^3c^4}{
|a_s|\hbar^2}R^3= -1.60\times 10^{-3}\frac{m^3c^4}{
|a_s|\hbar^2}R^3,
\end{equation}
\begin{equation}
\label{te3b}
E_{\rm tot}\sim -\frac{3(6\pi\zeta)}{64\pi}Mc^2=-1.79\times 10^{-2} M c^2.
\end{equation}
These relations can be compared with the exact results from Eq. (\ref{ord8}).

For $R\gg 1$ (nonattractive limit):
\begin{equation}
\label{te4}
E_{\rm tot}\sim-\frac{5R^9}{216\delta^2},\qquad E_{\rm
tot}\sim-\frac{5M^{9/5}}{6^{6/5}\delta^{1/5}}.
\end{equation}
This is also the asymptotic behavior of the general relations (\ref{mr4}) and
(\ref{te1}) for 
$R\rightarrow +\infty$ and $M\rightarrow +\infty$. We note that
$E_{\rm tot}\sim -(5/6)M^2/R$. Coming back to dimensional variables, we obtain 
\begin{equation}
E_{\rm tot}\sim -\frac{5\nu}{6} \frac{GM^2}{R}\sim -0.332  \frac{GM^2}{R}.
\end{equation}
This relation can be compared with the exact result
(\ref{pol8}) valid for a polytrope $n=1/2$.

{\it Remark:} Let us recover from our approach based on the Gaussian ansatz the
fact that the energy and the mass achieve their extrema at the same points.
The total energy can be written as $E_{\rm
tot}=V(M(R),R)$, where $M(R)$ corresponds to the mass-radius relation
(\ref{mr4}). Therefore
\begin{equation}
\frac{dE_{\rm tot}}{dR}=\frac{\partial V}{\partial
M}\frac{dM}{dR}+\frac{\partial V}{\partial
R}.
\end{equation}
Since $V'(R)=0$ for every state on the series of equilibria, there remains
\begin{equation}
\label{rql1}
\frac{dE_{\rm tot}}{dR}=\frac{\partial V}{\partial M}\frac{dM}{dR}.
\end{equation}
In general $\partial V/\partial M\neq 0$ at an equilibrium state. Therefore,
$E'_{\rm tot}(R)=0$ if, and only if, $M'(R)=0$.

\subsection{The eigenenergy}
\label{sec_vp}

\begin{figure}
\begin{center}
\includegraphics[clip,scale=0.3]{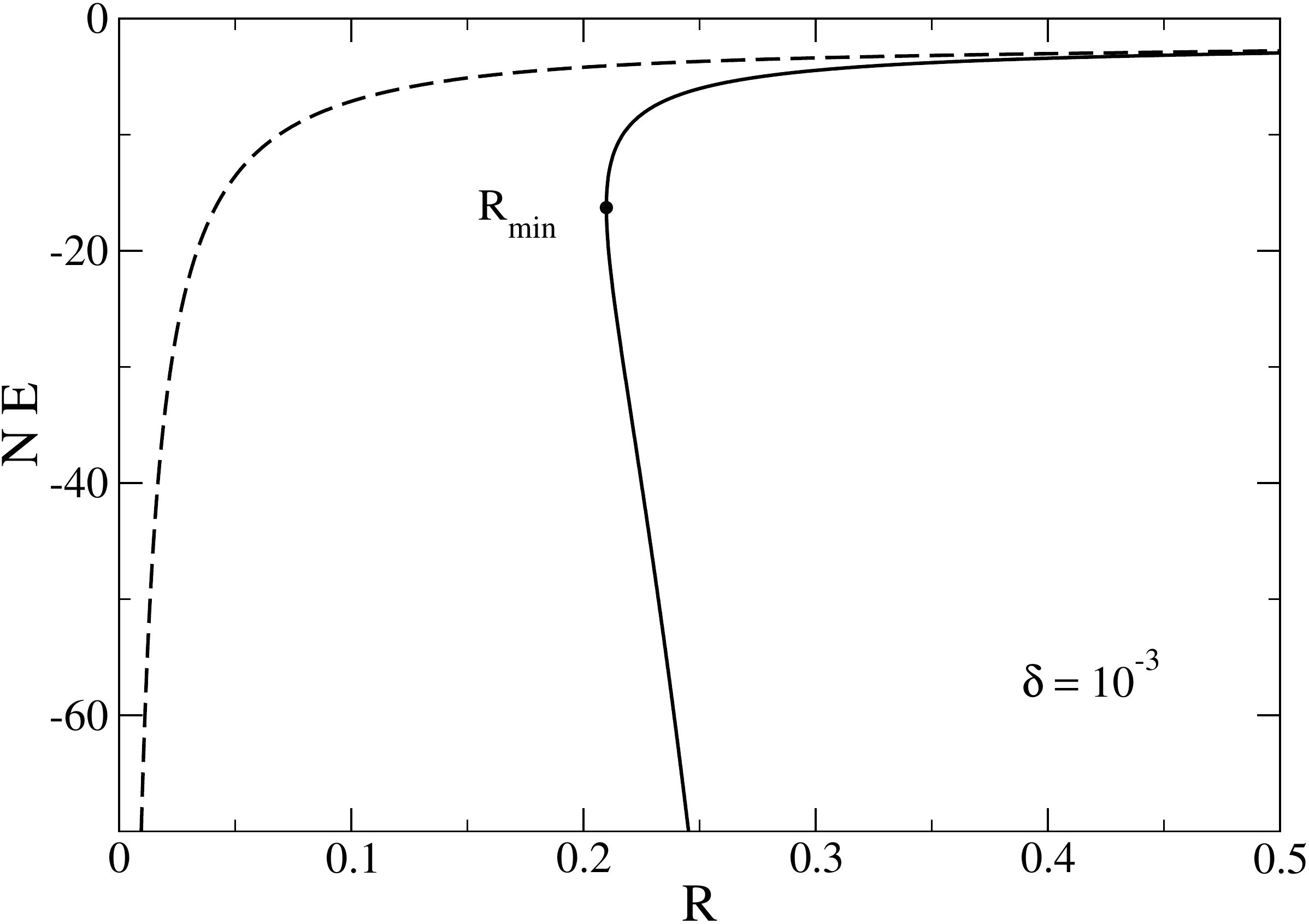}
\caption{Eigenenergy $NE$ as a function of the radius $R$ for 
axion stars with $\delta=10^{-3}$. The dashed line corresponds to dilute axion
stars with $\delta=0$.}
\label{reigen}
\end{center}
\end{figure}

\begin{figure}
\begin{center}
\includegraphics[clip,scale=0.3]{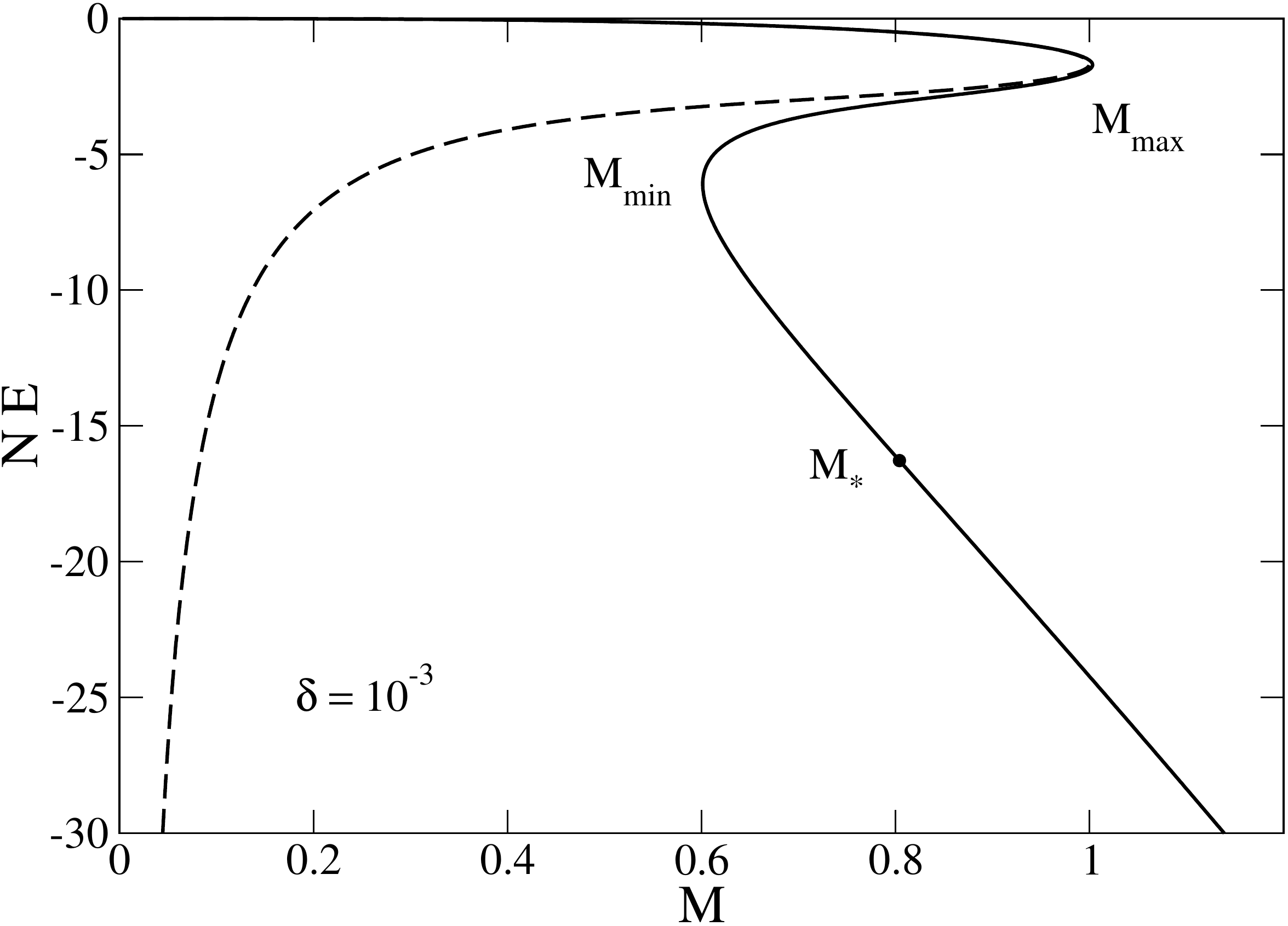}
\caption{Eigenenergy $NE$ as a function of the
mass $M$ for 
axion stars with $\delta=10^{-3}$. The dashed line corresponds to dilute axion
stars with $\delta=0$.}
\label{meigen}
\end{center}
\end{figure}

For a self-gravitating BEC described by a polytropic equation of state, the
quantity $NE$ (representing $N$ times the eigenenergy $E$) 
can be deduced from the total energy $E_{\rm tot}$ by
multiplying the gravitational term by $2$ and the internal energy by
$\gamma$ (see Sec. 5.2.1 of \cite{chavtotal}).
Therefore, according to Eq. (\ref{te1}), the eigenenergy of an axion star
with the equation of state (\ref{s2}) which is the sum of two polytropic
equations of state of index $\gamma=2$ and $\gamma=3$ is given
by 
\begin{equation}
\label{vp1}
NE=\frac{M}{R^2}-\frac{2M^2}{R}-\frac{2M^2}{3R^3}+3\delta\frac{M^3}{R^6}.
\end{equation}
We can obtain the functions $NE(R)$ and $NE(M)$ by combining Eqs. (\ref{mr4})
and (\ref{vp1}). These functions are plotted in Figs. \ref{reigen} and
\ref{meigen}.

For dilute axion stars with $\delta=0$
at the critical point
($M=R=1$), we get  
\begin{equation}
N E=-\frac{5}{3}.
\end{equation}
Coming back to dimensional variables, we obtain
\begin{equation}
N E=-\frac{5}{3}\frac{\sigma^2\nu^{1/2}}{(6\pi\zeta)^{3/2}}\frac{\hbar
m^{1/2}G^{1/2}}{|a_s|^{3/2}}=-0.452\, E_a
\end{equation}
or
\begin{equation}
E=-\frac{5}{3}\frac{\sigma\nu}{6\pi\zeta}\frac{Gm^2}{|a_s|}
=-0.417\, \frac{E_a}{N_a}.
\end{equation}
These relations can be compared with the exact results given in \cite{prd2}.
We note that $|E|\ll mc^2$ for $|a_s|\gg 2Gm/c^2$ (i.e.
$\delta\ll 1$) which justifies our nonrelativistic treatment. For QCD axions
with  $m=10^{-4}\,
{\rm eV}/c^2$ and
$a_s=-5.8\times 10^{-53}\, {\rm m}$, we get $N E=-1.175\times
10^{26}\, {\rm ergs}$ and $E=-1.52\times
10^{-31}\, {\rm ergs}$. For ULAs with $m=2.19\times
10^{-22}\, {\rm eV}/c^2$
and
$a_s=-1.11\times 10^{-62}\, {\rm fm}$, we get $N E=-2.085\times
10^{54}\, {\rm ergs}$ and $E=-3.83\times
10^{-42}\, {\rm ergs}$.

In the TF approximation, the first term in Eq. (\ref{vp1}) can be neglected.
Using Eq.  (\ref{tf1}), the eigenenergy can be written as 
\begin{equation}
\label{vp2}
NE=-\frac{R^3(1+R^2)^2}{216\delta^2}(9R^2+1).
\end{equation}

For $R\ll 1$ (nongravitational limit):
\begin{equation}
\label{vp5}
NE= E_{\rm tot}.
\end{equation}
This is an exact result independent of the Gaussian ansatz (see Appendix
\ref{sec_tfng}).

For $R\gg 1$ (nonattractive limit):
\begin{equation}
\label{vp3}
NE=\frac{9}{5}E_{\rm tot}.
\end{equation}
This is an exact result for a polytrope $n=1/2$ independent
of the Gaussian ansatz (see Appendix \ref{sec_pol}).
This is also the asymptotic behavior of the general relations
(\ref{mr4}) and (\ref{vp1}) for 
$R\rightarrow +\infty$ and $M\rightarrow +\infty$.

{\it Remark:} According to Eqs. (\ref{p2}) and (\ref{p4}), one
has $E/m=dE_{\rm
tot}/dM$ or, equivalently,
\begin{equation}
\label{rew1}
NE=M\frac{dE_{\rm tot}}{dM}.
\end{equation}
Let us check that this exact relation is satisfied by our approximate approach
based on the Gaussian ansatz. The total energy can be written as $E_{\rm
tot}=V(M,R(M))$, where $R(M)$ corresponds to the mass-radius relation
(\ref{mr4}). Therefore
\begin{equation}
\label{rew2}
\frac{dE_{\rm tot}}{dM}=\frac{\partial V}{\partial M}+\frac{\partial V}{\partial
R}\frac{dR}{dM}.
\end{equation}
Since $V'(R)=0$ for every state on the series of equilibria, there remains
\begin{equation}
\label{rew3}
\frac{dE_{\rm tot}}{dM}=\frac{\partial V}{\partial M}.
\end{equation}
This relation is equivalent to Eq. (\ref{rql1}). Therefore, we have to show
that ${E}/{m}={\partial V}/{\partial M}$ or, equivalently,
\begin{equation}
NE=M\frac{\partial V}{\partial M}.
\end{equation}
Taking the partial derivative of $V(R,M)$ given by Eq. (\ref{ep7}) with
respect to $M$, and comparing the result with Eq.  (\ref{vp1}), we find that
this relation is indeed satisfied.

\subsection{The pulsation}
\label{sec_w}

\begin{figure}
\begin{center}
\includegraphics[clip,scale=0.3]{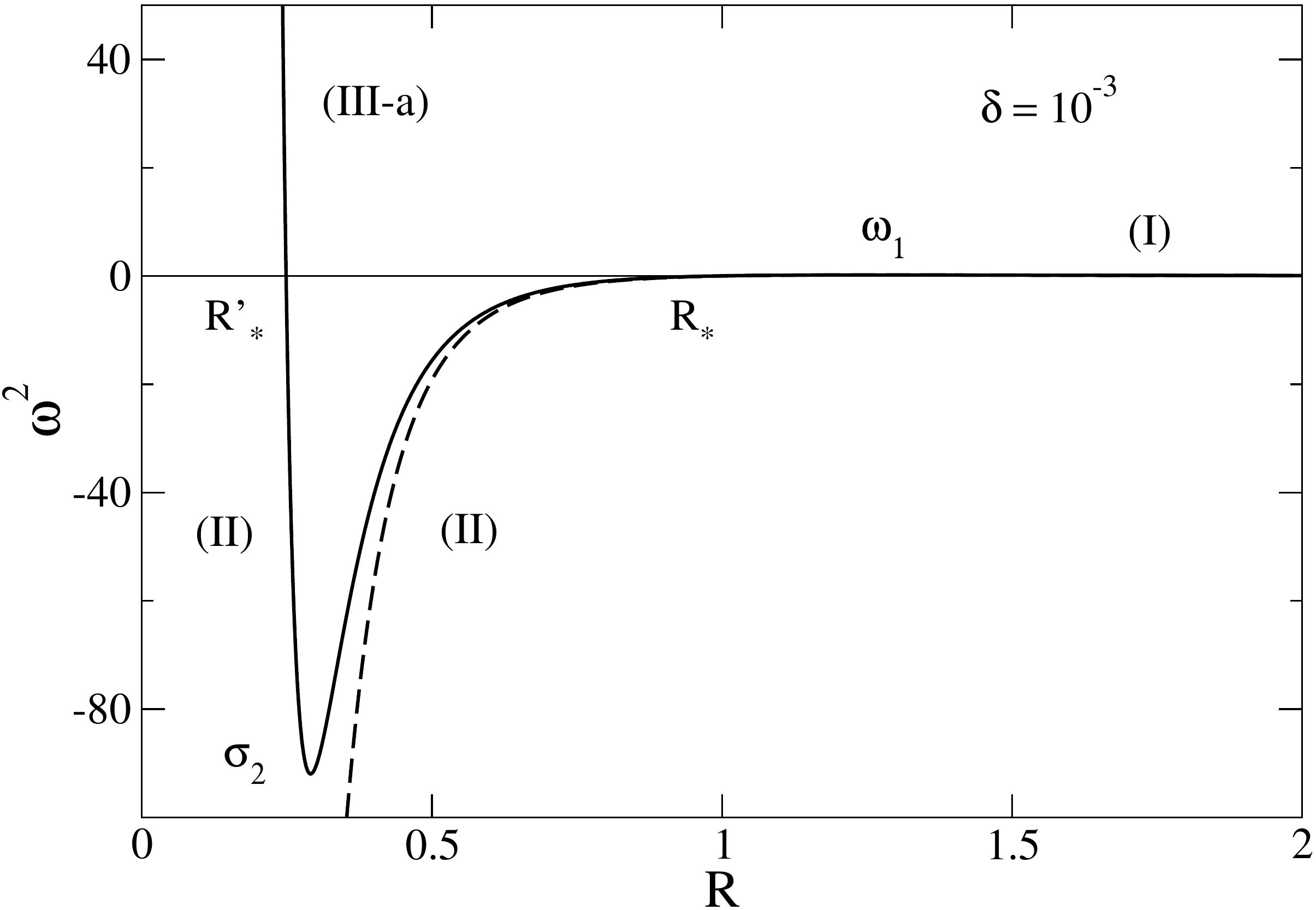}
\caption{Complex pulsation as a function of the radius $R$ for 
axion stars with $\delta=10^{-3}$. The dashed line corresponds to dilute axion
stars with $\delta=0$.}
\label{rpulsationZOOM}
\end{center}
\end{figure}

\begin{figure}
\begin{center}
\includegraphics[clip,scale=0.3]{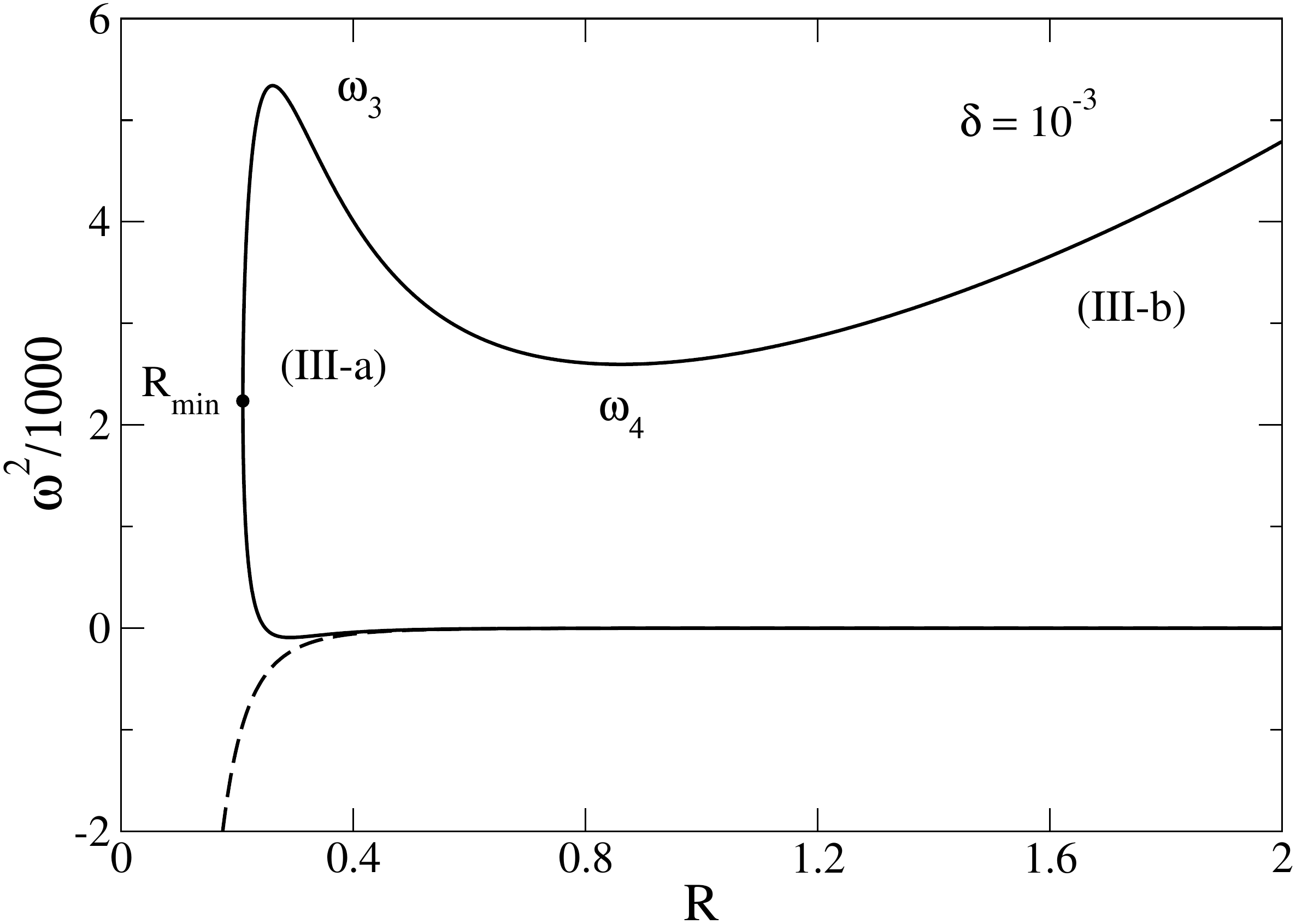}
\caption{Same as Fig. \ref{rpulsationZOOM} showing the branch of dense axion
stars with high pulsations ($\omega^2$ has been divided by $1000$).}
\label{rpulsation}
\end{center}
\end{figure}

\begin{figure}
\begin{center}
\includegraphics[clip,scale=0.3]{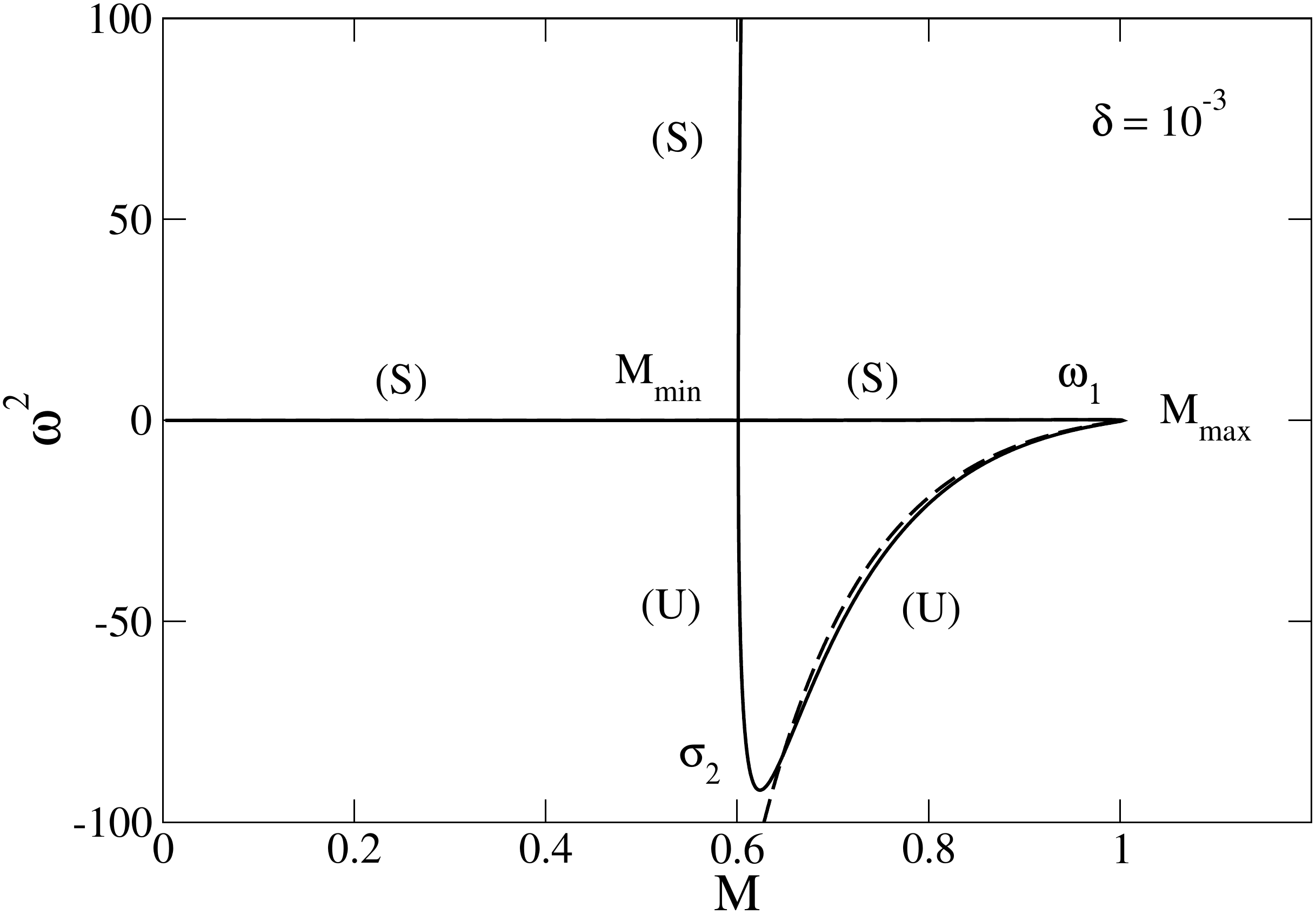}
\caption{Complex pulsation as a function of the mass $M$ for 
axion stars with $\delta=10^{-3}$. The dashed line corresponds to dilute axion
stars with $\delta=0$. }
\label{mpulsationZOOM}
\end{center}
\end{figure}

\begin{figure}
\begin{center}
\includegraphics[clip,scale=0.3]{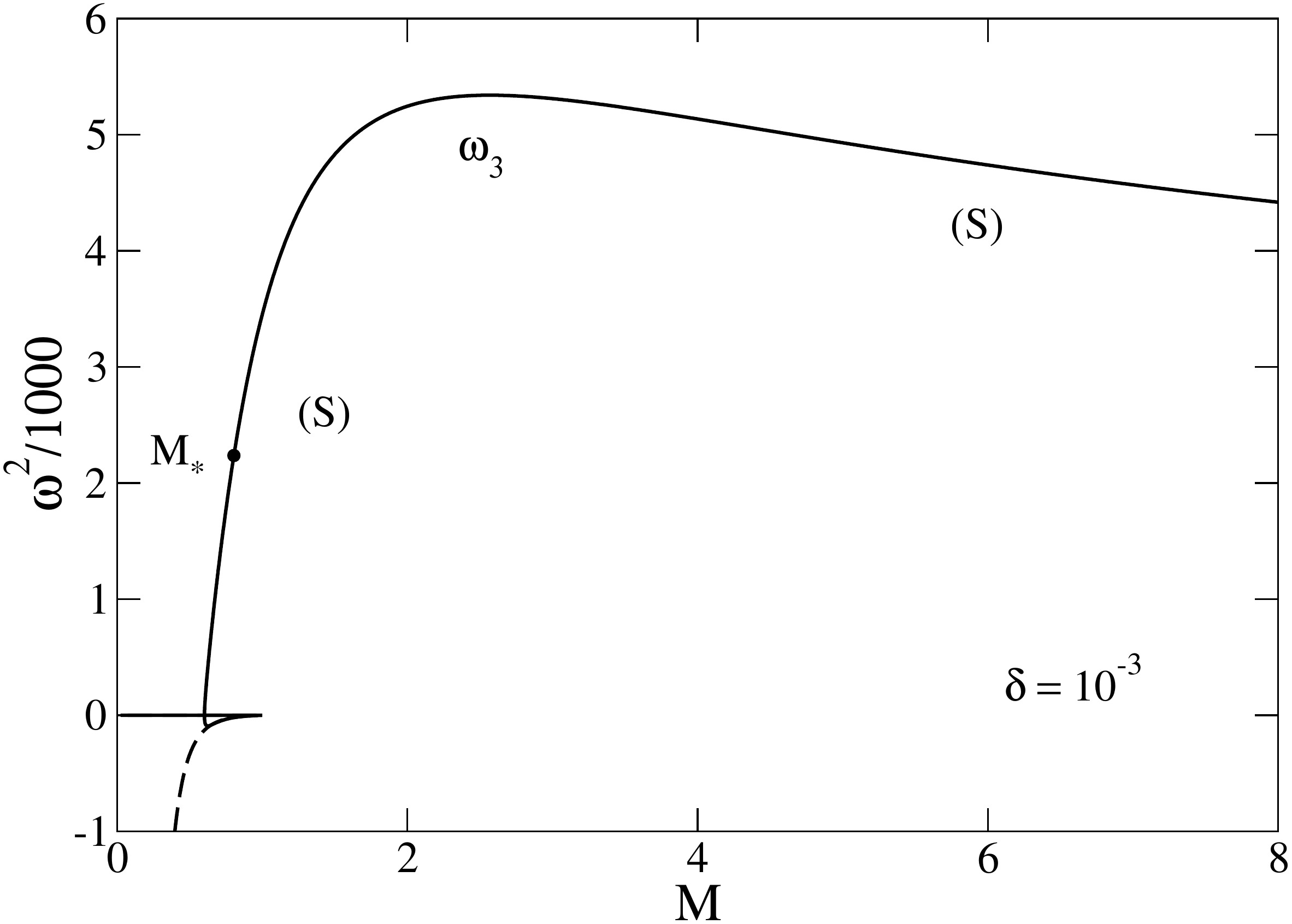}
\caption{Same as Fig. \ref{mpulsationZOOM} showing the branch of dense axion
stars with high pulsations ($\omega^2$ has been divided by $1000$). The minimum
pulsation $\omega_4$ corresponds to a large
mass $M_4=202$ that is not represented on the figure.}
\label{mpulsation}
\end{center}
\end{figure}

The complex pulsation of an axion star slightly displaced from equilibrium is
given by \cite{prd1,chavtotal}:
\begin{equation}
\label{w1}
\omega^2=\frac{V''(R)}{M}.
\end{equation}
Using Eq. (\ref{mr7}), we get
\begin{equation}
\label{w2}
\omega^2=\frac{6}{R^4}-\frac{2M}{R^3}-\frac{4M}{R^5}+42\delta\frac{M^2}{R^8
}.
\end{equation}
We can obtain the functions $\omega^2(R)$ and $\omega^2(M)$ by combining Eqs. 
(\ref{mr4}) and (\ref{w2}). These functions are plotted in Figs.
\ref{rpulsationZOOM}-\ref{mpulsation}. The pulsation vanishes at
the critical points
$(M_{\rm max}(\delta),R_*(\delta))$ and $(M_{\rm
min}(\delta),R'_*(\delta))$ where a change of stability occurs. The stable
branch of dilute axion stars (I) has a
positive square pulsation $\omega^2>0$. The pulsation achieves a maximum value
$\omega_1(\delta)$ at a point $(M_{1}(\delta),R_1(\delta))$. For $\delta=0$ this
point is characterized analytically in \cite{bectcoll} (see also Appendix
\ref{sec_maxw}). The unstable
branch of axion
stars (II) has a negative square pulsation $\omega^2<0$. The growth rate
$\sigma=\sqrt{-\omega^2}$ achieves a
maximum value $\sigma_2(\delta)$ at a
point $(M_{2}(\delta),R_2(\delta))$. The stable
branch of dense axion stars (III) has a
positive square pulsation $\omega^2>0$. The pulsation achieves a maximum value
$\omega_3(\delta)$ at a point $(M_{3}(\delta),R_3(\delta))$
and a minimum value $\omega_4(\delta)$ at a point 
$(M_{4}(\delta),R_4(\delta))$. This last point is  characterized analytically
below
in the TF approximation.

In the TF approximation, the first term in Eq. (\ref{w2}) can be neglected.
Using Eq.  (\ref{tf1}), the pulsation can be
written as
\begin{equation}
\label{w3}
\omega^2=\frac{1+R^2}{6\delta}\left (5+\frac{3}{R^2}\right ).
\end{equation}

For $R\ll 1$ (nongravitational limit):
\begin{equation}
\label{w8}
\omega^2\sim \frac{1}{2\delta R^2},\qquad \omega^2\sim
\frac{1}{2^{5/3}3^{2/3}}\frac{1}{\delta^{5/3}M^{2/3}}.
\end{equation}
Coming back to dimensional variables, we get
\begin{equation}
\omega=\left (\frac{27}{32\pi}\right )^{1/2}\left
(\frac{6\pi\zeta}{\alpha}\right )^{1/2}\frac{c}{R}=0.463\, \frac{c}{R}.
\end{equation}
This relation can be compared with the exact result  (\ref{tfng19}).

For $R\gg 1$ (nonattractive limit):
\begin{equation}
\label{w4}
\omega^2\sim \frac{5R^2}{6\delta},\qquad \omega^2\sim
\frac{5M^{2/5}}{(6\delta)^{3/5}}.
\end{equation}
This is also the asymptotic
behavior of the general relations (\ref{mr4}) and (\ref{w2}) for 
$R\rightarrow +\infty$ and $M\rightarrow +\infty$. Using Eq. (\ref{tf3}), we
find that $\omega^2\sim 5M/R^3$. Coming back to dimensional variables, we
obtain 
\begin{equation}
\omega^2\sim \frac{5\nu}{\alpha}\frac{GM}{R^3}\sim 1.33\frac{GM}{R^3}.
\end{equation}
This relation can be compared with the result  (\ref{led}) obtained from
the Ledoux formula for a  polytrope $n=1/2$.

In the TF approximation, the square pulsation $\omega^2$
has
a
minimum 
\begin{equation}
\label{w7}
\omega_4^2=\frac{5+\sqrt{15}}{6\delta}\left (1+\sqrt{\frac{3}{5}}\right )
\end{equation}
at the point 
\begin{equation}
\label{w6}
R_4=\left (\frac{3}{5}\right )^{1/4}, \qquad
M_4=\frac{\left (\frac{3}{5}\right )^{3/4}\left (1+\sqrt{\frac{3}{5}}\right
)}{6\delta}.
\end{equation}
We note that $M_4\sim M_0$. For such a large mass, general relativity must be
taken into account (see Secs. \ref{sec_tran} and \ref{sec_qgr}) so our
Newtonian results (\ref{w7}) and (\ref{w6}) may not be physically relevant.

{\it Remark:} We note that dilute axion stars have low pulsations while dense
axion stars have very high pulsations. Numerical applications are made in
Sec. \ref{sec_coll} for dense axion stars and in Appendix \ref{sec_maxw} for
dilute axion stars.

\subsection{The turning point argument}
\label{sec_turning}

In this section, we derive
a simplified version of Poincar\'e's turning point argument based on the
Gaussian ansatz. If we differentiate the mass-radius relation (\ref{mr2}) with
respect to $R$
and use Eq. (\ref{mr7}), we obtain
\begin{equation}
\label{ty1}
V''(R)+\left
(-\frac{2}{R^3}+\frac{2M}{R^2}+\frac{2M}{R^4}-18\delta\frac{M^2}{R^7}\right
)\frac{dM}{dR}=0.
\end{equation}
Simplifying the term in parenthesis with the aid of Eq. (\ref{mr2}), and
using Eq. (\ref{w1}), the foregoing equation can be rewritten as
\begin{eqnarray}
\label{ty2}
\omega^2(R)=-\frac{4R-M(R^2+1)}{MR^4}\frac{d M}{d R}.
\end{eqnarray}
It relates the square of the complex pulsation $\omega^2$ determining
the stability of the system to the slope of the mass-radius relation $M(R)$.

It is better to develop the following discussion at a general level. For an
arbitrary potential $V(R,M)$, the mass-radius relation $M(R)$ is given in
implicit form by
\begin{eqnarray}
\label{turning3}
\frac{\partial V}{\partial R}(R,M(R))=0.
\end{eqnarray}
Differentiating this relation with respect to $R$, and using Eq. (\ref{w1}), we
get
\begin{eqnarray}
\label{turning4}
M\omega^2=\frac{\partial^2 V}{\partial R^2}(R,M)=-\frac{\partial^2 V}{\partial
R\partial
M}(R,M)\frac{dM}{dR},
\end{eqnarray}
which is the generalization of Eq. (\ref{ty1}).  We first note that a
turning point of mass ($M'(R)=0$) corresponds to
$\omega^2=(1/M)\partial^2V/\partial R^2=0$. On the other
hand, a turning point of radius  ($R'(M)=0$ or $M'(R)=\pm\infty$)  corresponds
to $\partial^2V/\partial R\partial M=0$. A change of stability along the series
of equilibria  takes place
when $\omega^2=(1/M)\partial^2V/\partial R^2$ changes sign. According to Eq.
(\ref{turning4}), this happens at a
point  where $M'(R)$ changes sign
while $\partial^2V/\partial R\partial M$ does not change sign. Therefore, this 
happens
at a turning point of mass, {\it not} at a turning point of radius. At a
turning point of mass, both $dM/dR$ and $\omega^2$ change sign by passing
through
$0$ while  $\partial^2V/\partial R\partial M$ does not change sign. At a
turning point of radius, $dM/dR$ changes sign by passing through $\infty$ and
$\partial^2V/\partial R\partial M$ changes sign by passing through $0$ while
$\omega^2$ does not changes sign (it remains finite and nonzero). In conclusion,
a change of stability along the series of equilibria ($\omega^2=V''(R)=0$)
coincides with a turning point of mass ($M'(R)=0$). This is a simplified
version, for a potential $V(R,M)$, of the turning point argument obtained from
Poincar\'e's theory of linear
series of equilibria \cite{poincare} (see also the $M(R)$ theorem of Wheeler
\cite{htww} and the theory of catastrophes \cite{catastrophe}). We emphasize
that the stability of
the system is not directly related to the sign of the slope of the
mass-radius relation since there is no change of stability after a turning
point of radius although the slope changes. A change of slope is a necessary
(but not a sufficient) condition for a  change of stability.

Coming back to our particular problem, let us explicitly check that the
condition of marginal stability $\omega^2=(1/M)V''(R)=0$ returns Eq.
(\ref{mmm2})
corresponding to $M'(R)=0$ and that 
the condition $\partial^2V/\partial R\partial M=0$ returns Eq. (\ref{mir3})
corresponding to $R'(M)=0$. 

According to Eq. (\ref{mr7}), the
condition $V''(R)=0$ is equivalent to
\begin{equation}
\label{vpp}
\frac{6M}{R^4}-\frac{2M^2}{R^3}-\frac{4M^2}{R^5}+42\delta\frac{M^3}{
R^8}=0.
\end{equation}
Multiplying this equation by $R/7$ and adding the resulting expression
to Eq. (\ref{mr2}) in order to eliminate $\delta$, we recover Eq. (\ref{mmm2}).

According to Eq. (\ref{ep7}), we have
\begin{equation}
\label{id3}
\frac{\partial^2V}{\partial
M\partial R}=-\frac{2}{R^3}+\frac{2M}{R^2}+\frac{2M}{R^4} -18\delta\frac { M^2 }
{
R^7}.
\end{equation}
Multiplying Eq. (\ref{mr2}) by $-3/M$ and adding the resulting expression
to Eq. (\ref{id3}) in order to eliminate $\delta$, we obtain
\begin{equation}
\frac{\partial^2V}{\partial
M\partial R}=\frac{4R-M(R^2+1)}{R^4}.
\end{equation}
These equalities show the equivalence between Eqs. (\ref{ty1}), (\ref{ty2}) and
(\ref{turning4}). Furthermore, we can check that the condition  
$\partial^2V/\partial R\partial M=0$  returns Eq. (\ref{mir3}).

\subsection{The sign of the total energy}
\label{sec_tev}

\begin{figure}
\begin{center}
\includegraphics[clip,scale=0.3]{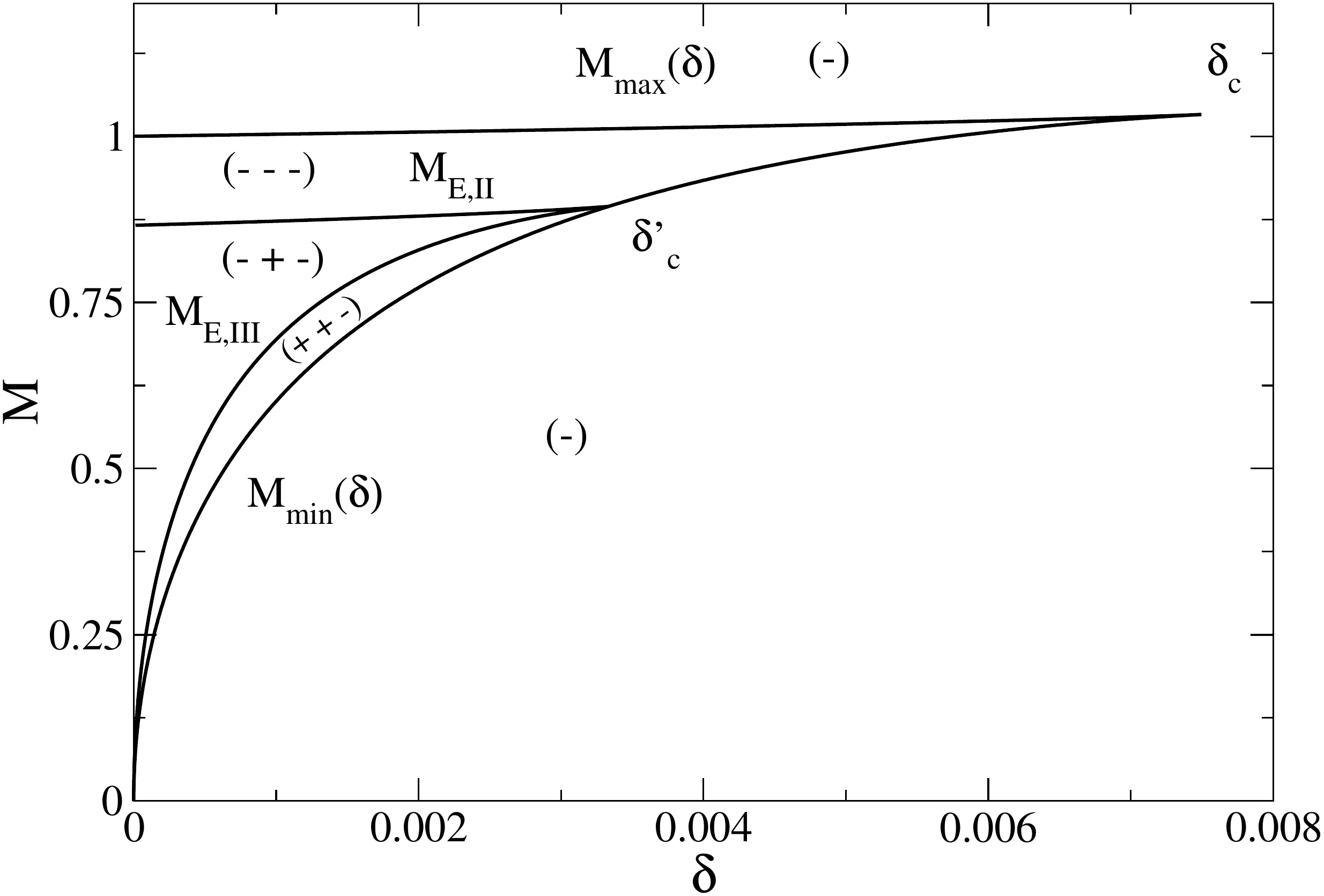}
\caption{The mass $M_{\rm E}(\delta)$ at which the
total energy vanishes as a function of $\delta$.  The upper branch  $M_{\rm
E,II}(\delta)$ corresponds to unstable axion stars and the lower branch  $M_{\rm
E,III}(\delta)$
corresponds to metastable dense axion stars. For
comparison, we have also plotted the maximum and
minimum mass. Finally, we have indicated the sign of the energy of the different
equilibrium states in
each domain. For example $(++-)$ means that the dense stable axion star
(left) has a positive energy, the unstable axion star (middle) has a positive
energy and the stable dilute axion (right) star has a negative
energy.}
\label{deltamzeroenergytotal}
\end{center}
\end{figure}

\begin{figure}
\begin{center}
\includegraphics[clip,scale=0.3]{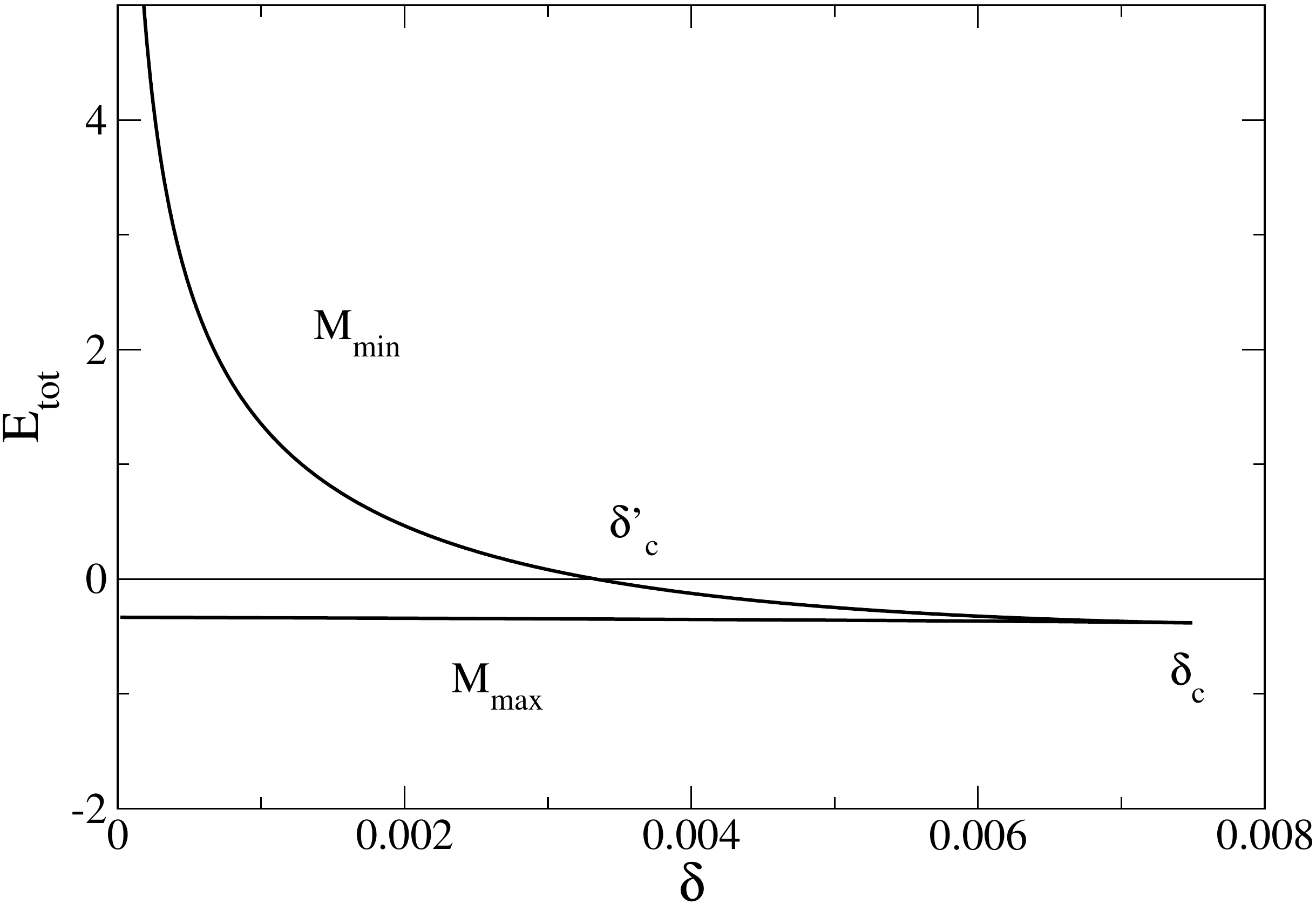}
\caption{Total energy at the point of maximum mass (lower branch) and at the
point of minimum mass (upper branch) as a function of $\delta$.}
\label{sign}
\end{center}
\end{figure}

We have seen that a change of stability in the series of
equilibria corresponds to a turning point of mass, or equivalently, to a turning
point of total energy. For future
discussions (see Sec. \ref{sec_unstable}), it is
useful to determine
the points in the series of equilibria where the total energy vanishes. Writing
$E_{\rm tot}=0$ in Eq. (\ref{te1}) and combining the resulting equation with the
mass-radius relation (\ref{mr3}), we find after simplification that the
mass $M_{\rm E}(\delta)$ of an equilibrium state
(stable or unstable) for which $E_{\rm tot}=0$ is determined by the parametric
equations
\begin{equation}
\label{me}
M_{\rm E}=\frac{4R_{\rm E}}{1+5R_{\rm E}^2},
\end{equation}
\begin{equation}
\label{deltae}
\delta=\frac{1}{48}R_{\rm E}^2(1+5R_{\rm E}^2)(1-3R_{\rm E}^2),
\end{equation}
where the parameter $R_{\rm E}$ is the radius. The function $M_{\rm E}(\delta)$
is plotted in
Fig. \ref{deltamzeroenergytotal}. It displays two branches   $M_{\rm
E,II}(\delta)$ and $M_{\rm
E,III}(\delta)$  whose meaning is explained below. From Eqs. (\ref{mmm1}),
(\ref{mmm2}) and
(\ref{te1}), we find that the total energy at the point of maximum mass is
always negative, going from $E_{\rm tot}=-1/3$ at $\delta=0$ to
$E_{\rm tot}=-(8/27)\sqrt{5/3}$ at $\delta=\delta_c$ (see Fig. \ref{sign}).
From Fig. \ref{metot}, we
conclude that the stable dilute axion stars
(branch I) always have a negative energy ($E_{\rm tot}<0$). On the other hand,
from Eqs. (\ref{mmm1}), (\ref{mmm2}), (\ref{te1}), (\ref{me}) and (\ref{deltae})
we find that the total energy at the point of minimum mass is  positive 
for $\delta<\delta'_c=1/300$  and negative for
$\delta>\delta'_c$, going from $E_{\rm tot}=+\infty$
at $\delta=0$ to
$E_{\rm tot}=-(8/27)\sqrt{5/3}$ at $\delta=\delta_c$  (at the transition  point
$\delta'_c=1/300$ we find $M=2/\sqrt{5}$ and
$R=1/\sqrt{5}$).
Considering  Fig. \ref{metot} again, we come to the following conclusions. When
$\delta<\delta'_c$, the unstable axion stars (branch II) have a positive
energy for $M_{\rm min}<M<M_{\rm E,II}$ and a negative energy for  $M_{\rm
E,II}<M<M_{\rm max}$ while the metastable dense axion stars (branch III) have a
positive
energy for $M_{\rm min}<M<M_{\rm E,III}<M_t$ and a negative energy for 
$M>M_{\rm E,III}$. When $\delta>\delta'_c$, the unstable axion stars and the  
stable dense axion stars have a negative energy.  For $\delta\rightarrow 0$, we
recover the result $(M_{\rm
E,II},R_{\rm E,II})=(\sqrt{3}/2,1/\sqrt{3})$  obtained in \cite{bectcoll}. We
also find that $M_{\rm E,III}\sim 16\sqrt{3\delta}$  and
$R_{\rm
E,III}\sim 4\sqrt{3\delta}$, so that ($R_{\rm
E,III},M_{\rm E,III})$ approches the point with the
minimum radius
($R_{\rm min},M_*$). More generally, $M_{\rm min}<M_{\rm
E,III}<M_*$ and $R_{\rm min}<R_{\rm
E,III}<R'_*$. Using the preceding  results, it is easy to
determine the sign of the
energy of each equilibrium state as indicated in Fig.
\ref{deltamzeroenergytotal}.

\section{The radius of the axion star resulting from a collapse or
an explosion at a critical point}
\label{sec_r}

\subsection{The different solutions as a function of $M$ depending on the value
of $\delta$}
\label{sec_ds}

We are now in a position to discuss the different equilibrium states of axion
stars (characterized by their radius $R$) as a function of their mass $M$ 
depending on the value of the interaction parameter $\delta$. Specifically, we
fix $\delta$ and determine the possible radii $R$ as a function of $M$ (see Fig.
\ref{mr}). 

(i) We first assume $\delta<\delta_c$. If
$M<M_{\rm min}(\delta)$, there is only one solution:
a stable dilute axion star (I). If $M_{\rm min}(\delta)<M<M_{\rm max}(\delta)$,
there are three solutions: a stable dilute axion star (I),
an unstable axion star (II), and a stable dense axion star (III). If  $M>M_{\rm
max}(\delta)$, there is only one solution: a stable dense axion
star (III).

(ii) We now assume $\delta>\delta_c$. In that case, there is only one solution:
a stable dilute axion star (I) if $M<M_*(\delta)$ and a
dense axion star (III) if $M>M_*(\delta)$.

\subsection{Collapse when $\delta\ll 1$ and $M\sim M_{\rm max}^+$}
\label{sec_coll}

Let us consider the situation where $\delta\ll 1$ and
$M\sim M_{\rm max}(\delta)^+\sim 1^+$. Physically, we assume that a stable axion
star gains mass (for example by accreting matter around it or through its
collision with another star) and passes above the critical point. In that case,
the dilute axion star
collapses and ultimately forms a dense axion star. Since $M_0\gg
1$ for $\delta\ll 1$, this dense axion star is of type (III-a). We assume
that there is no mass loss during the
collapse.\footnote{This assumption may not be correct and
will have to be revised in future works. Indeed, the collapse of a dilute axion
star is generally accompanied by an emission of relativistic axions (bosenova),
leaving a dense axion star as the remnant with a mass smaller than the initial
one \cite{braaten}. If the  mass loss is not too important, our
results should remain qualitatively correct. Similarly, white
dwarf stars above
the Chandrasekhar mass
collapse and form neutron stars. Since the maximum mass of white dwarf stars
$M_{\rm max}=1.44\, M_{\odot}$ \cite{chandra31} is larger than the maximum mass
of ideal neutron stars $M_{\rm max}=0.710\, M_{\odot}$ \cite{ov},
the collapse is necessarily accompanied by an ejection of mass (note, however,
that in more realistic models, the maximum mass of neutron stars, being in the
range
$2-2.4\, M_{\odot}$, is larger than the maximum mass of white dwarf stars).}
Using Eq.
(\ref{tf2}), the radius of the dense axion star is 
\begin{equation}
\label{r1}
R_{\rm coll}(\delta)= (6\delta)^{1/3}.
\end{equation}
We note that the scaling of $R_{\rm coll}(\delta)$ is different from the
scaling of $R_{\rm min}(\delta)$ given by Eq. (\ref{mir4}). Coming back to
dimensional variables we get
\begin{equation}
R_{\rm coll}\sim \left (\frac{96\pi\sigma}{27\nu}\right
)^{1/3}\left (\frac{\nu}{6\pi\zeta}\right
)^{1/6}\left (\frac{|a_s|\hbar^6}{Gm^7c^4}\right )^{1/6}=2.30 R_d,
\end{equation}
where the scale $R_d$ is defined in Appendix \ref{sec_oc}. This expression can
be compared with the exact result from Eq. (\ref{tfng14}).

The density of the dense axion star resulting from gravitational collapse
is $\rho_{\rm dense}=1/(8\pi\delta)$ [see Eq. (\ref{tf2b})]. In comparison, the
density of the dilute
axion star at the point of maximum mass is $\rho_{\rm dilute}=3/(4\pi)$. This
leads to a ratio  
\begin{equation}
\frac{\rho_{\rm dense}}{\rho_{\rm dilute}}=\frac{1}{6\delta}.
\end{equation}
For $\delta\ll 1$, the increase in density is great.

The pulsation of the dense axion star resulting from gravitational collapse is
given by [see Eq. (\ref{w8})]:
\begin{equation}
\label{r1b}
\omega_{\rm coll}(\delta)=\frac{1}{2^{5/6}3^{1/3}\delta^{5/6}}.
\end{equation}
For $\delta\ll 1$, the dense axion star has a very high
pulsation. In comparison, the maximum
pulsation of a dilute axion star is $\omega_{\rm max}=0.424$ (see Appendix
\ref{sec_maxw}). Coming back to
dimensional variables we get
\begin{eqnarray}
\label{r1c}
\omega_{\rm coll}&=&\frac{(\sigma/\alpha)^{1/2}}{2^{5/6}3^{1/3}}\left
(\frac{27\nu}{16\pi\sigma}\right )^{5/6}
\left (\frac{6\pi\zeta}{\nu}\right )^{2/3}\left
(\frac{Gm^7c^{10}}{|a_s|\hbar^6}\right )^{1/6}\nonumber\\
&=&0.202 \left
(\frac{Gm^7c^{10}}{|a_s|\hbar^6}\right )^{1/6}.
\end{eqnarray}
This expression can be compared with the
exact result from Eq. (\ref{tfng20}).

The total energy of the dense axion star resulting from gravitational collapse
is given by [see Eq. (\ref{te3})]:
\begin{equation}
\label{r1d}
E_{\rm tot}^{\rm coll}(\delta)= -\frac{1}{36\delta}.
\end{equation}
For $\delta\ll 1$, the dense axion star has a very negative energy. In
comparison, the energy of a
dilute axion star
with the  critical mass is $E_{\rm tot}^{\rm dilute}=-{1}/{3}$.
Coming back to dimensional variables we get
\begin{eqnarray}
\label{r1e}
E_{\rm tot}^{\rm coll}&=&-\frac{3\sigma}{64\pi}\left
(\frac{6\pi\zeta}{\nu}\right )^{1/2}\frac{\hbar
c^2}{(Gm|a_s|)^{1/2}}\nonumber\\
&=&-1.94\times 10^{-2}\frac{\hbar
c^2}{(Gm|a_s|)^{1/2}}.
\end{eqnarray}
We note that $E_{\rm tot}^{\rm
coll}=-(9\zeta/32)M_{\rm max}c^2=-1.79\times 10^{-2}\, M_{\rm max}c^2$. These
expressions can be compared with the
exact results from Eq. (\ref{mc2}). 

For QCD axions with   $\delta=8.88\times
10^{-16}$, assuming $M_{\rm coll}=M_{\rm max}=1$, we get $R_{\rm coll}=
1.75\times
10^{-5}$, $\rho_{\rm dense}=4.48\times 10^{13}$, $2\pi/\omega_{\rm
coll}=4.62\times 10^{-12}$ and $E_{\rm
tot}^{\rm coll}=-3.13\times
10^{13}$. Coming back to original variables, assuming $M_{\rm coll}=M_{\rm
max}=6.92\times
10^{-14}\, M_{\odot}$, we get $R_{\rm coll}=1.80\times
10^{-9}\, R_{\odot}=1.25\, {\rm m}$, $\rho_{\rm dense}=1.68\times 10^{19}\, {\rm
g/m^3}$, $2\pi/\omega_{\rm coll}=5.65\times
10^{-8}\, {\rm s}$ and
$E_{\rm
tot}^{\rm coll}=-2.21\times
10^{39}\, {\rm ergs}$ (in comparison $R_*=71.5\, {\rm
km}$, $\rho_{\rm dilute}=8.98\times 10^4\, {\rm g/m^3}$, $2\pi/\omega_{\rm
max}=50.3\, {\rm hrs}$, $E_{\rm
tot}^{\rm dilute}=-2.35\times
10^{25}\, {\rm ergs}$).

For ULAs with $\delta=1.02\times
10^{-8}$, assuming $M_{\rm coll}=M_{\rm max}=1$, we get 
$R_{\rm coll}= 3.94\times
10^{-3}$, $\rho_{\rm dense}=3.90\times 10^6$, $2\pi/\omega_{\rm coll}=3.54\times
10^{-6}$ and $E_{\rm
tot}^{\rm coll}=-2.72\times
10^{6}$. Coming back to original variables, assuming  $M_{\rm coll}=M_{\rm
max}=1.07\times
10^8\, M_{\odot}$, we get
$R_{\rm coll}= 1.23\, {\rm
pc}$, $\rho_{\rm dense}=9.21\times 10^{-10}\, {\rm g/m^3}$,
$2\pi/\omega_{\rm coll}=54.8\, {\rm yrs}$ and $E_{\rm
tot}^{\rm coll}=-3.40\times
10^{60}\, {\rm ergs}$ (in comparison $R_*=0.313\, {\rm kpc}$, $\rho_{\rm
dilute}=5.63\times 10^{-17}\, {\rm g/m^3}$, $2\pi/\omega_{\rm max}=229\, {\rm
Myrs}$,  $E_{\rm
tot}^{\rm dilute}=-4.17\times
10^{53}\, {\rm ergs}$).

\subsection{Explosion when $\delta\ll 1$ and $M\sim M_{\rm
min}^-$}
\label{sec_exp}

We now  consider the situation where $\delta\ll 1$ and
$M\sim M_{\rm min}(\delta)^-$. Physically, we assume that a stable dense
axion
star loses mass and passes below the critical point. In
that
case, the dense axion star
explodes and ultimately becomes a dilute axion star (I) with a very large
radius.\footnote{We consider this situation for curiosity and remain brief
because it may not be physically relevant. In view of the numerical applications
made at the end of this section, the explosion, if it really takes place, is
expected to form a dispersed cloud, not a star.} We
assume that
there is no mass loss during the explosion. Using Eqs.
(\ref{mrd3}) and (\ref{mmm3}), the radius and the average density of the dilute
axion star are
\begin{equation}
\label{r1f}
R_{\rm exp}(\delta)\sim \frac{3}{32\sqrt{\delta}}, \qquad
\rho_{\rm exp}(\delta)\sim \frac{524288}{27\pi}\delta^2.
\end{equation}
Coming back to
dimensional variables, we
get
\begin{equation}
\label{r2}
R_{\rm exp}\sim \frac{3}{128}\left (\frac{27}{\pi\sigma}\right
)^{1/2}\frac{(6\pi\zeta)^{3/2}}{\nu}\frac{|a_s|\hbar c}{Gm^2}=0.260\, R_e,
\end{equation}
\begin{equation}
\label{r3}
\rho_{\rm exp}\sim
\left (\frac{512}{27}\right )^3\frac{\pi\nu^3\sigma^3}{(6\pi\zeta)^6}\frac{
G^3m^6}{|a_s|^4\hbar^2c^4}=195\,\rho_e,
\end{equation}
where the scales $R_e$ and $\rho_e$ are defined in Appendix \ref{sec_oe}.

The pulsation of the dilute axion star resulting from the
explosion is given by
\begin{equation}
\label{r3b}
\omega_{\rm exp}\sim \frac{2048}{9\sqrt{2}}\delta=161\, \delta.
\end{equation}

The total energy of the dilute axion star resulting from the
explosion is given by 
\begin{equation}
\label{r3c}
E_{\rm tot}^{\rm exp}\sim -\frac{65536}{27}\delta^{3/2}=-2.43\times
10^3\, \delta^{3/2}.
\end{equation}
In comparison, the total energy  of the dense axion star with the minimum mass
is
\begin{equation}
\label{r3d}
E_{\rm tot}^{\rm dense}\sim
\frac{2}{27\sqrt{\delta}}=\frac{7.41\times 10^{-2}}{\sqrt{\delta}}.
\end{equation}
We note that it is positive. This is possible because the
dense axion star is metastable, not fully stable (see the Remark at the end of
Sec. \ref{sec_small2}).

For QCD axions with   $\delta=8.88\times
10^{-16}$, assuming $M_{\rm exp}=M_{\rm min}=6.36\times 10^{-7}$, we get $R_{\rm
exp}\sim 3.14\times
10^{6}$, $\rho_{\rm exp}=4.90\times 10^{-27}$,
$2\pi/\omega_{\rm exp}=4.40\times 10^{13}$ and $E_{\rm
tot}^{\rm exp}=-6.42\times
10^{-20}$  (in comparison $E_{\rm
tot}^{\rm dense}=2.49\times
10^{6}$). Coming back to original variables, assuming
$M_{\rm exp}=M_{\rm
min}=4.40\times
10^{-20}\, M_{\odot}$, we get $R_{\rm exp}=323\, R_{\odot}$,
$\rho_{\rm exp}=1.84\times 10^{-21}\, {\rm g/m^3}$,
$2\pi/\omega_{\rm exp}=17.1\, {\rm Gyrs}$ and
$E_{\rm tot}^{\rm exp}=-4.53\times
10^{6}\, {\rm ergs}$ (in comparison $E_{\rm
tot}^{\rm dense}=1.76\times
10^{32}\, {\rm ergs}$).

For ULAs with $\delta=1.02\times
10^{-8}$, assuming $M_{\rm exp}=M_{\rm min}=2.15\times 10^{-3}$, we get $R_{\rm
exp}\sim 928$, $\rho_{\rm exp}=6.42\times 10^{-13}$, $2\pi/\omega_{\rm
exp}=3.83\times 10^{6}$ and $E_{\rm
tot}^{\rm exp}=-2.50\times
10^{-9}$ (in comparison $E_{\rm
tot}^{\rm dense}=733$). Coming back to original variables,
assuming $M_{\rm exp}=M_{\rm
min}=2.30\times
10^{5}\, M_{\odot}$, we get $R_{\rm exp}=290\, {\rm
kpc}$, $\rho_{\rm exp}=1.52\times 10^{-28}\, {\rm g/m^3}$,
$2\pi/\omega_{\rm exp}=5.93\times 10^{4}\, {\rm Gyrs}$ and
$E_{\rm tot}^{\rm exp}=-3.12\times
10^{45}\, {\rm ergs}$ (in comparison $E_{\rm
tot}^{\rm dense}=9.16\times
10^{56}\, {\rm ergs}$).

\subsection{Radiation}
\label{sec_radiation}

Since energy is conserved, the collapse of a dilute axion star above the
maximum mass towards a dense axion star ($|E_{\rm tot}^{\rm dense}|\ll |E_{\rm
tot}^{\rm dilute}|$) must be accompanied by the emission of a form of radiation
that will carry energy (and mass) away. This is similar to the process of
gravitational cooling \cite{seidel94} or to the situation investigated by Levkov
{\it et al.} \cite{tkachevprl}. The final state could be a dense axion star
(remnant) with a smaller energy (and usually a smaller mass) than the energy of
the dilute axion star, plus a radiation. The same situation occurs for the
explosion of dense axion stars below the minimum mass. 

{\it Remark:} We note that the Gaussian ansatz is not able to describe this
complex evolution. For $M>M_{\rm max}$, we find that the radius $R(t)$ of the
star permanently oscillates about the equilibrium state $R_{\rm coll}$, at
fixed energy. An improvement of the model would be to heuristically introduce a
source of dissipation (damping) in the GPP equations in order to describe the
relaxation
process as proposed in \cite{chavtotal}.

\section{Stability analysis and phase transitions in axion stars}
\label{sec_sapt}

In this section we discuss the stability of the different solutions found
previously and the nature of phase transitions between dilute axion stars and
dense axion stars. This description is similar to the description of phase
transitions in the self-gravitating Fermi gas at finite temperature given in
Ref.  \cite{ijmpb}.

\subsection{Series of equilibria and stability of the equilibrium
states}
\label{sec_soe}

\begin{figure}
\begin{center}
\includegraphics[clip,scale=0.3]{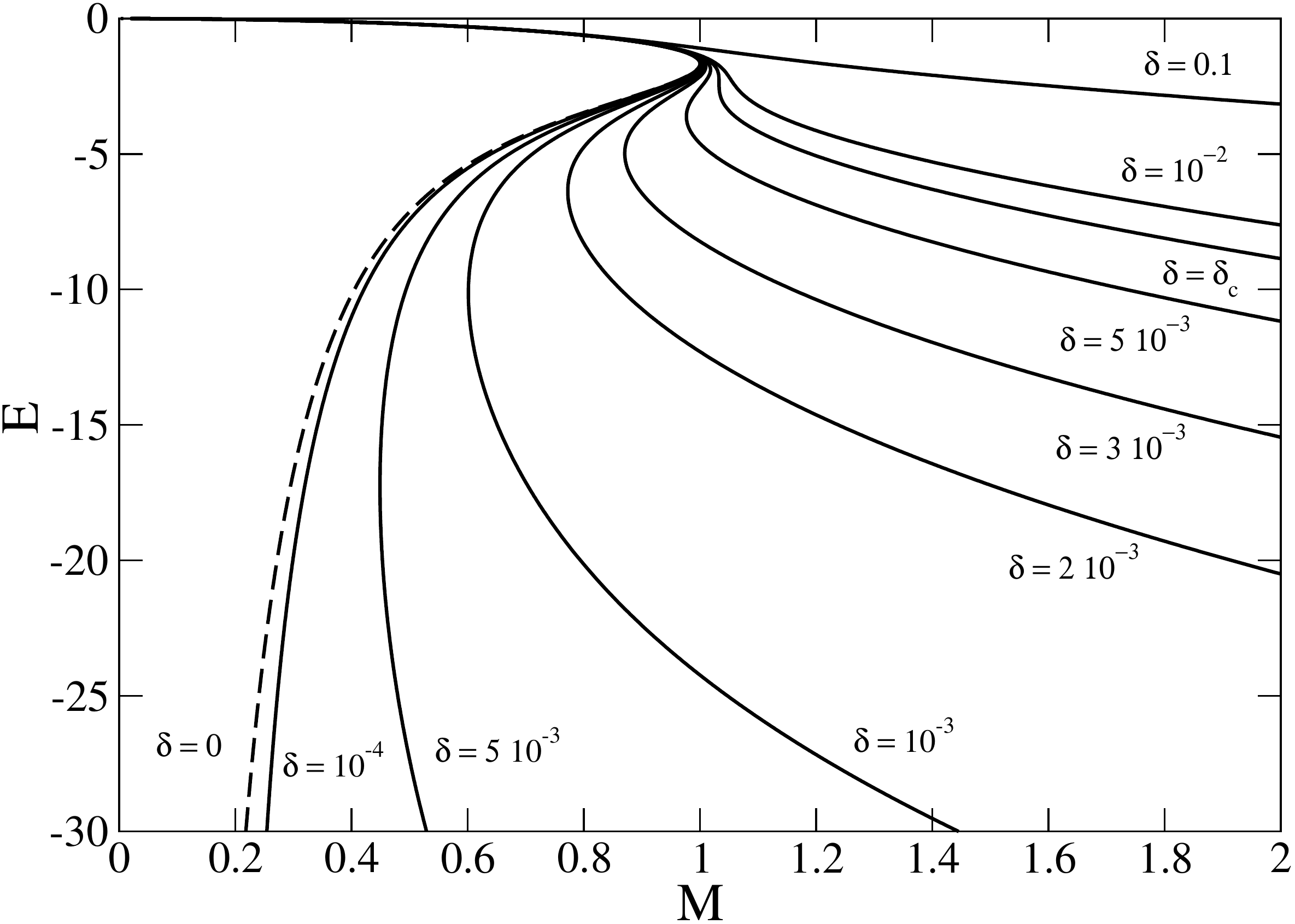}
\caption{Series of equilibria $E(M)$ for different values of the interaction
parameter $\delta$. The dashed line corresponds to dilute axion stars with
$\delta=0$.}
\label{multi}
\end{center}
\end{figure}

A stable axion star is a minimum of energy $E_{\rm tot}$ at fixed mass $M$. 
To determine the stability of axion stars, we can use the Poincar\'e theory
of linear series of equilibria (see Appendix \ref{sec_p}). 
The  variable conjugate to the mass $M$ (conserved quantity) with respect to
the energy $E_{\rm tot}$ (quantity to minimize) is the  eigenenergy
$E=\partial E_{\rm
tot}/\partial M$.\footnote{As before, we have introduced dimensionless
variables. We note that $E=NE/M$ where $NE$ is given by Eq. (\ref{vp1}).} To
apply the 
Poincar\'e theory to
the present situation, we have to plot the eigenenergy $E$ as a function of the
mass $M$  along the
series of
equilibria of axion stars.\footnote{As explained in
Appendix \ref{sec_p}, the correct series of equilibria to consider in
order to determine the stability of the solutions from the Poincar\'e theory and
study phase transitions is the eigenenergy-mass relation $E(M)$, not
the mass-radius relation $M(R)$. In particular, we cannot apply the Maxwell
constuction (see Appendix \ref{sec_m}) on the mass-radius relation
$M(R)$. The stability of the solutions can nevertheless be obtained from the
mass-radius relation by using Wheeler's theorem \cite{htww}.}
The series of equilibria $E(M)$ depends on the value of the interaction
parameter $\delta$ as shown in Fig. \ref{multi}.

\begin{figure}
\begin{center}
\includegraphics[clip,scale=0.3]{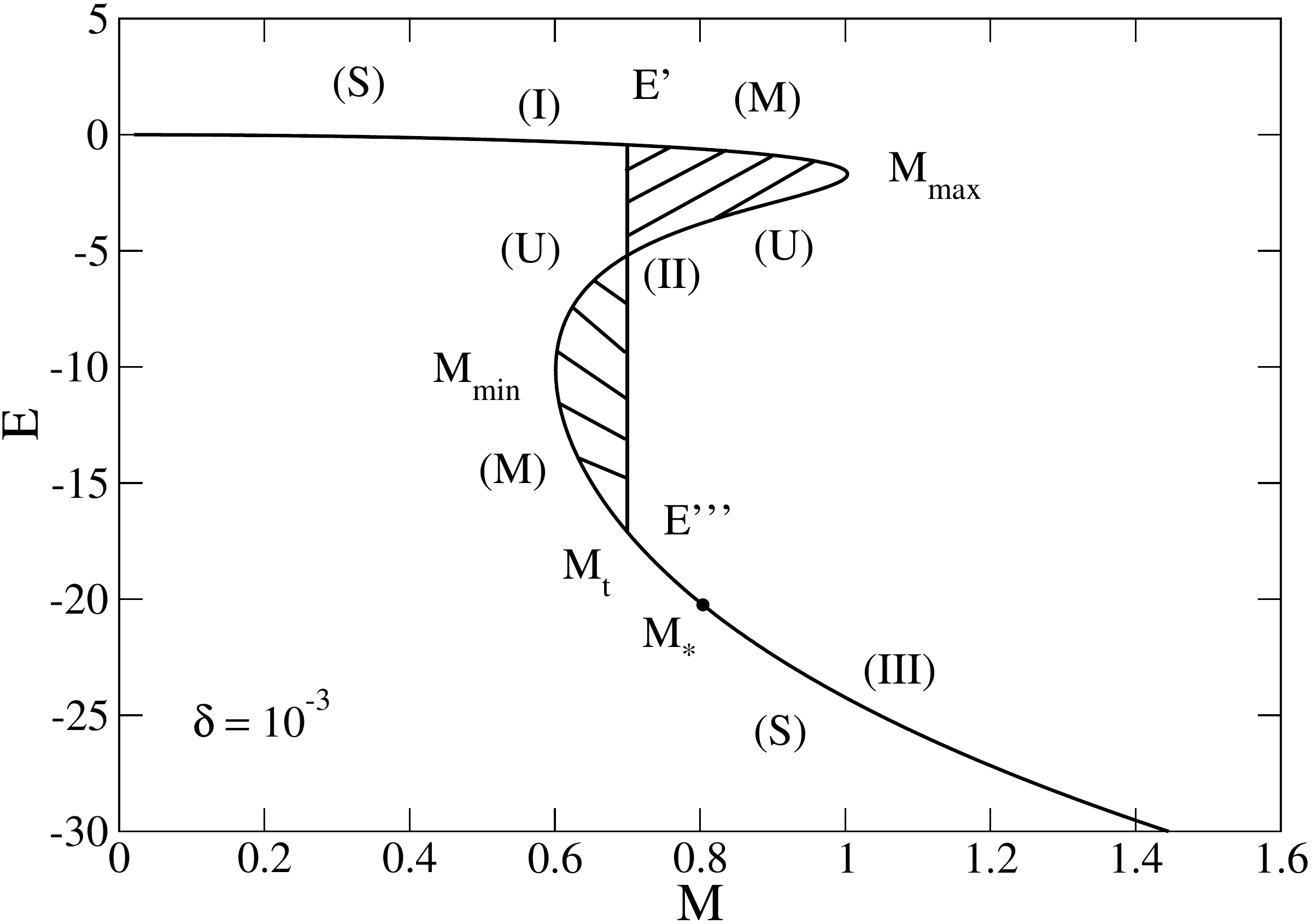}
\caption{Stability of the equilibrium states from the Poincar\'e theory of
linear series of equilibria. We have also performed the Maxwell
construction. The equilibrium states can be fully stable (S), metastable (M),
and unstable (U).}
\label{maxwell}
\end{center}
\end{figure}

In the following, for illustration, we shall consider the series of equilibria
$E(M)$ of Fig.
\ref{maxwell} corresponding to $\delta=10^{-3}$. This series of equilibria is
multi-valued. It has a $Z$-shape structure that gives rise to phase transitions.
The solutions on the upper branch (I) form the ``dilute phase''. They 
correspond to dilute axion stars. The solutions on the lower branch (III) form
the ``condensed phase''. They correspond to dense axion stars. As explained
below, the system has to cross a barrier of potential, played by the solutions
of the intermediate branch (II) corresponding to unstable axion stars, to pass
from the dilute phase to the
condensed phase (or inversely).

For $M\rightarrow 0$ and $E\rightarrow 0$, an axion
star is equivalent to a noninteracting Newtonian boson star in which
self-gravity is stabilized by the quantum potential (Heisenberg's
uncertainty principle) \cite{rb,membrado,prd1}. This structure is
known to be stable (it is a global minimum of energy $E_{\rm tot}$ at fixed mass
$M$). From the turning
point criterion, we conclude that the dilute axion stars on the branch (I) of
Fig. \ref{maxwell}  are  stable (energy minima) until
the first turning point of mass. At $M=M_{\rm max}(\delta)$, the
curve $E(M)$ rotates clockwise so that a mode of
stability is lost. Therefore, the axion stars on the intermediate branch (II) of
Fig. \ref{maxwell} are  unstable (maxima or saddle points of energy) until the
second
turning point of mass. They lie in the region where $dE/dM>0$ which
is inaccessible. This is the analogue of a region of negative
specific heats $dE/dT<0$ in thermodynamics \cite{ijmpb}. At $M=M_{\rm
min}(\delta)$, the
curve $E(M)$ rotates
anti-clockwise so that the mode of stability lost at $M=M_{\rm
max}(\delta)$ is regained. Therefore, the  dense
axion stars
on the branch (III) of
Fig. \ref{maxwell} are  stable.

{\it Remark:} We note that the turning point of radius at
($R_{\rm min},M_*)$ in the mass-radius relation $M(R)$ of Fig. \ref{mrQCD} does
not signal a change of stability as noted in Sec. \ref{sec_turning}.

\subsection{Fully stable and metastable states} \label{sec_small2}

We must now determine which configurations are fully stable (global minima of
energy)
and which configurations are metastable (local minima of energy). This can
be done by  plotting the energy $E_{\rm tot}$ of the two phases as a function of
the mass $M$ as done in Fig. \ref{metot}, comparing their
respective values, and determining at which mass 
$M_t(\delta)$ they
become equal. Alternatively, one can perform the Maxwell construction on the
curve $E(M)$ as described in Appendix \ref{sec_m}.  The Maxwell
construction is illustrated in Fig. \ref{maxwell}. The transition mass
$M_t(\delta)$
is such
that the two hatched areas are equal.

\begin{figure}
\begin{center}
\includegraphics[clip,scale=0.3]{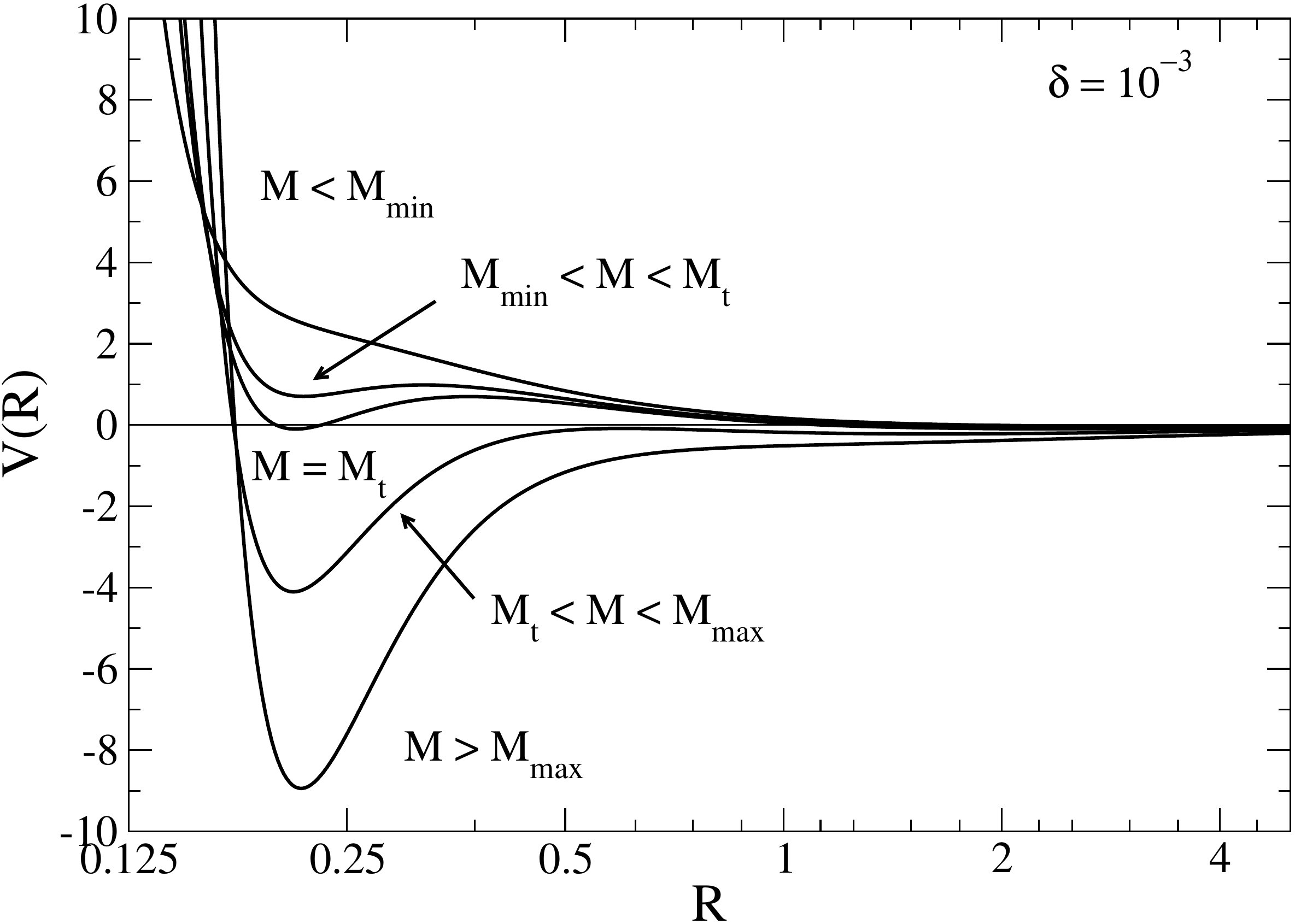}
\caption{Effective potential $V(R)$ as a function of the radius
for $\delta=10^{-3}$. For $M<M_{\rm min}=0.601$ (here $M=0.5$), the dilute axion
stars are fully stable and the dense axion stars are inexistent. For $M_{\rm
min}=0.601<M<M_t=0.700$ (here $M=0.65$), the dilute axion stars are fully stable
and the dense axion stars are metastable. For $M=M_t=0.700$, the dense axion
stars have the same energy as the dilute axion stars. For $M_{t}=0.700<M<M_{\rm
max}=1.00$ (here $M=0.9$), the dense axion stars are fully
stable and the dilute axion stars are metastable. For $M>M_{\rm max}=1.00$
(here $M=1.1$), the dense axion stars
are fully stable and the dilute axion stars are inexistent.}
\label{rvcomplet}
\end{center}
\end{figure}

When $M<M_{t}(\delta)$ the dilute axion stars (I) are fully stable (S) and
when $M_{t}(\delta)<M<M_{\rm max}(\delta)$ they are metastable (M).
Inversely, when
$M_{\rm min}(\delta)<M<M_{t}(\delta)$ the dense axion stars (III) are metastable
(M) 
and when $M>M_{t}(\delta)$ they are fully stable (S). The axion stars
on the branch (II) are unstable (U). These results are illustrated in Fig.
\ref{rvcomplet}. In a
strict sense, an equilibrium state
corresponds to a fully stable state (S).  Therefore, the
strict energy-mass and eigenenergy-mass relations are obtained from Figs.
\ref{metot} and \ref{maxwell} by keeping only the fully stable states (S).
From
these curves, we expect the occurrence of a first order phase
transition at $M=M_t(\delta)$ connecting the dilute phase
to the condensed phase.  It is accompanied by a discontinuity of
$E_{\rm tot}'(M)$  (the first derivative of $E_{\rm tot}(M)$)
at $M=M_t(\delta)$. Since $E_{\rm tot}'(M)=E(M)$,
this
corresponds to a discontinuity of the eigenenergy $E(M)$ at
$M_{t}(\delta)$. As a result,
the region where $dE/dM>0$ is replaced by a phase transition
(vertical
Maxwell plateau) that connects the dilute phase to the dense phase (see
Fig. \ref{maxwell}).

{\it Remark:} We can check in Fig. \ref{metot} that the stable branches
(I) and (III) have a lower energy $E_{\rm tot}$ than the unstable branch (II) as
it should, since a stable state corresponds to a (local) minimum of energy
$E_{\rm
tot}$ (see Figs. \ref{rv} and \ref{rvcomplet}). We also recall that a necessary
(but not sufficient)
condition
of nonlinear dynamical stability is that $E_{\rm tot}<0$. We can check in Fig.
\ref{metot} that the
configurations on the fully stable branches (S) corresponding to global energy
minima satisfy this condition, while this is
not
necessarily true for the configurations  on the metastable branches (M)
corresponding to local energy minima since they
are only linearly dynamically 
stable.\footnote{The stable dilute axion stars
always have a negative energy  while the unstable axion stars and the metastable
dense axion
stars may have negative or positive energies (see Sec.
\ref{sec_tev}). The reason why stable dilute axion stars have a negative
energy is due to the fact that $V(R)\rightarrow 0^{-}$ as
$R\rightarrow +\infty$ so the energy of the last minimum is necessarily
negative.} Finally, we note from the inspection of Fig.
\ref{maxwell} that
the eigenenergy $E$  of an unstable axion star (II) is higher
than the eigenenergy of a stable dense axion star (III) but lower than the
eigenenergy of a stable dilute axion star (I). There is no paradox, however,
since the eigenenergy $E$ represents the chemical potential $\mu$ (see Appendix
\ref{sec_p}), not the energy $E_{\rm
tot}$ that has to be a (local) minimum at equilibrium.

\subsection{Lifetime of metastable states}
\label{sec_lifetime}

The preceding discussion suggests the occurrence of a first order
 phase transition at $M_{t}(\delta)$. However, for systems with long-range
interactions such as self-gravitating systems, a first order phase transition
does not take place in practice. Indeed, the metastable states are long-lived
because the probability of a fluctuation
able to trigger the phase transition is extremely low. Indeed, to trigger a
phase transition from (I) to (III), or the converse, the system has to cross a
barrier of potential played by the
solutions on the intermediate
branch (II). This energetic barrier is illustrated in Figs.
\ref{rv} and \ref{rvcomplet}. For
$N\gg 1$, this barrier of potential rapidly increases as we depart
from the critical point $M_{\rm max}(\delta)$ and is usually very hard to cross.
Since the metastable
states are extremely robust, the first
order phase transition expected to occur at $M_t(\delta)$ does not take place in
practice. The system remains in the metastable phase past the
transition point. As a result, the physical  energy-mass and 
eigenenergy-mass relations must take metastable states into account. They are  
obtained from Figs. \ref{metot} and \ref{maxwell} by discarding only the
unstable states (U).

\begin{figure}
\begin{center}
\includegraphics[clip,scale=0.3]{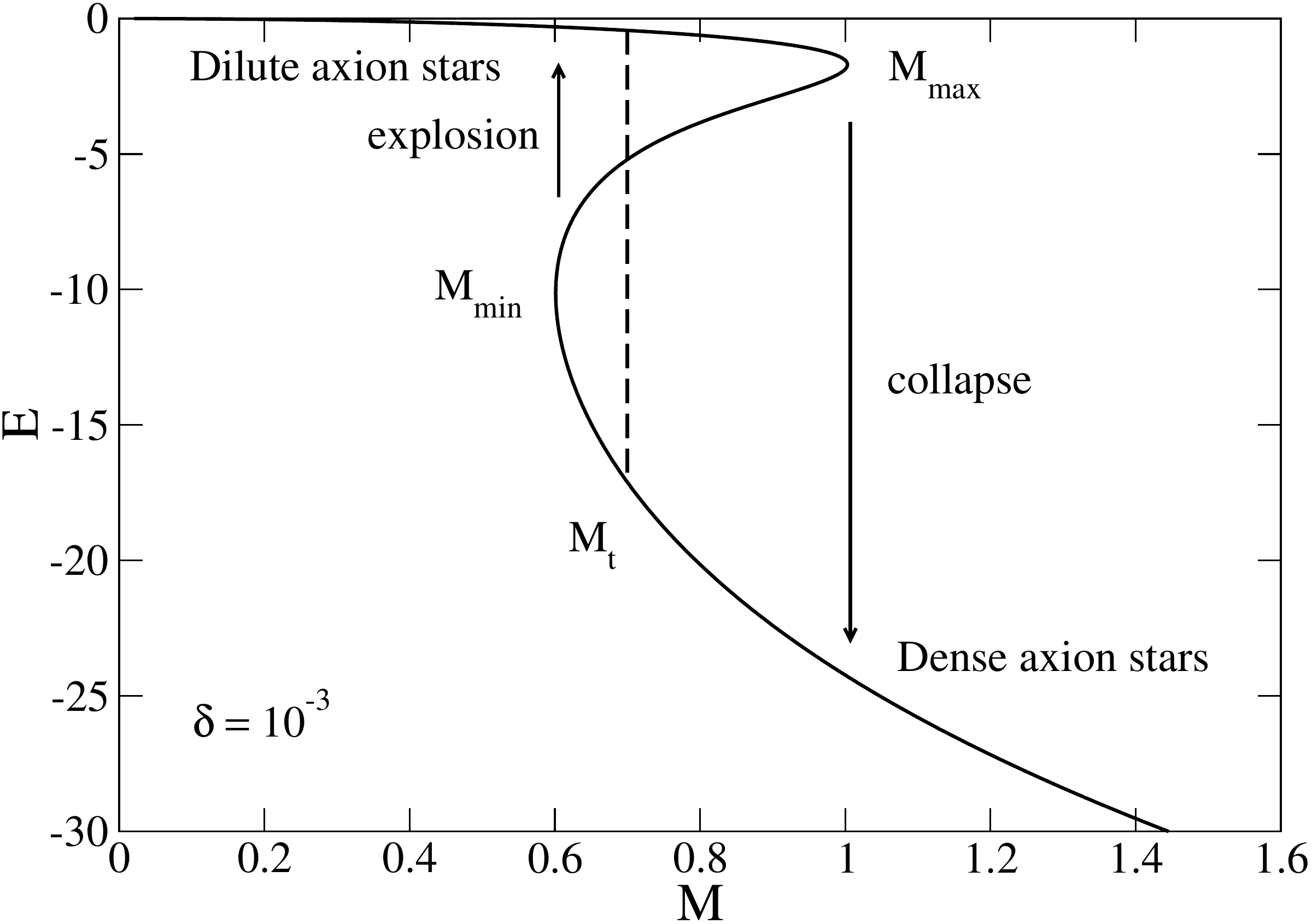}
\caption{Series of equilibria illustrating the collapse from a dilute axion
star to a dense axion star above $M_{\rm max}$ (see Sec. \ref{sec_coll}) or the
explosion from a dense
axion star to a dilute axion star below $M_{\rm min}$ (see Sec. \ref{sec_exp}).}
\label{summary}
\end{center}
\end{figure}

The true phase transition from dilute to dense axion stars occurs at the maximum
mass $M_{\rm max}(\delta)$ at which the metastable branch disappears (see
Fig. \ref{summary}). This point is similar to a 
spinodal point in thermodynamics. It also corresponds to a saddle node
bifurcation. If the mass of dilute axion stars increases above  $M_{\rm
max}(\delta)$ the system collapses due to the effect of self-gravity and the
attractive self-interaction of the bosons. However, when the repulsive
self-interaction is taken into account, the core of the system ceases to
shrink when it feels the repulsion of the bosons at sufficiently high densities,
and a dense axion star is
formed. Since this collapse is accompanied
by a discontinuous jump of energy $E_{\rm tot}$ (see Fig. \ref{metot}), this
is sometimes called a {\it zeroth order} phase transition.

For $M>M_t(\delta)$, the condensed states (dense axion stars) are fully stable
(S). If their mass decreases, the system remains
in the condensed phase past the transition mass $M_t(\delta)$. Indeed, for the
same reason as before, the condensed states with $M_{\rm
min}(\delta)<M<M_{t}(\delta)$ are
long-lived metastable states (M) so that the first order phase transition
from the dense phase to the dilute phase does not take
place in practice. However, below $M_{\rm min}(\delta)$ the condensed metastable
branch
disappears and the system undergoes a discontinuous transition
reversed to the collapse at $M_{\rm max}(\delta)$ (see Fig.
\ref{summary}). This
transition can be called an ``explosion'' since it transforms
 dense axion stars into dilute axion stars. Since the
collapse and the explosion occur at different values of mass,
due to the presence of metastable states, we can generate an {\it
hysteretic cycle} by varying the mass between $M_{\rm min}(\delta)$ and
$M_{\rm max}(\delta)$.

\subsection{The evolution of unstable equilibrium states}
\label{sec_unstable}

It is also interesting to determine the evolution of a slightly
perturbed unstable (U) equilibrium state. This evolution
will depend on the sign of its energy $E_{\rm tot}$ as determined in
Sec. \ref{sec_tev}. Recall that for $\delta<\delta_c$ and $M_{\rm
min}(\delta)<M<M_{\rm max}(\delta)$ the effective potential $V(R)$ has three
extrema (see Fig. \ref{rv}): a minimum corresponding to a
stable dense axion star
of small radius (left), a  maximum corresponding to an unstable axion star of
intermediate radius (middle), and another minimum corresponding to a stable
dilute axion
star of large radius (right). When $\delta'_c<\delta<\delta_c$
or when $\delta<\delta'_c$ and $M_{\rm E,II}(\delta)<M<M_{\rm
max}(\delta)$ the unstable axion star has a
negative total energy $E_{\rm tot}<0$ (see Fig. \ref{deltamzeroenergytotal}).
As a result, when slightly perturbed,
it either
relaxes towards a  stable dense axion star with a smaller radius or towards a
stable
dilute axion star with a larger radius (assuming an efficient dissipation
process
as explained in Sec. \ref{sec_radiation}). When
$\delta<\delta'_c$ and $M_{\rm
min}(\delta)<M<M_{\rm E,II}(\delta)$ the unstable axion star has a positive
total
energy  $E_{\rm tot}>0$ (see Fig. \ref{deltamzeroenergytotal}). As a
result, when slightly perturbed, it either
relaxes towards a dense axion star with a smaller radius (again assuming
dissipation) or explodes and disperses to infinity.

\subsection{Critical points and phase diagram}
\label{sec_cpt}

\begin{figure}
\begin{center}
\includegraphics[clip,scale=0.3]{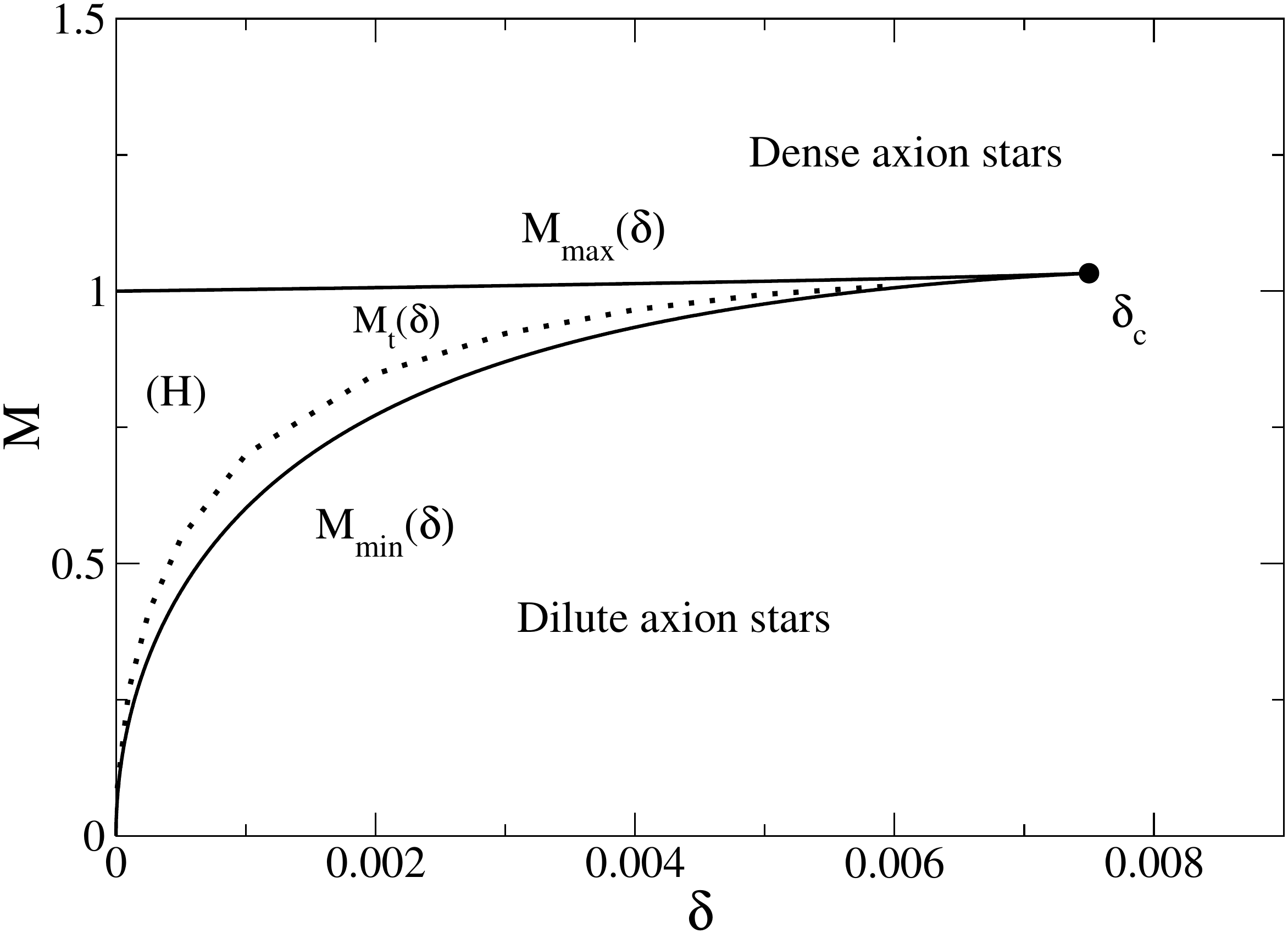}
\caption{Phase diagram of Newtonian axion stars. The $H$-zone corresponds to an
hysteretic zone where the actual phase depends on the history of the system. If
it is initially prepared in a dilute state, it will remain dilute until the
maximum mass $M_{\rm max}$ at which it will collapse and become dense.
Inversely, if the system  is initially prepared in a dense state, it will
remain dense until the
minimum mass $M_{\rm min}$ at which it will explode and become dilute.
We show in Appendix \ref{sec_abtm} that the transition mass
$M_{t}(\delta)$, which is by definition in the interval  $[M_{\rm
min}(\delta),M_{\rm max}(\delta)]$, approaches the mass 
$M_{*}(\delta)$ corresponding to the minimum radius as $\delta\rightarrow 0$.
We recall, however, that the
transition mass $M_{t}(\delta)$ is usually
not physically relevant because of the very long lifetime of metastable states
(see main text).}
\label{phasediagram}
\end{center}
\end{figure}

The deformation of the series of equilibria  $E(M)$ as a function of the
parameter $\delta$ is represented in Fig. \ref{multi}.

For $\delta=0$,
we recover the mass-eigenenergy relation of dilute axion stars
\cite{bectcoll} which presents a
maximum mass $M_{\rm max}=1$  that separates the branch of
stable dilute axion stars (I) and the branch of unstable axion stars (II). For
$M>M_{\rm max}=1$ the system collapses and is expected to form a Dirac peak (see
Appendix \ref{sec_dpbh}).

For $0<\delta<\delta_c$, a minimum mass $M_{\rm min}(\delta)$ exists in
addition to the maximum mass $M_{\rm max}(\delta)$. In that case, a
new branch appears
which corresponds to stable dense axion stars (III). The curve $E(M)$ becomes
multi-valued ($Z$-shape) so that
a  first order phase transition between dilute and dense axion stars is
expected to occur. If we keep only fully stable states (S), the $Z$-curve has to
be
replaced by a vertical Maxwell plateau. The extent of the
plateau decreases as $\delta$ increases towards $\delta_c$.

At the critical point $\delta_c$, the plateau disappears and the
series of
equilibria $E(M)$ presents an inflexion point. At that point, $dE/dM$ (the
equivalent of the specific heats in thermodynamics) is infinite.

For
$\delta>\delta_c$, the curve $E(M)$ is univalued so there is no phase
transition.  There is only one branch
of stable axion stars.

In summary, for $\delta=0$ the system exhibits a collapse
towards a Dirac peak at $M_{\rm max}=1$; for $0<\delta<\delta_c$ the system
exhibits a
first order phase transition at $M_t(\delta)$ characterized by a
vertical Maxwell plateau; for
$\delta>\delta_c$ there is no phase transition. We recall, however, that because
of the
presence of
long-lived
metastable states, the first order phase transition and the Maxwell plateau are
usually not physically relevant. Only the zeroth order phase
transitions that occur at $M_{\rm max}(\delta)$ and $M_{\rm min}(\delta)$ 
(spinodal points) are physically relevant.

The phase diagram of axion stars can be directly deduced from the properties of
the series of equilibria by identifying characteristic masses. We note $M_{\rm
max}(\delta)$ the end point of the metastable dilute phase (first
turning point of mass), $M_{\rm min}(\delta)$ the end point of the
metastable condensed phase (last turning point of mass), and
$M_{t}(\delta)$ the mass of transition determined by the
equality of the energies of the two phases.  The phase diagram is
represented in Fig. \ref{phasediagram}. It displays in
particular the critical point $\delta_c=3/400=0.0075$ at which the phase
transitions appear/disappear.

\subsection{The dilute limit $\delta\rightarrow 0$}
\label{sec_muinf}

It is of interest to discuss the limit $\delta\rightarrow 0$ specifically
so as to make the connection with the
results obtained for dilute axion stars ($\delta=0$) in \cite{bectcoll} when
there is no self-repulsion. For
$\delta\rightarrow 0$, the
transition mass $M_{t}(\delta)\rightarrow 0$ so that
the dilute axion stars on the branch (I)  are metastable (M)
while the dense axion stars on the branch (III) are fully stable (S).
However, the branch (III) is rejected at infinity ($E\rightarrow
-\infty$). It is made of Dirac
peaks (see Appendix \ref{sec_dpbh}). Therefore, the $\delta\rightarrow 0$ limit
of the
series of equilibria
$E(M)$ (Fig. \ref{multi}) is formed by
the metastable branch (I) and the unstable branch (II)  of the series of
equilibria $E(M)$ of dilute axion stars with $\delta=0$ plus a singular
stable branch (III)  at $E=-\infty$ made of Dirac peaks. Similarly, the
$\delta\rightarrow 0$ limit of the mass-radius relation 
$M(R)$ (Fig. \ref{mr}) is formed by
the metastable branch (I) and the unstable branch (II)  of the mass-radius
relation  $M(R)$ of dilute axion stars with $\delta=0$ plus a singular
stable branch (III) at $R=0$ made of Dirac peaks.

{\it Remark:} For $\delta\rightarrow 0$ but $\delta\neq 0$, the Dirac peaks on
the
singular branch (III)  are {\it not} black holes whatever the mass $M$ since the
typical mass at which general relativistic effects become important, of the
order of $M_0(\delta)$, tends to infinity for $\delta\rightarrow 0$ (see
Appendix \ref{sec_dpbh}).

\section{Qualitative description of general relativistic effects}
\label{sec_qgr}

\begin{figure}
\begin{center}
\includegraphics[clip,scale=0.3]{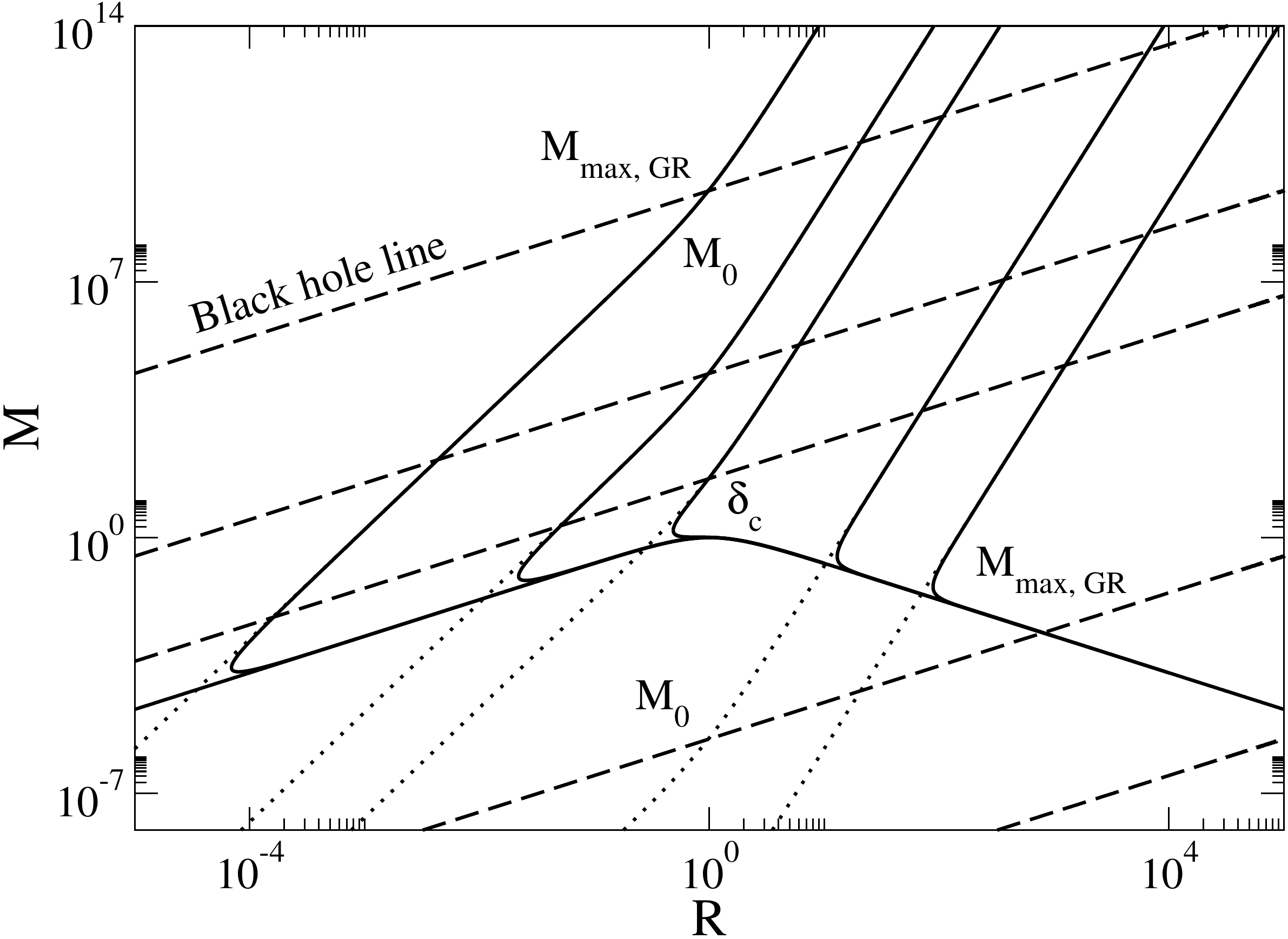}
\caption{Mass-radius relationship of Newtonian axion stars for different values
of the interaction parameter ($\delta=10^{-10}, 10^{-5}, \delta_c,
10^5, 10^{10}$). The intersection between the $M(R)$ curve and
the black hole line $M=Rc^2/2G$ ($M=0.310 R/\delta$ in
dimensionless form) determines the limit of validity of the Newtonian treatment
and the order of magnitude of the general relativistic maximum mass of axion
stars. The
mass-radius relation (solid line) and the black hole line (dashed line)
corresponding to the same $\delta$ can be easily identified by using the fact
that the black hole line   and the TF curve intersect each other at
$M\sim M_0$
(i.e. when the TF curve
presents a change of slope).
}
\label{mravecbh}
\end{center}
\end{figure}

Our Newtonian treatment is valid if the radius of an axion star of mass $M$
satisfies the condition $R\gg R_S=2GM/c^2$, where $R_S$ is the Schwarzschild
radius. In terms of dimensionless variables, this condition can be rewritten as
$R\gg R_S$ with $R_S=(81\zeta/4\nu)M\delta=3.23M\delta$ or,
equivalently, as
\begin{equation}
\label{qgr1}
M\ll \frac{4\nu}{81\zeta}\frac{R}{\delta}=0.310 \frac{R}{\delta} \qquad ({\rm
Newtonian}).
\end{equation}
When this condition is not satisfied, i.e. when $R\sim R_S$ or, in
dimensionless form, when
\begin{equation}
\label{qgr2}
M\sim \frac{4\nu}{81\zeta}\frac{R}{\delta}=0.310 \frac{R}{\delta}  \qquad ({\rm
GR}),
\end{equation}
general relativistic effects become important. Generalizing the arguments
developed in Appendix B of \cite{prd1}, the combination of the Newtonian
mass-radius relation $M(R)$ given by Eq. (\ref{mr4}) and the
relativistic criterion (\ref{qgr2}), sometimes called the black hole
line, gives a
qualitative  estimate of the general relativistic maximum mass $M_{\rm max,GR}$ 
of an axion star (see Fig. \ref{mravecbh} for an illustration). Although
this method can only give an order of magnitude of this maximum
mass,\footnote{It captures, however, the correct scaling of the maximum mass.}
we shall keep all the prefactors from our
analysis for future comparison.

\subsection{The general relativistic maximum mass of dilute axion stars}
\label{sec_grd}

Let us first determine the general relativistic maximum mass of dilute axion
stars.
Considering the
possible intersection between the stable part ($R>1$) of the Newtonian
mass-radius relation (\ref{mrd1}) and the black hole line 
(\ref{qgr2}), we obtain
\begin{equation}
\label{qgr3}
M_{\rm max,GR}^{\rm
dilute}=\frac{\delta_*}{\delta}\sqrt{\frac{2\delta}{\delta_*}-1},
\end{equation}
\begin{equation}
\label{qgr4}
R_{\rm *,GR}^{\rm
dilute}=\sqrt{\frac{2\delta}{\delta_*}-1},
\end{equation}
provided that $\delta\ge \delta_*$ with
\begin{equation}
\label{qgr5}
\delta_*=\frac{4\nu}{81\zeta}=0.310.
\end{equation}

These results lead to the following conclusions:

(i) If $\delta<\delta_*$ (strongly self-interacting axions), there is no
intersection. Therefore, the
Newtonian
treatment of dilute axion stars is always valid. In that case, the
maximum mass and the minimum radius of dilute axion stars are given by $M_{\rm
max}=1$ and $R_{*}=1$. We shall write them as
$M_{\rm max,N}^{\rm dilute}$ and $R_{\rm *,N}^{\rm dilute}$ to emphasize that
they have a nonrelativistic
(Newtonian) origin. We recall that $\delta\ll 1$ for QCD axions and typical
ULAs (such as those considered in Sec. \ref{sec_na}) so the Newtonian
approximation describes dilute axion stars very well.

\begin{figure}
\begin{center}
\includegraphics[clip,scale=0.3]{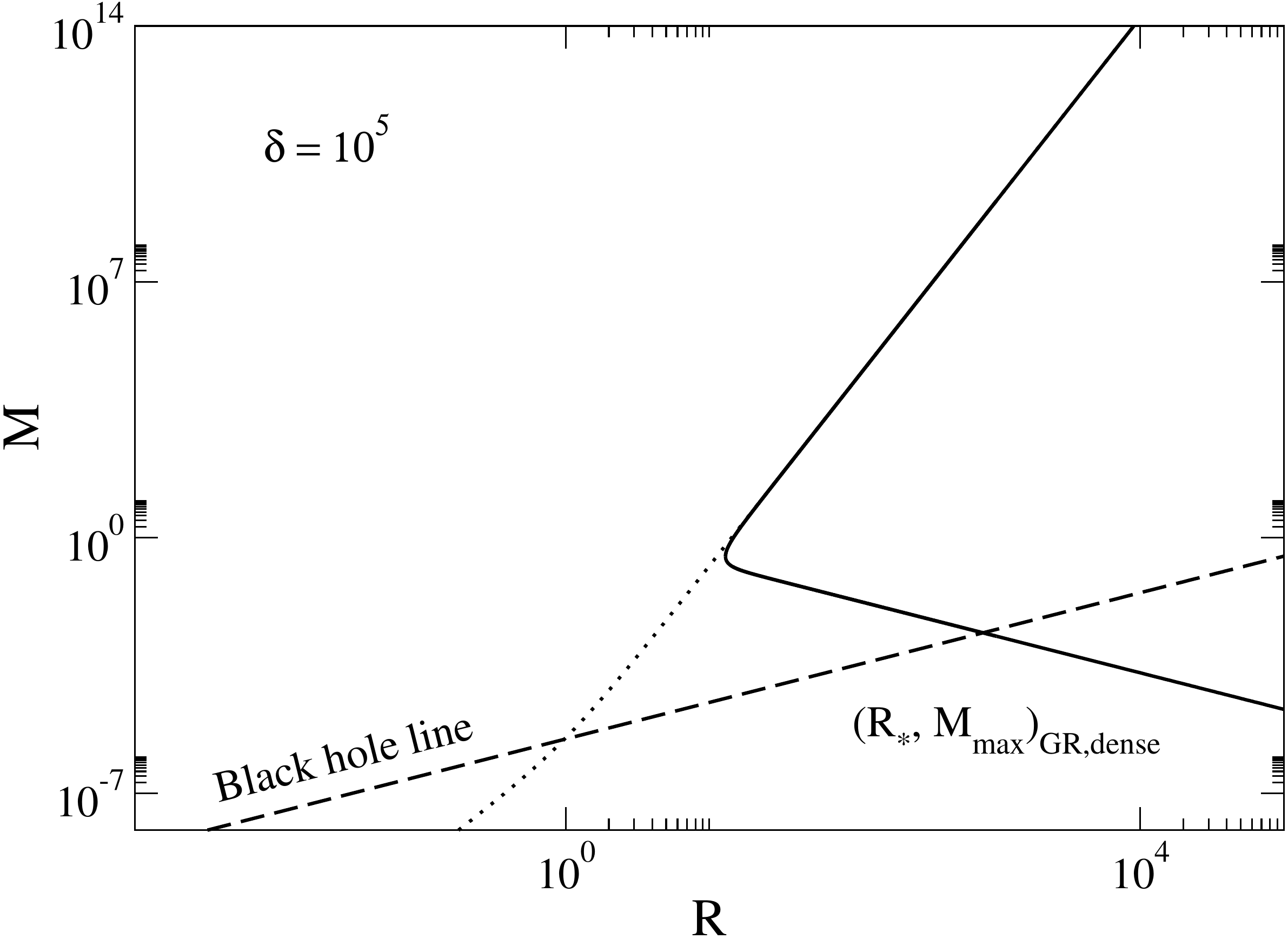}
\caption{Mass-radius relation of weakly self-interacting axion stars with
$\delta>\delta_*$ (for illustration we have taken
$\delta=10^5$). The Newtonian relation is only valid for $M\ll M_{\rm
max,GR}^{\rm
dilute}$ and $R\gg R_{\rm
*,GR}^{\rm dilute}$. Close to  $M_{\rm max,GR}^{\rm dilute}$, the real
(relativistic) mass-radius relation $M(R)$ should form a spiral like in the case
of noninteracting boson stars (see Appendix \ref{sec_cs}).}
\label{ex2}
\end{center}
\end{figure}

(ii) If $\delta>\delta_*$ (weakly self-interacting axions), there is one
intersection (see Fig. \ref{ex2}). In that case, the
Newtonian treatment is valid for $M\ll M_{\rm max,GR}^{\rm dilute}$ and
$R\gg R_{\rm *,GR}^{\rm dilute}$ but
ceases to be valid as we approach these values.\footnote{Note that the axion
stars become denser and denser when $M\rightarrow M_{\rm max,GR}^{\rm dilute}$
so they should not be called dilute axion stars anymore. However, we keep the
notation $M_{\rm max,GR}^{\rm dilute}$ to signify that it corresponds to the
general relativistic maximum mass of the branch of dilute axion stars.}. When
general relativity
is
taken into account, we expect that the mass-radius relation close to ($M_{\rm
max,GR}^{\rm
dilute}, R_{\rm *,GR}^{\rm dilute}$) forms a spiral (see below).
Above $M_{\rm max,GR}^{\rm dilute}$
the system is expected to collapse into a black hole.

If we come back to dimensional variables, we
obtain\footnote{These results are in agreement with previous estimates made in
Appendix B5 of
\cite{prd1} and in footnote 18 of \cite{bectcoll}.}
\begin{equation}
\label{qgr6}
M_{\rm max,GR}^{\rm
dilute}=\left
(\frac{\sigma}{2\nu}\right
)^{1/2}\sqrt{2-\frac{6\pi\zeta}{2\sigma}\frac{2|a_s|}{r_S}}\, \frac{M_P^2}{m},
\end{equation}
\begin{equation}
\label{qgr7}
R_{\rm *,GR}^{\rm
dilute}=\left
(\frac{2\sigma}{\nu}\right
)^{1/2}\sqrt{2-\frac{6\pi\zeta}{2\sigma}\frac{2|a_s|}{r_S}}\, \lambda_C,
\end{equation}
provided that $|a_s|\le |a_s|_*$ with
\begin{equation}
\label{qgr8}
|a_s|_*=\frac{\sigma}{6\pi\zeta}r_S=0.627\, r_S.
\end{equation}

For $a_s=0$ (noninteracting limit), we obtain
\begin{equation}
\label{qgr9}
M_{\rm max,GR}^{\rm
dilute}=\left
(\frac{\sigma}{\nu}\right
)^{1/2}\frac{M_P^2}{m}=1.37 M_c,
\end{equation}
\begin{equation}
\label{qgr10}
R_{\rm *,GR}^{\rm
dilute}=2\left
(\frac{\sigma}{\nu}\right
)^{1/2}\lambda_C=2.74\, R_c,
\end{equation}
where the scales $M_c$ and $R_c$ are defined in Appendix
\ref{sec_cs}. Our approximate relativistic treatment recovers the Kaup scales of
noninteracting boson stars. We
know that the mass-radius relation of general relativistic noninteracting boson
stars forms a spiral close to the maximum mass (the spiral being made of
unstable states), and we expect this
property to be preserved for weakly self-interacting axions with $|a_s|\le
|a_s|_*$.

For $|a_s|\ge |a_s|_*$, the maximum mass
and the minimum radius of dilute axion stars are given by Eqs. (\ref{cmmm1}) and
(\ref{cmmm2}). They can be written as
\begin{equation}
\label{qgr11}
M_{\rm max,N}^{\rm
dilute}=\left
(\frac{\sigma^2}{6\pi\zeta\nu}\right
)^{1/2}\left (\frac{r_S}{2|a_s|}\right )^{1/2}\frac{M_P^2}{m}=1.085 M_a,
\end{equation}
\begin{equation}
\label{qgr12}
R_{\rm *,N}^{\rm
dilute}=\left
(\frac{6\pi\zeta}{\nu}\right
)^{1/2}\left (\frac{2|a_s|}{r_S}\right )^{1/2}\lambda_C=1.732 R_a.
\end{equation}

We note from Eqs. (\ref{qgr6})-(\ref{qgr10}) that the general relativistic
maximum
mass $M_{\rm max,GR}^{\rm
dilute}$ and the general relativistic minimum radius $R_{\rm *,GR}^{\rm
dilute}$ of dilute axion stars change only weakly
with the  self-interaction for $|a_s|\le |a_s|_*$. By contrast, for $|a_s|\ge
|a_s|_*$, the Newtonian maximum mass $M_{\rm max,N}^{\rm dilute}$ decrease as
$|a_s|^{-1/2}$ and the Newtonian minimum radius $R_{\rm *,N}^{\rm dilute}$
increases as  $|a_s|^{1/2}$. At the transition $|a_s|=|a_s|_*$, we find that
\begin{equation}
\label{qgr11b}
M_{\rm max,N}^{\rm
dilute}=M_{\rm max,GR}^{\rm
dilute}=\left
(\frac{\sigma}{2\nu}\right
)^{1/2}\frac{M_P^2}{m}=0.969\, M_c,
\end{equation}
\begin{equation}
\label{qgr12b}
R_{\rm *,N}^{\rm
dilute}=R_{\rm *,GR}^{\rm
dilute}=\left
(\frac{2\sigma}{\nu}\right
)^{1/2}\lambda_C=1.94\, R_c.
\end{equation}

We note that the value of the scattering length $|a_s|_*$ that separates the
weakly self-interacting regime from the strongly self-interacting regime is of 
the order of the effective Schwarzschild radius $r_S$ of the axion. In Ref.
\cite{abrilphas}, we have
obtained a similar result in a cosmological
context. For QCD axions with $m=10^{-4}\,
{\rm eV}/c^2$ and
$|a_s|=5.8\times 10^{-53}\, {\rm m}$, we find
$|a_s|_*=1.66\times 10^{-67}\, {\rm m}$. For ULAs with $m=2.19\times 10^{-22}\,
{\rm eV}/c^2$ and $|a_s|=1.11\times 10^{-62}\, {\rm fm}$, we find
$|a_s|_*=3.64\times 10^{-85}\, {\rm m}$. Therefore, these types of axions are in
the strongly
self-interacting regime since $|a_s|\gg |a_s|_*$.

The results from Eqs. (\ref{qgr6})-(\ref{qgr12b}) can be expressed equivalently
in terms of $f$ and $\lambda$ by using Eq. (\ref{val4}). The critical
values of $f$ and $\lambda$ separating the  weakly
self-interacting regime from the strongly self-interacting regime are
\begin{equation}
\label{qgr13}
f_*=\left (\frac{6\pi\zeta}{64\pi\sigma}\right )^{1/2}M_Pc^2=8.91\times
10^{-2}M_Pc^2,
\end{equation}
\begin{equation}
\label{qgr14}
|\lambda|_*=\frac{16\pi\sigma}{6\pi\zeta}\left (\frac{m}{M_P}\right )^2=31.5
\left (\frac{m}{M_P}\right )^2.
\end{equation}
We note that $f_*$ is independent of the mass of the axion and is of
the order of $0.1 M_P c^2$. On the other hand,
the value of the
dimensionless self-interaction constant $|\lambda|_*$ separating the weakly
self-interacting regime from the strongly  self-interacting regime is 
$|\lambda|_*=2.11\times 10^{-63}$ for QCD axions with mass $m=10^{-4}\,
{\rm eV}/c^2$  and $|\lambda|_*=1.01\times 10^{-98}$ for ULAs with mass
$m=2.19\times 10^{-22}\, {\rm eV}/c^2$. Since QCD axions have
$\lambda=-7.39\times 10^{-49}$  and ULAs have $\lambda=-3.10\times
10^{-91}$, they are in the strongly self-interacting regime although
their values of $|\lambda|$ may seem very small at first sight. This is because
the critical parameter  $|\lambda|_*$ scales as $(m/M_P)^2$ which is an
extraordinarily small quantity.\footnote{In particle
physics and quantum field theory, one usually
considers that the
self-interaction is weak when $|\lambda|\ll 1$. In our situation, in which
gravity is in action, the
self-interaction is weak when 
$|\lambda|\ll (m/M_P)^2$, which is a very different criterion since the
ratio $m/M_P$ is extremely small (see Appendix \ref{sec_ws}). It is therefore
relevant to introduce a new
dimensionless self-interaction constant $\Lambda=\lambda (M_P/m)^2$ adapted to
our problem so that the weakly self-interacting
regime 
corresponds to $|\Lambda|\ll 1$ and the strongly self-interacting regime
corresponds to $|\Lambda|\gg 1$. We stress that the conditions $|\lambda|\ll 1$
and $|\Lambda|\ll 1$ have a very different physical meaning. We can be in the
strongly self-interacting  regime
$|\Lambda|\gg 1$ for our problem, even if we
are in the weakly self-interacting regime of quantum field theory $|\lambda|\ll
1$ (we have reached a similar conclusion in Sec. III.H. of \cite{abrilphas} in a
cosmological
context).}
It is therefore crucial to take the self-interaction of the axions into account
(see the Remark at the end of Sec. \ref{sec_na2}), although it is oftentimes
neglected \cite{ch2,marsh,hui}.

If we consider QCD axions with mass $m=10^{-4}\,
{\rm eV}/c^2$ and neglect their self-interaction (taking $a_s=0$), we obtain 
$M_{\rm max, GR}^{\rm dilute}=1.83\times 10^{-6}\, M_{\odot}$ and $R_{*,GR}^{\rm
dilute}=0.540\, {\rm cm}$. The maximum mass set by general relativity $M_{\rm
max, GR}^{\rm dilute}=1.83\times 10^{-6}\, M_{\odot}$ is much larger than the
Newtonian maximum mass $M_{\rm max,N}^{\rm dilute}=6.92\times 10^{-14}
M_{\odot}$ obtained when the attractive self-interaction of the axions
 is taken into account ($a_s=-5.8\times 10^{-53}\, {\rm m}$).

If we consider ULAs with mass $m=2.19\times 10^{-22}\, {\rm
eV}/c^2$
and neglect their self-interaction (taking $a_s=0$), we obtain $M_{\rm max,
GR}^{\rm dilute}=8.36\times
10^{11}\, M_{\odot}$ and $R_{\rm *,GR}^{\rm dilute}=0.0800\,
{\rm pc}$. The maximum mass set by general relativity $M_{\rm max,
GR}^{\rm dilute}=8.36\times
10^{11}\, M_{\odot}$ is much larger than the
Newtonian maximum mass $M_{\rm max,N}^{\rm dilute}=1.07\times 10^{8}
M_{\odot}$ obtained when the attractive self-interaction of the axions
 is taken into account ($a_s=-1.11\times
10^{-62}\, {\rm fm}$).

In this sense, an attractive self-interaction substantially
reduces the maximum mass of axion stars and make it nonrelativistic instead of
relativistic.

\subsection{The general relativistic maximum mass of dense axion stars}
\label{sec_fri}

We now consider the general relativistic maximum mass of dense axion stars.
Considering the
intersection between the Newtonian mass-radius relation (\ref{tf1}) in the TF
limit and the black hole line (\ref{qgr2}) we obtain
\begin{equation}
\label{qgr15}
M_{\rm max,GR}^{\rm
dense} =   \frac{\sqrt{-1+\sqrt{1+24\delta_*}}}{\sqrt{2}}  
\frac{\delta_*}{\delta}        =0.976\frac{\delta_*}{\delta},
\end{equation}
\begin{equation}
\label{qgr16}
R_{\rm *,GR}^{\rm
dense}=\frac{\sqrt{-1+\sqrt{1+24\delta_*}}}{\sqrt{2}}=0.976...
\end{equation}

\begin{figure}
\begin{center}
\includegraphics[clip,scale=0.3]{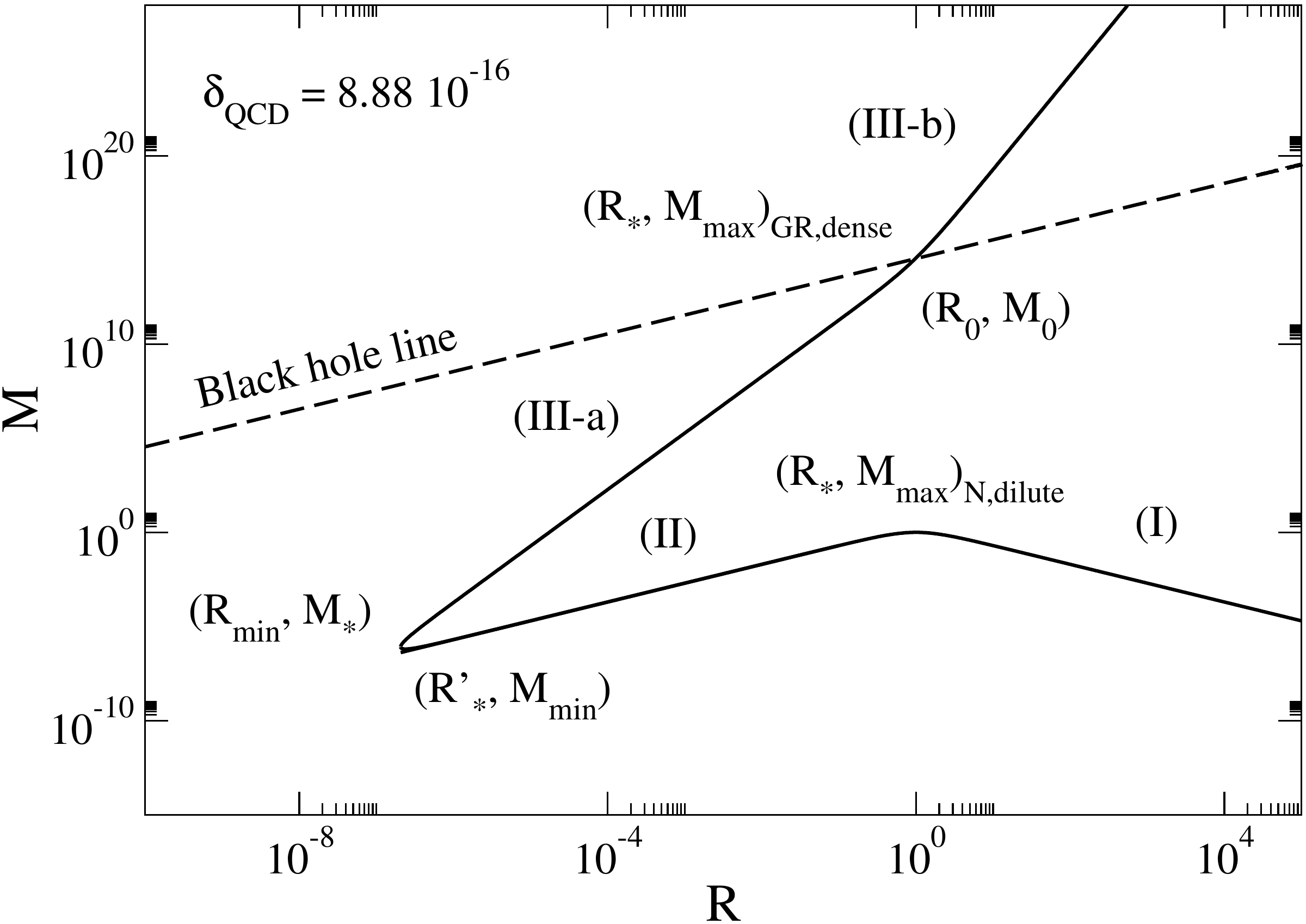}
\caption{Mass-radius relation of strongly self-interacting axion stars with
$\delta<\delta_*$ (for illustration we have taken
$\delta_{\rm QCD}=8.88\times 10^{-16}$). The Newtonian relation is only valid
for $M<M_{\rm max,GR}^{\rm dense}$. Close to  $M_{\rm max,GR}^{\rm
dense}$, the exact
(general relativistic) mass-radius relation $M(R)$ may form a spiral or stop
suddently
(see footnote 25).}
\label{ex1}
\end{center}
\end{figure}

These results lead to the following conclusions:

(i) If $\delta>\delta_*$ (weakly self-interacting axions)  there is no
intersection in the physical part of the TF curve  (the part that correctly
approximates the Newtonian mass-radius relation of dense axion stars) because
$R_{\rm *,GR}^{\rm
dense}<R_{\rm min}$. In that case, the maximum mass is $M_{\rm max,GR}^{\rm
dilute}$, the general relativistic maximum mass of dilute axion stars, and we
are led back to the discussion of point (ii) in Sec.
\ref{sec_grd}. Above $M_{\rm max,GR}^{\rm
dilute}$, the collapse of a dilute axion star is expected to form a black hole,
not a dense axion star.

(ii)  If $\delta<\delta_*$ (strongly self-interacting axions) there is one
intersection in the physical part of the TF curve since  $R_{\rm *,GR}^{\rm
dense}>R_{\rm min}$ (see Fig. \ref{ex1}). We can now refine the discussion of
point (i) in Sec. \ref{sec_grd}. If
$M<M_{\rm max,N}^{\rm dilute}=1$ and $R>R_{*,N}^{\rm dilute}=1$ the dilute
axion stars are stable and they can be treated by Newtonian gravity.  If
$M_{\rm max,N}^{\rm
dilute}<M<M_{\rm max,GR}^{\rm
dense}$, the system collapses into a dense axion star which can
again be treated by Newtonian gravity. If $M>M_{\rm
max,GR}^{\rm dense}$ the system collapses into a black hole. When general
relativity is taken into account, we
expect that the mass-radius relation close to ($M_{\rm max,GR}^{\rm
dense}, R_{\rm *,GR}^{\rm dense}$) forms a spiral or stops
suddently.\footnote{If the dense axion stars are described by a
polytropic equation
of state, as in Appendix \ref{sec_pol}, the relativistic
mass-radius relation is expected to form a spiral because polytropic spheres
become unstable above a maximum central density corresponding to the maximum
mass point. Alternatively, if the dense axion stars are
described by uniform density spheres, as in Appendix \ref{sec_tfng},
the relativistic mass-radius relation is expected to stop suddently at the point
of maximum
mass without forming a spiral because uniform density spheres, when they exist,
are always
stable. We will study this problem further in a specific paper.}
We recall that $\delta\ll 1$ for QCD axions
and typical
ULAs (such as those considered in Sec. \ref{sec_na}) so that $M_{\rm
max,GR}^{\rm
dense}\gg 1$ according to Eq. (\ref{qgr15}). Therefore, there is a wide range
of masses $[M_{\rm
max,N}^{\rm
dilute},M_{\rm max,GR}^{\rm dense}]$ in which a dilute axion star can
collapse
into a Newtonian dense axion star without forming a black hole.

If we come back to dimensional variables, we
obtain
\begin{equation}
\label{qgr17}
M_{\rm max,GR}^{\rm
dense}=0.976\left
(\frac{6\pi\zeta}{4\nu}\right
)^{1/2}\left (\frac{2|a_s|}{r_S}\right
)^{1/2}\frac{M_P^2}{m}=0.845 M_r,
\end{equation}
\begin{equation}
\label{qgr18}
R_{\rm *,GR}^{\rm
dense}=0.976\left
(\frac{6\pi\zeta}{\nu}\right
)^{1/2}\left (\frac{2|a_s|}{r_S}\right )^{1/2}\lambda_C=1.69
R_r,
\end{equation}
provided that $|a_s|\ge |a_s|_*$. We note that these scales are similar to the
scales characterizing the maximum mass and the minimum radius of boson stars
with a purely repulsive $|\varphi|^4$ self-interaction ($a_s>0$) (see Appendix
\ref{sec_is}). However, as far as we know, these scales are new in the
case of axion stars with an attractive  $|\varphi|^4$ self-interaction ($a_s<0$)
and a repulsive $|\varphi|^6$ self-interaction.

For QCD axions with  $m=10^{-4}\,
{\rm eV}/c^2$ and $a_s=-5.8\times 10^{-53}\, {\rm m}$,  we obtain
 $M_{\rm max, GR}^{\rm dense}=23.6\, M_{\odot}$ and $R_{\rm *,GR}^{\rm
dense}=69.8\,
{\rm
km}$, corresponding to an average density $\rho=3.30\times 10^{19}\, {\rm
g/m^3}$.

For ULAs with $m=2.19\times 10^{-22}\, {\rm eV}/c^2$
and $a_s=-1.11\times 10^{-62}\, {\rm fm}$, we obtain $M_{\rm max, GR}^{\rm
dense}=3.19\times 10^{15}\, M_{\odot}$ and $R_{\rm *,GR}^{\rm dense}=305\, {\rm
pc}$, corresponding to an average density $\rho=1.82\times 10^{-9}\, {\rm
g/m^3}$.

\begin{figure}
\begin{center}
\includegraphics[clip,scale=0.3]{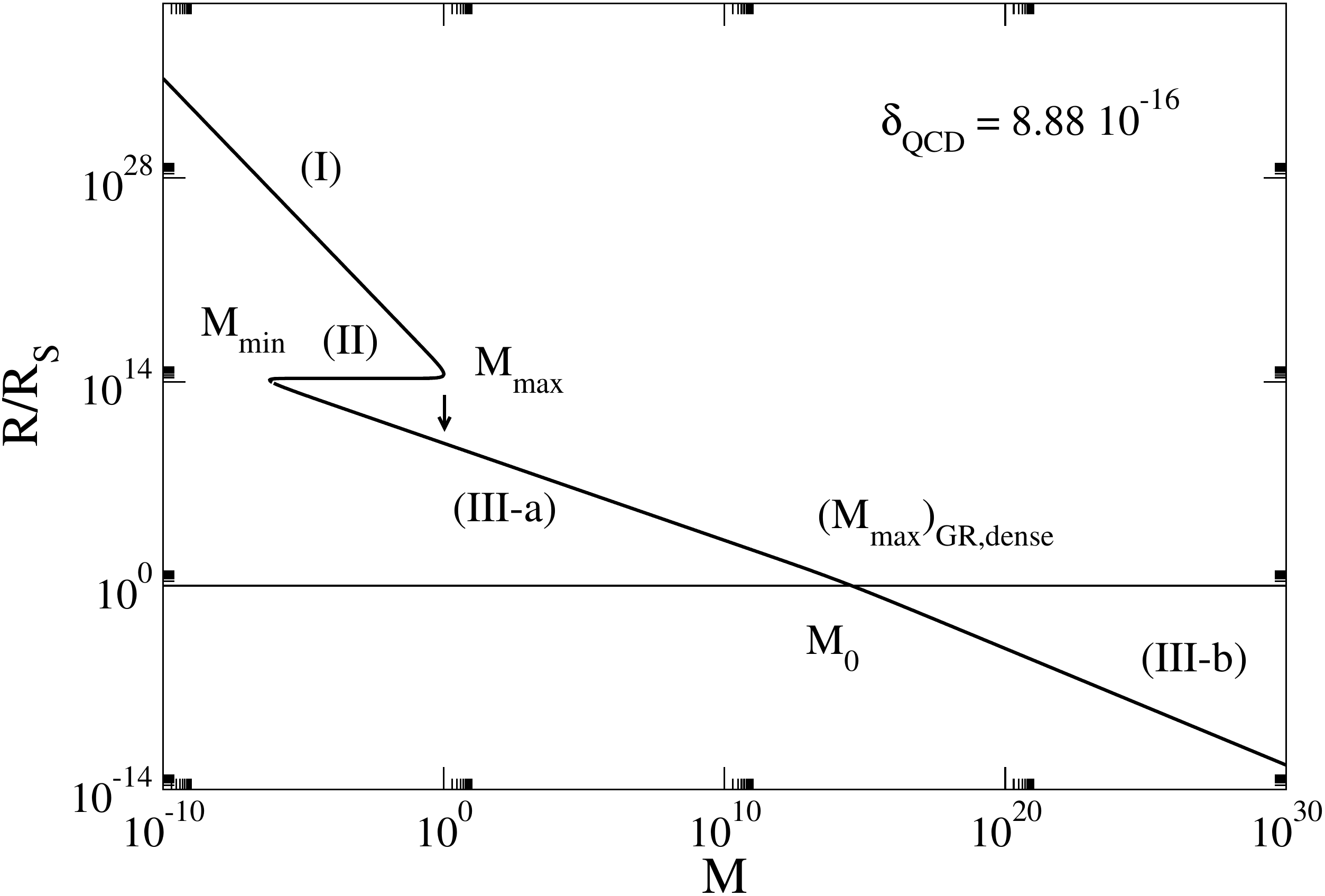}
\caption{Compactness $R/R_S=0.310 R/(M\delta)$ as a function of the
mass $M$ for QCD axion stars ($\delta=8.88\times 10^{-16}$). For
$\delta\rightarrow 0$, we find that $(R/R_S)(M_{\rm max})\sim 0.310/\delta$,
$(R/R_S)(M_{\rm min})\sim 0.116/\delta$, and $(R/R_S)_{\rm coll}\sim
0.563/\delta^{2/3}$.
We note that $R/R_S$ is constant ($R/R_S=0.155/\delta$) on the part of the
branch
(II) corresponding to unstable  axion stars where
$M\sim 2R$ (nongravitational limit). The axion stars are Newtonian when $M\ll
M_{\rm max,GR}^{\rm dense}$ (i.e. $R/R_S\gg 1$) and general relativistic
otherwise.}
\label{compacity}
\end{center}
\end{figure}

{\it Remark:} We note that the relativistic limit point $(M_{\rm max,GR}^{\rm
dense},R_{\rm *,GR}^{\rm
dense})$ is close to the point $(M_{0},R_{0})$ separating
nongravitational dense axion stars of type (III-a) and nonattractive dense
axion stars of type (III-b) in the TF regime. Therefore, we conclude
that dense axion stars of type (III-a) can be treated with a
Newtonian approach (up to about the end of this branch) while dense axion
stars of type (III-b) are always relativistic and cannot be treated with a 
Newtonian approach. Therefore, the Newtonian branch of type (III-b) is not
physically relevant, except, possibly, at its very begining. We also note
that $M_{\rm max,GR}^{\rm
dense}\sim M_{0}\gg M_{\rm max,N}^{\rm
dilute}$ when $\delta\ll 1$. Therefore, we conclude that the collapse of a
dilute axion star with mass  $M\gtrsim  M_{\rm max,N}^{\rm
dilute}$ leads to a dense axion star of type (III-b), not to a black hole.
Black holes are formed from the collapse of much heavier stars with a mass
$M>M_{\rm max,GR}^{\rm dense}\sim M_{0}\gg  M_{\rm max,N}^{\rm
dilute}$. To make things clear, we have plotted the mass-compactness relation
of axion stars in Fig. \ref{compacity}.

\subsection{Phase diagrams}

We can summarize the previous results on synthetic phase diagrams.

\begin{figure}
\begin{center}
\includegraphics[clip,scale=0.3]{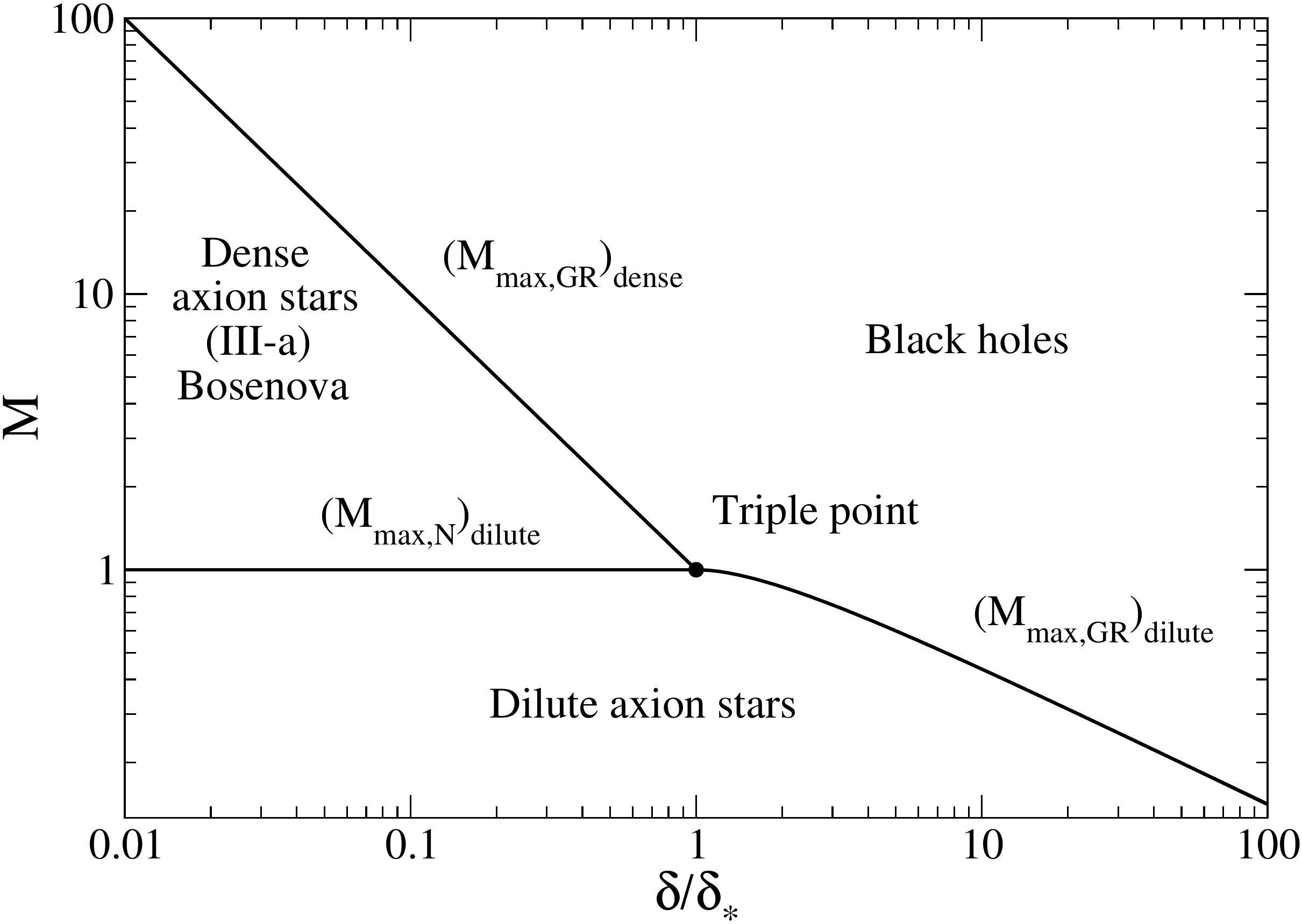}
\caption{Phase diagram of axion stars using dimensionless
variables.}
\label{dm}
\end{center}
\end{figure}

Figure \ref{dm} shows the phase diagram in terms of dimensionless variables. It
exhibits a triple point at $(\delta_*,M_{\rm max,N}^{\rm dilute})$ separating
three regions: dilute axion stars, dense axion stars of type (III-a), and black
holes.\footnote{We have represented the most natural phase diagram, obtained by
assuming that we start from a stable dilute axion star and that we 
progressively increase its mass. As reported in Fig. \ref{phasediagram}, in the
hysteretic zone (H), both dilute and dense axion stars are possible.
However, a dilute axion star is more natural.} We recall that the formation of
dense axion stars is usually accompanied by the emission of relativistic axions
(radiation),
so we have added the term ``bosenova'' in the phase diagram.

When $\delta>\delta_*$ (weakly self-interacting regime), we expect to observe
dilute axion stars
for $M<M_{\rm max,GR}^{\rm dilute}$  and
black holes for $M>M_{\rm max,GR}^{\rm dilute}$. 

When $\delta<\delta_*$ (strongly self-interacting regime), we expect to observe
dilute axion stars
for $M<M_{\rm max,N}^{\rm dilute}$, dense axion
stars for $M_{\rm max,N}^{\rm dilute}<M<M_{\rm max,GR}^{\rm dense}$, and
black holes for $M>M_{\rm max,GR}^{\rm dense}$.

\begin{figure}
\begin{center}
\includegraphics[clip,scale=0.3]{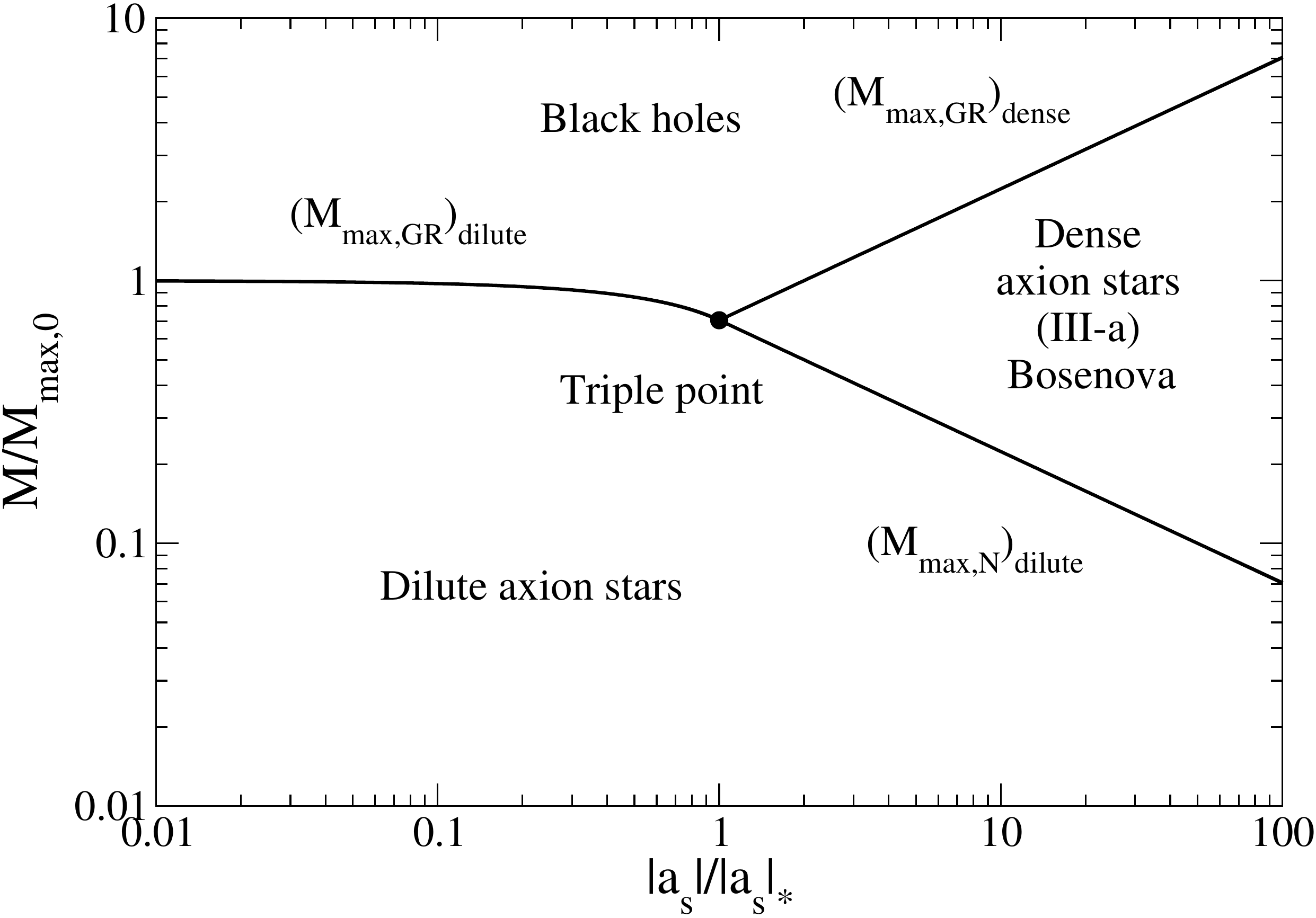}
\caption{Phase diagram of axion stars in terms of the scattering length
$|a_s|$.}
\label{dma}
\end{center}
\end{figure}

Figure \ref{dma} shows the phase diagram in terms of the scattering length
$|a_s|$. In
order to present a phase diagram that is not too sensitive on our
approximations, we
have normalized the scattering length by its transition value $|a_s|_*$ from Eq.
(\ref{qgr8}) and we have normalized the mass by the maximum general relativistic
mass $M_{\rm max,GR}^{\rm
dilute}(0)$ corresponding to noninteracting axion stars ($a_s=0$), that we
call $M_{\rm max,0}$ for brevity. It is given by Eq. (\ref{qgr9}). In this
way, the phase diagram is relatively universal while the exact values of
$|a_s|_*$ and $M_{\rm max,0}$ are expected to differ from the values
(\ref{qgr8}) and (\ref{qgr9}) obtained with our qualitative approach.  They
should be obtained from a rigorous calculation in general relativity. This will
be
considered in a future work.

\begin{figure}
\begin{center}
\includegraphics[clip,scale=0.3]{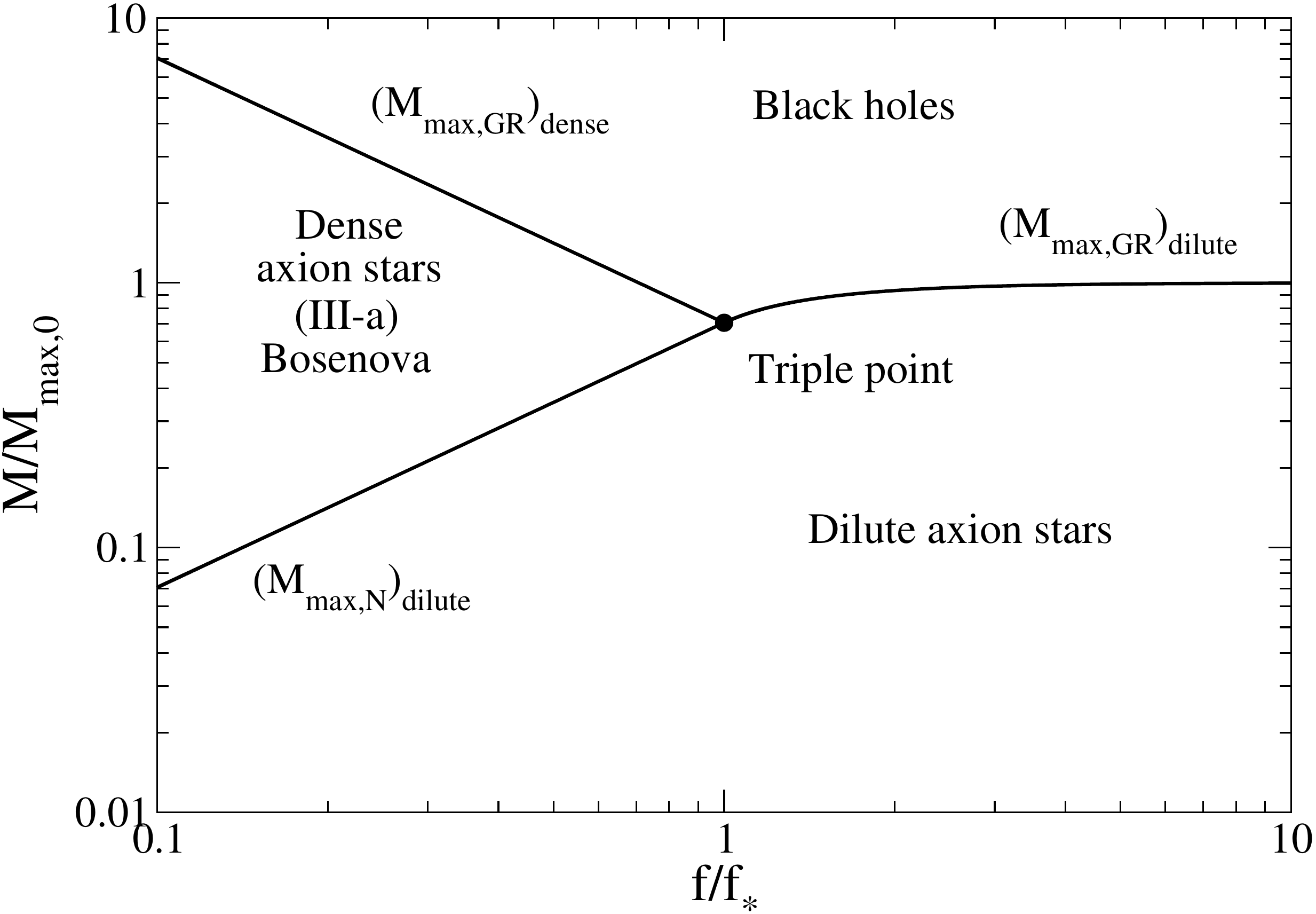}
\caption{Phase diagram of axion stars in terms of the  axion decay
constant $f$.}
\label{dmf}
\end{center}
\end{figure}

\begin{figure}
\begin{center}
\includegraphics[clip,scale=0.3]{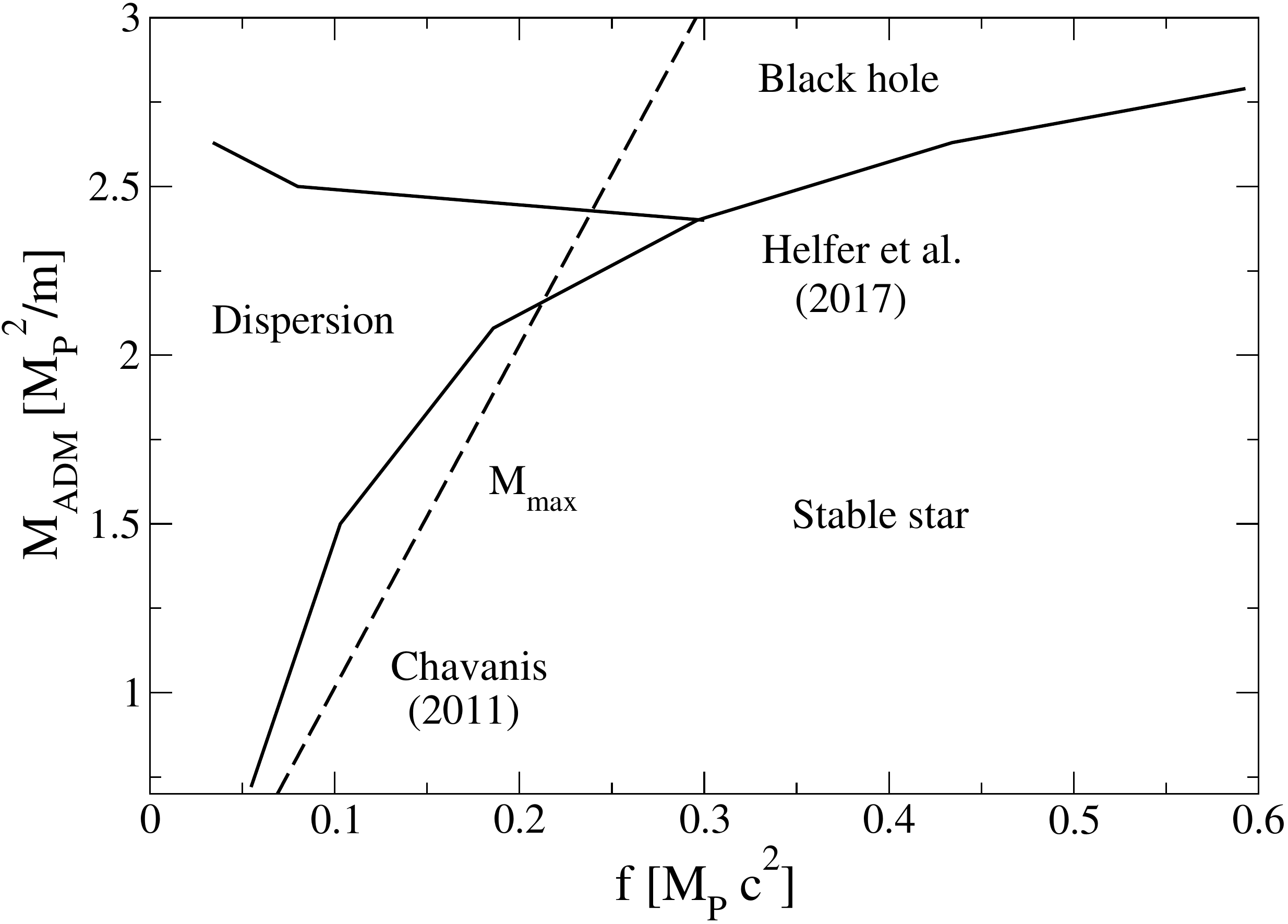}
\caption{Sketch of the phase diagram of axion stars obtained by Helfer {\it et
al.} \cite{helfer}. It can be compared with the phase diagram of Fig. \ref{dmf}
obtained from our analytical study.}
\label{marsh2}
\end{center}
\end{figure}

Using the relation
\begin{equation}
\label{qgr19}
\frac{\delta_*}{\delta}=\frac{|a_s|}{|a_s|_*}=\left (\frac{f_*}{f}\right
)^2=\frac{|\lambda|}{|\lambda|_*},
\end{equation}
we can easily obtain the phase diagrams in terms of $f$ and $\lambda$. The
normalized phase diagram in terms of $|\lambda|/|\lambda|_*$ coincides with Fig.
\ref{dma}. The phase diagram in terms of the axion decay constant
$f$ is presented in Fig. \ref{dmf}.

{\it Remark:} the triple point that we have obtained from our analytical
study is probably related to the one found numerically by Helfer {\it et
al.} \cite{helfer}. This is illustrated in Figs. \ref{dmf} and \ref{marsh2}
that have a similar qualitative behavior. We find that the triple point is
located at
$(f_*/M_Pc^2,M_*/(M_P^2/m))=(0.0891,0.969)$ while the exact values obtained by
Helfer {\it et al.} \cite{helfer} are $(f_*/M_Pc^2,M_*/(M_P^2/m))_{\rm
exact}=(0.3,2.4)$. One reason for this quantitative
difference is due to the approximations inherent to our analytical study:
Gaussian ansatz (see Sec. \ref{sec_mrdd}) and qualitative
treatment of relativistic effects (see Sec. \ref{sec_qgr}). Another
reason for this discrepency, that we believe to be more important, is due to the
fact that we are
not using the same axionic potential. Helfer {\it et al.} \cite{helfer} solve
the KGE equations for a
real SF described by the instantonic potential (\ref{inst1}) while we consider a
complex
SF described by the polynomial potential (\ref{inst9}) involving an attractive
$|\psi|^4$
term and a repulsive $|\psi|^6$ term. As an exemple of this intrinsic
difference, we note that, in our model, the maximum mass of dilute axion stars,
if it were calculated exactly, would be always smaller than the Kaup mass
$M_{\rm Kaup}=0.633 M_P^2/m$ (see Appendix \ref{sec_cs}), and would tends to the
Kaup mass in the noninteracting limit $f/M_Pc^2\rightarrow
+\infty$ (see Fig. \ref{dmf}). By contrast, the maximum mass obtained by Helfer
{\it
et al.} \cite{helfer} is always larger than the Kaup mass for the values of
$f/M_Pc^2$ considered, and apparently does not tend to the Kaup mass when
$f/M_Pc^2\rightarrow +\infty$  (see Fig. \ref{marsh2}).\footnote{Helfer {\it et
al.} \cite{helfer} do not consider the noninteracting limit $f/M_Pc^2\rightarrow
+\infty$ because they mention that high-particle physics generically imposes
$f<M_P c^2$.} This may be an effect of the periodicity
and anharmonicity of the instantonic axion potential that is not captured in our
polynomial approximation.

\section{Summary}
\label{sec_summary}

In this section, we summarize the main results of our study, and express the
physical quantities of axion stars (mass, radius, density) in terms of
dimensional
variables. We give the prefactors when we know
their exact values. Otherwise, we just give the scaling of the physical
quantities when the prefactor is uncertain (the prefactors obtained from the
Gaussian ansatz can be found in the main text if needed).

\subsection{Weakly self-interacting axions}

We first consider weakly self-interacting axions with
$|\lambda|<|\lambda|_*\sim (m/M_P)^2$.
For $M\ll M_{\rm max,GR}^{\rm dilute}\sim M_P^2/m$  and $R\gg R_{\rm *,GR}^{\rm
dilute}\sim \lambda_C$, we are on the branch (I) of the mass-radius relation 
corresponding to stable dilute axion stars. They can be described by
Newtonian gravity. Suppose that we increase their mass $M$
(physically, this can happen if the
axion stars accrete matter from the surrounding or collide with other stars). As
we approach the general relativistic maximum mass
$M_{\rm max,GR}^{\rm dilute}$, the axion stars become more and more
relativistic (they also become denser and denser). Close to $M_{\rm max,GR}^{\rm
dilute}$, the mass-radius relation 
forms a spiral.  For $M>M_{\rm max,GR}^{\rm dilute}$ there is no equilibrium
state
and  the  axion star collapses into a black hole.  For noninteracting
complex boson stars ($\lambda=0$), the maximum mass and the minimum radius are 
$M_{\rm max,GR}^{\rm dilute}=0.633
M_P^2/m$ and $R_{\rm *,GR}^{\rm
dilute}=6.03 \lambda_C$ \cite{kaup,rb}. For self-interacting axion stars, the
maximum mass
and
the minimum radius are not expected to change appreciably with
$|\lambda|$ when $\lambda<|\lambda|_*\sim (m/M_P)^2$
(see Sec. \ref{sec_grd}).

\subsection{Strongly self-interacting axions}

We now consider strongly self-interacting axions with $|\lambda|>|\lambda|_*\sim
(m/M_P)^2$. This is the regime of physical interest for QCD axions and ULAs.
For $M<M_{\rm max,N}^{\rm dilute}=5.073 M_P/\sqrt{|\lambda|}$  and
$R_{99}>R_{\rm *,N}^{\rm dilute}=1.1 \sqrt{|\lambda|}(M_P/m)\lambda_C$
\cite{prd1,prd2}, we are
on the branch (I) of the mass-radius relation corresponding to stable dilute
axion stars. They can be described by
Newtonian gravity up to the maximum mass $M_{\rm max,N}^{\rm dilute}$. Let us
assume that the mass progressively increases. For $M>M_{\rm
max,N}^{\rm dilute}$ the axion stars collapse and become denser and denser
until the collapse is stopped by the repulsive self-interaction. In that case,
the system reaches an equilibrium state corresponding to a dense axion star
located on the branch (III-a) of the mass-radius relation on which self-gravity
is weak. As discussed in \cite{tkachevprl}, this
collapse
may be associated with an emission of relativistic axions (bosenova phenomenon).
The star may also undergo a series of implosions and explosions. Assuming that
the mass is
approximately conserved during the collapse, the resulting dense axion star has
a 
radius $R_{\rm coll}=0.812 |\lambda|^{1/6}(M_P/m)^{1/3}\lambda_C$. For
sufficiently small masses and sufficiently small radii, the dense axion stars
on
the  branch (III-a)  of the mass-radius relation can be described by Newtonian
gravity. They have a constant
density $\rho_{\rm dense}=2.25 |\lambda|^{-1} m/\lambda_C^3$ independent of
their
mass. As  their mass increases further and approaches
$M_{\rm max,GR}^{\rm dense}\sim |\lambda|^{1/2}M_P^3/m^2$ (which is
of the order of $M_0$) corresponding to a radius $R_{\rm *,GR}^{\rm dense}\sim 
|\lambda|^{1/2}(M_P/m)\lambda_C$, the dense axion
stars become relativistic.\footnote{Depending on the exact
values of $M_{\rm max,GR}^{\rm dense}$ and
$M_0$, the dense axion stars may be in a Newtonian regime for $M_0<M<M_{\rm
max,GR}^{\rm dense}$ corresponding to the begining of
the branch
(III-b) of the mass-radius relation where the
self-attraction of the bosons is negligible, before becoming relativistic for
$M\sim M_{\rm
max,GR}^{\rm dense}$.}  Close to $M_{\rm max,GR}^{\rm dense}$, the mass-radius
relation forms a spiral or stops suddently (see footnote 25).  For $M>M_{\rm
max,GR}^{\rm
dense}$, 
there is no equilibrium state and the axion star collapses into a
black hole. Let us now assume that the mass decreases along the branch
(III-a) of the mass radius relation.  When
$M<M_{\rm min,N}^{\rm dense}\sim m/|\lambda|$ and $R<R_{\rm *,N}^{\rm
dense}\sim \lambda_C$, the dense
axion star explodes a forms a dilute axion
star of radius $R_{\rm exp}\sim |\lambda|(M_P/m)^2\lambda_C$ or, more
realistically, disperses to infinity.

It is interesting to see that the mass scales $M_{\rm max,GR}^{\rm
dilute}\sim M_c\sim M_P^2/m$, $M_{\rm max,N}^{\rm dilute}\sim M_a\sim
M_P/\sqrt{|\lambda|}$, $M_{\rm max,GR}^{\rm
dense}\sim M_r\sim \sqrt{|\lambda|} M_P^3/m^2$ and $M_{\rm min,N}^{\rm
dense}\sim M_i\sim m/|\lambda|$ presenting the scalings
$M_P^2/m$, $M_P/\sqrt{|\lambda|}$, $\sqrt{|\lambda|} M_P^3/m^2$ and 
$m/|\lambda|$ play a fundamental role in the problem.

\subsection{Analogy with white dwarfs, neutron stars and black holes}

\begin{figure}
\begin{center}
\includegraphics[clip,scale=0.3]{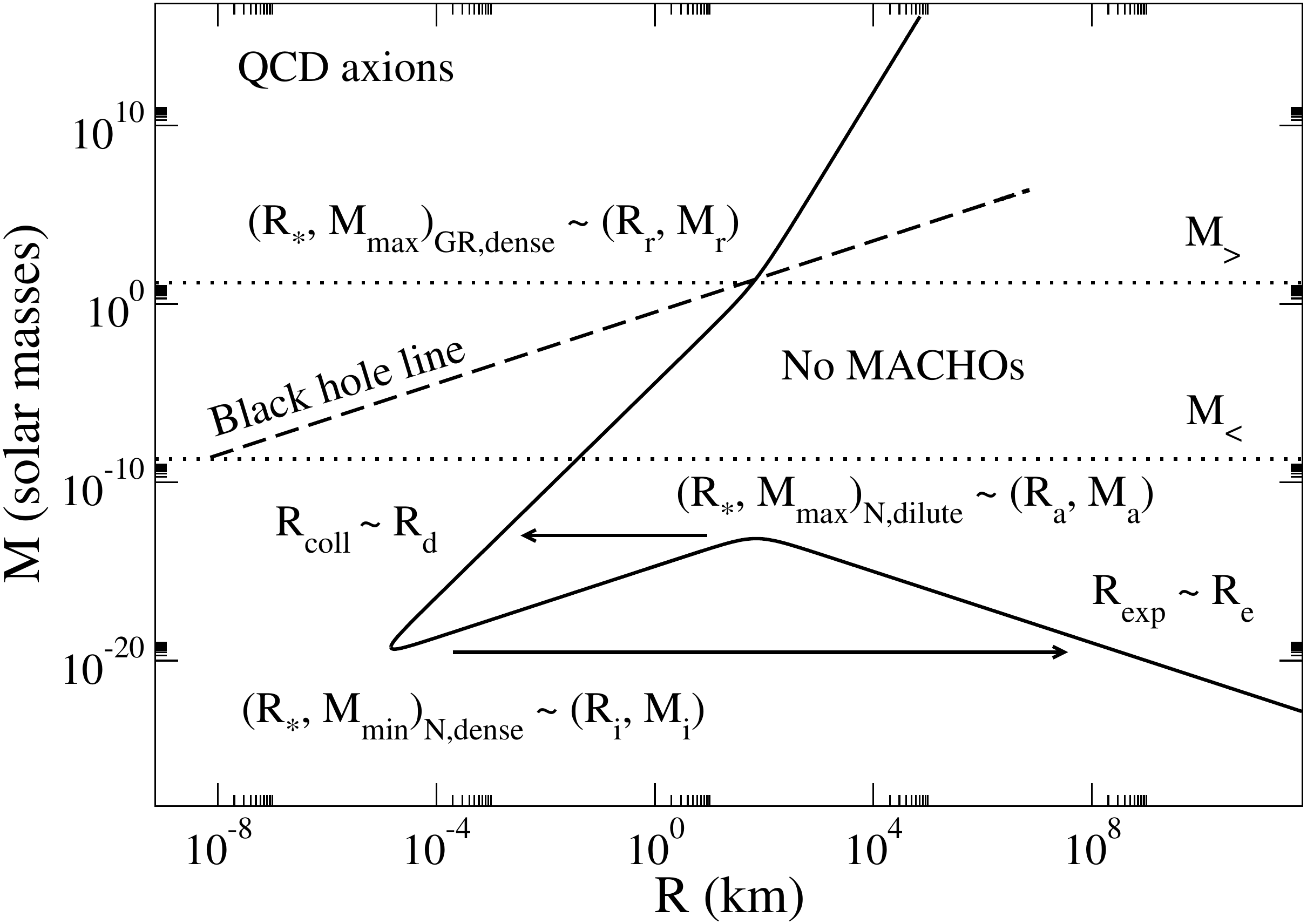}
\caption{Mass-radius relation of QCD axion
stars with dimensional
variables obtained from the Gaussian ansatz. The Newtonian maximum mass and
the corresponding radius of dilute axion
stars are 
$M_{\rm max,N}^{\rm dilute}=6.92\times 10^{-14} M_{\odot}$ and $R_{\rm *,N}^{\rm
dilute}=71.5 \, {\rm km}$. The radius of the dense axion star resulting from
the collapse of a dilute axion star with the maximum mass is $R_{\rm
coll}=1.25 \, {\rm m}$. The general relativistic maximum mass and
the corresponding
radius of dense axion stars are 
$M_{\rm max,GR}^{\rm dense}=23.6\, M_{\odot}$ and $R_{\rm
*,GR}^{\rm
dense}=69.8 \, {\rm km}$. Close to  $M_{\rm max,GR}^{\rm
dense}$, the  mass-radius relation $M(R)$ may form a spiral or stop suddently.
The Newtonian
minimum  mass and the corresponding radius
of dense
axion stars are 
$M_{\rm min,N}^{\rm dense}=4.40\times 10^{-20}\, M_{\odot}$ and $R_{\rm
*,N}^{\rm
dense}=1.70 \, {\rm cm}$. The radius of the dilute axion star resulting from
the explosion of a dense axion star with the minimum mass is $R_{\rm
exp}=2.25\times 10^{8} \, {\rm km}$ suggesting that the star is dispersed away.
From the
exact results of \cite{prd2}, we get  $M_{\rm max,N}^{\rm
dilute}=6.46\times 10^{-14} M_{\odot}$ and $R_{\rm *,99,N}^{\rm
dilute}=227 \, {\rm km}$. From the exact results of \cite{braaten},
we get $M_{\rm min,N}^{\rm dense}=1.2\times 10^{-20}\,
M_{\odot}$ and $R_{\rm
*,N}^{\rm dense}=1.81 \, {\rm cm}$. The authors of \cite{braaten} find no
solution with
$M_{\rm max}^{\rm dense}= 1.9\, M_{\odot}$ (see Appendix \ref{sec_braa}).
From the exact
result of Appendix \ref{sec_tfng}, we get $R_{\rm coll}=0.756\, {\rm m}$. We
have also plotted the lines
$M_{<}=2\times 10^{-9}\, M_{\odot}$ and $M_{>}=15\, M_{\odot}$
between which MACHOs are excluded from observations
\cite{eros,griest}. We see that axionic DM can  be in the
form of
gases of axions,
or dilute axion stars, or dense axion stars with $M<2\times 10^{-9}\,
M_{\odot}$, or dense axion stars (or axionic black holes) with $M>15\,
M_{\odot}$.}
\label{axionsqcd}
\end{center}
\end{figure}

\begin{figure}
\begin{center}
\includegraphics[clip,scale=0.3]{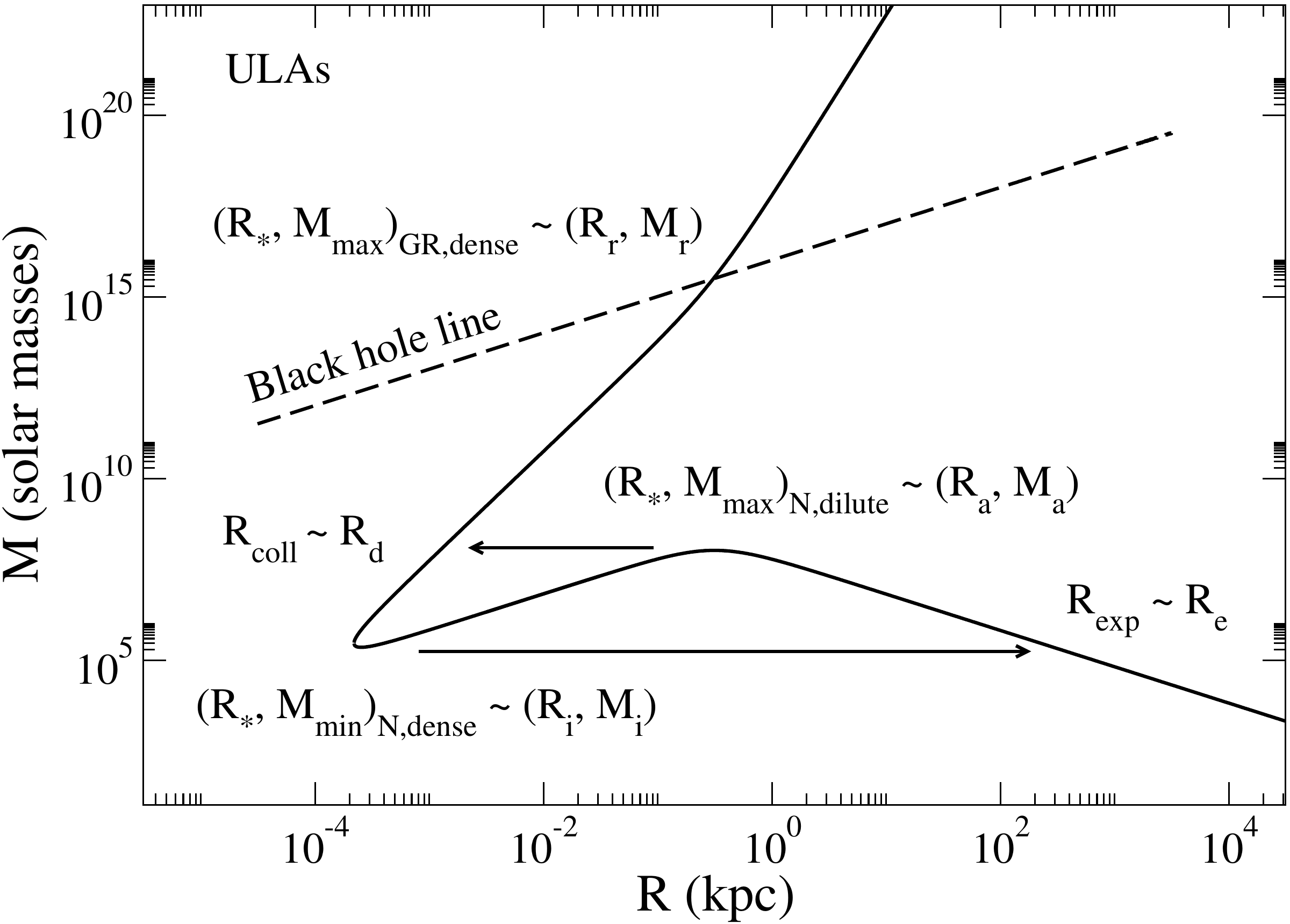}
\caption{Mass-radius relation of ULA
clusters (this concerns more precisely their solitonic core) with dimensional
variables obtained from the Gaussian ansatz. The Newtonian maximum mass and
the corresponding radius of dilute clusters are 
$M_{\rm max,N}^{\rm dilute}=1.07\times 10^{8} M_{\odot}$ and $R_{\rm *,N}^{\rm
dilute}=0.313 \, {\rm kpc}$. The radius of the dense cluster
resulting from
the collapse of a dilute cluster with the maximum mass is $R_{\rm
coll}=1.23 \, {\rm pc}$. The general relativistic maximum mass
and corresponding radius of dense clusters are 
$M_{\rm max,GR}^{\rm dense}=3.19\times 10^{15}\, M_{\odot}$ and $R_{\rm
*,GR}^{\rm
dense}=0.305 \, {\rm kpc}$.  Close to  $M_{\rm max,GR}^{\rm
dense}$, the  mass-radius relation $M(R)$ may form a spiral or stop
suddently. However, the  general relativistic  maximum mass $M_{\rm max,GR}^{\rm
dense}=3.19\times 10^{15}\, M_{\odot}$  is too large to describe the
solitonic core of ULA
clusters suggesting that ULA clusters
never form
black holes. The Newtonian minimum  mass and
corresponding radius of dense clusters are 
$M_{\rm min,N}^{\rm dense}=2.30\times 10^{5}\, M_{\odot}$ and $R_{\rm
*,N}^{\rm
dense}=0.253 \, {\rm pc}$.  The radius of the dilute cluster
resulting from
the explosion of a dense cluster with the minimum mass is $R_{\rm
exp}=290 \, {\rm kpc}$  suggesting that the cluster is dispersed away. From the
exact results of \cite{prd2}, we get  $M_{\rm max,N}^{\rm
dilute}=10^{8} M_{\odot}$ and $R_{\rm *,N}^{\rm
dilute}=1 \, {\rm kpc}$. From the exact
result of Appendix \ref{sec_tfng}, we find $R_{\rm coll}=0.745\, {\rm pc}$.
}
\label{ula}
\end{center}
\end{figure}

\begin{figure}
\begin{center}
\includegraphics[clip,scale=0.3]{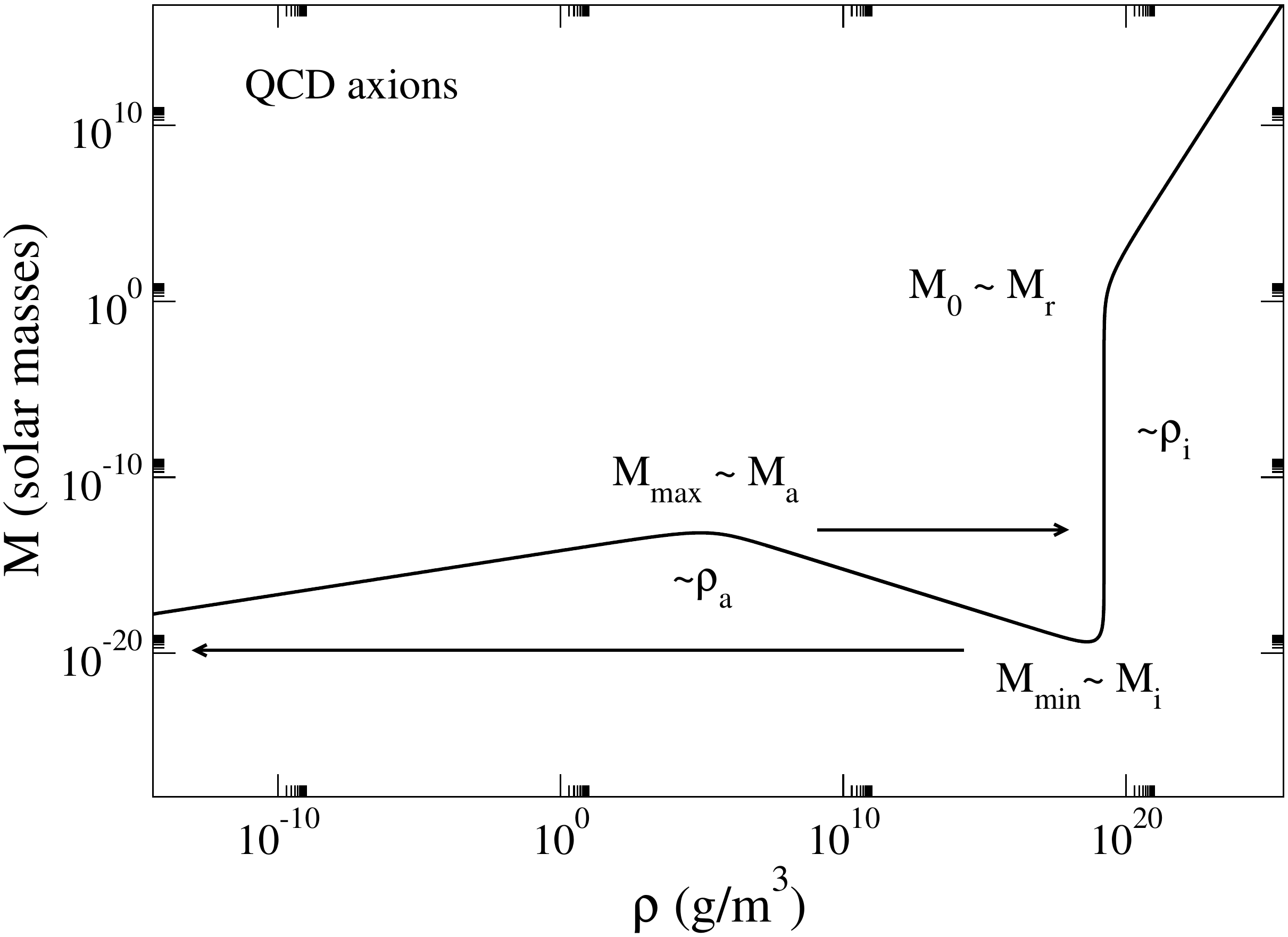}
\caption{Mass-density relation of QCD axion
stars with dimensional
variables obtained from the Gaussian ansatz. The density of dilute axion stars
at the Newtonian maximum mass is $\rho=8.98\times 10^4\, {\rm
g/m^3}$ (the exact value obtained from the results of \cite{prd2} is
$\rho=2.62\times
10^3\, {\rm
g/m^3}$). The constant density of dense axion stars in
the TF $+$ nongravitational regime, which  includes the dense axion star
resulting from the collapse of a dilute axion star with the maximum mass, is
$\rho_{\rm dense}=1.68\times 10^{19}\, {\rm g/m^3}$ (the exact value obtained 
from the results of Appendix
\ref{sec_tfng} is $\rho_{\rm dense}=7.07\times
10^{19}\, {\rm g/m^3}$). The density of dense axion stars
at the general relativistic maximum mass is $\rho=3.30\times 10^{19}\, {\rm
g/m^3}$. The density of dense axion stars at the Newtonian
minimum  mass is $\rho=4.21\times
10^{18}\, {\rm g/m^3}$ (the exact value obtained from the results of
\cite{braaten} is $\rho=9.61\times
10^{17}\, {\rm g/m^3}$). The density  of the dilute axion star resulting
from
the explosion of a dense axion star with the minimum mass is
$\rho=1.84\times
10^{-21}\, {\rm g/m^3}$ suggesting that the star is dispersed away.
}
\label{rhomassdimQCD}
\end{center}
\end{figure}

\begin{figure}
\begin{center}
\includegraphics[clip,scale=0.3]{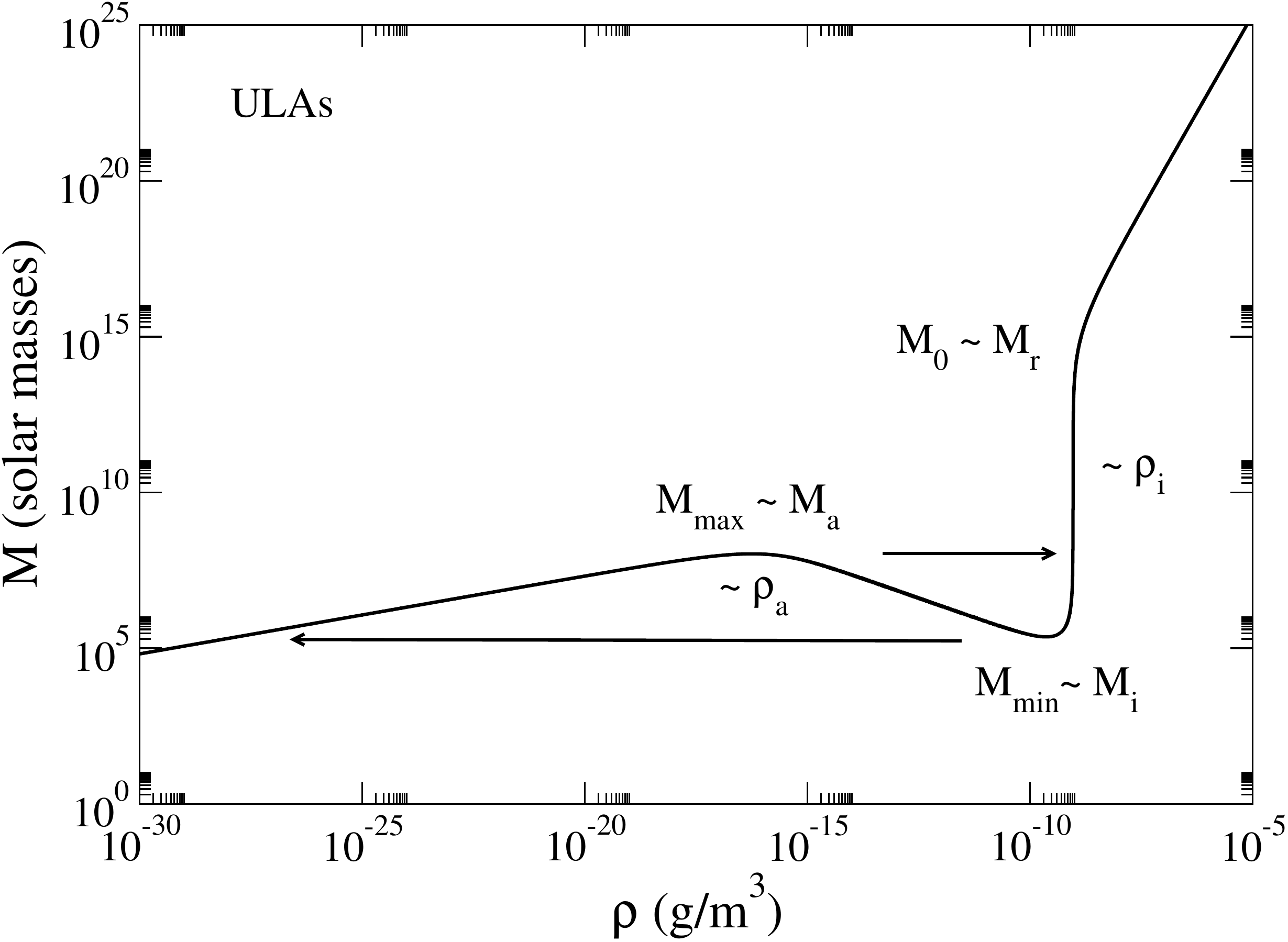}
\caption{Mass-density relation of ULAs clusters (this concerns more
precisely their solitonic
core) with dimensional
variables obtained from the Gaussian ansatz. The density of dilute axionic
clusters
at the Newtonian maximum mass is $\rho=5.63\times 10^{-17}\, {\rm
g/m^3}$ (the exact value obtained  from the results of \cite{prd2} is
$\rho=1.62\times
10^{-18}\, {\rm
g/m^3}$). The constant density of dense axionic clusters in
the TF $+$ nongravitational regime, which  includes the dense axionic cluster
resulting from the collapse of a dilute axionic cluster with the maximum mass,
is
$\rho=9.21\times 10^{-10}\, {\rm g/m^3}$ (the exact value obtained from the
results of Appendix
\ref{sec_tfng} is $\rho=3.88\times
10^{-9}\, {\rm g/m^3}$). The density of dense axion stars
at the general relativistic maximum mass is $\rho=1.82\times 10^{-9}\, {\rm
g/m^3}$. The density of dense axion stars at the Newtonian
minimum  mass is $\rho=2.30\times
10^{-10}\, {\rm g/m^3}$. The density  of
the dilute axionic clusters resulting
from
the explosion of a dense axionic cluster with the minimum mass is
$\rho=1.52\times
10^{-28}\, {\rm g/m^3}$ suggesting that the cluster is dispersed away.
}
\label{rhomassdimULA}
\end{center}
\end{figure}

We have represented the mass-radius relations of QCD axion stars and ULA
clusters in Figs. \ref{axionsqcd} and \ref{ula} using dimensional variables.
Numerical values of the characteristic masses and radii  are given in the
captions. We note that the mass-radius relation of QCD
axion stars reported in Fig. \ref{axionsqcd}, and obtained analytically in our
approximate study, is in good agreement with the exact mass-radius
relation of QCD axion stars obtained numerically by Braaten {\it et al.}
\cite{braaten} (see their Fig. 1). We also note that Figs.
\ref{axionsqcd} and \ref{ula}
look the same except for a change of scales. However, this impression
is misleading in the sense that there is no scale-invariance in the
general mass-radius relation (\ref{mr4}) of axion stars: the self-interaction
parameter $\delta$ cannot be eliminated from this relation by a simple
rescaling. Only the branch of dilute axion stars is invariant for sufficiently
small values of $\delta$ (see footnote 14).

The mass-radius relation of axion stars shares some analogies with the
mass-radius relation of fermion stars at $T=0$ such as white dwarf stars and
neutron stars based on the Harrison-Wakano-Wheeler (HWW)
equation of state (compare
Figs. \ref{axionsqcd} and \ref{ula} with Fig. 7 of \cite{htww}, Fig. 2 of
\cite{mt}, or Fig. 11.2 of \cite{weinberg}). For sufficiently
small
masses and sufficiently large
radii, a star at $T=0$ (more properly at $T\ll T_F$ where $T_F$ is the Fermi
temperature) is in a Newtonian nonrelativistic white dwarf stage where the
gravitational attraction is balanced by the quantum pressure of the degenerate
electron gas on account of the Pauli exclusion
principle. As we approach the Newtonian Chandrasekhar maximum mass $M_{\rm
max,N}^{\rm WD}=0.776 M_P^3/H^2=1.44\, M_{\odot}$ (where $H$ is the proton
mass),
special relativistic effects come into play and the radius  $R_{\rm *,N}^{\rm
WD}$ tends to zero.\footnote{If we take general relativity into account, we
obtain a minimum radius $R_{\rm
*,GR}^{\rm WD}=102\, {\rm km}$.} Above the Chandrasekhar maximum mass, white
dwarf stars collapse, the electrons and the protons combine together to form
neutrons, and the system becomes a neutron star in which the gravitational
attraction
is balanced by the quantum
pressure of the degenerate neutron gas (some mass is
ejected during the collapse).
Above the Oppenheimer-Volkoff general relativistic maximum mass $M_{\rm
max,GR}^{\rm NS}=0.384 M_P^3/m^2=0.710\, M_{\odot}$ (where $m$ is the
neutron mass), corresponding to a radius $R_{\rm *,GR}^{\rm NS}=9.18\, {\rm
km}$, the neutron star collapses into a black hole.

In the analogy between axion stars and fermion stars, the dilute axion stars are
the counterpart of the white dwarf
stars and the dense axion stars are the counterpart of the neutron stars.
Similarly, the Newtonian maximum mass of dilute axion stars  $M_{\rm max,N}^{\rm
dilute}$ is the counterpart of the Newtonian Chandrasekhar
maximum mass of white dwarf stars $M_{\rm max,N}^{\rm WD}$  and the
general relativistic  maximum mass of dense axion stars $M_{\rm max,GR}^{\rm
dense}$ is the counterpart of the  general relativistic  Oppenheimer-Volkoff
maximum mass of neutron stars $M_{\rm max,GR}^{\rm NS}$. Close
to the general relativistic  maximum mass, the mass-radius relation of
dense axion stars is expected to form a spiral (see footnote 25 for more
details) similarly to the mass-radius relation of
neutron stars.

There
is, however, a difference in the sense that the Chandrasekhar mass is larger
than the Oppenheimer-Volkoff maximum mass while, in the present context, the
Newtonian maximum mass of dilute axion stars is smaller than the general
relativistic maximum mass of   dense axion stars. In
this sense, the mass-radius relation of axion stars is more similar to the
mass-radius relation of compact stars obtained with the Skyrme-Cameron-Saakyan
(SCS) equation of state (see Fig. 3 of \cite{mt}). Indeed, in more
realistic models of neutron stars, the maximum mass is
substantially larger than the ideal Oppenheimer-Volkov
limit, being in the range
$2-2.4\, M_{\odot}$, and overcomes the Chandrasekhar limiting mass of white
dwarf stars.
This is why Shapiro and Teukolsky \cite{shapiroteukolsky} draw a mass-radius
relation (see their Fig. 6.3) in
which the
maximum mass of neutron stars is larger than the maximum mass of white dwarf
stars. This figure, which can be seen as a synthesis
of Figs. 2 and 3 of \cite{mt} is similar to Figs. \ref{axionsqcd} and \ref{ula}
of axion stars.

We finally note that
we can determine the stability of the different configurations of 
fermion stars represented in Fig. 7 of \cite{htww}, in Figs. 2 and 3 of
\cite{mt}, in Fig. 11.2 of \cite{weinberg} and in Fig. 6.3
of
\cite{shapiroteukolsky} by using the
Poincar\'e theory of linear series of equilibria, as shown here in the
case of axion stars. The Poincar\'e theory provides an
alternative, and a generalization, of the $M(R)$ theorem devised by
Wheeler \cite{htww} in the physics of compact objects (white dwarfs and neutron
stars) and used in \cite{mt,weinberg,shapiroteukolsky}.

For completeness, we have also represented the mass-density
relations of QCD axion stars and ULA
clusters in Figs. \ref{rhomassdimQCD} and \ref{rhomassdimULA}  using
dimensional variables. Again, they present analogies with the mass-central
density relations of fermion stars (compare Figs.
\ref{rhomassdimQCD} and \ref{rhomassdimULA} with Fig. 5 of \cite{htww} or
Figs. 6.2 and 9.1 of
\cite{shapiroteukolsky}). If the mass-radius relation of dense axion stars forms
a spiral close to $M_{\rm max,GR}^{\rm
dense}$ (for the reason explained previously), then the mass-density relation
will display damped oscillations as
in the case of neutron stars (see the previous references and Fig. 1 of
\cite{mz}).

\subsection{Analogy with self-gravitating fermions at finite temperature}

We have seen that axion stars are expected to undergo a phase
transition between a dilute phase and a dense phase when the self-interaction
parameter $\delta$ is below a critical value $\delta_c=0.0075$.

This phase transition is similar to the Van der Waals phase transition between a
gas and a liquid  when the temperature $T$ is below a critical value $T_c$.
However,  in our case,  the phase transition is not of thermodynamical origin
since we work at $T=0$. On the other hand, the metastable axionic stars
generally have very long lifetimes so that the first order phase transition that
is expected to occur at $M_t$ does not take place in practice. Only the zeroth
order phase transitions at $M_{\rm max}$ and $M_{\rm min}$ are physical. By
contrast, in the Van der Waals fluid, the metastable states are less robust
and the liquid/gas coexistence usually exists at a unique pressure
$P_t$ determined by the Maxwell construction. Therefore, a 
first order phase transition really occurs in that case.

The nature of phase transitions in axion stars at $T=0$ is also similar to the
nature of
phase transitions in the Newtonian self-gravitating Fermi gas at finite
temperature studied in \cite{ijmpb}. Above a critical size, self-gravitating
fermions\footnote{They may represent
electrons in white dwarf stars, neutrons in neutron stars, or massive neutrinos
in DM models.} are expected to display a first order phase
transition between a gaseous phase and a condensed phase with a core-halo
structure. In the gaseous phase, the gravitational attraction is equilibrated by
the thermal pressure. In the condensed phase, the gravitational attraction is
counterbalanced by the quantum pressure of the degenerate Fermi gas arising from
the Pauli exclusion principle. In the self-gravitating Fermi gas at finite
temperature, metastable states have very long lifetimes so that the
predicted first order phase transition does not take place  in practice, like in
the case of axion stars. Only the zeroth order phase transitions at the
spinodal points take place. The mass-radius relations of Fig.
\ref{mr} can be compared with the caloric curves of Fig. 14 in  \cite{ijmpb} and
the phase
diagram of Fig. \ref{phasediagram} can
be compared with the phase diagrams
of Figs. 34 and 35 in  \cite{ijmpb}. The phenomenology of the phase
transitions in the 
self-gravitating Fermi gas at finite temperature is, however, richer than the
phenomenology of the phase transitions in axion stars because this system
displays two critical points, one in each statistical 
ensembles (microcanonical and canonical). This is related to the phenomenon
of ensembles inequivalence that characterizes the thermodynamics of systems with
long-range interactions like self-gravitating systems \cite{paddy,ijmpb}.

\section{Conclusion}

In this paper, we have studied the mass-radius relation of dilute and
dense axion stars by using an analytical approach based on a Gaussian ansatz
like in our previous works \cite{prd1,bectcoll,chavtotal}.
Our results are in good qualitative, and even quantitative, agreement with the
exact results of Chavanis and Delfini \cite{prd2} for dilute axion stars, and
the exact results of Braaten {\it et al.} \cite{braaten} for dense axion stars,
obtained by solving
the GPP equations numerically (see the caption of Fig. \ref{axionsqcd} for a
comparison
between our approximate analytical results and the exact numerical ones). An
interest of our
analytical study is that it allows us to vary more easily the parameters
and to isolate characteristic mass, length and density scales that
play an important role in the problem.

We have also taken into account relativistic effects in a qualitative
manner and obtained an estimate of the maximum mass of axion stars set by
general relativity. Above that mass, the axion star is expected to collapse into
a black hole. We have obtained a phase diagram presenting
a triple point separating dilute axion stars, dense
axion stars (that may be the remnants of a bosenova) and
axionic
black holes. This phase diagram is in qualitative agreement
with the phase diagram obtained  numerically by Helfer {\it et al.}
\cite{helfer} by solving the KGE equations.  The quantitative disagreement in
the position of
the triple point may be due to our approximations or to the fact that we use a
different
potential. Indeed, Helfer {\it et al.} \cite{helfer} solve the KGE equations
for a real SF described by the instantonic potential (\ref{inst1}) while we
consider a
complex SF described by a polynomial potential (\ref{inst9}) involving an
attractive 
$|\psi|^4$ term and a repulsive $|\psi|^6$ term. 

The evolution of axion stars above the maximum mass  is still
unclear and deserves further research. As we have indicated, the collapse of a
dilute axion star can
lead to a dense axion star (possibly associated with an emission of relativistic
axions - bosenova) when $M\gtrsim M_{\rm max}$, or to a black hole when $M\gg
M_{\rm
max}$. The mass of the dense axion star, or the mass of the black
hole, that
results from the collapse of a dilute axion star is not precisely known and
should be studied in detail through numerical simulations. As shown by
Levkov \cite{tkachevprl}, the collapse of a dilute axion star with $M\gtrsim
M_{\rm max}$ may be accompanied by an important emission of radiation (bosenova)
that carries
away a large part of the initial mass (about $30\%$).

A first domain of application of these ideas concerns the formation and the
evolution of mini QCD
axion stars. Dilute QCD axion stars have a maximum mass $M_{\rm
max}^{\rm exact}=6.46\times
10^{-14} M_{\odot}$ [see Eq. (\ref{maxmr1})], a minimum radius
$(R_{99}^*)^{\rm
exact}=227 \, {\rm
km}$ [see Eq. (\ref{maxmr2})], a maximum density ${\rho}_{\rm
dilute}^{\rm
exact}=2.62\times 10^3\,
{\rm g/m^3}$ [see Eq. (\ref{maxmr2b})], and a typical
minimum energy $E_{\rm
tot}\sim -2.35\times 10^{25}\, {\rm
erg}$ [see Eq. (\ref{etotcrit})] while dense
QCD
axion stars that result from the collapse of a dilute QCD axion star with the
maximum mass $M_{\rm
max}^{\rm exact}=6.46\times
10^{-14} M_{\odot}$ have a radius $R_{\rm coll}^{\rm exact}=0.756 \,
{\rm m}$ [see Eq. (\ref{tfng14})] and 
a density $\rho_{\rm dense}^{\rm exact}=7.07\times 10^{19}\, {\rm
g/m^3}$ [see Eq. (\ref{tfng3rho})]. They are far from the black hole limit since
$R/R_S=3.96\times 10^9\gg 1$. They
have a low
pulsation period $2\pi/\omega_{\rm coll}^{\rm exact}=6.70\times 10^{-9}\, {\rm
s}$ [see Eq. (\ref{tfng20})], and a very negative energy $(E_{\rm
tot}^{\rm
coll})_{\rm exact}=-3.25\times 10^{40}\,
{\rm erg}$ [see Eq. (\ref{mc2})]. We can
check that self-gravity is weak in dense QCD axion stars since 
$GM^2/R=1.46\times 10^{31}\,
{\rm ergs}\ll |(E_{\rm tot}^{\rm
coll})_{\rm exact}|$. The formation
time of a dense QCD axion star (collapse time) is $t_{\rm
coll}=2.90\gamma^{-1/4}
t_D=9.86 \gamma^{-1/4}\, {\rm hrs}$ where $\gamma=M/M_{\rm max}-1$
(see
Eq. (185)
of \cite{bectcoll}). For $\gamma=10^{-4}$ this gives $t_{\rm
coll}=98.6\, {\rm hrs}$. Both dilute and dense QCD axion stars could be the
constituents of DM in the form of MACHOs. Axionic black
holes with a mass larger than $15 M_{\odot}$ could possibly exist (see Fig.
\ref{axionsqcd}) but their
mechanism of formation (if relevant) is uncertain and remains to be
established.

On the other hand, the collisions between dilute axion stars
and neutron stars possessing a strong
magnetic field \cite{tkachev2015,iwazaki,raby,bai}, the collisions
between dilute axion stars and the magnetized  accretion disk of a black
hole \cite{iwazakinew}, or the collapse of dilute axion
stars above their maximum mass \cite{tkachevprl}, can produce a radiation (of
different origin). If
the mass of the axion star is rapidly
transformed into electromagnetic radiation, for example during the collision
with a neutron star  (magnetar) or an accretion disk, this radiation should be
detectable.
Interestingly, as noted by \cite{tkachev2015,iwazaki,raby,iwazakinew,bai}, the
characteristics of mini QCD axion stars are
consistent
with
the observed properties of recently detected unusual radio pulses known as
FRBs: The observed frequency of the bursts ($\sim 1.4\, {\rm GHz}$) is
comparable to
the proper frequency $mc^2/(2\pi \hbar)$ of the QCD axion with a mass $m\sim
10^{-5}\, {\rm eV/c^2}$, the observed rate of the bursts in a galaxy ($\sim
10^{-3}$ per year) gives a mass of axion stars $\sim 10^{-12}\, M_{\odot}$
which is comparable to their maximum mass,\footnote{Iwazaki
\cite{iwazaki} identifies this mass with the mass of axitons $M_{\rm
axiton}\sim 10^{-12}\, M_{\odot}$ \cite{kt}
while it would be more relevant to identify it with the maximum mass 
$M_{\rm max}=6.46\times 10^{-14}\,
M_{\odot}$ of dilute axion stars \cite{prd1,prd2}.} the size of the emitting
source must be
less than $300\, {\rm km}$ which is
consistent with the radius of axion stars $R_{99}^*$, and the large
amount of radiation energy of the bursts ($\sim 10^{38}-10^{40}\,
{\rm erg}$) is of the
order of $M_{\rm max} c^2$. Therefore, FRBs can be
matched in a model which assumes an explosive decay of axion stars as a result
of collisions with neutron stars and accretion disks. In the case of a
collapse above the maximum mass, the spectrum obtained in
\cite{tkachevprl} may serve as a distinctive signature of the axion star
collapse.

Another domain of application of these ideas concerns the formation and the
evolution of DM halos.
It has been proposed that DM halos could be giant BECs made of ULAs. In this
context, an important issue is whether the DM particle has a self-interaction or
not. In many works, the self-interaction of the boson is neglected because the
self-interaction parameter $|\lambda|\sim 10^{-91}$ is extremely small. However,
we have shown in this paper and in previous ones
\cite{prd2,bectcoll,abrilphas} that even for
apparently small values of  $|\lambda|$ we can be in a regime of strong
self-interaction. This is because, for the self-interaction to be weak in the
context of DM halos, 
$|\lambda|$ has to be small with respect to $(M_P/m)^2$, not with respect to
$1$.
Now, for ULAs,  $(M_P/m)^2$ is a very small quantity of the order of $10^{-98}$.
Therefore, we claim that the self-interaction of the axions has to be taken
into account, and that ULAs are actually in a strong self-interaction regime
(see Appendix \ref{sec_ws} for a more detailed discussion).

The sign of the scattering length is also of crucial importance. A repulsive
self-interaction has the tendency to stabilize a DM halo. In that case, the
mass of the ULA can be substantially larger than the value it should
have to account for the observations if it were noninteracting  (see Appendix D
of \cite{abrilphas}). For example, a fiducial
model of self-repulsive bosons corresponds to a mass $m=3\times 10^{-21}\, {\rm
eV/c^2}$ and a self-interaction constant $\lambda/8\pi=1.69\times
10^{-87}$ \cite{shapiro,abrilphas}. The boson mass  is about one order of
magnitude larger than
the value $m=2.92\times 10^{-22}\, {\rm
eV/c^2}$ required in the absence of self-interaction.
This is precisely what we need to alleviate some tensions with the observations
of
the Lyman-$\alpha$ forest \cite{hui} or some tensions with the observations of
the
abundance of
ultrafaint lensed galaxies at redshift $z\simeq 6$ in the Hubble Frontier Fields
(HFF) \cite{menci}. Therefore, a repulsive self-interaction may solve
these problems (see the Remark at the end of Appendix D.4 of \cite{abrilphas}).
We also note that cosmological constraints from CMB and
big bang nucleosynthesis (BBN) exclude the possibility that the bosons are
noninteracting \cite{shapiro,abrilphas}. In these models, which are based on a
complex SF, a repulsive
self-interaction between the bosons ($a_s>0$) seems to be the most likely in
order to account for the observations.

However, if the DM particle is an
axion (a real SF), the 
self-interaction is generally attractive ($a_s<0$). Now, an
attractive self-interaction has the tendency to destabilize a DM halo.
A DM halo can be stable only below a maximum mass
\cite{prd1,prd2}. Recently,
Diez-Tejedor and Marsh \cite{dtm} have obtained some
constraints on the value of the axion decay constant. They argue that if the
axion mass is temperature-independent, then $m\sim 10^{-22}\, {\rm eV/c^2}$ and
$f\gtrsim 10^{16} \, {\rm GeV}$. They also consider the possibility that 
the axion mass is temperature-dependent and discuss the case where
$m\sim 10^{-22}\, {\rm eV/c^2}$ and $f\sim 10^{15} {\rm GeV}$,
corresponding to $a_s=-1.96\times 10^{-64}\, {\rm fm}$ and $\lambda=-2.50\times
10^{-93}$. This leads to dilute axion stars with a maximum mass $M_{\rm
max}^{\rm exact}=1.11\times 10^9 (f/m)\, M_{\odot}$ [see Eq.
(\ref{maxmr1})], a minimum radius $(R_{99}^*)^{\rm exact}=428/(fm)\, {\rm pc}$
[see Eq. (\ref{maxmr2})], a maximum density $\rho_{\rm
dilute}^{\rm
exact}=2.29\times 10^{-16} (f^4 m^2)\, {\rm g/m^3}$ [see Eq.
(\ref{maxmr2b})], and a typical minimum energy $E_{\rm tot}\sim -1.19\times
10^{56} (f^3/m)\,
{\rm ergs}$
[see Eq. (\ref{etotcrit})], where the axion
mass
$m$ is
measured in units of $10^{-22}\, {\rm eV/c^2}$ and the axion decay constant $f$
is measured in units of $10^{15}{\rm GeV}$.
We should regard $M_{\rm
max}^{\rm exact}=1.11\times 10^9 (f/m)\, M_{\odot}$ as the maximum mass of the
solitonic core of
the DM halo, not the maximum mass of the DM halo itself (see footnote 1).
If some DM
halos have an axionic solitonic  core of mass $M_{\rm
core}>M_{\rm max}^{\rm
exact}=1.11\times 10^9 (f/m)\, M_{\odot}$, they should undergo
core collapse.
One possibility is that they form a dense axionic nucleus (axion star)
of mass
$M_{\rm
max}^{\rm exact}=1.11\times 10^9 (f/m)\, M_{\odot}$,
radius $R_{\rm coll}^{\rm exact}=0.949/(mf^{1/3})\, {\rm
pc}$ [see Eq. (\ref{tfng14})], and density
$\rho_{\rm dense}^{\rm exact}=2.10\times 10^{-8}
(m^2f^2)\, {\rm g/m^3}$ [see Eq. (\ref{tfng3rho})]. This
nucleus is not a black hole, not even a relativistic object, because
$R/R_S=8930f^{-4/3}\gg
1$. 
This dense
nucleus
has a very low pulsation period $2\pi/\omega_{\rm
coll}^{\rm exact}=8.24/(mf^{1/3})\, {\rm
yrs}$ [see Eq. (\ref{tfng20})] and a very
negative energy $(E_{\rm
tot}^{\rm coll})_{\rm exact}=-5.59\times 10^{62} (f/m)\, {\rm
ergs}$ [see Eq. (\ref{mc2})]. Note that this energy
is not of gravitational origin since $GM^2/R\sim 1.11\times 10^{59}
(f^{7/3}/m)\, {\rm ergs} \ll |E_{\rm
tot}^{\rm coll}|_{\rm exact}$ for $f\lesssim 1$. The formation time of a dense
axionic nucleus (collapse time) is
$t_{\rm coll}=2.90\gamma^{-1/4}
t_D=3.80/(\gamma^{1/4}mf^2)\, {\rm Myrs}$ where $\gamma=M/M_{\rm max}-1$
[see Eq. (185) of \cite{bectcoll}]. For $\gamma=10^{-4}$ this
gives
$t_{\rm
coll}=38.0/(mf^2)\, {\rm Myrs}$.

We therefore suggest that DM halos may harbor
a dense axionic nucleus (supermassive axion star). We note that part (or all?)
of the mass and energy of
the dilute axionic solitonic core may be lost during
the collapse and could be converted in a radiation as
studied by Levkov {\it et al.} \cite{tkachevprl}. Indeed, the unstable
solitonic core
of a DM halo with a mass $M_{\rm core}>M_{\rm max}$ could undergo a series of
collapses and explosions and emit  waves with a very characteristic spectrum
[see Fig 3(a) of \cite{tkachevprl}]. In that case, the dense axionic
nucleus would be the remant of a bosenova. It would have a mass $\sim 10^9
\, M_{\odot}$ if there is no mass loss during the collapse or smaller if there
is mass loss (e.g. by
radiation) during the
collapse.

Interestingly, the mass $\sim 10^9 \, M_{\odot}$ is of the order of the
maximum mass of the dark objects that are reported to exist at the center of
the galaxies. It is also interesting to note that certain characteristics of
dense axionic nuclei are
similar to the characteristics of quasistellar radio sources (quasars and active
galactic nuclei) in strong radio galaxies. They both present large energies
$\sim 10^{60}\, {\rm ergs}$ and cyclic variations with periods of the order of
$10$ years. At present, it is believed that a quasar consists of a supermassive
black hole
with a mass $\sim 10^6-10^{9}\, M_{\odot}$ surrounded by an accretion disk of
gas. Their source of
energy and their large
luminosity is the
result of the immense friction caused by the gas falling 
towards the black hole. The energy is released in the form of electromagnetic
radiation. The active galactic nucleus radiates in the direction of a jet
directed by the lines of a magnetic field. The jet velocity is close to the
speed of light. Quasars cease to shine when the the supermassive black hole has
consumed all the gas and dust around it. Our galaxy may have been active in the
past but is now quiet since its central black hole has nothing to ``eat'' to
produce radiation.

Instead of a central black hole, other possibilities have been contemplated.
In a first attempt to understand the source of the energy
emitted by the radio galaxies, Hoyle and Fowler \cite{hf}  suggested the
possibility that a mass of the order of $10^8\, M_{\odot}$ has
condensed in the galactic nucleus into a
supermassive star in which nuclear-energy generation takes place.
Such an object
would be supported in hydrostatic equilibrium almost entirely by radiation
pressure. In
the Hoyle-Fowler theory, the energy corresponds to
the gravitational energy of the supermassive star.  Assuming a radius $\sim
10^{-3}\,
{\rm pc}$ (corresponding to a superstar with a density $\sim 1 \, {\rm
g/m^3}$ and $R/R_S\sim 104$), this gives a gravitational energy $GM^2/R\sim
10^{60}\, {\rm ergs}$. 
The gravitational energy $GM^2/R$ may be made available for external dissipation
necessary to transfer angular momentum from the supermassive star to
the surrounding material. Hoyle and Fowler \cite{hf}  imagined a magnetized
field torroidally wound between the central star and a surrounding  disk and
argued that the energy stored in the magnetic field could be of the order of
$GM^2/R$. This powerful field may explode throwing jets from the center of
galaxies.  The superstars of mass $10^8\, M_{\odot}$ imagined by Hoyle and
Fowler \cite{hf} were found to be unstable from the viewpoint of general
relativity, but Fowler \cite{wf} showed that stability is gained if they are
rotating.

More recently, it has been proposed that a supermassive neutrino star
(fermion ball) may form at the center of a galaxy. It may arise through a
phase transition in a self-gravitating gas of neutrinos with mass $17.2\, {\rm
keV/c^2}$ from a dilute gaseous phase to a condensed phase when the temperature
passes below a critical value 
\cite{bilicviollierN,bilicviollierRG,ijmpb,kingfermionic,ruffini}. These
compact
dark objects with a mass $10^6-10^9\, M_{\odot}$ and a radius $\sim 10^{-3}-
10^{-2}\, {\rm pc}$ could mimic the effect of a supermassive black hole
purported to exist in the core of galaxies to power active galactic
nuclei and quasars.

Similarly, we could  consider the possibility that the central dark object in a
galaxy is a boson star 
\cite{torres2000,guzmanbh} or a
supermassive dense axion star. Realistic models should take into account the
formation of an accretion disc around the boson star.  An important
difference between dense axion stars and black holes resides in the fact that 
dense axion stars (like certain types of fermion balls and
boson stars) are
nonrelativistic objects.\footnote{We note that the compact object associated
with Sagittarius $A^*$ has a mass $M=4.31\times
10^6\, M_{\odot}$ and a radius $R\le
10^{10}\, {\rm km}$ implying $R/R_S\le 785$. Therefore, this
object is not necessarily  a black hole unless its radius
is much smaller than $R\le
10^{10}\, {\rm km}$.}
Still they can have a huge mass $M\sim 10^9 \, M_{\odot}$ and a huge energy
$E_{\rm tot}\sim  -M c^2\sim - 10^{60}\, {\rm ergs}$.

In conclusion, the formation of dense axion stars, the radiation emitted
during the collapse of a dilute axion star above the maximum mass (bosenova),
or the electromagnetic radiation created during the collision of a dilute axion
star with a neutron star or an accretion disk, may have
different  applications of great astrophysical interest such as FRBs and,
possibly, the formation of supermassive galactic nuclei accompanied by intense
radiation. These topics will be the object of future research.

\appendix

\section{Characteristic scales}
\label{sec_chs}

\subsection{The scales associated with self-interacting nonrelativistic
self-gravitating
BECs}
\label{sec_cis}

We introduce the mass and length scales associated with nonrelativistic
self-gravitating BECs with a $|\psi|^4$ self-interaction:
\begin{eqnarray}
M_a=\left (\frac{\hbar^2}{Gm|a_s|}\right )^{1/2}, \qquad R_a=\left
(\frac{|a_s|\hbar^2}{Gm^3}\right
)^{1/2}.
\label{cis1}
\end{eqnarray}
In the context of our paper, these scales characterize the maximum mass
$M_{\rm max,N}^{\rm dilute}$ and the
minimum
radius $R_{\rm *,N}^{\rm dilute}$ of nonrelativistic dilute axion stars. We
also introduce the
density scale 
\begin{equation}
\rho_{a}=\frac{M_a}{R_a^3}=\frac{Gm^4}{\hbar^2a_s^2},
\label{cis2}
\end{equation}
the timescale
\begin{eqnarray}
t_a=\frac{1}{\sqrt{G\rho_a}}=\frac{|a_s|\hbar}{Gm^2},
\label{cis9}
\end{eqnarray}
the global energy scale
\begin{eqnarray}
E_a=\frac{GM_a^2}{R_a}=\left
(\frac{Gm\hbar^2}{|a_s|^3}\right )^{1/2},
\label{cis10}
\end{eqnarray}
the individual energy scale\footnote{Dimensionally, the eigenenergy scale
is equal to
the gravitational energy of two bosons separated by a distance equal to their
scattering length. This is also the gravitational energy of a boson of ``size''
$|a_s|$.}
\begin{eqnarray}
\frac{E_a}{N_a}=\frac{Gm^2}{|a_s|},
\end{eqnarray}
where $N_a=M_a/m$, and the pressure scale
\begin{equation}
\label{cis11}
P_a=\frac{G^2m^5}{|a_s|^3\hbar^2}.
\end{equation}
These scales
can be expressed in terms of $f$ and
$\lambda$ as
\begin{eqnarray}
M_a=\left (\frac{32\pi f^2\hbar}{Gm^2c^3}\right )^{1/2}=\left
(\frac{8\pi\hbar c}{|\lambda| G}\right )^{1/2},
\label{cis4}
\end{eqnarray}
\begin{eqnarray}
R_a=\left (\frac{\hbar^3c^3}{32\pi f^2
Gm^2}\right )^{1/2}=\left (\frac{|\lambda|\hbar^3}{8\pi Gm^4 c}\right )^{1/2},
\label{cis3}
\end{eqnarray}
\begin{eqnarray}
\rho_a=\frac{(32\pi f^2)^2Gm^2}{\hbar^4c^6}=\frac{(8\pi)^2Gm^6c^2}{
\lambda^2\hbar^4}, 
\label{cis5}
\end{eqnarray}
\begin{eqnarray}
t_a=\frac{\hbar^2c^3}{32\pi
f^2Gm}=\frac{|\lambda|\hbar^2}{8\pi Gm^3c},
\label{cis9b}
\end{eqnarray}
\begin{equation}
E_a=(32\pi f^2)^{3/2}\left
(\frac{G}{\hbar m^2c^9}\right )^{1/2}
=\left (\frac{8\pi}{|\lambda|}\right )^{3/2}\left (\frac{Gm^4c^3}{\hbar}\right
)^{1/2},
\label{cis10b}
\end{equation}
\begin{equation}
\frac{E_a}{N_a}=32\pi f^2 \frac{Gm}{\hbar c^3}
=\frac{8\pi}{|\lambda|}\frac{Gm^3c}{\hbar},
\end{equation}
\begin{equation}
\label{cis11b}
P_a=(32\pi f^2)^3\frac{G^2m^2}{\hbar^5 c^9}=\left
(\frac{8\pi}{|\lambda|}\right )^3\frac{G^2m^8c^3}{\hbar^5}.
\end{equation}
They can also be written as
\begin{eqnarray}
M_a&=&\left (\frac{r_S}{2|a_s|}\right
)^{1/2}\frac{M_P^2}{m}
=\sqrt{32\pi}\frac{f}{M_Pc^2}\frac{M_{P}^2}{m}\nonumber\\
&=&\left (\frac{8\pi}{|\lambda|}\right )^{1/2}M_P,
\label{cis6}
\end{eqnarray}
\begin{eqnarray}
\label{cis7}
R_a&=&\left (\frac{2|a_s|}{r_S}\right
)^{1/2}\lambda_C
=\frac{1}{\sqrt{32\pi}}\frac{M_Pc^2}{f}\lambda_C\nonumber\\
&=&\left (\frac{|\lambda|}{8\pi}\right )^{1/2}\frac{M_P}{m}\lambda_C,
\end{eqnarray}
\begin{eqnarray}
\label{cis8}
\rho_a&=&\left (\frac{r_S}{2|a_s|}\right
)^{2}\frac{M_P^2}{m\lambda_C^3}
=1024\pi^2 \left (\frac{f}{M_Pc^2}\right )^4
\frac{M_P^2}{m\lambda_C^3}\nonumber\\
&=&\left (\frac{8\pi}{|\lambda|}\right )^{2}\left (\frac{m}{M_P}\right
)^4\frac{M_P^2}{m\lambda_C^3}=\left
(\frac{8\pi}{|\lambda|}\right )^{2}\frac{m^3}{M_P^2\lambda_C^3},\nonumber\\
\end{eqnarray}
\begin{eqnarray}
t_a&=&\frac{2|a_s|}{r_S}\frac{\lambda_C}{c}=\frac{1}{32\pi}\left (\frac{
M_Pc^2 } { f } \right )^2\frac{\lambda_C}{c}\nonumber\\
&=&\frac{|\lambda|}{8\pi}\left (\frac{M_P}{m}\right
)^2\frac{\lambda_C}{c},
\label{cis12}
\end{eqnarray}
\begin{eqnarray}
E_a&=&\left
(\frac{r_S}{2|a_s|}\right
)^{3/2}\frac{M_P^2c^2}{m}=(32\pi)^{3/2}\left (\frac{f}{
M_Pc^2 }  \right )^3\frac{M_P^2c^2}{m}\nonumber\\
&=&\left (\frac{8\pi}{|\lambda|}\right )^{3/2}\left (\frac{m}{M_P}\right
)^3\frac{M_P^2c^2}{m}=\left (\frac{8\pi}{|\lambda|}\right
)^{3/2}\frac{m^2c^2}{M_P},\nonumber\\
\label{cis13}
\end{eqnarray}
\begin{eqnarray}
\frac{E_a}{N_a}&=&
\frac{r_S}{2|a_s|}mc^2=32\pi\left (\frac{f}{
M_Pc^2 }  \right )^2 mc^2\nonumber\\
&=&\frac{8\pi}{|\lambda|}\left (\frac{m}{M_P}\right
)^3 M_P c^2=\frac{8\pi}{|\lambda|}\frac{m^3c^2}{M_P^2},\nonumber\\
\end{eqnarray}
\begin{eqnarray}
\label{cis14}
P_a&=&\left (\frac{r_S}{2|a_s|}\right
)^{3}\frac{M_P^2c^2}{m\lambda_C^3}=(32\pi)^{3}\left (\frac{f}{
M_Pc^2 }  \right )^6\frac{M_P^2c^2}{m\lambda_C^3}\nonumber\\
&=&\left (\frac{8\pi}{|\lambda|}\right )^{3}\left (\frac{m}{M_P}\right
)^6\frac{M_P^2c^2}{m\lambda_C^3}=\left (\frac{8\pi}{|\lambda|}\right
)^{3}\frac{m^5c^2}{M_P^4\lambda_C^3}.\nonumber\\
\end{eqnarray}

These scales appear in our works
\cite{prd1,prd2,bectcoll} on
self-gravitating BECs with attractive or repulsive self-interactions. For a
repulsive
self-interaction ($a_s>0$), they determine the transition between the
noninteracting limit and
the TF limit. For an attractive
self-interaction ($a_s<0$), $M_a$
is of the order of the maximum mass of a self-gravitating BEC and $R_a$ is
of the order of the corresponding radius. The mass-radius relation
$M(R)$ (parametrized by the central density $\rho_0$)
of nonrelativistic self-gravitating
BECs with a repulsive self-interaction is represented in Fig. 4 of \cite{prd2}.
In the noninteracting limit,  the mass-radius
relation is given by Eq. (\ref{maxmr2bb}). In the TF limit, the
radius of the system is independent of its mass and given by Eq.
(\ref{maxmr2pos}). At the transition, combining Eqs. 
(\ref{maxmr2bb}) and (\ref{maxmr2pos}), we obtain the scalings of Eq.
(\ref{cis1}). The mass-radius relation $M(R)$ (parametrized by
the central density $\rho_0$)
of nonrelativistic
self-gravitating
BECs with an attractive self-interaction is represented in Fig. 6 of
\cite{prd2}.
In the noninteracting limit,  the mass-radius
relation is given by Eq. (\ref{maxmr2bb}). In the nongravitational limit, the
mass-radius relation is given by Eq. (\ref{maxmr2cc}).\footnote{We recall
that these configurations are unstable.} At the transition,
combining Eqs. (\ref{maxmr2bb}) and (\ref{maxmr2cc}), we obtain the scalings of
Eqs.
(\ref{cis1}). They determine the maximum
mass of Newtonian self-attracting boson stars above which there is no
equilibrium state. By numerically solving the GPP
equations, or the equivalent hydrodynamic equations, one gets the
exact
values $M_{\rm max}=1.012 M_a$, $R_{\rm
min}=5.5\lambda_a$ and $(\rho_0)_{\rm
max}=0.5\rho_a$ \cite{prd2}.

\subsection{The scales associated with self-attractive and self-repulsive
nongravitational BECs}
\label{sec_ngsi}

We introduce the mass and length scales associated with
nongravitational BECs with an attractive $|\psi|^4$ self-interaction ($a_s<0$)
and a repulsive $|\psi|^6$ self-interaction:
\begin{eqnarray}
M_i=\frac{\hbar}{|a_s|c},\qquad R_i=\frac{\hbar}{mc}.
\label{ngsi1}
\end{eqnarray}
We note that $R_i$ coincides with the Compton
wavelength $\lambda_C$ of the bosons.
To our knowledge, these
scales have not been introduced
previously. In the context of our paper, they characterize
the minimum mass $M_{\rm min,N}^{\rm
dense}$ and the corresponding radius $R_{\rm *,N}^{\rm dense}$ of dense axion
stars, as well as  their minimum radius $R_{\rm min,N}^{\rm dense}$ 
and the corresponding mass $M_{\rm *,N}^{\rm dense}$.   We
also introduce the
density scale
\begin{eqnarray}
\rho_i=\frac{M_i}{R_i^3}=\frac{m^3c^2}{|a_s|\hbar^2}.
\label{ngsi2}
\end{eqnarray}
It characterizes the constant density $\rho_{\rm dense}$ of nongravitational
dense axion stars
in the TF limit.
These scales can be expressed in terms of $f$
and $\lambda$ as
\begin{equation}
\label{ngsi6}
M_i=\frac{32\pi
f^2}{mc^4}=\frac{8\pi m}{|\lambda|},
\end{equation}
\begin{equation}
\label{ngsi7}
\rho_i=\frac{32\pi m^2f^2}{\hbar^3 c}=\frac{8\pi
m^4c^3}{|\lambda|\hbar^3}. 
\end{equation}
They can also be written as
\begin{eqnarray}
\label{ngsi4}
M_i&=&\frac{r_S}{2|a_s|}\frac{M_P^2}{m}=32\pi\left (\frac{f}{M_P c^2}\right
)^2\frac{M_P^2}{m}\nonumber\\
&=&\frac{8\pi}{|\lambda|}\left
(\frac{m}{M_P}\right
)^2\frac{M_P^2}{m}=\frac{8\pi m}{|\lambda|},
\end{eqnarray}
\begin{equation}
\label{ngsi3}
R_i= \lambda_C,
\end{equation}
\begin{eqnarray}
\label{ngsi5}
\rho_i&=&\frac{r_S}{2|a_s|}\frac{M_P^2}{m\lambda_C^3}=32\pi\left (\frac{f}{M_P
c^2}\right
)^2\frac{M_P^2}{m\lambda_C^3}\nonumber\\
&=&\frac{8\pi}{|\lambda|}\left
(\frac{m}{M_P}\right
)^2\frac{M_P^2}{m\lambda_C^3}=\frac{8\pi}{|\lambda|}\, \frac{m
}{\lambda_C^3}.
\end{eqnarray}

\subsection{The scales associated with dense axion stars resulting from a
collapse at  $M_{\rm max,N}^{\rm dilute}$}
\label{sec_oc}

We introduce the mass and length scales associated with dense axion
stars
resulting from the collapse of dilute axion stars at the critical point:
\begin{eqnarray}
M_d=\frac{\hbar}{\sqrt{Gm|a_s|}}, \qquad R_d=\left
(\frac{|a_s|\hbar^6}{Gm^7c^4}\right )^{1/6}.
\label{oc1}
\end{eqnarray}
They characterize $M_{\rm max,N}^{\rm dilute}$ and $R_{\rm coll}$.
We also introduce the
density scale 
\begin{equation}
\rho_{d}=\frac{M_d}{R_d^3}=\frac{m^3c^2}{|a_s|\hbar^2}.
\label{oc2}
\end{equation}
We note that
\begin{eqnarray}
\label{oc3}
M_d=M_a,\qquad \rho_d=\rho_i.
\end{eqnarray}
These scales can be expressed in terms of $f$
and $\lambda$ as in Eqs. (\ref{cis4}), (\ref{ngsi7}) and
\begin{eqnarray}
\label{oc6}
R_d=\left (\frac{\hbar^7}{32\pi Gf^2m^6c}\right )^{1/6}=\left
(\frac{|\lambda|\hbar^7}{8\pi G m^8c^5}\right )^{1/6}.
\end{eqnarray}
They can also be written as in Eqs. (\ref{cis6}), (\ref{ngsi5}) and 
\begin{eqnarray}
\label{oc4}
R_d&=&\left
(\frac{2|a_s|}{r_S}\right
)^{1/6}\lambda_C=\frac{1}{(32\pi)^{1/6}}\left (\frac{M_Pc^2}{f}\right
)^{1/3}\lambda_C\nonumber\\
&=&\left (\frac{|\lambda|}{8\pi}\right
)^{1/6}\left (\frac{M_P}{m}\right )^{1/3}\lambda_C.
\end{eqnarray}

\subsection{The scales associated with dilute axion stars resulting from an
explosion at  $M_{\rm min,N}^{\rm dense}$ }
\label{sec_oe}

We introduce the mass and length scales associated with dilute axion
stars
resulting from the explosion of dense axion stars at the critical point:
\begin{eqnarray}
M_e=\frac{\hbar}{|a_s|c}, \qquad R_e=\frac{|a_s|\hbar c}{Gm^2}.
\label{oe1}
\end{eqnarray}
They characterize $M_{\rm min,N}^{\rm dense}$ and $R_{\rm exp}$.
We also introduce the
density scale 
\begin{equation}
\rho_{e}=\frac{M_e}{R_e^3}=\frac{G^3m^6}{|a_s|^4\hbar^2c^4}.
\label{oe2}
\end{equation}
We note that
\begin{eqnarray}
M_e=M_i.
\end{eqnarray}
These scales  can be expressed in terms of $f$
and $\lambda$ as in Eq. (\ref{ngsi6}) and
\begin{eqnarray}
\label{oe5}
R_e=\frac{\hbar^2c^4}{32\pi f^2 G m}=\frac{|\lambda|\hbar^2}{8\pi G
m^3},
\end{eqnarray}
\begin{equation}
\rho_{e}=\frac{(32\pi f^2)^4 G^3m^2}{\hbar^6c^{16}}=\frac{(8\pi)^4G^3m^{10}}{
\lambda^4\hbar^6}.
\label{oe6}
\end{equation}
They can also be written as in Eq. (\ref{ngsi4}) and
\begin{eqnarray}
\label{oe3}
R_e&=&
\frac{2|a_s|}{r_S}\lambda_C=\frac{1}{32\pi}\left
(\frac{M_Pc^2}{f}\right
)^{2}\lambda_C\nonumber\\
&=&\frac{|\lambda|}{8\pi}\left (\frac{M_P}{m}\right )^{2}\lambda_C,
\end{eqnarray}
\begin{eqnarray}
\label{oe4}
\rho_e&=&\left (\frac{r_S}{2|a_s|}\right
)^4\frac{M_P^2}{m\lambda_C^3}=(32\pi)^4\left (\frac{f}{M_P
c^2}\right
)^8\frac{M_P^2}{m\lambda_C^3}\nonumber\\
&=&\left (\frac{8\pi}{|\lambda|}\right )^4\left
(\frac{m}{M_P}\right
)^8\frac{M_P^2}{m\lambda_C^3}=\left (\frac{8\pi}{|\lambda|}\right
)^4\frac{m^7}{M_P^6\lambda_C^3}.\nonumber\\
\end{eqnarray}

\subsection{The scales associated with noninteracting relativistic
self-gravitating
 BECs}
\label{sec_cs}

We introduce the mass and length scales associated with noninteracting
relativistic self-gravitating  BECs: 
\begin{eqnarray}
M_c=\frac{\hbar c}{Gm},\qquad R_c=\frac{\hbar}{mc}.
\label{cs1}
\end{eqnarray}
They are connected to each other by the relativistic scaling $M_c=c^2R_c/G$. We
note that $R_c$ coincides with the Compton
wavelength $\lambda_C$ of the bosons. In
the context of our paper,
these scales characterize the maximum mass $M_{\rm max,GR}^{\rm dilute}$ and
the corresponding radius $R_{\rm *,GR}^{\rm dilute}$ of relativistic dilute
axion stars. We also introduce the
density scale  
\begin{equation}
\rho_c=\frac{M_c}{R_c^3}=\frac{m^2c^4}{G\hbar^2}.
\label{cs2}
\end{equation}
These scales can be rewritten as
\begin{eqnarray}
\label{cs3}
M_c=\frac{M_P^2}{m},\qquad R_c=\lambda_C,\qquad
\rho_c=\frac{M_P^2}{m\lambda_C^3}.
\end{eqnarray}

These scales appear in the
works of Kaup \cite{kaup} and Ruffini and
Bonazzola  \cite{rb} on boson stars. $M_c$ is of the order of
the maximum mass of a noninteracting relativistic self-gravitating  BEC and
$R_c$ is
of the order of the corresponding radius. The
mass-radius relation $M(R)$ of noninteracting
self-gravitating
BECs in the context of general relativity is represented in Fig. 3
of \cite{seidel90}. In the nonrelativistic limit, the
mass-radius
relation is given by Eq.  (\ref{maxmr2bb}). This relation is
obtained by balancing the repulsive quantum pressure against the
attractive gravitational force (see Sec. II. G of \cite{prd1}). The radius of
the system is of the order of the particle's de Broglie wavelength $R\sim
\hbar/mv$ associated with the virial velocity $v\sim (GM/R)^{1/2}$.
The system becomes
relativistic when its radius
$R$ approaches the Schwarzschild radius $R_S$. Combining Eqs.
(\ref{maxmr2bb}) and
(\ref{sw}) we obtain the scalings of Eq. (\ref{cs1}) (see Appendix B of
\cite{prd1}). They
determine the
maximum mass and minimum radius of noninteracting boson stars beyond which there
is no equilibrium
state. By
numerically solving the KGE equations, one gets the exact values $M_{\rm
max}=0.633 M_c$ and $(R_{95})_{\rm min}=6.03R_c$ (leading
to $R_{95}/R_S=4.76$) \cite{kaup,rb}. 
The mass-radius relation forms a spiral but the configurations  with a central
density $\epsilon_0\ge
(\epsilon_0)_{\rm max}$ where  $(\epsilon_0)_{\rm max}$ is the density
corresponding to the configuration with the maximum mass are unstable.

\subsection{The scales associated with self-attractive and self-repulsive
relativistic self-gravitating BECs}
\label{sec_is}

We introduce the mass and length scales associated
with  relativistic self-gravitating BECs with an attractive $|\psi|^4$
self-interaction ($a_s<0$)
and a repulsive $|\psi|^6$ self-interaction:
\begin{eqnarray}
M_r=\left
(\frac{|a_s|\hbar^2 c^4}{G^3m^3}\right )^{1/2}, \qquad R_r=\left
(\frac{|a_s|\hbar^2}{Gm^3}\right )^{1/2}.
\label{is1}
\end{eqnarray}
They are connected to each other by the relativistic scaling $M_r=c^2R_r/G$. 
In the context of our paper, these scales characterize the maximum mass
$M_{\rm
max,GR}^{\rm
dense}$ and the corresponding radius $R_{\rm *,GR}^{\rm dense}$ of general
relativistic dense axion stars as well as  the mass $M_0$ and
the corresponding radius $R_0$ marking the transition between
nongravitational and nonattractive Newtonian dense axion stars. 
We also introduce the  density
scale
\begin{equation}
\rho_r=\frac{M_r}{R_r^3}=\frac{m^3c^2}{|a_s|\hbar^2}.
\label{is2}
\end{equation}
We note that
\begin{equation}
R_r= R_a, \qquad \rho_r=\rho_i
\end{equation}
These scales can be expressed in terms of $f$ and $\lambda$ as in
Eqs. (\ref{cis3}), (\ref{ngsi7}) and
\begin{equation}
M_r=\left (\frac{\hbar^3c^7}{32\pi f^2 G^3m^2}\right )^{1/2}=\left
(\frac{|\lambda|\hbar^3c^3}{8\pi G^3m^4}\right )^{1/2}.
\label{is5}
\end{equation}
They can also be written as in Eqs. (\ref{cis7}), (\ref{ngsi5}) and
\begin{eqnarray}
\label{is3}
M_r&=&\left
(\frac{2|a_s|}{r_S}\right )^{1/2}\frac{M_P^2}{m}
=\frac{1}{\sqrt{32\pi}}\frac{M_Pc^2}{f}\frac{M_P^2}{m}\nonumber\\
&=&\left
(\frac{|\lambda|}{8\pi}\right
)^{1/2}\frac{M_P^3}{m^2}.
\end{eqnarray}

To the best of our knowledge, these scales have not been
introduced previously. However, apart from the presence of
the
absolute value, $R_r$ and $M_r$ have the same expressions
as the scales associated with relativistic BECs with a purely repulsive
$|\psi|^4$ self-interaction ($a_s>0$). These scales appear in the works of Colpi
{\it et al.}
\cite{colpi} and Tkachev \cite{tkachev} on boson
stars and in the work of Chavanis and Harko \cite{chavharko} on neutron
stars with a superfluid core (BEC stars). In that context, $M_r$ is of the order
of the maximum
mass of a relativistic self-gravitating BEC in the TF limit and $R_r$ is
of the order of the corresponding radius. The mass-radius relation
$M(R)$ (parametrized by the central density $\epsilon_0$)
of self-gravitating
BECs with a repulsive self-interaction in the context of general relativity is
represented in  Fig. 9 of \cite{chavharko}. In the nonrelativistic limit, the
radius of the system is independent of its mass and given by Eq.
(\ref{maxmr2pos}).  It is
obtained by balancing the repulsive pressure force coming from the
self-interaction against the attractive gravitational force (see
Sec. II. G of \cite{prd1}). The system
becomes relativistic when its radius
$R$ approaches the Schwarzschild radius $R_S$. Combining Eqs. 
(\ref{maxmr2pos}) and (\ref{sw}), we obtain the scalings of Eq.
(\ref{is1})  (see Appendix B
of \cite{prd1}).  They
determine the
maximum mass and the minimum radius of self-repulsive boson stars beyond which
there
is no equilibrium
state. By numerically solving
the KGE equations \cite{colpi}
or the equivalent hydrodynamic equations \cite{chavharko}, one gets the exact
values $M_{\rm max}=0.307 M_r$, $R_{\rm
min}=1.92 R_r$ and $(\epsilon_0)_{\rm
max}=1.19\rho_r c^2$ (leading to $R/R_S=3.13$). Chavanis and Harko
\cite{chavharko} have argued that, because
of their superfluid core, neutron stars could actually be BEC stars. Indeed, the
neutrons could form Cooper pairs and behave as bosons of mass $2m_n$ (where
$m_n=0.940\, {\rm GeV/c^2}$ is the mass of the neutron). By adjusting the
value of the self-interaction constant, the maximum mass of these BEC stars
could account for the abnormal mass (in the range $2-2.4\, M_{\odot}$) of
certain neutron stars \cite{Lat,Dem,black1,black2,black3,antoniadis} that
is much larger than the Oppenheimer-Volkoff  limit $M_{\rm OV}=0.384\, 
M_P^3/m^2=0.709\,  M_\odot$  based on the assumption
that neutron stars are ideal fermion stars \cite{ov}.

\section{On the collapse time of axion stars}
\label{sec_tco}

When the self-interaction is purely attractive, the collapse time $t_{\rm
coll}(M,R_0)$ of a dilute axion star with mass
$M>M_{\rm max}$ and  initial radius $R_0$ towards a singular
state of vanishing radius ($R=0$) has been studied in
detail in our previous paper \cite{bectcoll}
within the Gaussian ansatz. Eby
{\it et al.} \cite{ebycollapse} used our formalism to
study the collapse time of a  dilute axion star towards a dense axion star,
taking
into account the fact that the dense axion star has a nonzero radius $R>0$.
We show below that this refinement is not necessary. Since
the radius of a dense axion star is
extremely small, the same result for $t_{\rm coll}(M,R_0)$ is
obtained by taking $R=0$ as in \cite{bectcoll} instead of $R=0^+$ as in
\cite{ebycollapse}. The results obtained in  \cite{bectcoll} are also relevant
if, instead of forming a dense axion star, the collapse of a dilute axion star
is followed by an explosion and an emission of relativistic axions as in the
simulations of Levkov {\it et al.} \cite{tkachevprl}. For a purely attractive
self-interaction, $t_{\rm coll}(M,R_0)$ represents the moment at which the
self-similar collapse leads to a finite time singularity. When collisions
between axions are taken into account, it gives a
good estimate of the time at which the explosion takes place. Again, the value
of $t_{\rm coll}$ is not expected to be very sensitive on the exact value
of $R_e$ at which the explosion takes place as long as $R_e\ll R_0$.

\begin{figure}
\begin{center}
\includegraphics[clip,scale=0.3]{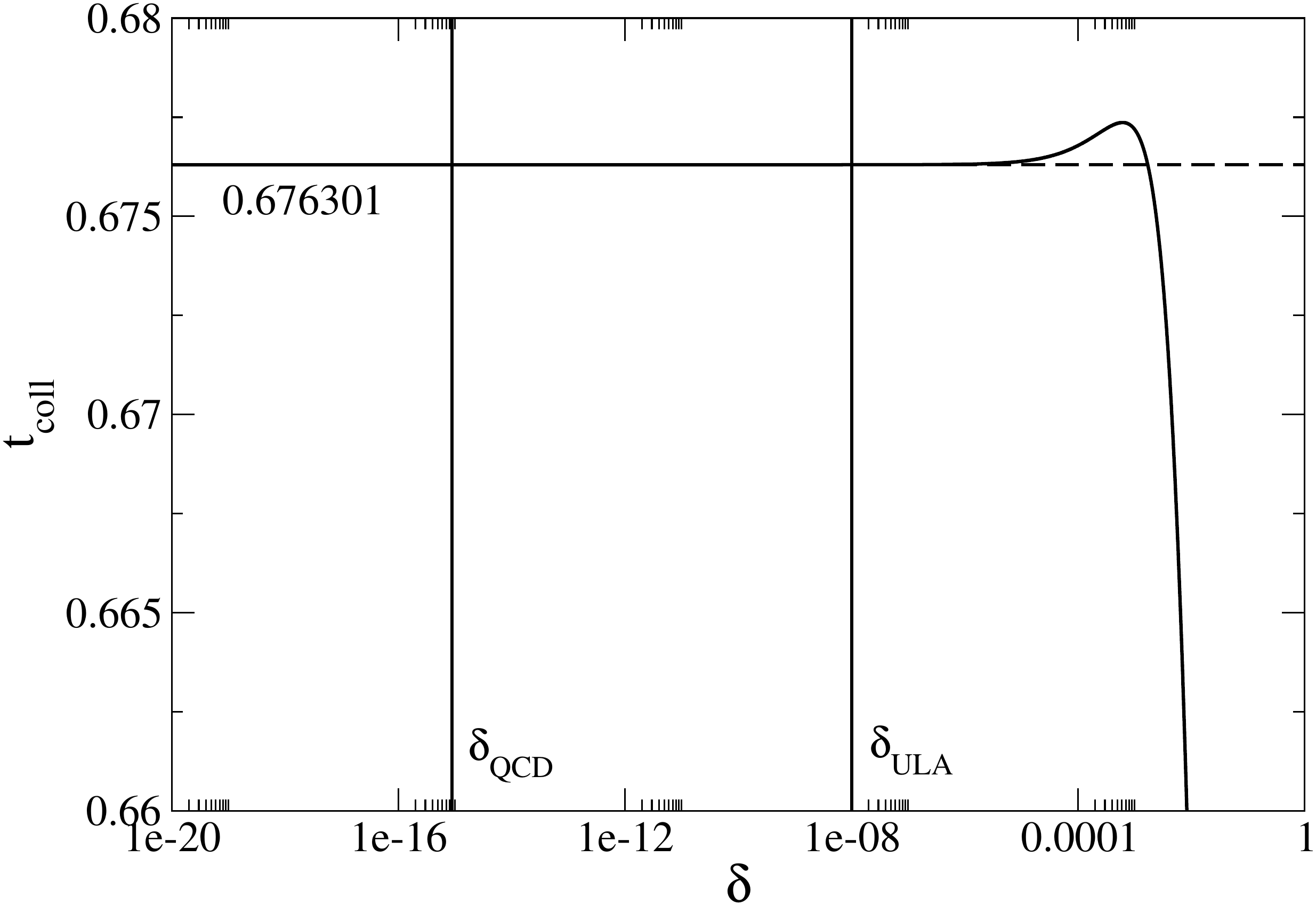}
\caption{Collapse time of axion stars as a function of $\delta$ for $M=2$ and
$R_0=1$. We note that it presents a small maximum at
$\delta=6.16\times 10^{-4}$ before decreasing. The dashed line corresponds to
the result obtained in \cite{bectcoll} by taking $\delta=0$. It provides an
excellent approximation of the collapse time of QCD axions
($\delta_{\rm QCD}=8.88\times
10^{-16}$) and ULAs ($\delta_{\rm ULA}=1.02\times
10^{-8}$).}
\label{tcoll}
\end{center}
\end{figure}

Within the Gaussian ansatz, the equation of motion of a self-gravitating  BEC
with mass $M$ and initial radius $R_0$ is
given by \cite{bectcoll}:
\begin{eqnarray}
\int_{R_0}^{R(t)}\frac{dR}{\sqrt{E_{\rm tot}-V(R)}}=\pm\left (\frac{2}{M}\right
)^{1/2}t,
\label{tco1}
\end{eqnarray}
where the sign $+$ corresponds to an explosion/expansion and the sign $-$ to a
collapse/contraction. In the following, we assume $\delta<\delta_c$ and
$M>M_{\rm
max}(\delta)$ so that a collapse takes place. If the BEC starts from $R_0$
without initial velocity ($\dot R_0=0$), its total energy is given by $E_{\rm
tot}=V(R_0)$. In the case where the collapse is stopped by
the self-repulsion of the bosons, the collapse time is given by
\begin{eqnarray}
t_{\rm coll}=\left (\frac{M}{2}\right
)^{1/2}\int_{R}^{R_0}\frac{dR'}{\sqrt{V(R_0)-V(R')}},
\label{tco2}
\end{eqnarray}
where $R$ is the radius of the resulting dense axion star with mass $M$. It is
given by
Eq.
(\ref{mr2}), which can be rewritten in the more convenient form
\begin{equation}
\label{tco3}
\delta=\frac{R^3(1+R^2)M-2R^4}{6M^2}.
\end{equation}
For $\delta\ll 1$, we have in good approximation
\begin{equation}
\label{tco4}
R\sim (6M\delta)^{1/3}.
\end{equation}
For given $M$ and $R_0$, we can plot $t_{\rm coll}(\delta)$ by proceeding as
follows (for simplicity, we do not explicitly write $M$ and $R_0$ in the
function $t_{\rm coll}(M,R_0,\delta)$). Instead of prescribing
$\delta$, we prescribe $R$ and calculate $\delta(R)$ from Eq. (\ref{tco3}). We
can then determine $t_{\rm coll}(R)$ from Eq. (\ref{tco2}). By running $R$
between $0$ and $1$, we can  obtain $t_{\rm
coll}(\delta)$ in parametric form, with parameter $R$. The result of this
calculation is reported in Fig. \ref{tcoll} for   $M=2$ and $R_0=1$.

This ``exact'' result can be compared with the ``approximate'' expression of
$t_{\rm coll}$ obtained in \cite{bectcoll} which amounts to taking
$\delta=R=0$ in Eq. (\ref{tco2}).  For   $M=2$ and $R_0=1$, we have
found that  $t_{\rm coll}=0.676301...$ \cite{bectcoll}. As expected, this
``approximate'' result provides an excellent approximation of the ``exact''
result
for the values of $\delta$ corresponding to QCD axions and ULAs.

For QCD axions with   $\delta=8.88\times
10^{-16}$, we obtain $R=2.20\times 10^{-5}$ and $t_{\rm coll}=0.676$ (in
terms of dimensional variables, this corresponds to $R_{\rm
min}=1.57\, {\rm m}$ and $t_{\rm coll}=2.30\, {\rm hrs}$).

For ULAs with $\delta=1.02\times
10^{-8}$, we obtain $R=4.96\times 10^{-3}$ and $t_{\rm coll}=0.676$ (in terms of
dimensional variables, this corresponds to $R=1.55\, {\rm pc}$ and
$t_{\rm coll}=10.5\, {\rm Myrs}$).

\section{Dirac peaks and black holes}
\label{sec_dpbh}

\subsection{Absence of global minimum of energy in the case of a purely
attractive
self-interaction}
\label{sec_pf}

Let us consider a self-gravitating BEC with a purely attractive
$|\psi|^4$ self-interaction as in \cite{bectcoll}. The energy functional $E_{\rm
tot}[\rho]$ is given by Eq. (\ref{pm1}), where $V(\rho)$ is given by the first
term of Eq. (\ref{s1}). To show the
absence of a global energy minimum $E_{\rm tot}$ at fixed mass $M$, it is
sufficient to construct a particular
family of distributions $\rho_{\epsilon}({\vec x})$, depending on a parameter
$\epsilon$,
that conserve the mass $M$, and for which the energy $E_{\rm
tot}[\rho_{\epsilon}]$ tends to $-\infty$ when
$\epsilon\rightarrow 0$ (we use a strategy similar to the one developed in
Appendixes  A and B of \cite{sp2} in relation to the thermodynamics of
self-gravitating systems). In the present
context, it is convenient to take a Gaussian density profile  of size
$R=\epsilon$ \cite{prd1,bectcoll,chavtotal}. We also take ${\bf u}={\bf 0}$
since we are interested in equilibrium states. In that case, the total energy
$E_{\rm tot}=V(R)=V({\epsilon})$ is given by Eq. (\ref{ep2}) without the last
term (corresponding to a repulsive self-interaction). For
$R=\epsilon\rightarrow 0$, the total energy clearly goes to $-\infty$ implying
the absence of a global minimum of energy for any mass $M$.

The previous argument proves the absence of a global minimum of energy at fixed
mass. However, it says nothing about the possible existence of local energy
minima (metastable states). As we have seen, for $M<M_{\rm max}$, there exists a
local energy minimum corresponding to a dilute axion star \cite{prd1}. As
discussed in Sec. \ref{sec_lifetime}, this metastable state can have a very long
lifetime so it is physically relevant. However, for
$M>M_{\rm max}$ there is no equilibrium state at all and the system is expected
to collapse.\footnote{This assumes that the system remains compact. If not, a
star of mass $M>M_{\rm max}$ can break into different pieces of mass $M'<M_{\rm
max}$ and form a stable system \cite{cotner}.}

The family of distributions contructed above forms a Dirac peak
$\rho(\vec x)=M\delta(\vec x)$ when $R=\epsilon\rightarrow 0$. This Dirac peak
has an infinitely negative energy ($E_{\rm tot}\rightarrow -\infty$). Therefore,
we
expect that  a self-gravitating BEC with a purely attractive
$|\psi|^4$ self-interaction will ultimately form a Dirac peak if $M>M_{\rm
max}$. As discussed in
the Introduction, this Dirac peak may appear in the post-collapse regime of
the dynamics. Since a Dirac peak has a finite mass and a vanishing radius, it
may be regarded as a form of ``extreme'' black hole. However,
to make things
rigorous, and prove that a
self-gravitating BEC with a purely attractive self-interaction really collapses 
into a black hole, one must take general relativity into account and study the
KGE
equations.

\subsection{The case of self-gravitating BECs with attractive and repulsive
self-interactions}

We now consider a self-gravitating BEC with an 
attractive
$|\psi|^4$ self-interaction and a repulsive $|\psi|^6$ self-interaction.
In that case, we have seen that there exists a global minimum of energy for
any mass $M$. There also exists a local minimum of energy for $M_{\rm
min}< M < M_{\rm max}$ as described in Sec. \ref{sec_sapt}. In the limit
$\delta\rightarrow 0$, it seems that the general mass-radius relation
(\ref{mr4}) reduces to the mass-radius relation (\ref{mrd1}) of dilute axion
stars, suggesting that above $M_{\rm max}$ the system forms a Dirac peak
according to the arguments given in Appendix \ref{sec_pf}.  However, when
repulsive interactions are included, we have shown that dense axion stars with
$M_{\rm max}<M<M_0$ are {\it not} black holes,
whatever the value of $\delta>0$. We note furthermore that
$M_0\rightarrow +\infty$ when
$\delta\rightarrow 0$ implying that this result is true for any mass $M$. How
can we
solve this apparent paradox?

We must be careful that the limit
$\delta\rightarrow 0$ is not equivalent to $\delta=0$. Indeed, for
$\delta\rightarrow 0$ the mass-radius relation
tends to the mass-radius relation of dilute axion stars only for sufficiently
large radii. At very small radii we recover the branch of dense
axion stars that is pushed towards the vertical axis $R=0$ as $\delta\rightarrow
0$. To be specific, let us consider a dense axion star with a mass 
$M\in [M_{\rm max},M_0]$ resulting from the collapse of a dilute axion star.
It is a global minimum of energy (fully stable state). For a fixed mass
$M$ and
$\delta\rightarrow 0$ its radius tends to zero as $R\sim (6\delta M)^{1/3}$
[see Eq. (\ref{ray1})] and its energy tends to $-\infty$ as $E_{\rm tot}\sim
-M/(36\delta)$ [see Eq. (\ref{te3})]. 
Therefore, in the limit $\delta\rightarrow 0$, a dense axion star tends towards
a
Dirac peak $\rho({\vec x})\rightarrow M\delta({\vec x})$. This is compatible
with the case
$\delta=0$ considered in Appendix \ref{sec_pf}. However, for $\delta>0$, the
dense axion
star is never a black hole, whatever small $\delta$ may be (the
reason is that, in dimensionless variables, the Schwarzschild radius depends on
$\delta$). Indeed,
it would be
a black hole if $R\sim R_S$. Since $R\sim (6\delta M)^{1/3}$ [see Eq.
(\ref{ray1})] and $R_S=3.23
M\delta$ [see Eq. (\ref{qgr1})], we find that $R\gg R_S$ for $\delta\rightarrow
0$
(assuming  $M\ll
M_0=1/6\delta$).  Therefore, the black hole criterion is
never satisfied for dense axion
stars although they tend to Dirac peaks as $\delta\rightarrow 0$. On the
contrary, such structures are highly nonrelativistic. In conclusion, for
$\delta\rightarrow 0$ the dense axion star tends towards a Dirac peak, but it is
never a black hole.

Another way to resolve this apparent paradox is to
notice that the limit $\delta\rightarrow 0$ corresponds to $c\rightarrow
+\infty$ [see Eq. (\ref{ep8})], the other parameters being fixed. In this
limit, the Schwarzschild radius
$R_S=2GM/c^2$ tends to zero as $1/c^2$ while, according to Eq.
(\ref{tf2c}),
the radius of the dense axion star tends to zero as $1/c^{2/3}$. Therefore,
$R\gg R_S$ implying that the dense axion star is not a black hole.

{\it Remark:} Things would be different if the coefficient of
the repulsive term in Eq. (\ref{s2}) involved a new parameter $\chi$,
independent of the other parameters, measuring the
strength of the repulsive self-interaction. In that case, when
$\chi\rightarrow 0$, the radius of the star would tend to
zero ($R\rightarrow 0$) and, at some point, $R$ would be of the order of $R_S$
implying that the star is a black hole. This situation does not
occur for axion stars because the coefficient of the repulsive term in Eq.
(\ref{s2}) depends on $a_s$, $m$, $c$ and $\hbar$ so that it is completely
constrained.

\section{The Poincar\'e theory of linear series of equilibria}
\label{sec_p}

In this Appendix, we briefly explain how one can apply the
Poincar\'e \cite{poincare}  theory of linear series of equilibria to our
problem in order to
determine the stability of axion stars.

In a static Universe ($a=1$ and $H=0$), the GPP equations (\ref{gpe6}) and
(\ref{gpe7}), or the
equivalent quantum EP equations (\ref{mad3})-(\ref{mad6}), conserve the
total energy
\begin{equation}
\label{pm1}
E_{\rm tot}=\int\rho \frac{{\bf u}^2}{2}\, d{\bf r}+\frac{1}{m}\int \rho Q\,
d{\bf r}+\int
V(\rho)\, d{\bf r}
+\frac{1}{2}\int\rho\Phi\, d{\bf r}
\end{equation}
and the mass
\begin{eqnarray}
\label{p0}
M=\int\rho\, d{\bf r},
\end{eqnarray}
where we have expressed these functionals in terms of the hydrodynamic
variables $\rho$ and ${\bf u}$ (see Ref. \cite{chavtotal} for details).
As a result, a stable  equilibrium state
 is a (local) minimum of energy $E_{\rm tot}$ at fixed
mass $M$.\footnote{From  general considerations, one can show that a
global minimum of energy at fixed mass is nonlinearly dynamically stable
\cite{holm}. However, we shall also be interested in local minima of energy at
fixed mass that are linearly  dynamically stable. These states are
metastable but they have very long lifetimes (see Sec. \ref{sec_lifetime}) so
they are
fully relevant for our problem.} We therefore have to consider the
minimization problem 
\begin{eqnarray}
\label{p1}
E_{\rm tot}(M)=\min_{\rho,{\bf u}} \left\lbrace E_{\rm tot}[\rho,{\bf u}]\quad
|\quad
M\quad {\rm
fixed}\right\rbrace.
\end{eqnarray}
An equilibrium state, being a critical
point of energy at fixed mass, is
determined by the variational
principle
\begin{eqnarray}
\label{p2}
\delta E_{\rm tot}-\frac{\mu}{m}\delta M=0,
\end{eqnarray}
where $\mu$ is a Lagrange multiplier that takes  into account the mass
constraint.
Using the results of \cite{chavtotal}, this variational problem gives
${\bf u}={\bf 0}$ (the equilibrium state is static) and the eigenvalue equation
\begin{eqnarray}
\label{p3}
m\Phi+m h(\rho)+Q=\mu,
\end{eqnarray}
where we recall that $Q=-(\hbar^2/2m)\Delta\sqrt{\rho}/\sqrt{\rho}$
and $\Phi({\bf r})=-G\int
{\rho({\bf r}')}{|{\bf r}-{\bf r}'|}^{-1}\,
d{\bf r}'$. The density can be written formally as
\begin{eqnarray}
\label{p5}
\rho=h^{-1}\left (\frac{\mu}{m}-\frac{Q}{m}-\Phi\right ).
\end{eqnarray}
We note, however, that Eq. (\ref{p5}) is an integrodifferential equation.
Taking the gradient of Eq. (\ref{p3}), and using Eq. (\ref{mad8}), 
we obtain the condition of hydrostatic equilibrium 
\begin{eqnarray}
\label{p3b}
\nabla P+\rho\nabla\Phi+\frac{\rho}{m}\nabla Q={\bf 0}.
\end{eqnarray}
This returns the steady state of the quantum Euler equation (\ref{mad5}).
Equation
(\ref{p3}) is also equivalent to the time-independent GP equation obtained by
substituting into Eq. (\ref{gpe6}) a wavefunction of the form $\psi({\bf
r},t)=\phi({\bf r})e^{-iEt/\hbar}$ and making the
identification
\begin{eqnarray}
\label{p4}
\mu=E.
\end{eqnarray}
This shows that the Lagrange multiplier 
(chemical potential) in the variational problem associated with Eq. (\ref{p1})
is equal to the eigenenergy $E$.  Inversely, the eigenenergy $E$ may be
interpreted as a chemical potential. 

We call critical point (or equilibrium state) any solution of the variational
principle (\ref{p2}).
The ensemble of these critical points forms a series of equilibria. An
equilibrium state is determined  by the differential equation (\ref{p3}). This
equation  is equivalent to the fundamental equation of hydrostatic
equilibrium obtained by coupling Eq. (\ref{p3b}) to the Poisson equation
(\ref{mad6}). There can be
several equilibrium states with the same mass $M$ because the variational
problem
(\ref{p2}) can have several solutions: a global minimum of energy (fully
stable state), a local minimum of  energy
(metastable state), and
a maximum or saddle point of energy (unstable state). It is convenient to
represent all these solutions on the series of equilibria (see, e.g., Fig.
\ref{maxwell})
in order to see its global structure and apply the Poincar\'e theory (see
below). On the other hand, as already indicated, local minima of  energy
(metastable states) can be as much
physical, or even more physical, than global minima of  energy for the
timescales achieved in astrophysics (see  Sec.
\ref{sec_lifetime} and Refs. \cite{metastable,ijmpb} in the context of stellar
systems).

Stable equilibrium states correspond to (local or global) minima of energy at
fixed mass. The stability of an equilibrium state can be determined by studying
the sign of the second order variations of the energy functional $E_{\rm tot}$
whose general expression is given in Ref. \cite{chavtotal}. An
equilibrium state is stable if $\delta^2E_{\rm
tot}>0$ for all perturbations that conserve mass. In general,
it is difficult to
study the sign of $\delta^2E_{\rm tot}$ because we have to
solve a complicated eigenvalue equation. Fortunately, the
stability of the system can be directly deduced from topological properties
of continuous series of equilibia. This graphical method just requires to solve
the first order
variational problem (\ref{p2}).  This is the so-called Poincar\'e theory of
linear series of equilibria. It
works as follows. According to Eqs. (\ref{p2}) and (\ref{p4}), the eigenenergy
$E/m=\partial
E_{\rm tot}/\partial M$ is the variable conjugate to the mass $M$ (constraint)
with respect to the total energy $E_{\rm tot}$ (potential to minimize).
Therefore, if we plot $E$ as a function of $M$ along the series of equilibria,
we have the
following results:

(i) A change
of stability can occur only at a turning point of mass. 

(ii) One mode of stability
is lost is the curve $E(M)$ turns clockwise and one mode of stability is gained
if the curve $E(M)$ rotates anti-clockwise. 

Therefore, if we know a limit in which the
configuration is stable, we can use the Poincar\'e theory to deduce the
stability of the whole series of equilibria.

This theory does not tell us if an equilibrium state is fully stable (global
minimum of  $E_{\rm tot}$ at fixed $M$) or metastable (local minimum of 
$E_{\rm tot}$ at fixed $M$). To settle this issue we can plot $E_{\rm tot}(M)$
along the series of equilibria and compare the total energy of the different
equilibrium states for the same
mass $M$. Alternatively, we can perform a Maxwell construction (see Fig.
\ref{maxwell}) as explained in Appendix \ref{sec_m}. 

A general property can be noted. According to Eq. (\ref{p2}), the total energy
$E_{\rm tot}$ and the mass $M$ reach their
extrema along the series of equilibria at the same points since
\begin{eqnarray}
\delta M=0
\Leftrightarrow \delta E_{\rm
tot}=0.
\end{eqnarray}
This implies that the curve  $E_{\rm tot}(M)$ presents cusps at these
extrema. This is illustrated in Fig. \ref{metot}.

The Poincar\'e theory of linear series of equilibria is very general. For
example, it was applied by Lynden-Bell and Woods \cite{lbw} and Katz
\cite{katzpoincare} in relation to the thermodynamics of classical
self-gravitating
systems and by Chavanis \cite{ijmpb} in relation to the thermodynamics of
fermionic self-gravitating systems. In that context, the role of $E_{\rm tot}$
is played by the entropy $S$
in the microcanonical ensemble (MCE) and by the free energy $F$ in the canonical
ensemble (CE); the role of $M$ is played by the energy $E$ and the mass $M$
in MCE and by the mass $M$ in CE; the role of $\mu/m=E/m$ is played by the
inverse temperature $\beta$ and the chemical potential $\mu/m$
in MCE and by the chemical potential $\mu/m$ in CE. An interesting feature of
the statistical mechanics of self-gravitating systems is the notion of ensembles
inequivalence \cite{paddy} leading to various types of phase transitions
\cite{ijmpb}. The Poincar\'e theory of linear series of equilibria was also
applied to dilute axion stars in \cite{prd1,prd2,bectcoll} and
to general relativistic white dwarf stars and neutron stars. In the
present paper, it has been  applied to dense axion stars.

{\it Remark:} An alternative to the  Poincar\'e theory of linear series of
equilibria is the $M(R)$ theorem devised by Wheeler in the
physics of compact objects \cite{htww}. To apply it, one plots the mass-radius
relation $M(R)$ parametrized by the central density $\rho_0$. We know from
classical theory that the configurations of very low central density are stable.
As we move along the series of equilibria  $M(R)$ in the direction of increasing
central density all configurations remain stable until the first extremal
value (maximum or minimum) of $M$ is reached. At the first extremal point the
fundamental mode of radial oscillation becomes unstable. All configurations
between the first extremal point and the second have one unstable radial mode.
At the second extremal point (and also at each succeeding extremal point) one
more mode of radial oscillation changes stability. The direction of stability
change is determined by the way in which the $M(R)$-curve bends as it passes
through the extremal point. If the direction of bend is counterclockwise,
then one previously stable radial mode becomes unstable at the turning point;
if the direction bend is clockwise, then one previously unstable mode becomes
stable at the turning point. This method is illustrated in
Fig. 7 of \cite{htww}, in Figs. 2 and 3 of
\cite{mt}, in Fig. 11.2 of \cite{weinberg} and in Fig. 6.3
 of \cite{shapiroteukolsky}. We note that the $M(R)$
theorem of 
Wheeler is less general than the Poincar\'e theory. For example, it cannot be
directly applied in the context of the thermodynamics of self-gravitating
systems
\cite{lbw,katzpoincare,ijmpb}.

\section{Nongravitational dense axion stars in the TF limit}
\label{sec_tfng}

We have seen that dense axion stars are described by a mixed polytropic equation
of state (\ref{s2}) involving an attractive self-interaction
(first term with $a_s<0$) and a repulsive  self-interaction (second term). More
generally, let us consider a system described by a mixed polytropic equation of
state of the form
\begin{equation}
\label{tfng1}
P(\rho)=K_1\rho^{\gamma_1}+K_2\rho^{\gamma_2},
\end{equation}
where the first term is attractive ($K_1<0$) and
the second term is repulsive ($K_2>0$).\footnote{We assume that $\gamma_1>0$ and
$\gamma_2>0$.} We
consider the situation in which the quantum potential and the self-gravity are
negligible (TF $+$ nongravitational limit). For the equation
of state (\ref{s2}), this corresponds to the branch
(III-a) of the mass-radius relation of dense axion
stars.

\subsection{Constant density}
\label{sec_cd}

In the TF $+$ nongravitational limit, the equilibrium state of the system
results from the
balance between the attractive self-interaction and the repulsive
self-interaction. This is a situation where the system is spatially homogeneous
and the total pressure  is equal to zero ($P=0$):
\begin{equation}
\label{tfng2}
-|K_1|\rho^{\gamma_1}+K_2\rho^{\gamma_2}=0.
\end{equation}
The resulting density is
\begin{equation}
\label{tfng3}
\rho=\left (\frac{|K_1|}{K_2}\right
)^{1/(\gamma_2-\gamma_1)}.
\end{equation}
Specializing on the equation of state (\ref{s2}), we find that the
axion star has a constant density
\begin{equation}
\label{tfng3rho}
\rho_{\rm dense}=\frac{9 m^3c^2}{32\pi|a_s|\hbar^2}=8.95\times
10^{-2}\rho_i.
\end{equation}

For QCD axions with  $m=10^{-4}\,
{\rm eV}/c^2$ and $a_s=-5.8\times 10^{-53}\, {\rm m}$, we obtain
$\rho_{\rm dense}=7.07\times 10^{19}\, {\rm g/m^3}$.

For ULAs with $m=2.19\times 10^{-22}\, {\rm eV}/c^2$
and $a_s=-1.11\times 10^{-62}\, {\rm fm}$, we obtain $\rho_{\rm
dense}=3.88\times
10^{-9}\, {\rm g/m^3}$.

{\it Remark:} we note that $\rho_{\rm dense}$ given by Eq.
(\ref{tfng3rho}) corresponds to the density scale appearing in the potential of
Eq. (\ref{g1}). As a result, we directly recover the fact that the dilute limit
corresponds
to $\rho\ll\rho_{\rm dense}$. The scaling of Eq. (\ref{tfng3rho}) 
can also be obtained from the KG equation (\ref{kge1}) by equating the rest-mass
term $(m^2c^2/\hbar^2)\varphi^2$ and the self-interaction term
$(m^2c^3/f^2\hbar)\varphi^4$ [see Eq.
(\ref{inst3})], or by equating the self-interaction terms
$(m^2c^3/f^2\hbar)\varphi^4$ and $(m^2c^4/f^4)\varphi^6$. In the  dilute limit,
one has $(m^2c^2/\hbar^2)\varphi^2\ll (m^2c^3/f^2\hbar)\varphi^4\ll
(m^2c^4/f^4)\varphi^6$. Finally, we note that $\rho_{\rm dense}$ depends only on
$|a_s|/m^3$ which
is fixed for QCD axions (see the Remark at the end of Sec. \ref{sec_na1}).

\subsection{Mass-radius relation}
\label{sec_braa}

Since the density is uniform, the mass-radius relation is
\begin{equation}
\label{tfng5}
M=\frac{3m^3c^2}{8\hbar^2|a_s|}R^3.
\end{equation}
For QCD axions, we obtain
\begin{equation}
\label{tfng6}
\frac{R}{R_{\odot}}=2.72\times 10^{-5}\left (\frac{M}{M_{\odot}}\right )^{1/3}.
\end{equation}
This relation can be compared with the exact  expression
$R/R_{\odot}=2.1\times 10^{-5}(M/M_{\odot})^{0.305}$ found
numerically by Braaten {\it et al.} \cite{braaten}. The
exponent $0.305$ instead of $1/3$ takes into account the fact 
that the density is not exactly constant on the branch (III-a), implying
that self-gravity plays a certain role on this branch. Indeed, the
pressure
cannot be exactly zero otherwise the dense axion star would collapse under its
own gravity. Therefore, the star must be (slightly)  inhomogeneous so that the
pressure gradient arising from the
self-interaction can equilibrate the self-gravity.
Its exact structure
can be obtained by solving the generalized Lane-Emden equation of Appendix
\ref{sec_gle}. However,
gravitational effects are subdominant on the branch (III-a) so we expect that
even if the density profile of the dense axion
stars is not uniform, their central (or average) density should be
almost independent of the mass and approximately given by Eq. (\ref{tfng3rho}).
On the other hand, the different value
of the prefactor comes in part from the different axionic potential considered.
Braaten
{\it et al.} \cite{braaten}
use the effective potential (\ref{inst6}) while we use the polynomial
potential (\ref{inst9}). The value of the density at which the pressure $P$
given by Eq. (\ref{g3}) vanishes is $\rho=0.3025\, m^3c^2/|a_s|\hbar^2$ instead
of $\rho=8.95\times 10^{-2}\, m^3c^2/|a_s|\hbar^2$ corresponding to the
vanishing of the pressure $P$
given by Eq. (\ref{s2}).
Assuming that the density is uniform (with the warning made previously), we
obtain a mass-radius relation of the
form of Eq. (\ref{tfng6}) with a prefactor $1.81\times 10^{-5}$ which is closer
to the value found by Braaten {\it et al.} \cite{braaten}, showing that the
exact form of the potential has some  importance. Another
consequence of using the
polynomial
potential (\ref{inst9}) instead of the effective potential (\ref{inst6})
 is that we find solutions for any mass $M$ in the Newtonian approximation
while Braaten {\it et al.}
\cite{braaten} find solutions only for $M<1.9\, M_{\odot}$.
This is because, with the effective potential (\ref{inst6}),
new (secondary) branches appear at high densities as explained in Sec.
\ref{sec_g}. There
are no such branches in our simplified model based on the polynomial potential
(\ref{inst9}). It is easy to show from simple scaling arguments that the maximum
mass
found in \cite{braaten} scales as $A (|a_s|\hbar^2c^4/G^3m^3)^{1/2}$. Using
their numerical value $1.9\, M_{\odot}$ obtained for QCD axions, we obtain the
general value of the prefactor: $A=0.0679$. It is about one order
of magnitude smaller than the prefactor of $M_0$ and $M_{\rm max,GR}^{\rm
dense}$ found in our approach. This means that for certain potentials, such as
the one considered by Braaten {\it et al.} \cite{braaten}, the
fundamental Newtonian branch of dense axion stars 
may end before general relativistic effects come into play.

\subsection{Energy}

In the TF $+$ nongravitational limit, the virial theorem, the total energy and
the eigenenergy of a system described by the equation of state
(\ref{tfng1}) are [see Eqs. (169)-(171) of  \cite{chavtotal}]:
\begin{equation}
\label{ord1}
3(\gamma_1-1)U_1+3(\gamma_2-1)U_2=0,
\end{equation}
\begin{eqnarray}
\label{ord2}
E_{\rm tot}=U_1+U_2,
\end{eqnarray}
\begin{equation}
\label{ord3}
NE=\gamma_1 U_1+\gamma_2 U_2,
\end{equation}
where $U=[K/(\gamma-1)]\int\rho^{\gamma}\, d{\bf r}$ is the
internal energy of a polytrope of index $\gamma$. From
these equations we obtain
\begin{equation}
\label{ord4}
E_{\rm tot}=NE.
\end{equation}
Since this relation has been derived without having to evaluate the
functionals explicitly, the same exact relation can be obtained from the
Gaussian ansatz (see Sec. \ref{sec_vp})  since the formulae obtained with
the Gaussian ansatz are consistent with Eqs. (\ref{ord1})-(\ref{ord3})
\cite{chavtotal}.

If we now evaluate the functionals for a  spatially homogeneous system (see Sec.
\ref{sec_cd}), using $U=[K/(\gamma-1)]\rho^{\gamma}V$ where $V=(4/3)\pi R^3$
is the volume of the system, we first note that the virial theorem (\ref{ord1})
returns
Eq. (\ref{tfng2}). On the other hand, Eqs. (\ref{ord2}) and (\ref{ord3})
become 
\begin{equation}
\label{ord5}
E_{\rm tot}=\left
(\frac{K_1}{\gamma_1-1}\rho^{\gamma_1}+\frac{K_2}{\gamma_2-1}\rho^{\gamma_2}
\right )V,
\end{equation}
\begin{equation}
\label{ord6}
NE=\left
(\frac{\gamma_1
K_1}{\gamma_1-1}\rho^{\gamma_1}+\frac{\gamma_2 K_2}{\gamma_2-1}\rho^{ \gamma_2
}\right )V.
\end{equation}
Using Eq. (\ref{tfng2}), we
obtain
\begin{equation}
\label{ord7}
E_{\rm
tot}=NE=-\frac{\gamma_2-\gamma_1}{(\gamma_1-1)(\gamma_2-1)}|K_1|\rho^{\gamma_1}
V,
\end{equation}
where $\rho$ is given by Eq. (\ref{tfng3}). Specializing on the equation
of state (\ref{s2}), and using Eqs. (\ref{tfng3rho}) and (\ref{tfng5}),  we get
\begin{equation}
\label{ord8}
E_{\rm tot}=NE=-\frac{27m^3c^4}{256\hbar^2|a_s|}R^3=-\frac{9}{32}Mc^2.
\end{equation}

\subsection{Pulsation}

The pulsation of a nongravitational dense axion star in the TF limit can be
obtained as follows. If we neglect the self-gravity and the quantum potential,
the hydrodynamic equations (\ref{mad3}) and (\ref{mad5}) reduce to
\begin{eqnarray}
\frac{\partial\rho}{\partial t}+\vec\nabla\cdot (\rho
{\vec v})=0,
\label{mad3q}
\end{eqnarray}
\begin{eqnarray}
\frac{\partial {\vec v}}{\partial t}+({\vec v}\cdot
\vec\nabla){\vec v}=-\frac{1}{\rho}\vec\nabla P.
\label{mad5q}
\end{eqnarray}
Linearizing these equations about an homogeneous sphere of density $\rho$ given
by Eq. (\ref{tfng3rho}) and writing the time evolution of the perturbation
under the form $\delta\rho\sim e^{-i\omega t}$, we obtain the eigenvalue
equation
\begin{eqnarray}
\label{tfng15}
\Delta\delta\rho+\frac{\omega^2}{c_s^2}\delta\rho=0,
\end{eqnarray}
where $c_s^2=P'(\rho)$ is the square of the speed of sound. For radial
perturbations, the solution of
Eq. (\ref{tfng15}) that is regular at the origin is
\begin{eqnarray}
\label{tfng16}
\delta\rho=A\frac{\sin\left (\frac{\omega}{c_s}r\right )}{r}.
\end{eqnarray}
The boundary condition $\delta\rho=0$ at $r=R$ implies
\begin{eqnarray}
\label{tfng17}
\omega=\pi\frac{c_s}{R}.
\end{eqnarray}
From the equation of state (\ref{s2}), we find that
\begin{eqnarray}
\label{tfng18}
c_s=\frac{3}{4}c.
\end{eqnarray}
Therefore, the pulsation of the axion star is
\begin{eqnarray}
\label{tfng19}
\omega=\frac{3\pi c}{4R}.
\end{eqnarray}
The pulsation period $2\pi/\omega$ is of the order of the time
taken by a photon to cross the axion star.

{\it Remark:} For a dense  axion star of radius 
$R\sim R_{\rm min}$ [see Eq. (\ref{mir6})], we find that $\omega\sim
mc^2/\hbar$, corresponding to the proper pulsation $\omega_{\rm
proper}=mc^2/\hbar$ of the axion. More generally, we can write
$\omega/\omega_{\rm proper}=R_{\rm min}/R$ so the pulsation of the axion stars
is smaller than the proper pulsation of the axion.

\subsection{The dense axion star resulting from the collapse of a dilute axion
star at the critical point}

The dense axion star resulting from the collapse of a dilute axion star with
mass $M_{\rm
max,N}^{\rm dilute}$ has a radius
\begin{eqnarray}
\label{tfng14}
R_{\rm coll}=1.39\left (\frac{|a_s|\hbar^6}{Gm^7c^4}\right )^{1/6}=1.39 R_d,
\end{eqnarray}
a pulsation\footnote{We note that the pulsation satisfies the relation
$\omega_{\rm
coll}/\omega_{\rm proper}=1.99 \, \delta^{1/6}$. Since the exponent $1/6$ is
small, $\omega_{\rm coll}$ is relatively close to $\omega_{\rm proper}$.
This may question the validity of the fast oscillation approximation made in
Sec. \ref{sec_gpe}.}
\begin{eqnarray}
\label{tfng20}
\omega_{\rm coll}=1.70\left (\frac{Gm^7c^{10}}{|a_s|\hbar^6}\right )^{1/6},
\end{eqnarray}
and an energy\footnote{It is interesting to note that the
energy $|E_{\rm tot}^{\rm coll}|$ of the dense axion star that results from
gravitational collapse  is of the
order of the rest-mass energy $M_{\rm max} c^2$ of the dilute axion star that
collapses. This is rather coincidental since we are using purely
Newtonian gravity in conformity with the fact that $R\gg R_S$ so there is no
rest-mass energy in our problem. We obtain this
result
because we have kept relativistic contributions  in the
self-interaction potential (see footnote 9). Indeed, the speed
of light occurs in the repulsive self-interaction term. In
this connection, we note that Tkachev \cite{tkachev2015} argues that the energy
of FRBs is of the order of $M_{\rm max} c^2$ [see his Eq. (3)]. This is
consistent with our results if we assume that  $|E_{\rm
tot}^{\rm coll}|$ corresponds to the energy
released during the collapse - possibly in the form of radiation (see Sec.
\ref{sec_radiation}).}
\begin{eqnarray}
\label{mc2}
E_{\rm tot}^{\rm coll}=-\frac{9}{32}M_{\rm max}c^2=-0.285\frac{\hbar
c^2}{\sqrt{Gm|a_s|}}.
\end{eqnarray}

For QCD axions with  $m=10^{-4}\,
{\rm eV}/c^2$ and $a_s=-5.8\times 10^{-53}\, {\rm m}$, we obtain $M_{\rm
max,N}^{\rm dilute}=6.46\times 10^{-14}\, M_{\odot}$, $R_{\rm coll}=75.6\, {\rm
cm}$, $\rho_{\rm coll}=7.07\times 10^{19}\, {\rm g/m^3}$, 
$2\pi/\omega_{\rm coll}=6.70\times 10^{-9}\, {\rm s}$, and $E_{\rm tot}^{\rm
coll}=-3.25\times 10^{40}\, {\rm erg}$ (for comparison $2\pi/\omega_{\rm
proper}=4.13\times 10^{-11}\, {\rm s}$).

For ULAs with $m=2.19\times 10^{-22}\, {\rm eV}/c^2$
and $a_s=-1.11\times 10^{-62}\, {\rm fm}$, we obtain $M_{\rm
max,N}^{\rm dilute}=10^{8}\, M_{\odot}$, $R_{\rm
coll}=0.745\, {\rm pc}$, $\rho_{\rm coll}=3.88\times
10^{-9}\, {\rm g/m^3}$, $2\pi/\omega_{\rm coll}=6.47\, {\rm yrs}$, and $E_{\rm
tot}^{\rm
coll}=-5.03\times 10^{51}\, {\rm erg}$ (for comparison $2\pi/\omega_{\rm
proper}=0.599\, {\rm yrs}$).

{\it Remark:} These results can be compared with the minimum
period of white dwarfs and neutron stars (see the Remark at the end of Appendix
\ref{sec_maxw}). We stress that the pulsation period of dense axion stars
resulting from
the collapse of a dilute axion star with the maximum mass is very small. We
also note that the density of dense QCD axion stars is comparable to the maximum
density of neutron stars, but their mass and radius are smaller.

\subsection{The dense axion star resulting from the collapse of an axion
minicluster}
\label{sec_qpo}

As recalled in the Introduction, QCD axion stars are expected to result
from the gravitational collapse of axion miniclusters
(axitons) \cite{tkachev,tkachevrt}. However, the
mass of axitons $M_{\rm
axiton}\sim 10^{-12}\, M_{\odot}$ is larger than the maximum mass $M_{\rm
max}^{\rm
exact}=6.46\times
10^{-14}\, M_{\odot}$
of dilute axion
stars \cite{prd1}. Therefore,
when axion miniclusters gravitationally collapse, they can either (i) form a
dilute
axion star of mass $M<M_{\rm max}$ by sheding mass, (ii) fragment
into
several axion drops with mass $M'<M_{\rm max}$ (this is possible because the
size
 of  axion miniclusters, $R_{\rm axiton}\sim 10^{6}\, {\rm km}$, is
significantly larger than the size $(R_{99}^{*})^{\rm exact}=227 \,
{\rm km}$ of dilute
axion stars), or  (iii) directly form a
dense
axion star of mass $M_{\rm axiton}$. 
Let
us consider this third possibility. For QCD axions with 
$m=10^{-4}\,
{\rm eV}/c^2$ and $a_s=-5.8\times 10^{-53}\, {\rm m}$, 
using the foregoing results and assuming $M_{\rm coll}=M_{\rm
axiton}=10^{-12}\, M_{\odot}$, we get $R_{\rm coll}=1.89\, {\rm m}$, $\rho_{\rm
dense}=7.07\times 10^{19}\, {\rm
g/m^3}$, $2\pi/\omega_{\rm coll}=1.68\times
10^{-8}\, {\rm s}$, and
$E_{\rm
tot}^{\rm coll}=-5.03\times
10^{41}\, {\rm ergs}$ (in comparison $R_{\rm axiton}\sim 10^6\, {\rm
km}$ and $\rho_{\rm axiton}=4.75\times 10^{-7}\, {\rm g/m^3}$).
On the other hand, using the results of \cite{bectcoll}, we find that a
minicluster of mass $M_{\rm axiton}/M_{\rm max}^{\rm exact}=15.5$ and radius
$R_{\rm axiton}/(R_{99}^{*})^{\rm exact}=4405$ collapses towards a dense
axion star on a time $t_{\rm coll}=82483 t_D=32.0\, {\rm yrs}$.

\subsection{The maximum mass of general relativistic dense axion stars}
\label{sec_mmd1}

We have seen that nongravitational dense axion stars in the TF limit,
corresponding to the branch (III-a) of the general mass-radius relation, have a
constant density given by Eq. (\ref{tfng3rho}) and a vanishing pressure
($P=0$). When the mass $M$ of the star approaches the value $M_0$ (see
Appendix \ref{sec_tranA}), self-gravity starts to become important but we may
still assume that the density approximately keeps its value given by Eq.
(\ref{tfng3rho}). On the other hand, general relativity must be taken
into
account (see Sec. \ref{sec_fri}). In that case, we have to deal with
a uniform sphere of energy density $\epsilon=\rho c^2$ in general relativity.
We know that the radius of a uniform sphere of mass $M$ is restricted
by the inequality\footnote{Uniform spheres exist for any $R\ge (9/8)R_S$ and
they are stable. Buchdahl \cite{buchdahl} has shown that the
lower limit to $R$
set by the inequality (\ref{tfng21}) is absolute in the sense that it is
satisfied by any star, whatever its equation of state.
However, such a lower bound is
usually not reached. The radius of a star is usually
substantially larger than $(9/8)R_S$. For example, the
minimum radius of ideal $^4_2{\rm He}$ white dwarfs satisfies $R_{\rm
min}=247\, R_S$ (corresponding to $M=1.39\, M_{\odot}$, $R_{\rm min}=1.02\times
10^{3}\, {\rm km}$, and 
$\rho_0=2.35\times 10^{16}\,
{\rm g/m^3}$) and the minimum radius of ideal neutron stars satisfies
$R=4.37\, R_S$ (corresponding to $M=0.710\, M_{\odot}$, $R=9.16\,
{\rm km}$, and 
$\rho_0=3.54\times 10^{21}\,
{\rm g/m^3}$).}
\begin{equation}
\label{tfng21}
R\ge \frac{9}{8}\frac{2GM}{c^2}.
\end{equation}
Combining Eqs. (\ref{tfng5}) and (\ref{tfng21}), we obtain the following
expressions for the general relativistic maximum mass and corresponding
radius of dense axion
stars 
\begin{equation}
\label{tfng22}
M_{\rm max, GR}^{\rm dense}=\frac{16}{27}\sqrt{\frac{2}{3}}\left
(\frac{|a_s|\hbar^2c^4}{G^3m^3}\right )^{1/2}=0.484 M_r,
\end{equation}
\begin{equation}
\label{tfng23}
R_{*,GR}^{\rm dense}=\frac{4}{3}\sqrt{\frac{2}{3}}\left
(\frac{|a_s|\hbar^2}{Gm^3}\right
)^{1/2}=1.09 R_r.
\end{equation}

If we use the qualitative argument of Sec. \ref{sec_qgr}, which amounts
to replacing Eq. (\ref{tfng21}) by $R>2GM/c^2$, we get
\begin{equation}
\label{tfng12}
M_{\rm max, GR}^{\rm dense}=\frac{1}{\sqrt{3}}\left
(\frac{|a_s|\hbar^2c^4}{G^3m^3}\right )^{1/2}=0.577 M_r,
\end{equation}
\begin{equation}
\label{tfng13}
R_{*,GR}^{\rm dense}=\frac{2}{\sqrt{3}}\left (\frac{|a_s|\hbar^2}{Gm^3}\right
)^{1/2}=1.15 R_r.
\end{equation}

For QCD axions with  $m=10^{-4}\,
{\rm eV}/c^2$ and $a_s=-5.8\times 10^{-53}\, {\rm m}$,  we obtain
 $M_{\rm max, GR}^{\rm
dense}=13.5\, M_{\odot}$ and $R_{\rm *,GR}^{\rm dense}=45.0\, {\rm
km}$ (with the qualitative argument of Sec. \ref{sec_qgr}, we
get $M_{\rm max, GR}^{\rm
dense}=16.1\, M_{\odot}$ and $R_{\rm *,GR}^{\rm dense}=47.5\, {\rm
km}$).

For ULAs with $m=2.19\times 10^{-22}\, {\rm eV}/c^2$
and $a_s=-1.11\times 10^{-62}\, {\rm fm}$, we obtain $M_{\rm max, GR}^{\rm
dense}=1.83\times 10^{15}\, M_{\odot}$ and $R_{\rm *,GR}^{\rm dense}=197\, {\rm
pc}$ (with the qualitative argument of Sec. \ref{sec_qgr}, we get $M_{\rm max,
GR}^{\rm
dense}=2.18\times 10^{15}\, M_{\odot}$ and $R_{\rm *,GR}^{\rm dense}=208\, {\rm
pc}$).

\section{Nonattractive dense axion stars in the TF limit}
\label{sec_pol}

We consider dense axion stars for which the quantum potential and the attractive
self-interaction can be neglected  (TF $+$ nonattractive
limit). This corresponds to the branch (III-b) of the mass-radius relation of
Newtonian dense axion stars. We recall that only the begining of this branch is
valid (in
the best case) because the value $M_0$ of the mass at which this branch
appears is precisely of the order of the maximum mass $M_{\rm max,GR}^{\rm
dense}$ set by general relativity (see Sec. \ref{sec_fri}).

\subsection{Polytrope $n=1/2$}

In the TF $+$ nonattractive
limit, the axion star is equivalent
to a barotropic star with an equation of state
\begin{equation}
\label{pol1}
P=\frac{64\pi^2\hbar^4a_s^2}{9m^6c^2}\rho^3.
\end{equation}
This is the equation of state of a polytrope with polytropic constant
$K={64\pi^2\hbar^4a_s^2}/{9m^6c^2}$ and polytropic index $\gamma=3$ (i.e.
$n=1/2$).

\subsection{Mass-radius relation}

The general mass-radius relation of polytropic spheres is \cite{chandrabook}:
\begin{equation}
\label{pol2}
M^{(n-1)/n}R^{(3-n)/n}=\frac{K(1+n)}{(4\pi)^{1/n}G}\omega_n^{(n-1)/n}, 
\end{equation}
where $\omega_n$ is a constant that can be obtained from the Lane-Emden
equation (some values of
$\omega_n$ are tabulated in \cite{chandrabook}).
Specializing on the equation of state (\ref{pol1}), we obtain
\begin{equation}
\label{pol3}
M=\frac{3Gm^6c^2}{2\hbar^4a_s^2}\omega_{1/2}R^5=0.0323 \frac{Gm^6c^2}{
\hbar^4a_s^2}R^5,
\end{equation}
where we have used $\omega_{1/2}=0.02156...$.

\subsection{Energy}

In the TF limit, the virial theorem, the total energy and
the eigenenergy of a polytrope are [see Eqs. (169)-(171) of 
\cite{chavtotal}]:
\begin{equation}
\label{pol6}
3(\gamma-1)U+W=0,
\end{equation}
\begin{equation}
\label{pol6q}
E_{\rm tot}=U+W,
\end{equation}
\begin{equation}
\label{pol12}
N E=\gamma U+2W.
\end{equation}
From these equations we obtain
\begin{equation}
\label{pol7}
E_{\rm tot}=\left (1-\frac{n}{3}\right )W,
\end{equation}
\begin{equation}
\label{pol21}
U=-\frac{n}{3}W,
\end{equation}
\begin{equation}
\label{pol13}
NE=\frac{1}{3}(5-n)W.
\end{equation}
Eliminating the gravitational energy $W$ from these relations, we get
\begin{equation}
\label{pol14}
E_{\rm tot}=\frac{3-n}{5-n}NE.
\end{equation}
This relation shows that the total energy $E_{\rm tot}$ is not equal to $N$
times the
eigenenergy $E$, as one might naively expect. Specializing on the polytrope
$n=1/2$, we obtain
\begin{equation}
\label{pol15}
E_{\rm tot}=\frac{5}{9}NE.
\end{equation}
Since this relation has been derived without having to evaluate the
functionals explicitly, the same exact relation can be obtained from the
Gaussian ansatz (see Sec. \ref{sec_vp})  since the formulae obtained with
the Gaussian ansatz are consistent with Eqs. (\ref{pol6})-(\ref{pol12})
\cite{chavtotal}.

In the TF limit, the gravitational energy of a polytropic sphere of index
$n<5$ is given by the Betti-Ritter
formula \cite{chandrabook}:
\begin{equation}
\label{pol5}
W=-\frac{3}{5-n}\frac{GM^2}{R}.
\end{equation}
Substituting for $W$ from Eq. (\ref{pol5}) into Eqs.
(\ref{pol7})-(\ref{pol13}), we obtain
\begin{equation}
\label{pol7j}
E_{\rm tot}=-\frac{3-n}{5-n}\frac{GM^2}{R},
\end{equation}
\begin{equation}
\label{pol21j}
U=\frac{n}{5-n}\frac{GM^2}{R},
\end{equation}
\begin{equation}
\label{pol13j}
NE=-\frac{GM^2}{R}.
\end{equation}
For a polytrope of index $n=1/2$, the first two equations become
\begin{equation}
\label{pol8}
E_{\rm tot}=-\frac{5}{9}\frac{GM^2}{R}, \qquad U=\frac{1}{9}\frac{GM^2}{R}.
\end{equation}
On the other hand, we note that the third relation is independent of the
index of the polytrope. Actually, in the TF limit, this relation is valid for an
arbitrary
equation of state provided that the enthalpy vanishes on the boundary of the
star. It can be
directly obtained from Eqs. (\ref{p3}) and
(\ref{p4}) by applying this relation at $r=R$
and using $h(R)=0$ and $\Phi(R)=-GM/R$.

\subsection{Pulsation}

An approximate expression of the complex pulsation of a polytrope is given by
the Ledoux formula \cite{ledoux}:
\begin{equation}
\label{pol17}
\omega^2=(4-3\gamma)\frac{W}{I},
\end{equation}
where $I$ is the moment of inertia of the star. For a polytrope $n=1/2$, we have
$I=0.489\,
MR^2$ and $W=-(2/3)GM^2/R$, giving
\begin{equation}
\label{led}
\omega^2=6.82 \frac{GM}{R^3}.
\end{equation}
Specializing on the equation of state (\ref{pol1}), the pulsation is given by
Eq. (\ref{led}), the mass and the radius being related to each other by
Eq. (\ref{pol3}).

{\it Remark:} It can be shown that polytropes of index $0<n<3$ are dynamically
stable while  polytropes of index $3<n<5$ are dynamically unstable. In
particular, the polytrope $n=1/2$ is stable. We note that stable polytropes have
a negative energy ($E_{\rm tot}<0$) while unstable polytropes have a positive
energy ($E_{\rm tot}>0$).

\subsection{The maximum mass of general relativistic dense axion stars}
\label{sec_mmd2}

An estimate of the maximum mass of general relativistic
dense axion stars is obtained by combining the Newtonian mass-radius relation
(\ref{pol3}) with the constraint $R\ge R_S$, where $R_S$ is the Schwarzschild
radius defined in Eq. (\ref{sw}).\footnote{We could also use 
the Buchdahl inequality $R\ge (9/8)R_S$ \cite{buchdahl} but this refinement is
not necessary since our
procedure can only give a rough estimate of the maximum mass anyway. A more
detailed
calculation of the maximum mass of general relativistic dense axion stars will
be reported elsewhere.} This
gives a maximum general relativistic
mass 
\begin{equation}
\label{pol24}
M_{\rm max, GR}^{\rm dense}=\frac{1}{2(3\omega_{1/2})^{1/4}}\left
(\frac{|a_s|\hbar^2c^4}{G^3m^3}\right )^{1/2}=0.991 M_r
\end{equation}
and a corresponding radius
\begin{equation}
\label{pol25}
R_{\rm *,GR}^{\rm dense}=\frac{1}{(3\omega_{1/2})^{1/4}}\left
(\frac{|a_s|\hbar^2}{Gm^3}\right )^{1/2}=1.98 R_r.
\end{equation}
As expected, these values are of the same order of magnitude as those obtained 
in Appendix
\ref{sec_mmd1}. This is because general relativistic effects set in precisely
at the transition between nongravitational and nonattractive dense axion stars
(see Sec. \ref{sec_fri}).

For QCD axions with  $m=10^{-4}\,
{\rm eV}/c^2$ and $a_s=-5.8\times 10^{-53}\, {\rm m}$,  we obtain
 $M_{\rm max, GR}^{\rm dense}=27.7\, M_{\odot}$ and $R_{\rm *,GR}^{\rm
dense}=81.9\, {\rm
km}$.

For ULAs with $m=2.19\times 10^{-22}\, {\rm eV}/c^2$
and $a_s=-1.11\times 10^{-62}\, {\rm fm}$, we obtain $M_{\rm max, GR}^{\rm
dense}=3.74\times 10^{15}\, M_{\odot}$ and $R_{\rm *,GR}^{\rm dense}=358\, {\rm
pc}$.

\section{Transition between nongravitational and nonattractive dense axion
stars in the TF limit}
\label{sec_tranA}

We have seen that nongravitational dense axion stars in the TF limit,
corresponding to the branch (III-a),  have a constant density given by Eq.
(\ref{tfng3rho})
and a mass-radius relation given by Eq. (\ref{tfng5}). 
On the other hand, nonattractive dense axion stars in the TF limit,
corresponding to the branch (III-b), are
equivalent to polytropes of index $n=1/2$ with a mass-radius relation given by
Eq. (\ref{pol3}). From these results, we find that the transition  between
nongravitational and nonattractive dense axion
stars in the TF limit corresponds to
\begin{equation}
\label{ext1}
M_0=\frac{3}{64\omega_{1/2}^{3/2}}\left
(\frac{|a_s|\hbar^2c^4}{G^3m^3}\right )^{1/2}=14.8 M_r,
\end{equation}
\begin{equation}
\label{ext2}
R_0=\frac{1}{2(\omega_{1/2})^{1/2}}\left
(\frac{|a_s|\hbar^2}{Gm^3}\right )^{1/2}=3.41 R_r.
\end{equation}

For QCD axions with  $m=10^{-4}\,
{\rm eV}/c^2$ and $a_s=-5.8\times 10^{-53}\, {\rm m}$,  we obtain
 $M_{0}=414\, M_{\odot}$ and $R_{0}=141\, {\rm
km}$.

For ULAs with $m=2.19\times 10^{-22}\, {\rm eV}/c^2$
and $a_s=-1.11\times 10^{-62}\, {\rm fm}$, we obtain $M_{0}=5.60\times
10^{16}\, M_{\odot}$ and $R_{0}=615\, {\rm
pc}$.

{\it Remark:} We note that the value of $M_0$ given by Eq. (\ref{ext1})
is substantially larger than the value of  $M_{\rm max, GR}^{\rm dense}$ given
by Eqs. (\ref{tfng22}) and (\ref{pol24}). This suggests that the Newtonian
branch (III-b) is not physically
relevant, even marginally. However, since our approach is qualitative, we
should remain cautious about this claim.

\section{Generalized Lane-Emden equation}
\label{sec_gle}

We have seen that dense axion stars are described  by an equation of state of
the form of Eq. (\ref{s3}) which is the sum of two polytropic equations of state
of
index $\gamma=2$ and  $\gamma=3$ and polytropic constants $K_2<0$ and
$K_3>0$. More generally, we consider a self-gravitating system described by an
equation of state of the form
\begin{eqnarray}
P=K\rho^{\gamma}+K_*\rho^{\gamma_*},
\label{gle1}
\end{eqnarray}
which is the sum of two polytropic equations of state of index $\gamma=1+1/n$
and  $\gamma_*=1+1/n_*$ and polytropic constants $K$ and $K_*$. The fundamental
differential equation of hydrostatic
equilibrium for a spherically symmetric barotropic self-gravitating
system writes \cite{chandrabook}:
\begin{eqnarray}
\frac{1}{r^2}\frac{d}{dr}\left (\frac{r^2}{\rho}\frac{dP}{dr}\right )=-4\pi
G\rho.
\label{gle2}
\end{eqnarray}
Substituting Eq. (\ref{gle1}) into Eq. (\ref{gle2}), and rearranging some terms,
we
obtain
\begin{equation}
\frac{1}{r^2}\frac{d}{dr}\left (r^2 \frac{d}{dr}\left\lbrack
\frac{K(n+1)}{4\pi G}\rho^{\frac{1}{n}}+\frac{K_*(n_*+1)}{4\pi
G}\rho^{\frac{1}{n_*}}\right \rbrack\right )=-\rho.
\label{gle3}
\end{equation}
Defining $\theta$ and $\xi$  through the relations
\begin{eqnarray}
\rho=\rho_0\theta^n,\qquad \xi=\left\lbrack \frac{4\pi
G\rho_0^{1-1/n}}{K(n+1)}\right \rbrack^{1/2} r,
\label{gle4}
\end{eqnarray}
where $\rho_0$ is the central density, we can rewrite the foregoing equation
under the form
\begin{equation}
\frac{1}{\xi^2}\frac{d}{d\xi}\left \lbrack \xi^2
\frac{d}{d\xi}(\theta+\chi\theta^q) \right \rbrack=-\theta^n,
\label{gle5}
\end{equation}
where we have introduced the abbreviations
\begin{eqnarray}
\chi=\frac{K_*(n_*+1)}{K(n+1)}\rho_0^{1/n_*-1/n},\qquad q=\frac{n}{n_*}. 
\label{gle6}
\end{eqnarray}
Equation (\ref{gle5}) must be solved with the boundary conditions at
the origin
\begin{eqnarray}
\theta(0)=1,\qquad \theta'(0)=0.
\label{gle7}
\end{eqnarray}
The differential equation (\ref{gle5}) is a generalization of the Lane-Emden
equation for  a spherically symmetric self-gravitating system described by a
polytropic equation of state $P=K\rho^{\gamma}$ \cite{chandrabook}. The
traditional Lane-Emden equation is recovered from Eq. (\ref{gle5}) by setting
$\chi=0$. 

The generalized Lane-Emden equation (\ref{gle5}) 
describes dense axion stars with the equation of state (\ref{s3})
in the TF limit (in that case $\gamma=2$, $K=K_2<0$, $\gamma=3$, and
$K_*=K_3>0$). This corresponds to the configurations forming the branch
(III) of the general mass-radius relation of axion stars. The nongravitational
limit, corresponding to the branch (III-a), is described by the constant density
solution $\theta=1$ studied in Appendix \ref{sec_tfng}.\footnote{We note that
a constant density profile is not an exact solution of the  generalized
Lane-Emden equation (\ref{gle5}) as discussed after Eq. (\ref{tfng6}).
However, we expect that on the  the branch (III-a) the typical (central
or average) density is independent of the mass, as for a uniform sphere, so the
results of Appendix \ref{sec_tfng} should provide correct orders of magnitude. A
detailed study of Eq. (\ref{gle5})  will be reported elsewhere.}  The
nonattractive limit, corresponding to
the branch (III-b), is described by the standard Lane-Emden equation associated
with a polytrope of index $n=1/2$.

\section{The maximum pulsation of dilute axion stars}
\label{sec_maxw}

Using the Gaussian ansatz, we have shown in Ref. \cite{bectcoll} that the
pulsation of stable dilute axion stars
with $\delta=0$ presents a maximum
\begin{eqnarray}
\label{maxw1}
\omega_{\rm
max}=\frac{\sqrt{8}(\sqrt{5}-1)^{1/2}}{(\sqrt{5}+3)^{1/2}(1+\sqrt{5})}
=0.4246...
\end{eqnarray}
at
\begin{eqnarray}
\label{maxw2}
R=\left (\frac{1+\sqrt{5}}{2}\right )^{1/2}=1.272...
\end{eqnarray}
\begin{eqnarray}
\label{maxw3}
M=\frac{[8(1+\sqrt{5})]^{1/2}}{3+\sqrt{5}}=0.9717...
\end{eqnarray}
Coming back to dimensional variables, we get 
\begin{equation}
\label{maxw4}
\omega_{\rm max}=0.425 \frac{\nu}{6\pi\zeta}\left (\frac{\sigma}{\alpha}\right
)^{1/2}\frac{Gm^2}{|a_s|\hbar}=0.100\, \omega_a,
\end{equation}
where $\omega_a=1/t_a$ (see Appendix \ref{sec_cis}).

For QCD
axions with  $m=10^{-4}\,
{\rm eV}/c^2$ and
$a_s=-5.8\times 10^{-53}\, {\rm m}$,  we
obtain a minimum period $2\pi/\omega_{\rm max}=50.3\, {\rm hrs}$ (it
corresponds to an axion star with $R=90.9\, {\rm km}$, $M=6.72\times 10^{-14}\,
M_{\odot}$, and $\rho=4.25\times 10^4\, {\rm g/m^3}$).

For ULAs with $m=2.19\times 10^{-22}\, {\rm eV}/c^2$
and
$a_s=-1.11\times 10^{-62}\, {\rm fm}$, we obtain a
minimum period
$2\pi/\omega_{\rm max}=229\, {\rm Myrs}$ (it corresponds to an axionic DM halo
with  $R=0.398\, {\rm kpc}$, $M=1.04\times 10^{8}\, M_{\odot}$, and 
$\rho=2.67\times 10^{-17}\,
{\rm g/m^3}$).

{\it Remark:} These results can be compared with the minimum
period $T_{\rm min}=2.13\, {\rm s}$ of ideal $^4_2{\rm He}$ white dwarfs
(corresponding to $R=1.47\times 10^{3}\, {\rm km}$,
$M=1.39\, M_{\odot}$, and 
$\rho_0=6.58\times 10^{15}\,
{\rm g/m^3}$) and with the minimum period
$T_{\rm min}=9.00\times 10^{-4}\, {\rm
s}$ of ideal neutron stars (corresponding to $R=11.5\, {\rm km}$,
$M=0.670\, M_{\odot}$, and 
$\rho_0=1.48\times 10^{21}\,
{\rm g/m^3}$).

\section{Maxwell construction}
\label{sec_m}

In a first order phase transition, the point of transition where the two phases
have the same probability can be determined by the Maxwell construction.
Applied to our problem, the Maxwell construction tells us that the transition
mass $M_t$ at which the two phases (dilute and dense axion stars) have the same
energy $E_{\rm tot}$ is such that the two hatched areas in Fig. \ref{maxwell}
are equal.
This result can be proven as follows.

The equal area Maxwell condition $A_1=A_2$ can be expressed as
\begin{equation}
\label{m1}
\int_{E'}^{E'''}(M-M_t) dE=0,
\end{equation}
where $E'$ is the eigenenergy of the dilute axion star and $E'''$ is the
eigenenergy of the dense axion star at the transition mass $M_t$. 
We introduce the grand potential 
\begin{equation}
\label{m2}
G=E_{\rm tot}-\frac{E}{m}M=E_{\rm tot}-NE
\end{equation}
which is the Legendre transform of the energy $E_{\rm tot}$ with respect to
the mass $M$, with conjugate parameter $E/m$. From
Eq. (\ref{p2}), we get
\begin{eqnarray}
\label{m3}
\delta G&=&\delta E_{\rm tot}-\frac{E}{m}\delta M-\frac{\delta E}{m}
M\nonumber\\
&=&\frac{E}{m}\delta M-\frac{E}{m}\delta M-\frac{\delta
E}{m}M\nonumber\\
&=&-\frac{\delta E}{m}M.
\end{eqnarray}
Using this relation, we can integrate Eq. (\ref{m1}), thereby obtaining
\begin{equation}
\label{m4}
-m(G'''-G')-M_t(E'''-E')=0.
\end{equation}
Using Eq. (\ref{m2}), this equation  can be rewritten as
\begin{equation}
\label{m5}
E_{\rm tot}'''=E_{\rm tot}'.
\end{equation}
This shows that the Maxwell construction is equivalent to the equality of
the total energy of the two phases at the transition.

\section{Asymptotic behavior of the transition mass}
\label{sec_abtm}

In this Appendix, we determine the asymptotic behavior of the transition mass
$M_t$ as $\delta\rightarrow 0$.

Let us first consider the dilute phase. For $M_t\rightarrow 0$, we can neglect
the self-interaction terms in the energy [last two terms in  Eq.
(\ref{ep7})]. We can also use the asymptotic mass-radius relation $M_t\sim
2/R_t^{\rm dilute}$ [see Eq. (\ref{mrd3})].
Therefore, we get \cite{prd2}:
\begin{eqnarray}
(E_{\rm tot})_t^{\rm dilute}\sim -\frac{M_t^3}{4}.
\label{abtm1}
\end{eqnarray}
We note that $(E_{\rm tot})_t^{\rm dilute}\rightarrow
0$ when $\delta\rightarrow 0$.

We now consider the condensed phase. For $M_t\rightarrow 0$, we can neglect
the self-gravity in the energy [second term in  Eq.
(\ref{ep7})]. Assuming the scalings $M_t\sim a\sqrt{\delta}$
 and $R_t^{\rm dense}\sim b\sqrt{\delta}$ when
$\delta\rightarrow 0$, 
we obtain
\begin{eqnarray}
(E_{\rm tot})_t^{\rm dense}\sim \left
(\frac{a}{b^2}-\frac{a^2}{3b^3}+\frac{a^3}{b^6}\right )\frac{1}{\sqrt{\delta}}.
\label{abtm2}
\end{eqnarray}
At the transition point, the energy of the two phase must be equal: $(E_{\rm
tot})_t^{\rm dilute}=(E_{\rm tot})_t^{\rm dense}$. Comparing Eqs.
(\ref{abtm1}) and (\ref{abtm2}), we see that the term in brackets in
Eq. (\ref{abtm2}) must necessarily vanish otherwise the total energy would tend
to infinity, instead of zero, when $\delta\rightarrow 0$. Therefore,
${a}/{b^2}-{a^2}/{3b^3}+{a^3}/{b^6}=0$. On the other hand, substituting
the scalings $M_t\sim a\sqrt{\delta}$  and $R_t^{\rm dense}\sim b\sqrt{\delta}$
into the mass-radius relation (\ref{mr2}) we obtain a second equation
$a^2/b^4-a/6b+1/3=0$. Solving these two equations, we get $a=16\sqrt{3}$
and $b=4\sqrt{3}$  implying $M_t\sim 16\sqrt{3\delta}$ and
$R_t^{\rm dense}\sim 4\sqrt{3\delta}$ when $\delta\rightarrow 0$. Comparing
these results with Eqs.
(\ref{mir4}), we conclude that the transition point ($R_t$,$M_t$)
asymptotically approaches the point ($R_*$,$M_*$) corresponding to the minimum
radius when $\delta\rightarrow 0$.

\section{Weak and strong self-interaction limits}
\label{sec_ws}

In this Appendix, we try to clarify the notion of weak and strong
self-interaction
limits
for axions.

\subsection{$|\lambda|\ll 1$}

The strength of the self-interaction can be measured by the dimensionless
parameter $\lambda$. This parameter is usually very small with
respect to unity ($|\lambda|\ll 1$) as
required by particle physics and quantum field theory, meaning that axions are
in a weak self-interaction regime. For QCD axions, one has $\lambda_{\rm
QCD}=-7.39\times 10^{-49}$ and for typical ULAs, one has  $\lambda_{\rm
ULA}=-3.10\times
10^{-91}$. The condition  $|\lambda|\ll 1$ corresponds to (i) $|a_s|\ll
\hbar/mc=\lambda_C$, i.e., the scattering length of the axions is small as
compared to their
Compton wavelength; (ii) $f\gg mc^2$, i.e., the axion decay constant is large
as compared to their rest mass energy. We shall say that the
condition $|\lambda|\ll 1$ corresponds to the weak self-interaction limit from
the
viewpoint of particle physics. We note that gravity $(G)$ does not enter into
this criterion.

\subsection{$1/\delta\gg 1$}

The study of this paper, focusing on self-gravitating axion stars, shows that
the
strength of the self-interaction can be measured by another dimensionless
parameter $1/\delta$. This parameter is usually very large ($1/\delta\gg 1$),
meaning that axions are
in a strong self-interaction regime. For QCD axions, one has $\delta^{-1}_{\rm
QCD}=1.13\times 10^{15}$ and for typical ULAs, one has  $\delta^{-1}_{\rm
ULA}=9.80\times
10^{7}$. The condition  $1/\delta\gg 1$ corresponds to (i) $|a_s|\gg 
2Gm/c^2=r_S$, i.e., the scattering length of the axions is large as compared to
their
effective Schwarzschild radius; (ii) $f\ll M_P c^2$, i.e., the
axion decay constant is small
as compared to the Planck energy; (iii) $|\lambda|\gg (m/M_P)^2$. We note that
gravity ($G$) enters into this criterion. We
shall say that the
condition $1/\delta\gg 1$ corresponds to the strong self-interaction limit from
the
viewpoint of gravity. 

In the strongly self-interacting regime $1/\delta\gg 1$, we have shown that (i)
axion stars can be treated with
Newtonian gravity (except for dense axion stars with a very large mass
$M\sim M_{\rm max,GR}^{\rm dense}$); (ii) there is a critical mass $M_{\rm
max,N}^{\rm dilute}$ marking the onset
of a phase
transition between dilute and dense axion stars.

In the weakly self-interacting regime $1/\delta\ll 1$, axion stars are Newtonian
when $M\ll M_{\rm max,GR}^{\rm dilute}$ and general relativistic when $M\sim
M_{\rm max,GR}^{\rm dilute}$.

{\it Remark:} We have already introduced the parameter
$1/\delta$ (denoted $\sigma$) in  \cite{abrilphas} in a different, cosmological,
context. In this context, the conditions $1/\delta\ll
1$ (weak self-interaction) and $1/\delta\gg 1$ (strong
self-interaction) determine two different regimes in the evolution of a
Universe dominated by a self-interacting complex SF with a purely quartic
potential (see the phase diagrams of Figs. 8, 18 and 19 of  \cite{abrilphas}).

\subsection{$|\lambda|\gg (m/M_P)^2$}

We note that the two conditions $|\lambda|\ll 1$ and $1/\delta\gg 1$ are both
satisfied for axions, i.e., the axions are weakly self-interacting from the
viewpoint of
particle physics and strongly self-interacting from the viewpoint of gravity.
It may seem strange that a particle with a coupling constant
as small as $|\lambda|_{\rm QCD}=7.39\times 10^{-49}$ or $|\lambda|_{\rm
ULA}=3.10\times
10^{-91}$  is
considered to be strongly self-interacting (from the viewpoint of gravity). The
reason is that $|\lambda|$ is small with respect to $1$ but large with
respect to $(m/M_P)^2$ which is itself extremely small: 
$(m/M_P)_{\rm QCD}^2=6.71\times 10^{-65}$ and   $(m/M_P)_{\rm
ULA}^2=3.22\times 10^{-100}$. In this sense, there is no paradox.

\subsection{$|\lambda|\gg (M_P/M_{\rm ground})^2\sim 10^{-92}$}

The previous criteria are relativistic in essence since they depend on the speed
of light. In this section, we consider another criterion that does not depend on
$c$.

Let us consider the most compact halo that we know  and let us assume
that it corresponds to the ground state ($T=0$) of a self-gravitating BEC (see
Appendix D of \cite{abrilphas}). To be specific, we identify this halo with
Fornax which has a mass $M_{\rm ground}\sim 10^{8}\, M_{\odot}$ and a radius 
$R_{\rm ground}\sim 1\, {\rm kpc}$. As shown in \cite{prd2}, the condition
to be in the noninteracting regime is that $M_{\rm ground}\ll
M_{a}={\hbar}/\sqrt{Gm|a_s|}$ or, equivalently,  $M_{\rm ground}\ll
{M_P}/\sqrt{|\lambda|}$.\footnote{For a repulsive self-interaction
($a_s>0$), the noninteracting limit corresponds to $M\ll
M_{\rm TF}\sim {\hbar}/\sqrt{Gm a_s}$  and the TF limit corresponds  to $M\gg
M_{\rm TF}\sim {\hbar}/\sqrt{Gm a_s}$.
For an attractive self-interaction ($a_s<0$), the noninteracting limit
corresponds to $M\ll M_{\rm max}\sim {\hbar}/\sqrt{Gm|a_s|}$; the
halo becomes unstable when $M\sim M_{\rm max}\sim {\hbar}/\sqrt{Gm|a_s|}$.} This
condition can be rewritten as
$|\lambda|\ll (M_P/M_{\rm ground})^2\sim 10^{-92}$. Therefore, in
order to be able to
neglect the self-interaction of the bosons, $|\lambda|$ has to be small with
respect to $1.20\times 10^{-92}$ (!), not only small with
respect to $1$. This striking condition was stressed in Appendix A.3 of
\cite{prd2}.  In many case, one has $10^{-92}\ll |\lambda|\ll
1$. This shows one more time that the self-interaction of the bosons
(either
attractive or repulsive) is of considerable importance in gravitational
problems (axion stars, axionic DM halos...) even if their self-interaction 
parameter 
$|\lambda|$ looks very small at first sight.

{\it Remark:} As mentioned above, one cannot ignore the
self-interaction of the axions when considering mini QCD axion stars and ULA
clusters that form in the nonlinear regime of the cosmic evolution. Things are
different, however, when considering the formation of structures by Jeans
instability in the linear regime of the cosmic evolution as investigated
in \cite{suarezchavanisprd3}. In that case, we find that: (i) the attractive
self-interaction
of QCD axions cannot be neglected in the ultrarelativistic era;
(ii) the self-interaction of QCD axions can be neglected in the
matter era (they behave as CDM); (ii) the self-interaction of ULAs cannot be
generally neglected in the matter era.

\section{Condensation temperature}
\label{sec_tc}

The GP equation is valid at $T=0$, or
for $T\ll T_c$, where
\begin{eqnarray}
T_c=\frac{2\pi\hbar^2\rho^{2/3}}{m^{5/3}k_B\zeta(3/2)^{2/3}}
\label{tc1}
\end{eqnarray}
is
the condensation temperature of the bosons (here $\zeta(3/2)=2.6124...$). For
self-gravitating BECs with attractive self-interaction close to the critical
point ($M\sim M_{\rm max}$ and $R\sim R_*$), using Eqs. (\ref{maxmr1})
and (\ref{maxmr2}), we find that the order of magnitude of the condensation
temperature is
\begin{eqnarray}
k_B T_c\sim m \left (\frac{\hbar G}{a_s^2}\right )^{2/3}.
\label{tc2}
\end{eqnarray}
It can be expressed in terms of $f$ and
$\lambda$ as
\begin{eqnarray}
k_B T_c\sim m \left (\frac{G f^4}{\hbar m^2 c^6}\right
)^{2/3}\sim m \left (\frac{G m^2 c^2}{\lambda^2\hbar}\right
)^{2/3}.
\label{tc3}
\end{eqnarray}
It can also be written as
\begin{eqnarray}
k_B T_c&\sim& \left (\frac{r_s}{2|a_s|}\right )^{4/3}\left (\frac{M_P}{m}\right
)^{1/3}M_P c^2\nonumber\\
&\sim& (32\pi)^{4/3}\left (\frac{f}{M_P c^2}\right )^{8/3}\left
(\frac{M_P}{m}\right
)^{1/3}M_P c^2\nonumber\\
&\sim& \left (\frac{8\pi}{|\lambda|}\right )^{4/3}\left
(\frac{m}{M_P}\right )^{8/3} \left
(\frac{M_P}{m}\right
)^{1/3}M_P c^2.\qquad 
\label{tc4}
\end{eqnarray}

For QCD axions  with 
$m=10^{-4}\, {\rm eV}/c^2$ and $a_s=-5.8\times
10^{-53}\, {\rm m}$,  we find
$T_c\sim 2.11\times 10^{23}\, {\rm K}$ and $k_B T_c\sim 1.82\times
10^{10}\,
{\rm GeV}$.

For ULAs with
$m=2.19\times
10^{-22}\, {\rm eV}/c^2$ and
$a_s=-1.11\times 10^{-62}\, {\rm fm}$, we find
$T_c\sim 4.19\times 10^{38}\, {\rm K}$ and  $k_B T_c\sim 3.61\times 10^{25}\,
{\rm GeV}$.

These values of the condensation temperature are considerable. Whatever the
physical origin of the temperature $T$, it is clear that the condition $T\ll
T_c$ will be fulfilled by many orders of magnitude. This implies that excited
(thermal) states are completely negligible for self-gravitating BECs. Therefore,
in excellent approximation, bosonic DM can be considered to be at zero
thermodynamic temperature ($T=0$) described by the GP equation. Still, DM halos
may have an effective nonzero temperature $T_{\rm eff}\neq 0$ responsible for
their atmosphere as discussed in \cite{chavtotal,prep,kingfermionic}.

\end{document}